\documentclass[twoside,12pt]{article}
\usepackage{epsfig}
\usepackage{graphicx}
\usepackage{wrapfig}
\usepackage{amsmath}
\usepackage{latexsym}
\usepackage{amssymb}
\usepackage{color}
\usepackage{multirow}
\usepackage{url}
\usepackage{wrapfig}
\usepackage{longtable}
\usepackage{array}
\usepackage{hyperref}
\usepackage[normalem]{ulem}
\usepackage{bbold}
 
\def\Journal#1#2#3#4{{#1} {#2} (#4) #3 }

\def\NPA{{\em Nucl. Phys.} A}

\def\PLB{{\em Phys. Lett.} B}

\def\PRL{\em Phys. Rev. Lett.}

\def\PREP{\em Phys. Rep.}

\def\PRD{{\em Phys. Rev.} D}
\def\PRC{{\em Phys. Rev.} C}

\newcommand{\be}{\begin{equation}}
\newcommand{\ee}{\end{equation}}
\newcommand{\bea}{\begin{eqnarray}}
\newcommand{\eea}{\end{eqnarray}}

\newcommand{\open}{\sphericalangle}
\begin{document}
\bibliographystyle{elsart-num}
\title{ \vspace{1cm} Parton Fragmentation Functions}
\author{A.\ Metz,$^{1}$ A.\ Vossen$^2$\\
\\
$^1$Department of Physics, Temple University, Philadelphia, PA 19122, USA\\
$^2$Center for Exploration of Energy and Matter, Indiana University, Bloomington, IN 47408}

\maketitle
\begin{abstract} 
The field of fragmentation functions of light quarks and gluons is  reviewed.
In addition to integrated fragmentation functions, attention is paid to the dependence of fragmentation functions on transverse momenta and on polarization degrees of freedom.
Higher-twist and di-hadron fragmentation functions are considered as well.
Moreover, the review covers both theoretical and experimental developments in hadron production in electron-positron annihilation, deep-inelastic lepton-nucleon scattering, and proton-proton collisions.
\end{abstract}
\eject
\tableofcontents

\section{Introduction}
\label{sec:introduction}
Quantum chromodynamics (QCD) is the generally accepted fundamental theory of the strong interaction.
Thanks to the asymptotic freedom of QCD~\cite{Gross:1973id,Politzer:1973fx} many high-energy scattering processes can be analyzed using perturbation theory.
In most cases such analyses are in the form of factorization theorems, which separate the perturbatively calculable part of the cross section from the non-perturbative part~\cite{Collins:1989gx}.
If specific particles are identified in the final state, parton fragmentation functions (FFs) appear frequently as non-perturbative ingredient of QCD factorization formulas.
(In the older literatue the term parton decay functions instead of FFs was often used.)
Generally, FFs describe how the color-carrying quarks and gluons transform into color-neutral particles such as hadrons or photons. 

The concept of FFs was already used shortly after the parton model~\cite{Bjorken:1969ja,Drell:1970yt,Feynman:book} had been introduced~\cite{Berman:1971xz}.
This was in the pre-QCD era, at a time when key characteristic features of the partons, like their interactions, were heavily under debate.
Early on FFs were considered as counterpart of parton distribution functions (PDFs).
While PDFs were understood as probability densities for finding partons, with a given momentum, inside color-neutral particles, FFs were understood as probability densities for finding color-neutral particles inside partons~\cite{Berman:1971xz,Feynman:book}.

FFs are intimately connected with another type of non-perturbative objects, the so-called (time-like) cut-vertices~\cite{Mueller:1978xu,Mueller:1981sg}.
More precisely, cut-vertices correspond to certain moments of FFs~\cite{Collins:1981uw}.
They were introduced in order to obtain a formulation of processes with identified hadrons that is similar to, for instance, Wilson's operator product expansion~\cite{Wilson:1969zs}.
Nowadays FFs are more frequently used than cut-vertices.

The best studied FF is what we denote by $D_1^{h/i}(z)$ in this paper.
It describes the fragmentation of an unpolarized parton of type $i$ into an unpolarized hadron of type $h$, where the hadron carries the fraction $z$ of the parton momentum.
Here one has in mind the longitudinal momentum of the hadron, that is, the component of the momentum along the direction of motion of the parton.
Therefore, $D_1^{h/i}(z)$ is often called a collinear FF or also an integrated FF, since the transverse momentum of the hadron $\vec{P}_{hT}$ relative to the parton is integrated over. 
To be now more specific about the meaning of this function we note that the quantity $D_1^{h/i}(z) \, dz$ is the number of hadrons $h$ inside parton $i$ in the momentum fraction range $[z,z+dz]$.
When taking into account higher-order QCD effects this parton model interpretation of FFs gets distorted~\cite{Collins:2011zzd}.
\begin{figure}[t]
\begin{center}
\includegraphics[width=16.0cm]{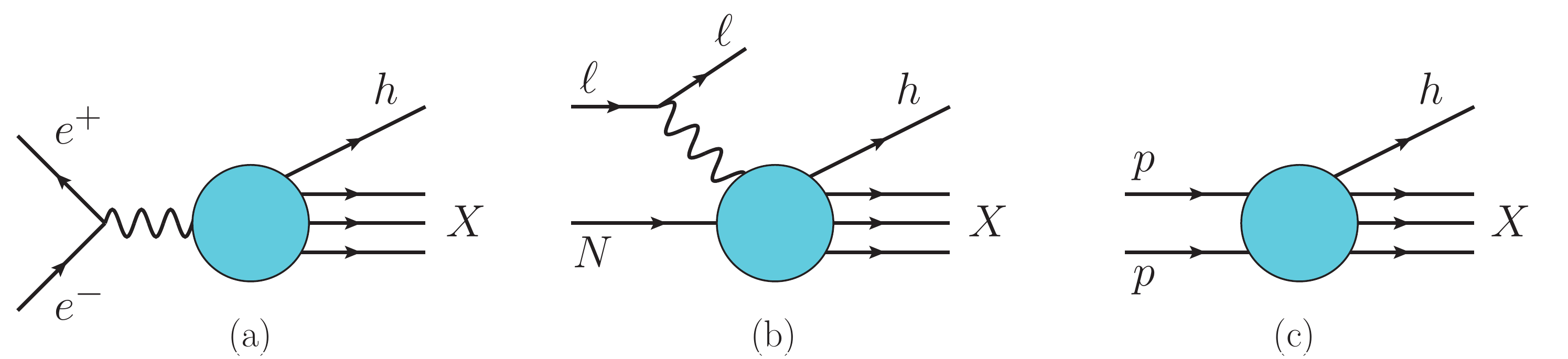} 
\end{center}
\vspace{-0.4cm}
\caption{Representation of the scattering amplitude for processes whose QCD description involves FFs:
Single-inclusive hadron production in~(a) $e^+ e^-$ annihilation,~(b) deep-inelastic lepton-nucleon scattering,~(c) proton-proton scattering.}
\label{f:fundamental_proc}
\end{figure}

In general, the following processes have played and continue to play a crucial role in studies of FFs:
\begin{itemize}
\item single-inclusive hadron production in electron-positron annihilation, $e^+ + e^- \to h + X$. 
Often this process is simply denoted as single-inclusive annihilation (SIA).
\item semi-inclusive deep-inelastic lepton-nucleon scattering (SIDIS), $\ell + N \to \ell + h + X$.
\item single-inclusive hadron production in proton-proton collisions, $p + p \to h + X$.
Related processes like proton-antiproton ($p\bar{p}$) collisions have been studied as well.
\end{itemize}

\begin{figure}[t]
\begin{center}
\includegraphics[width=18.0cm]{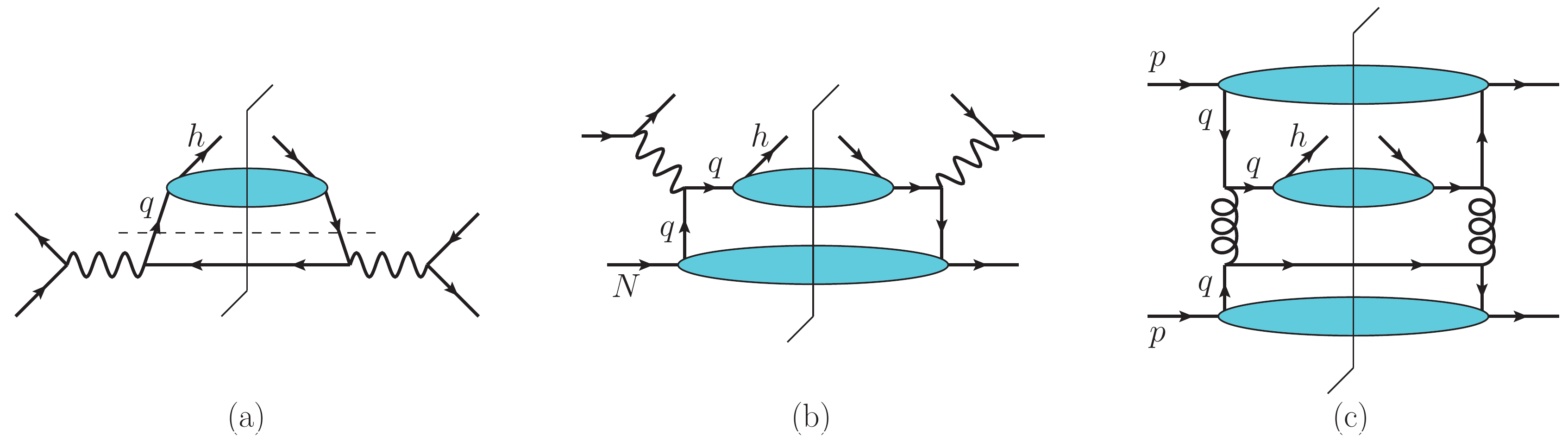} 
\end{center}
\vspace{-0.4cm}
\caption{Parton-model representation of the cross section for single-inclusive hadron production in~(a) $e^+ e^-$ annihilation,~(b) deep-inelastic lepton-nucleon scattering,~(c) proton-proton scattering.
In the case of SIA the factorization of the cross section into the hard process $e^+ e^- \to q \bar{q}$ and the quark fragmentation part is indicated by the horizontal dashed line.
Processes~(b) and~(c) depend on both FFs and PDFs.
For proton-proton scattering only one of the leading-order diagrams is shown.}
\label{f:fundamental_proc_PM}
\end{figure}

The scattering amplitudes for these reactions are displayed in Fig.~\ref{f:fundamental_proc}.
In these cases QCD factorization theorems schematically read~\cite{Collins:1989gx,Brock:1993sz}
\begin{eqnarray} \label{e:fact_epem}
\sigma^{e^+ e^- \to h X} & = & \hat{\sigma} \otimes FF \,,
\\  \label{e:fact_sidis}
\sigma^{\ell N \to \ell h X} & = & \hat{\sigma} \otimes PDF \otimes FF \,, 
\phantom{\frac{1}{1}}
\\  \label{e:fact_pp}
\sigma^{pp \to h X} & = & \hat{\sigma} \otimes PDF \otimes PDF \otimes FF \,,
\end{eqnarray}
where $\hat{\sigma}$ indicates the respective process-dependent partonic cross section that can be computed in perturbation theory.
The parton-model representation of the cross section for the three processes is shown in Fig.~\ref{f:fundamental_proc_PM}.
Using the parton model, or in other words leading order perturbative QCD (pQCD), it is often straight forward to write down a factorization formula.
However, in full QCD it is typically challenging to analyze and factorize radiative corrections to arbitrary order in the strong coupling~\cite{Collins:1989gx,Brock:1993sz}.
Factorization theorems only hold if specific kinematic conditions are satisfied, where the minimum requirement is the presence of a hard scale that allows one to use pQCD.
For SIA that scale is provided by the center-of-mass ({\it cm}) energy $\sqrt{s}$.
For SIDIS it is the momentum transfer between the leptonic and the hadronic part of the process, while in the case of hadronic collisions it is the transverse momentum of the final state hadron relative to the collision axis.
The specific form of the factorization theorem, including the precise meaning of the ``multiplication" $\otimes$, also depends on the kinematics of the process.
In addition, it can depend on the polarization state of one or more of the involved particles.
More information on this point will be given later in this paper and the references quoted there.
We also mention that the factorization theorems in Eqs.~(\ref{e:fact_epem})-(\ref{e:fact_pp}) hold in the sense of a Taylor expansion in powers of $1/Q$, where here $Q$ denotes the hard scale of a process.
The term on the r.h.s.~of these equations then represents the leading contribution.
Factorization theorems have been written down for certain subleading terms as well, but in most such cases all-order proofs do not exist.

An interesting and important early application of FFs in the 1970s was for the production of large-transverse-momentum hadrons in hadronic collisions, where FFs are needed according to~(\ref{e:fact_pp}).
Data for this process had been obtained in {\it pp} collisions at the ISR (Intersecting Storage Ring) collider at CERN~\cite{Alper:1973nv,Banner:1973nu,Busser:1973hs}, and in fixed-target proton-tungsten collisions at Fermi-Lab~\cite{Cronin:1974zm}.
The data could not be explained unless one takes subprocesses involving gluons into account~\cite{Cahalan:1974tp,Sivers:1975dg,Combridge:1977dm,Cutler:1977qm,Owens:1977sj}.
This observation can be considered the first phenomenological indication for the existence of gluons, as it was made before the discovery of the 3-jet events by the TASSO, PLUTO, MARK-J and JADE collaborations at DESY~\cite{Brandelik:1979bd,Berger:1979cj,Barber:1979yr,Bartel:1979ut} which are often quoted in this context.

In addition to $D_1^{h/i}(z)$, one can consider a number of other FFs by including (i) the spin of the parton and/or hadron, (ii) the transverse momentum of the hadron relative to the parton, (iii) higher-twist effects, and (iv) fragmentation into more than one hadron.
All the FFs that emerge when making such generalizations are important in their own as they contain a lot of information about non-perturbative QCD dynamics in general, and the hadronization process in particular.
In a number of cases they are simply needed to describe existing data.
Very often they are also considered and used just as a tool for studying PDFs of hadrons, most notably of the nucleon.
Especially in that regard generalizations of $D_1^{h/i}(z)$ received a lot of attention over the past two decades.
In this article we will discuss the following four types of FFs:
\begin{itemize}
\item {\bf Integrated FFs:} The major focus for this type of FFs will be on $D_1^{h/i}(z)$.
There exists an enormous body of experimental and theoretical work for this function.
Its phenomenology is very rich, not the least due to the large variety of hadrons that has been studied.
But we also discuss the corresponding spin-dependent integrated FFs $G_1^{h/i}(z)$ and $H_1^{h/i}(z)$, which are needed when the hadron is polarized parallel or perpendicular to its momentum, respectively.
The three functions listed here are leading-twist (twist-2) objects.
\item {\bf Transverse-momentum dependent (TMD) FFs:} One can also study FFs that depend on the hadron's transverse momentum $\vec{P}_{hT}$, in addition to their dependence on $z$.
Such TMD FFs allow one, for instance, to probe the transverse-momentum dependence of PDFs in a process like SIDIS.
By keeping the transverse momentum of the fragmentation process one also finds new FFs.
Two of them already became important in high-energy spin physics, with the Collins function for quarks $H_1^{\perp \, h/q}$ representing the most prominent example~\cite{Collins:1992kk}. 
At present people are interested in this function mainly because in SIDIS it couples to the transversity PDF~\cite{Collins:1992kk}, which is one of the three leading-twist collinear quark distributions of the nucleon~\cite{Ralston:1979ys,Jaffe:1991kp,Cortes:1991ja}. 
Because of its chiral-odd nature transversity cannot be measured in inclusive DIS.
As a result the present knowledge about this function is still poor compared to the other leading-twist PDFs, that is, the unpolarized distribution and the helicity distribution of partons.
\item {\bf Higher-twist FFs:} Here we limit ourselves to the most important case, namely twist-3 FFs.
In general, higher-twist FFs comprise different subclasses~\cite{Balitsky:1990ck,Kanazawa:2015ajw}. 
Analogous to twist-2 FFs, one subclass is defined through two-parton correlation functions, yet these twist-3 FFs do not have a density interpretation.
Another subclass parameterizes quark-gluon-quark correlation functions which can be considered as quantum interference effects.
Such three-parton FFs are very important for describing several twist-3 observables like the large transverse single-spin asymmetries (SSAs) that were observed for instance in $p^{\uparrow} p \to h X$.
Also certain moments of TMD FFs are related to three-parton FFs.
\item {\bf Di-hadron FFs (DiFFs):} These objects parameterize the fragmentation of a parton into two hadrons.
They were introduced when trying to get a detailed understanding of the structure of jets~\cite{Konishi:1978yx,Konishi:1979cb}.
The DiFFs used in the original papers depend on the longitudinal hadron momenta only and describe unpolarized fragmentation.
Later it was realized that DiFFs can also serve as crucial tools for measuring polarization of partons~\cite{Collins:1993kq}, especially if one keeps the dependence on the relative momentum between the two hadrons.
In this context we mention in particular the function $H_1^{\open \, h_1 h_2 / q}$ which is relevant for the fragmentation of a transversely polarized quark, and therefore allows one to address the transversity distribution of the nucleon~\cite{Collins:1993kq}.
Sometimes this FF is considered a cleaner ``analyzer" of transversity than the aforementioned Collins function $H_1^{\perp \, h/q}$ because one can use collinear factorization rather than TMD factorization.
Other DiFFs have attracted attention too, but presently not at the same level as $H_1^{\open \, h_1 h_2 / q}$.
\end{itemize}
Many new developments related to these different FFs appeared over the last years.
In this review we make an attempt to summarize the main findings.
Certain parts of the material presented below are also discussed in some detail in a number of other papers~\cite{Barone:2001sp,D'Alesio:2007jt,Albino:2008aa,Arleo:2008dn,Albino:2008gy,Burkardt:2008jw,Barone:2010zz,Radici:2011ix,Aidala:2012mv,Angeles-Martinez:2015sea,Rogers:2015sqa,Perdekamp:2015vwa,Pisano:2015wnq,Boglione:2015zyc,Aschenauer:2015ndk,Bacchetta:2016ccz,Avakian:2016rst}. 
There are some important related topics that we cannot cover at all.
One of them is parton fragmentation in a strongly interacting medium such as nuclear matter or the quark-gluon plasma.
More information on this research area can be found, e.g., in Refs.~\cite{Arleo:2008dn,Accardi:2009qv,Burke:2013yra,Aschenauer:2016our}.
Also, we do not consider heavy-quark FFs.
The fragmentation into a heavy quark can be computed perturbatively~\cite{Mele:1990cw,Arleo:2008dn}, but the transition from the heavy quark into a heavy-flavored hadron contains non-perturbative effects.
Several parameterizations of such effects exist in the literature~\cite{Kartvelishvili:1977pi,Bowler:1981sb,Peterson:1982ak,Andersson:1983ia,Collins:1984ms,Colangelo:1992kh,Braaten:1993jn,Braaten:1994bz}.

The rest of this paper is organized as follows: In Sec.~\ref{sec:properties}, we review the definitions of the various FFs and their most important properties.
The observables that can be used to extract the FFs are summarized in Sec.~\ref{sec:observables}.
This is followed by Sec.~\ref{sec:experiments} which contains an overview of the different experiments and datasets that are relevant for the defined observables, along with the most important results.
In Sec.~\ref{sec:global_fits}, we discuss the efforts to extract FFs from the experimental results through global fits, and in Sec.~\ref{sec:models} models for FFs are briefly presented.
Section~\ref{sec:otherTopics} contains several topics that are important for the field of FFs, but cannot be discussed at length in this review.
This applies in particular to FFs for polarized hadrons, which mainly matter for hyperon production.
While in Sec.~\ref{sec:properties} we summarize most of the properties of the FFs relevant in this case, any additional discussion on this topic is limited to a few paragraphs in Sec.~\ref{sec:otherTopics}.
The conclusions and an outlook are given in Sec.~\ref{sec:conclusions}.

\section{Properties of Fragmentation Functions}
\label{sec:properties}
\begin{figure}[t]
\begin{center}
\includegraphics[width=9.0cm]{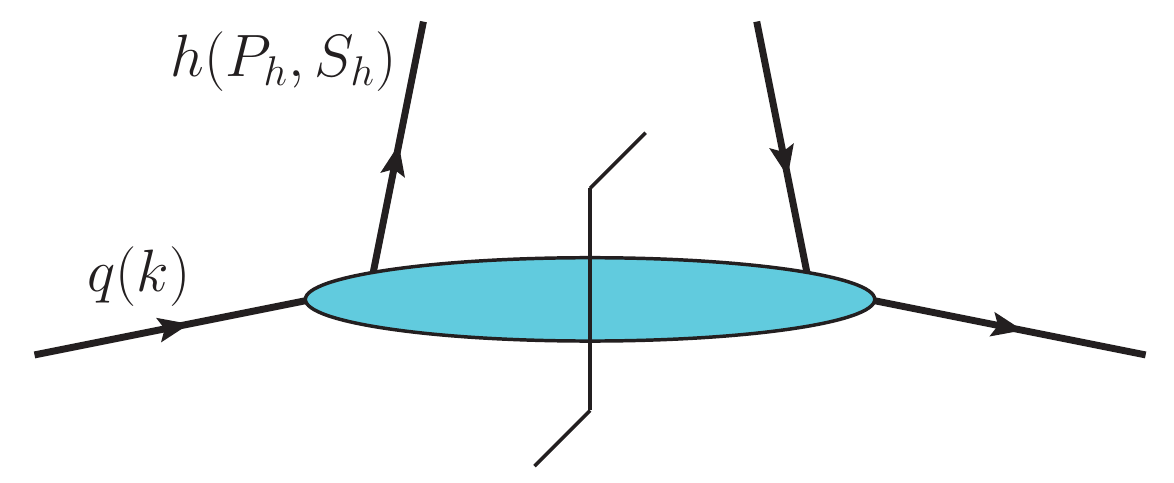} 
\end{center}
\vspace{-0.4cm}
\caption{Representation of the correlator $\Delta^{h/q}$ in~(\ref{e:corr_quark_int}), which describes the fragmentation of a quark into a hadron.
The figure contains no reference to the Wilson lines.}
\label{f:quark_frag_correlator}
\end{figure}

\subsection{Definition of FFs}
\label{sec:definition}
Here we summarize the definitions of integrated FFs, TMD FFs, three-parton FFs, as well as DiFFs --- see, in particular, also Refs.~\cite{Collins:1981uw,Mulders:1995dh,Bianconi:1999cd,Bacchetta:2006tn,Collins:2011zzd,Kanazawa:2015ajw}.
We largely follow the so-called Amsterdam notation for FFs~\cite{Mulders:1995dh,Mulders:2000sh,Bacchetta:2006tn,Meissner:2007rx}.
The relation to the so-called Torino notation, which has also been used frequently, is discussed for instance in~\cite{Bacchetta:2004jz,Anselmino:2005sh}.  
The focus is on fragmentation into spin-$\frac{1}{2}$ hadrons, which of course also covers the important case of fragmentation into spin-$0$ hadrons.
In the case of DiFFs we only consider spin-$0$ hadrons.
Work on the classification of FFs for spin-$1$ hadrons can be found in~\cite{Ji:1993vw,Bacchetta:2000jk,Chen:2016moq}.

\subsubsection{Definition of integrated FFs}
\label{sec:definition_integrated}
We begin with the field-theoretic definition of integrated FFs for quarks, antiquarks, and gluons.
Let us first specify the kinematics of the final-state hadron and the fragmenting parton.
The hadron is characterized by its 4-momentum $P_h$ and the covariant spin vector $S_h$, while $k$ denotes the momentum of the parton.
In a reference frame in which the hadron has no transverse momentum one can write
\begin{eqnarray}  \label{e:Ph}
P_h & = & (P_h^+, P_h^-, \vec{0}_T) = \bigg( \frac{M_h^2}{2 P_h^-}, P_h^-, \vec{0}_T \bigg) \,,
\\  \label{e:Sh}
S_h & = & (S_h^+, S_h^-, \vec{S}_{hT}) = \bigg( - \Lambda_h \frac{M_h}{2 P_h^-}, \Lambda_h \frac{P_h^-}{M_h}, \vec{S}_{hT} \bigg) \,,
\\  \label{e:k}
k & = & (k^+, k^-, \vec{k}_T) = \bigg( z \, \frac{k^2 + \vec{k}_T^{\, 2}}{2 P_h^-}, \frac{P_h^-}{z}, \vec{k}_T \bigg) \,,
\end{eqnarray}
where $M_h$ is the mass of the hadron, while $\Lambda_h$ and $\vec{S}_T$ describes longitudinal and transverse polarization of the hadron, respectively.
(We use $a^{\pm} = (a^0 \pm a^3) / \sqrt{2}$ for the light-cone plus- and minus-components of a generic 4-vector $a$.)
Obviously one has $P_h^2 = M_h^2$, $P_h \cdot S_h = 0$, and $S_h^2 = - \Lambda_h^2 - \vec{S}_{hT}^2$. 
We assume that both the parton and the hadron have a large minus-momentum, with the hadron carrying the fraction $z$ of the parton's momentum.
For later convenience we also introduce two light-like vectors through
\begin{equation}
n = (n^+, n^-, \vec{n}_T) = (0, 1, \vec{0}_T) \,, \qquad
\bar{n} = (\bar{n}^+, \bar{n}^-, \vec{\bar{n}}_T) = (1, 0, \vec{0}_T) \,,
\phantom{\frac{1}{1}}
\end{equation}
which satisfy $n^2 = \bar{n}^2 = 0$ and $n \cdot \bar{n} = 1$.
This definition implies $n \cdot a = a^+$ and $\bar{n} \cdot a = a^-$ for a generic 4-vector $a$.
Note that, upon neglecting mass effects, one has $P_h = P_h^- \, n$.

It is well known that ordinary FFs are specified through (bilocal) fragmentation correlators, where for integrated FFs one deals with light-cone correlators. 
(See Fig.~\ref{f:quark_frag_correlator} for a representation of the quark-quark ({\it qq}) fragmentation correlator.) 
In the case of a quark with flavor $q$ fragmenting into a hadron $h$ the field-theoretic expression of the fragmentation correlator reads~\cite{Collins:1981uw,Mulders:1995dh,Bacchetta:2006tn,Collins:2011zzd}
\begin{eqnarray} \label{e:corr_quark_int}
\Delta^{h/q}(z;P_h,S_h) & = & \sum_X \hspace{-0.5cm} \int \; \int \frac{d \xi^+}{2\pi} \, e^{i k^- \xi^+} \,
\langle 0 | \, {\cal W}(\infty^+ , \xi^+) \, \psi_q(\xi^+, 0^-, \vec{0}_T) \, | P_h, S_h; X \rangle  
\nonumber \\[0.1cm]
& & \times \; \langle P_h, S_h; X | \, \bar{\psi}_q(0^+, 0^-, \vec{0}_T) \, {\cal W}(0^+ , \infty^+) \, | 0 \rangle \,,
\\[0.1cm]
& & \textrm{with} \;\, \sum_X \hspace{-0.5cm} \int \, \equiv \, \sum_X \int \frac{d^3 \vec{P}_X}{(2\pi)^3 \, 2P_X^0} \,.
\nonumber
\end{eqnarray}
This correlator is a $(4 \! \times \! 4)$-matrix in Dirac space, and an average in color space is understood.  
It depends on the (longitudinal) momentum fraction $z$ through $k^- = P_h^- / z$, and on $P_h$ as well as $S_h$.
Since $P_h^-$ and $k^-$ are large, the two quark fields are separated along the (conjugate) light-cone plus-direction.
Wilson lines are included in~(\ref{e:corr_quark_int}) in order to ensure color gauge invariance of the correlator $\Delta^{h/q}$. 
We define a Wilson line connecting two points $a$ and $b$ according to
\begin{equation} \label{e:wilson_quark}
[a^+, a^-, \vec{a}_T; b^+, b^-, \vec{b}_T] = {\cal P} \, \textrm{exp} \bigg[ -i g \int_a^b ds_{\mu} \, A^{\mu}(s) \bigg] \,,
\phantom{\frac{1}{1}}
\end{equation}
where ${\cal P}$ indicates path-ordering, and $g$ is the strong coupling.
The Wilson line in~(\ref{e:wilson_quark}) is in the fundamental representation, i.e., $A^{\mu} = A_a^{\mu} \, T_a$, with $T_a = \lambda_a / 2$ and $\lambda_a$ denoting the Gell-Mann matrices.
It is understood that the line integral in~(\ref{e:wilson_quark}) runs along a straight line.
Using this notation we define 
\begin{equation}
{\cal W}(a^+,b^+) = [a^+, 0^-, \vec{0}_T; b^+, 0^-, \vec{0}_T] \,,
\end{equation}
which specifies the Wilson lines that appear in~(\ref{e:corr_quark_int}).
Note that in~(\ref{e:corr_quark_int}) each quark field is associated with a Wilson line (phase factor) which runs to light-cone infinity.
These two Wilson lines can be combined into a single Wilson line which directly connects the two quark fields through a straight line.
The same situation appears for integrated two-parton PDFs.
Keeping the two Wilson lines in~(\ref{e:corr_quark_int}) makes the notation somewhat more ``symmetric" and, in particular, reveals the similarity to the definition of TMD FFs discussed below. 
Additionally, we point out that in particular through the Wilson lines, $\Delta^{h/q}$ in~(\ref{e:corr_quark_int}) also depends on the vector $\bar{n}$.
For simplicity this dependence is not displayed.
Moreover, we have suppressed the dependence of $\Delta^{h/q}$ on a renormalization scale $\mu$, which is caused by the renormalization of the two quark field operators and a divergence arising from the integration upon the transverse quark momentum~\cite{Collins:2011zzd}.
We will return to this point in Sec.~\ref{subsec:evolution} when discussing the evolution of FFs.

For the following discussion it is convenient to introduce a shorthand notation for the trace of $\Delta^{h/q}$ with an arbitrary Dirac matrix $\Gamma$~\cite{Mulders:1995dh},
\begin{equation} \label{e:corr_quark_trace}
\Delta^{h/q \, [\Gamma]}(z; P_h, S_h) \equiv \frac{1}{4z} \, \textrm{Tr} \, [\Delta^{h/q}(z, P_h, S_h) \, \Gamma] \,.
\end{equation}
Then $\Delta^{h/q}$ may be written in terms of 16 independent Dirac matrices according to
\begin{eqnarray} \label{e:corr_quark_decomp}
\Delta^{h/q}(z;P_h,S_h) & = & z \, \Big( 
   \Delta^{h/q \, [\gamma^-]} \, \gamma^+
 - \Delta^{h/q \, [\gamma^- \gamma_5]} \, \gamma^+ \gamma_5
 +\Delta^{h/q \, [i \sigma^{i-} \gamma_5]} \, i \sigma^{i+} \gamma_5
\nonumber \\
& & \hspace{0.2cm}
 + \, \Delta^{h/q \, [\mathbb{1}]} \, \mathbb{1}
 - \Delta^{h/q \, [i \gamma_5]} \, i \gamma_5
 - \Delta^{h/q \, [\gamma^i]} \, \gamma^i
 \phantom{\Big(}
\nonumber \\ 
& & \hspace{0.2cm}
 + \, \Delta^{h/q \, [\gamma^i \gamma_5]} \, \gamma^i \gamma_5
 -\tfrac{1}{2} \, \Delta^{h/q \, [i \sigma^{ij} \gamma_5]} \, i \sigma^{ij} \gamma_5
 + \Delta^{h/q \, [i \sigma^{-+} \gamma_5]} \, i \sigma^{-+} \gamma_5
 \phantom{\Big(}
\nonumber \\
& & \hspace{0.2cm}
 + \, \Delta^{h/q \, [\gamma^+]} \, \gamma^-
 - \Delta^{h/q \, [\gamma^+ \gamma_5]} \, \gamma^- \gamma_5
 +\Delta^{h/q \, [i \sigma^{i+} \gamma_5]} \, i \sigma^{i-} \gamma_5
 \Big) \,,
\end{eqnarray}
where $\sigma_{\mu\nu} = \frac{i}{2} [\gamma_{\mu}, \gamma_{\nu}]$, $i,j$ denote transverse Lorentz indices, and a summation over repeated indices is understood.
Integrated quark FFs are directly given by the traces in~(\ref{e:corr_quark_decomp}).
If $P_h^-$ is large, the first three traces on the r.h.s.~ of~(\ref{e:corr_quark_decomp}) dominate.
They define the leading-twist (twist-2) FFs according to~\cite{Collins:1981uw,Mulders:1995dh,Bacchetta:2006tn,Collins:2011zzd}
\begin{eqnarray} \label{e:D1_quark_z}
\Delta^{h/q \, [\gamma^-]}(z; P_h, S_h) & = & \frac{1}{z^2} \, \Big[ D_1^{h/q}(z) \Big] \,, 
\\[0.2cm] \label{e:G1_quark_z}
\Delta^{h/ q \, [\gamma^- \gamma_5]}(z; P_h, S_h) & = & \frac{1}{z^2} \, \Big[ \Lambda_h \, G_1^{h/q}(z) \Big] \,,
\\[0.2cm] \label{e:H1_quark_z}
\Delta^{h/q \, [i \sigma^{i-} \gamma_5]}(z; P_h, S_h) & = & \frac{1}{z^2} \, \Big[ S_{hT}^i \, H_1^{h/q}(z) \Big] \,.
\end{eqnarray}
Here $D_1^{h/q}$ is the well-known unpolarized FF which describes the number density of unpolarized hadrons in an unpolarized quark.
Note that the definition of $D_1^{h/q}$ is appropriate for a spin-0 hadron.
For spin-$\frac{1}{2}$ hadrons this function gets multiplied by 2 if one sums over the hadron spins.
The FF $G_1^{h/q}$ describes the density of longitudinally polarized hadrons in a longitudinally polarized quark, whereas $H_1^{h/q}$ describes the density of transversely polarized hadrons in a transversely polarized quark.
In Sec.~\ref{subs:interpretation} below we will come back to the physical interpretation of leading-twist FFs.
Using Eqs.~(\ref{e:D1_quark_z}),~(\ref{e:corr_quark_trace}),~(\ref{e:corr_quark_int}) one immediately obtains the operator definition for $D_1^{h/q}(z)$,
\begin{eqnarray} \label{e:D1_quark_def}
D_1^{h/q}(z) & = & \frac{z}{4} \, \sum_X \hspace{-0.5cm} \int \; \int \frac{d \xi^+}{2\pi} \, e^{i k^- \xi^+} \,
\textrm{Tr} \, \Big[ \langle 0 | \, {\cal W}(\infty^+ , \xi^+)  \, \psi_q(\xi^+, 0^-, \vec{0}_T) \, | P_h, S_h; X \rangle  
\nonumber \\[0.1cm]
& & \times \; \langle P_h, S_h; X | \, \bar{\psi}_q(0^+, 0^-, \vec{0}_T) \, {\cal W}(0^+ , \infty^+) \, | 0 \rangle \, \gamma^- \Big] \,.
\end{eqnarray}
It is straightforward to write down the corresponding definitions for $G_1^{h/q}(z)$ and $H_1^{h/q}(z)$.

Let us now proceed to the twist-3 (two-parton) FFs, which are suppressed by a factor $M_h / P_h^-$ relative to the leading FFs.
A total of six twist-3 {\it qq} FFs can be identified~\cite{Mulders:1995dh,Boer:1997mf,Goeke:2005hb,Bacchetta:2006tn},
\begin{eqnarray} \label{e:E_quark_z}
\Delta^{h/q \, [\mathbb{1}]}(z; P_h, S_h) & = & \frac{M_h}{z^2 P_h^-} \, \Big[ E^{h/q}(z) \Big] \,,
\\[0.2cm]  \label{e:EL_quark_z}
\Delta^{h/ q \, [i \gamma_5]}(z; P_h, S_h) & = & \frac{M_h}{z^2 P_h^-} \, \Big[ \Lambda_h \, E_L^{h/q}(z) \Big] \,,
\\[0.2cm]  \label{e:DT_quark_z}
\Delta^{h/q \, [\gamma^i ]}(z; P_h, S_h) & = & \frac{M_h}{z^2 P_h^-} \, \Big[ - \varepsilon_T^{ij} \, S_{hT}^j \, D_T^{h/q}(z) \Big] \,,
\\[0.2cm]  \label{e:GT_quark_z}
\Delta^{h/q \, [\gamma^i \gamma_5]}(z; P_h, S_h) & = & \frac{M_h}{z^2 P_h^-} \, \Big[ S_{hT}^i \, G_T^{h/q}(z) \Big] \,,
\\[0.2cm]  \label{e:H_quark_z}
\Delta^{h/ q \, [i \sigma^{ij} \gamma_5]}(z; P_h, S_h) & = & \frac{M_h}{z^2 P_h^-} \, \Big[ \varepsilon_T^{ij} \, H^{h/q}(z) \Big] \,,
\\[0.2cm]  \label{e:HL_quark_z}
\Delta^{h/q \, [i \sigma^{-+} \gamma_5]}(z; P_h, S_h) & = & \frac{M_h}{z^2 P_h^-} \, \Big[ \Lambda_h \, H_L^{h/q}(z) \Big] \,,
\end{eqnarray}
where obviously two FFs appear for an unpolarized target, two for a longitudinally polarized target, and two for a transversely polarized target.
We have used $\varepsilon_T^{ij} = \varepsilon^{\mu \nu ij} \, \bar{n}_{\mu} n_{\nu} = \varepsilon^{-+ij}$, and the sign convention $\varepsilon_T^{12} = 1$.
Higher-twist FFs are not necessarily smaller than twist-2 FFs, but the (small) factor $M_h/P_h^-$ on the r.h.s.~of~(\ref{e:E_quark_z})-(\ref{e:HL_quark_z}) reduces their impact on observables.
For completeness we also include the twist-4 case~\cite{Goeke:2005hb},
\begin{eqnarray} \label{e:D3_quark_z}
\Delta^{h/q \, [\gamma^+]}(z; P_h, S_h) & = & \frac{M_h^2}{z^2 (P_h^-)^2} \, \Big[ D_3^{h/q}(z) \Big] \,, 
\\[0.2cm] \label{e:G3_quark_z}
\Delta^{h/ q \, [\gamma^+ \gamma_5]}(z; P_h, S_h) & = & \frac{M_h^2}{z^2 (P_h^-)^2} \, \Big[ \Lambda_h \, G_3^{h/q}(z) \Big] \,,
\\[0.2cm] \label{e:H3_quark_z}
\Delta^{h/q \, [i \sigma^{i+} \gamma_5]}(z; P_h, S_h) & = & \frac{M_h^2}{z^2 (P_h^-)^2} \, \Big[ S_{hT}^i \, H_3^{h/q}(z) \Big] \,.
\end{eqnarray}
The structures of the traces in~(\ref{e:D1_quark_z})-(\ref{e:H1_quark_z}), (\ref{e:E_quark_z})-(\ref{e:HL_quark_z}), and~(\ref{e:D3_quark_z})-(\ref{e:H3_quark_z}) follow from parity invariance.
(Some additional structures that appear when relaxing the parity constraint have been discussed in Ref.~\cite{Kang:2010qx}.)
Moreover, hermiticity implies that all the FFs are real.
Time-reversal does not impose a constraint on the number of allowed FFs because the state $| P_h, S_h; X \rangle$ in~(\ref{e:corr_quark_int}) is an out-state which includes all the interactions between the particles~\cite{Collins:1992kk}.
Time-reversal, however, does transform out-states into in-states of non-interacting particles, and therefore the correlator is transformed into an unrelated object.
If one ignored the difference between in-states and out-states, three of the twist-3 FFs above ($E_L$, $D_T$, $H$) were forbidden due to time-reversal invariance of the strong interaction~\cite{Collins:1992kk,Goeke:2005hb}.
Therefore, these functions are called na\"ive T-odd.
Even though na\"ive T-odd FFs can exist because of interactions between particles, they are nonzero only if the amplitude describing the fragmentation $q \to h X$ has a minimum of two different components with at least one of them having an imaginary part.
This feature is also clearly seen in model-calculations~\cite{Collins:1992kk,Bacchetta:2001di}.
Additional discussion on the application of time-reversal to FFs can be found in~\cite{Barone:2001sp,Collins:2011zzd}.

Now we consider the case of antiquarks.
One can define antiquark FFs by means of (Dirac traces of) the correlator~\cite{Mulders:1995dh,Boer:1997mf,Pitonyak:2013dsu}
\begin{eqnarray} \label{e:corr_antiquark_int}
\bar{\Delta}^{h/q}(z;P_h,S_h) & = & \sum_X \hspace{-0.5cm} \int \; \int \frac{d \xi^+}{2\pi} \, e^{i k^- \xi^+} \,
\langle 0 | \, {\cal W}(\infty^+, \xi^+) \, \bar{\psi}_q(\xi^+, 0^-, \vec{0}_T) \, | P_h, S_h; X \rangle  
\nonumber \\[0.1cm]
& & \times \; \langle P_h, S_h; X | \, \psi_q(0^+, 0^-, \vec{0}_T) \, {\cal W}(0^+, \infty^+) \, | 0 \rangle \,,
\end{eqnarray}
which appears naturally in calculations with antiquarks.
One finds that $\bar{\Delta}^{h/q}$ can be expressed in terms of the correlator $\Delta^{h/\bar{q}}$ via~\cite{Mulders:1995dh,Boer:1997mf,Pitonyak:2013dsu}
\begin{eqnarray} \label{e:antiquark_tr1}
\bar{\Delta}^{h/q \, [\Gamma]} & = &  + \, \Delta^{h/\bar{q} \, [\Gamma]} \quad \textrm{for} \;\, \Gamma = \gamma^{\mu} \,, \; i \sigma^{\mu\nu} \gamma_5 \,,  
\\[0.2cm]  \label{e:antiquark_tr2}
\bar{\Delta}^{h/q \, [\Gamma]} & = &  - \, \Delta^{h/\bar{q} \, [\Gamma]} \quad \textrm{for} \;\, \Gamma = \mathbb{1} \,, \; i \gamma_5 \,, \; \gamma^{\mu} \gamma_5 \,,
\end{eqnarray}  
where $\Delta^{h/\bar{q}}$, in comparison to the {\it qq}-correlator in~(\ref{e:corr_quark_int}), involves charge-conjugated quark fields and therefore defines FFs for antiquarks, in full analogy to Eqs.~(\ref{e:D1_quark_z})-(\ref{e:H1_quark_z}), (\ref{e:E_quark_z})-(\ref{e:HL_quark_z}), and~(\ref{e:D3_quark_z})-(\ref{e:H3_quark_z}).

Finally, we discuss the definition of integrated leading-twist FFs for gluons.
They can be specified through the gluon-gluon ({\it gg}) fragmentation correlator~\cite{Collins:1981uw,Mulders:2000sh,Collins:2011zzd}
\begin{eqnarray} \label{e:corr_gluon_int}
\Delta^{h/g,ij}(z;P_h,S_h) & = & \sum_X \hspace{-0.5cm} \int \; \int \frac{d \xi^+}{2\pi} \, e^{i k^- \xi^+} \,
\langle 0 | \, {\cal W}_{ba}(\infty^+ , \xi^+) \, F_a^{- i} (\xi^+, 0^-, \vec{0}_T) \, | P_h, S_h; X \rangle  
\nonumber \\[0.1cm]
& & \times \; \langle P_h, S_h; X | \, F_c^{- j}(0^+, 0^-, \vec{0}_T) \, {\cal W}_{cb}(0^+ , \infty^+) \, | 0 \rangle \,,
\end{eqnarray}
with the gluons represented by components of the gluon field strength tensor $F_a^{\mu\nu}$.
Here the Wilson lines are in the adjoint representation.
We decompose the correlator in~(\ref{e:corr_gluon_int}) according to 
\begin{equation}  \label{e:corr_gluon_decomp}
\Delta^{h/g,ij}(z;P_h,S_h) = \frac{1}{2} \, \delta_T^{ij} \, \Big( \delta_T^{kl} \Delta^{h/g,kl}(z;P_h,S_h) \Big)
 - \frac{i}{2} \, \varepsilon_T^{ij} \, \Big( i \varepsilon_T^{kl} \Delta^{h/g,kl}(z;P_h,S_h) \Big)
 + \hat{S} \, \Delta^{h/g,ij}(z;P_h,S_h) \,,
\end{equation}
where the three terms on the r.h.s.~of this equation correspond, in order, to unpolarized, circularly polarized, and linearly polarized gluons.
We have used $\delta_T^{ij} = - g_T^{ij}$ with $g_T^{\mu\nu} = g^{\mu\nu} - n^{\mu} \bar{n}^{\nu} - n^{\nu} \bar{n}^{\mu}$, and the symmetrization operator $\hat{S}$ which is defined through~\cite{Diehl:2001pm,Diehl:2003ny,Meissner:2007rx}
\begin{equation}
\hat{S} \, O^{ij} \equiv \frac{1}{2} \, \Big( O^{ij} + O^{ji} - \delta_T^{ij} \, O^{kk} \Big)
\end{equation}
for a generic tensor $O^{ij}$.
The integrated gluon FFs can then be specified via~\cite{Collins:1981uw,Mulders:2000sh,Collins:2011zzd}
\begin{eqnarray} \label{e:D1_gluon_z}
\delta_T^{ij} \, \Delta^{h/g,ij}(z;P_h,S_h) & = & \frac{2 P_h^-}{z^2} \, \Big[ D_1^{h/g}(z) \Big] \,,
\\ \label{e:G1_gluon_z}
i \varepsilon_T^{ij} \, \Delta^{h/g,ij}(z;P_h,S_h) & = & \frac{2 P_h^-}{z^2} \, \Big[ \Lambda_h \, G_1^{h/g}(z) \Big] \,,
\\[0.2cm]  \label{e:H1_gluon_z}
\hat{S} \, \Delta^{h/g,ij}(z;P_h,S_h) & = & 0 \,,
\end{eqnarray}
where $D_1^{h/g}$ is the unpolarized gluon FF, while $G_1^{h/q}$ describes the density of longitudinally polarized hadrons in a circularly polarized gluon.
There is no integrated gluon FF for linearly polarized gluons fragmenting into a spin-$\frac{1}{2}$ hadron, which is a consequence of conservation of angular momentum.
This is like for PDFs where no (integrated) gluon transversity exists for a spin-$\frac{1}{2}$ target.
We refrain from listing integrated gluon FFs of higher twist, but the interested reader finds more information on their classification in Ref.~\cite{Mulders:2000sh}.
Currently the phenomenology of such functions is unexplored.

We also want to comment briefly on the support properties of integrated FFs. 
Translating in Eqs.~(\ref{e:corr_quark_int}), (\ref{e:corr_antiquark_int}), (\ref{e:corr_gluon_int}) the field operators that are not located at the origin allows one to remove the $\xi^+$ dependence from the matrix elements, such that the $\xi^+$ integral can be performed easily.
Then one finds that, for all partons, the FFs vanish unless $0 \le z \le 1$  (see also~\cite{Collins:1981uw}).
This also implies that there is no (formal) relation between quark FFs and antiquark FFs, which is in contrast to PDFs, where antiquark PDFs with positive momentum fractions can be expressed through quark PDFs with negative momentum fractions (and vice versa).

\subsubsection{Definition of TMD FFs}
\label{sec:definition_TMD}
In full analogy to integrated FFs, one defines TMD FFs by means of a parton fragmentation correlator, where now one keeps the dependence on the transverse momentum $\vec{k}_T$ of the parton.
The TMD {\it qq} correlator reads~\cite{Collins:1981uw,Collins:1992kk,Mulders:1995dh,Boer:1997mf,Boer:2003cm,Bacchetta:2006tn,Collins:2011zzd,Pitonyak:2013dsu}
\begin{eqnarray} \label{e:corr_quark_kT}
\Delta^{h/q}(z,\vec{k}_T;P_h,S_h) & = & \sum_X \hspace{-0.5cm} \int \; \int \frac{d \xi^+ \, d^2\vec{\xi}_T}{(2\pi)^3} \, e^{i (k^- \xi^+ - \vec{k}_T \cdot \vec{\xi}_T)} \,
\langle 0 | \, {\cal W}_1(\infty , \xi) \, \psi_q(\xi^+, 0^-, \vec{\xi}_T) \, | P_h, S_h; X \rangle  
\nonumber \\[0.1cm]
& & \times \; \langle P_h, S_h; X | \, \bar{\psi}_q(0^+, 0^-, \vec{0}_T) \, {\cal W}_2(0, \infty) \, | 0 \rangle \,,
\end{eqnarray}
which obviously is not a light-cone correlator since the quark fields are not only separated along the $\xi^+$ direction but also along the transverse direction $\xi_T$. 
The Wilson lines in~(\ref{e:corr_quark_kT}) are given by
\begin{eqnarray} \label{e:wilson_1}
{\cal W}_1(\infty, \xi) & = & [\infty^+, 0^-, \vec{\infty}_T; \infty^{+}, 0^-, \vec{\xi}_T] \times
[\infty^+, 0^-, \vec{\xi}_T; \xi^+, 0^-, \vec{\xi}_T] \,,
\\[0.2cm] \label{e:wilson_2}
{\cal W}_2(0, \infty) & = & [0^+, 0^-, \vec{0}_T; \infty^{+}, 0^-, \vec{0}_T] \times
[\infty^{+}, 0^-, \vec{0}_T; \infty^{+}, 0^-, \vec{\infty}_T] \,,
\end{eqnarray}
with the individual Wilson lines specified in~(\ref{e:wilson_quark}).
Transverse gauge links like those on the r.h.s.~in~(\ref{e:wilson_1}), (\ref{e:wilson_2}) are crucial for a proper gauge-invariant definition of TMD correlation functions as was first pointed out in Refs.~\cite{Ji:2002aa,Belitsky:2002sm}.
The future-pointing Wilson lines in~(\ref{e:wilson_1}), (\ref{e:wilson_2}) naturally appear for the correlation function  when deriving factorization in $e^+ e^-$ annihilation. 
For SIDIS, {\it a priori,} factorization seemed to require past-pointing Wilson lines instead, i.e.,  the replacement of all occurrences of $\infty^{+}$ in the gauge links by $-\infty^{+}$~\cite{Boer:2003cm}.
However, later on in Ref.~\cite{Collins:2004nx} it was shown that factorization can actually be derived for both $e^+ e^-$ annihilation and SIDIS using, for instance, future-pointing Wilson lines.
This result is intimately connected with the universality of TMD FFs, a point which will be discussed in a bit more detail in Sec.~\ref{sec:universality}.
Also, like for the integrated correlator in~(\ref{e:corr_quark_int}), we have suppressed a dependence of $\Delta^{h/q}$ in~(\ref{e:corr_quark_kT}) on the light-cone vector $\bar{n}$ and on a renormalization scale $\mu$, where here the latter merely is due to renormalization of the quark fields~\cite{Collins:2011zzd}.
An additional complication in the case of TMDs are the infamous light-cone singularities which show up when evaluating the TMD correlator in~(\ref{e:corr_quark_kT}) (or the corresponding correlator for TMD PDFs) in perturbation theory by including gauge field degrees of freedom --- see for instance~\cite{Collins:1981uw,Collins:2003fm}.
These singularities arise as a consequence of the kinematical (eikonal) approximation that is made when deriving factorization for higher-order diagrams and which leads to the Wilson line in the parton correlator.
In fact, they already appear at the one-loop level.
For $k_T$ integrated parton correlators the light-cone singularties cancel after summing over real and virtual loop corrections~\cite{Collins:1981uw}, however there is no such cancellation for TMDs.
Like UV divergences, these divergences need to be regularized, but this cannot be achieved using dimensional regularization~\cite{Collins:1992tv}.
 The light-cone singualrities, plus a proper treatment of leading-twist soft-gluon effects, necessitate some modification of the TMD correlator in~(\ref{e:corr_quark_kT}).
However, these issues have no influence on the main focus of this subsection, namely the classification of TMD FFs. 
We briefly come back to this discussion in Sec.~\ref{sec:tmd_evolution}.

In the TMD case, the Dirac traces defined in~(\ref{e:corr_quark_trace}) are parameterized in terms of eight leading-twist FFs according to~\cite{Collins:1981uw,Collins:1992kk,Mulders:1995dh,Boer:1997mf,Boer:2003cm,Bacchetta:2006tn,Collins:2011zzd,Pitonyak:2013dsu}
\begin{eqnarray} \label{e:D1_quark_kT}
\Delta^{h/q \, [\gamma^-]}(z, \vec{k}_T; P_h, S_h) & = &  
D_1^{h/q}(z, z^2 \vec{k}_T^{\,2}) 
+ \frac{\varepsilon_T^{ij} \, k_T^i \, S_{hT}^j}{M_h} \, D_{1T}^{\perp \, h/q}(z, z^2 \vec{k}_T^{\,2}) \,,
\\[0.1cm]  \label{e:G1_quark_kT}
\Delta^{h/q \, [\gamma^- \gamma_5]}(z, \vec{k}_T; P_h, S_h) & = &  
\Lambda_h \, G_{1L}^{h/q}(z, z^2 \vec{k}_T^{\,2}) 
+ \frac{\vec{k}_T \cdot \vec{S}_{hT}}{M_h} \, G_{1T}^{h/q}(z, z^2 \vec{k}_T^{\,2}) \,,
\\[0.1cm]  \label{e:H1_quark_kT}
\Delta^{h/q \, [i \sigma^{i-} \gamma_5]}(z,\vec{k}_T; P_h, S_h) & = & 
S_{hT}^i \, H_{1T}^{h/q}(z, z^2 \vec{k}_T^{\,2}) 
- \frac{\varepsilon_T^{ij} \, k_T^j}{M_h} \, H_{1}^{\perp \, h/q}(z, z^2 \vec{k}_T^{\,2}) 
\nonumber \\
& & + \, \frac{k_T^i}{M_h} \, \bigg(
\Lambda_h \, H_{1L}^{\perp \, h/q}(z, z^2 \vec{k}_T^{\,2}) 
+ \frac{\vec{k}_T \cdot \vec{S}_{hT}}{M_h} \, H_{1T}^{\perp \, h/q}(z, z^2 \vec{k}_T^{\,2}) \bigg) \,.
\end{eqnarray}
The FF $D_{1T}^{\perp \, h/q}$ and the Collins function $H_1^{\perp \, h/q}$ are na\"ive T-odd.
Also, note that the density interpretation of FFs is typically understood in a frame of reference in which the fragmenting parton has no transverse momentum, and the hadron has the transverse momentum $\vec{P}_{hT}$.
One can show that $\vec{P}_{hT} = - z \vec{k}_T$~\cite{Collins:1981uw}, and therefore $z^2 \vec{k}_T^{\,2} = \vec{P}_{hT}^{\,2}$.

Integrating the {\it qq} correlator in~(\ref{e:corr_quark_kT}) upon $k_T$ one obtains the correlator in~(\ref{e:corr_quark_int}). 
This provides the following relations between the integrated FFs and the TMD FFs,
\begin{eqnarray}  \label{e:D1_int}
D_1^{h/q}(z) & = & z^2 \int d^2 \vec{k}_T \, D_1^{h/q}(z, z^2 \vec{k}_T^{\,2}) =
\int d^2 \vec{P}_{hT} \, D_1^{h/q}(z,\vec{P}_{hT}^{\,2}) \,,
\\[0.2cm]  \label{e:G1_int}  
G_1^{h/q}(z) & = & z^2 \int d^2 \vec{k}_T \, G_{1L}^{h/q}(z, z^2 \vec{k}_T^{\,2}) \,,
\\  \label{e:H1_int}
H_1^{h/q}(z) & = & z^2 \int d^2 \vec{k}_T \, \bigg[ H_{1T}^{h/q}(z, z^2 \vec{k}_T^{\,2}) 
+ \frac{\vec{k}_T^{\,2}}{2 M_h^2} \, H_{1T}^{\perp \, h/q}(z, z^2 \vec{k}_T^{\,2}) \bigg] \,.
\end{eqnarray}
A few comments are in order at this point.
The aforementioned light-cone singularities in the TMD FFs as defined above do cancel between real and virtual diagrams when performing the $k_T$ integral, which implies that the integrated FFs are free of such divergences~\cite{Collins:1981uw,Collins:2003fm}.
On the other hand, if one first regularizes the light-cone singularities of TMD FFs, and afterwards performs the $k_T$ integration one no longer has simple relations of the type (\ref{e:D1_int})-(\ref{e:H1_int}) --- see for instance~\cite{Ji:2004wu,Collins:2011zzd}.
Moreover, the transverse momentum integrals in~(\ref{e:D1_int})-(\ref{e:H1_int}) contain UV divergences.
We understand that UV divergences on the l.h.s.~and on the r.h.s.~of these equations are dealt with in exactly the same manner.

For later convenience we also define the following $k_T$ moments of a generic FF $X^{h/q}$,
\begin{equation}  \label{e:kT_moment}
X^{(n)\, h/q}(z) = z^2 \int d^2 \vec{k}_T \, \bigg( \frac{\vec{k}_T^{\,2}}{2 M_h^2} \bigg)^n X^{h/q}(z, z^2 \vec{k}_T^{\,2}) \,.
\end{equation}
The case $n=1$ for the functions $D_{1T}^{\perp \, h/q}$, $G_{1T}^{h/q}$, $H_1^{\perp \, h/q}$, $H_{1L}^{\perp \, h/q}$ is of particular interest as these objects can appear in certain twist-3 observables that are described in collinear factorization.

One defines TMD FFs for antiquarks as in the case of integrated FFs by using the TMD counterpart of the correlator in Eq.~(\ref{e:corr_antiquark_int}) and the relations~(\ref{e:antiquark_tr1}), (\ref{e:antiquark_tr2}) which also hold in the TMD case.

The relevant correlator for gluon TMD FFs reads~\cite{Collins:1981uw,Mulders:2000sh,Collins:2011zzd}
\begin{eqnarray} \label{e:corr_gluon_kT}
\Delta^{h/g,ij}(z,\vec{k}_T;P_h,S_h) & = & \sum_X \hspace{-0.5cm} \int \; \int \frac{d \xi^+ \, d^2\vec{k}_T}{(2\pi)^3} \, e^{i (k^- \xi^+ - \vec{k}_T \cdot \vec{\xi}_T)} \,
\langle 0 | \, {\cal W}_{1, ba}(\infty, \xi) \, F_a^{- i} (\xi^+, 0^-, \vec{\xi}_T) \, | P_h, S_h; X \rangle  
\nonumber \\[0.1cm]
& & \times \; \langle P_h, S_h; X | \, F_c^{- j}(0^+, 0^-, \vec{0}_T) \, {\cal W}_{2, cb}(0, \infty) \, | 0 \rangle \,,
\end{eqnarray}
where we use Wilson lines as specified in~(\ref{e:wilson_1}), (\ref{e:wilson_2}), but in the adjoint representation.
With the decomposition~(\ref{e:corr_gluon_decomp}) of the gluon fragmentation correlator one can define eight leading-twist TMD FFs of gluons through~\cite{Collins:1981uw,Mulders:2000sh,Collins:2011zzd}
\begin{eqnarray}  \label{e:D1_gluon_kT}
\delta_T^{ij} \, \Delta^{h/g,ij}(z,\vec{k}_T;P_h,S_h) & = & 2 P_h^- \, \bigg[
D_1^{h/g}(z, z^2 \vec{k}_T^{\,2}) 
+ \frac{\varepsilon_T^{ij} \, k_T^i \, S_{hT}^j}{M_h} \, D_{1T}^{\perp \, h/g}(z, z^2 \vec{k}_T^{\,2}) \bigg] \,,
\\[0.1cm]  \label{e:G1_gluon_kT}
i \varepsilon_T^{ij} \, \Delta^{h/g,ij}(z,\vec{k}_T;P_h,S_h) & = & 2 P_h^- \, \bigg[ 
\Lambda_h \, G_{1L}^{h/g}(z, z^2 \vec{k}_T^{\,2}) 
+ \frac{\vec{k}_T \cdot \vec{S}_{hT}}{M_h} \, G_{1T}^{h/g}(z, z^2 \vec{k}_T^{\,2}) \bigg] \,,
\\[0.1cm]  \label{e:H1_gluon_kT}
\hat{S} \, \Delta^{h/g,ij}(z,\vec{k}_T;P_h,S_h) & = & 2 P_h^- \, \hat{S} \, \bigg[
\frac{k_T^i \, \varepsilon_T^{jk} \, S_{hT}^k}{2 M_h} \, H_{1T}^{h/g}(z, z^2 \vec{k}_T^{\,2}) 
+ \frac{k_T^i \, k_T^j}{2 M_h^2} \, H_1^{\perp \, h/g}(z, z^2 \vec{k}_T^{\,2}) 
\nonumber \\
& & + \, \frac{k_T^i \, \varepsilon_T^{jk} \, k_T^k}{2 M_h^2} \, \bigg(
\Lambda_h \, H_{1L}^{\perp \, h/g}(z, z^2 \vec{k}_T^{\,2}) 
+ \frac{\vec{k}_T \cdot \vec{S}_{hT}}{M_h} \, H_{1T}^{\perp \, h/g}(z, z^2 \vec{k}_T^{\,2}) \bigg) \bigg] \,,
\end{eqnarray}
where $D_{1T}^{\perp \, h/g}$, $H_{1T}^{h/g}$, $H_{1L}^{\perp \, h/g}$, $H_{1T}^{\perp \, h/g}$ are na\"ive T-odd functions.
One readily verifies that the relations~(\ref{e:D1_int}), (\ref{e:G1_int}) also hold for gluon FFs.
Our notation of the gluon TMD FFs is inspired by the notation used in Ref.~\cite{Meissner:2007rx} for gluon TMD PDFs. 
In order to fully see the close analogy of the notation for quark and gluon FFs one has to consider the correspondence
\begin{equation}
\hat{S} \, \Delta^{h/g,ij} \; \longleftrightarrow \; P_h^- \, \hat{S} \, \frac{k_T^i}{M_h} \, \varepsilon_T^{jk} \, \Delta^{h/q \, [i \sigma^{k-} \gamma_5]}
\end{equation}
between the expressions in Eqs.~(\ref{e:H1_quark_kT}) and (\ref{e:H1_gluon_kT}).
For convenience we list the relations between the FFs defined in~(\ref{e:D1_gluon_kT}), (\ref{e:G1_gluon_kT}), (\ref{e:H1_gluon_kT}), and those of Ref.~\cite{Mulders:2000sh}, where a full set of gluon TMD FFs was defined for the first time,
\begin{eqnarray}
&& 
D_1^{h/g} = + \, \hat{G} \,, \hspace{0.4cm}
D_{1T}^{\perp \, h/g} = - \, \hat{G}_T \,, \hspace{0.4cm}
G_{1L}^{\perp \, h/g} = - \, \Delta \hat{G}_L \,, \hspace{0.4cm}
G_{1T}^{\perp \, h/g} = - \, \Delta \hat{G}_T \,, 
\nonumber \\
&& 
H_{1T}^{h/g} = - \bigg( \Delta \hat{H}_T - \frac{\vec{k}_T^2}{2 M_h^2} \, \Delta \hat{H}_T^\perp  \bigg) \,, \hspace{0.4cm}
H_1^{\perp \, h/g} = + \, \hat{H}^\perp \,, \hspace{0.4cm}
H_{1L}^{\perp \, h/g} = - \, \Delta \hat{H}_L^\perp \,, \hspace{0.4cm}
H_{1T}^{\perp \, h/g} = - \, \Delta \hat{H}_T^\perp \,. 
\phantom{aaaa}
\end{eqnarray}
Finally, for the definition of TMD FFs of higher twist we refer to the literature~\cite{Mulders:1995dh,Boer:1997mf,Mulders:2000sh,Bacchetta:2004zf,Goeke:2005hb,Bacchetta:2006tn,Chen:2016moq}.
Let us also mention that presently the status of TMD factorization beyond leading twist is unclear~\cite{Gamberg:2006ru,Bacchetta:2008xw}.

\subsubsection{Definition of three-parton FFs}
\label{sec:definition_3parton}
So far in this section we exclusively considered fragmentation correlators with two parton fields.
However, for a complete description of twist-3 observables it is mandatory to also include three-parton fragmentation correlators.
We call the functions that parameterize such correlators three-parton FFs.
As we explain below in a bit more detail, after taking into account certain relations among twist-3 FFs one can actually express twist-3 observables entirely through three-parton FFs~\cite{Balitsky:1987bk,Balitsky:1990ck,Kanazawa:2015ajw}.
At twist-3 level, three-parton FFs can therefore be considered the truly fundamental objects. 
There exist two types of three-parton FFs: quark-gluon-quark ({\it qgq}) and tri-gluon ({\it ggg}) FFs. 
Here we will limit ourselves to a brief discussion of the former type.
Only if the final state hadron is transversely polarized the {\it ggg} FFs do matter.
\begin{figure}[t]
\begin{center}
\includegraphics[width=9.0cm]{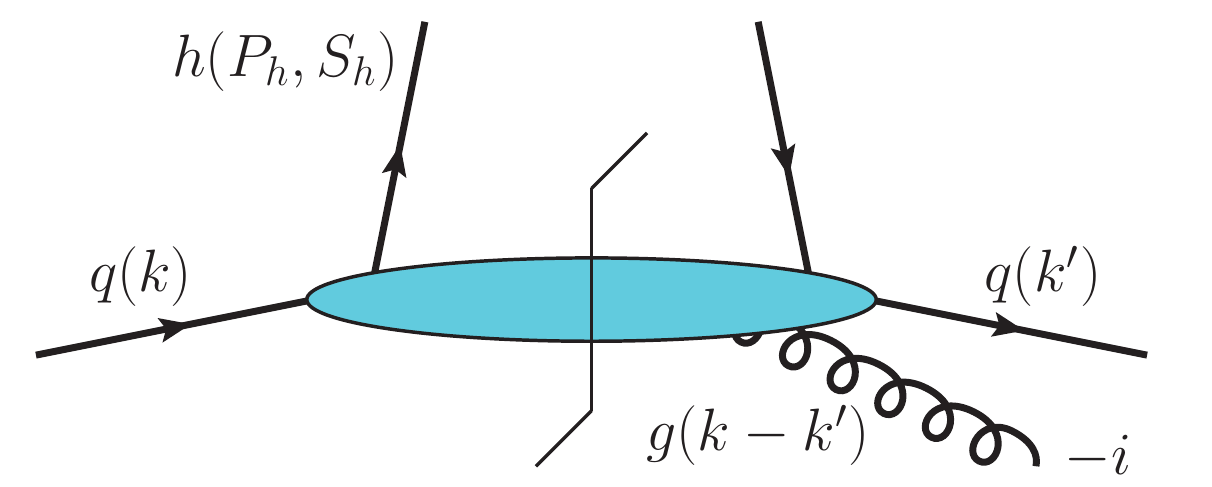} 
\end{center}
\vspace{-0.4cm}
\caption{Representation of the correlator $\Delta_F^{h/q;\, i}$ in~(\ref{e:corr_qgq}), which defines three-parton F-type FFs.
The figure contains no reference to the Wilson lines.}
\label{f:qgq_frag_correlator}
\end{figure}

The so-called F-type {\it qgq} FFs are defined through the correlator (see also Fig.~\ref{f:qgq_frag_correlator})~\cite{Koike:2001zw,Eguchi:2006qz,Meissner:thesis} 
\begin{eqnarray} \label{e:corr_qgq}
\Delta_F^{h/q;\,i}(z,z';P_h,S_h) & = & \frac{1}{P_h^-} \, \sum_X \hspace{-0.5cm} \int \; \int \frac{d \xi^+ d \zeta^+}{(2\pi)^2} \, e^{i k'^- \xi^+} \, e^{i(k^- - k'^-) \zeta^+} \,
\nonumber \\[0.1cm]
& & \times \; \langle 0 | \, {\cal W}(\infty^+, \zeta^+) \, i g F^{-i}(\zeta^+, 0^-, \vec{0}_T) \, {\cal W}(\zeta^+, \xi^+) \, \psi_q(\xi^+, 0^-, \vec{0}_T) \,  | P_h, S_h; X \rangle  
\nonumber \\[0.1cm]
& & \times \; \langle P_h, S_h; X | \, \bar{\psi}_q(0^+, 0^-, \vec{0}_T) \, {\cal W}(0^+, \infty^+) \, | 0 \rangle \,,
\phantom{\frac{1}{1}}
\end{eqnarray}
where the gluon is represented by the component $F^{-i}$ of the field strength tensor, and $g$ is the strong coupling constant.
Note that in~(\ref{e:corr_qgq}) all parton fields are located on the light-cone. 
This specific situation is sufficient for the twist-3 observables considered in this review.
The correlator in Eq.~(\ref{e:corr_qgq}) can be parameterized in terms of four three-parton FFs~\cite{Koike:2001zw,Eguchi:2006qz,Meissner:thesis} where, using the trace notation from~(\ref{e:corr_quark_trace}), one finds~\cite{Meissner:thesis}
\begin{eqnarray} \label{e:DeltaF_F}
\Delta_F^{h/q;\, i \, [\gamma^-]}(z,z'; P_h, S_h) & = & \frac{M_h}{P_h^-} \, \Big[ i \varepsilon_T^{ij} \, S_{hT}^j \, \hat{D}_{FT}^{h/q}(z,z') \Big] \,, 
\\[0.2cm] \label{e:DeltaF_G}
\Delta_F^{h/q;\, i \, [\gamma^- \gamma_5]}(z,z'; P_h, S_h) & = & \frac{M_h}{P_h^-} \, \Big[ \delta_T^{ij} \, S_{hT}^j \, \hat{G}_{FT}^{h/q}(z,z') \Big] \,,
\\[0.2cm] \label{e:DeltaF_H}
\Delta_F^{h/q;\, i \, [i \sigma^{j-} \gamma_5]}(z,z'; P_h, S_h) & = & \frac{M_h}{P_h^-} \, \Big[ i \varepsilon_T^{ij} \, \hat{H}_{FU}^{h/q}(z,z') + \Lambda_h \, \delta_T^{ij} \, \hat{H}_{FL}^{h/q}(z,z') \Big] \,.
\end{eqnarray}
The three-parton FFs depend on two arguments.
It is also important that these functions have both a real and imaginary part~\cite{Meissner:thesis,Yuan:2009dw,Kang:2010zzb,Metz:2012ct,Kanazawa:2013uia} which we denote by~\cite{Metz:2012ct}
\begin{equation}
\hat{D}_{FT}^{h/q,\, \Re} \equiv {\rm Re} \, \hat{D}_{FT}^{h/q} \,, \qquad
\hat{D}_{FT}^{h/q,\, \Im} \equiv {\rm Im} \, \hat{D}_{FT}^{h/q} \,, \; {\rm etc.}
\end{equation}
Instead of using the field strength tensor like in~(\ref{e:corr_qgq}), the gluon may also be represented through a component of the covariant derivative. 
The corresponding D-type {\it qgq} correlator is also parameterized through four independent three-parton FFs --- the D-type functions $\hat{D}_{DT}^{h/q}$, $\hat{G}_{DT}^{h/q}$, $\hat{D}_{DU}^{h/q}$, $\hat{D}_{DL}^{h/q}$~\cite{Meissner:thesis,Yuan:2009dw,Kang:2010zzb,Metz:2012ct,Kanazawa:2013uia,Kanazawa:2015ajw}.
Depending on the situation either F-type or D-type functions are more convenient.
The two types of three-parton FFs are related, and we refer to the literature for more information~\cite{Eguchi:2006qz,Meissner:thesis,Yuan:2009dw,Metz:2012ct,Kanazawa:2013uia}.

QCD equations of motion allow one to obtain relations between twist-3 integrated FFs given in~(\ref{e:E_quark_z})-(\ref{e:HL_quark_z}), certain $k_T$ moments of TMD FFs, and three-parton FFs.
Here we just list one such relation that will be relevant for the discussion of fits for higher-twist FFs in Sec.~\ref{sec:fit_higher_twist}~\cite{Metz:2012ct,Kanazawa:2014dca},  
\begin{equation} \label{e:relation_twist3}
H^{h/q}(z) = - \,  2 z H_1^{\perp(1) h/q}(z) + 2 z^3 \int_{z}^{\infty} \frac{dz_1} {z_1^2} \, \frac{1} {\frac{1} {z}-\frac{1} {z_{1}}} \, \hat{H}_{FU}^{h/q,\,\Im}(z,z_{1}) \,,
\end{equation}
where the moment of the Collins function $H_1^{\perp(1) \, h/q}$ is defined according to Eq.~(\ref{e:kT_moment}).
Additional constraints among twist-3 FFs arise from so-called Lorentz invariance relations (LIRs)~\cite{Bukhvostov:1984xf,Bukhvostov:1984as,Balitsky:1987bk,Balitsky:1990ck,Belitsky:1997zw,Belitsky:1997ay,Ball:1998sk,Kodaira:1998jn,Kundu:2001pk,Accardi:2009au,Kanazawa:2014tda,Kanazawa:2015ajw}, such that all twist-3 FFs can be expressed through three-parton FFs. 
For the FFs in Eq.~(\ref{e:relation_twist3}) one finds~\cite{Kanazawa:2015ajw}
\begin{eqnarray} \label{e:H_twist3}
H^{h/q}(z) & = & 2 \int_z^1 dz_1 \int_{z_1}^\infty \frac{dz_2}{z_2^2} \,
\frac{\big[ 2 \big( \frac{2}{z_1} - \frac{1}{z_2} \big) + \frac{1}{z_1} \big( \frac{1}{z_1} - \frac{1}{z_2} \big) \delta \big( \frac{1}{z_1} - \frac{1}{z} \big) \big] \hat{H}_{FU}^{h/q,\, \Im}(z_1,z_2)}{\big( \frac{1}{z_1} - \frac{1}{z_2} \big)^2} \,,
\\  \label{e:H1perp_twist3}
H_1^{\perp(1) h/q}(z) & = & - \, \frac{2}{z} \int_z^1 dz_1 \int_{z_1}^\infty \frac{dz_2}{z_2^2} \,
\frac{\big( \frac{2}{z_1} - \frac{1}{z_2} \big) \hat{H}_{FU}^{h/q,\, \Im}(z_1,z_2)}{\big( \frac{1}{z_1} - \frac{1}{z_2} \big)^2} \,.
\end{eqnarray}
The functions $H^{h/q}$, $H_1^{\perp (1) \, h/q}$, $\hat{H}_{FU}^{h/q,\, \Im}$ play an important role in the QCD description of transverse SSAs in processes like $p^{\uparrow} p \to h X$~\cite{Metz:2012ct,Kanazawa:2014dca}.
Additional relations of the type~(\ref{e:H_twist3}),~(\ref{e:H1perp_twist3}), which matter for a polarized final-state hadron, can be found in Ref.~\cite{Kanazawa:2015ajw}.

\subsubsection{Definition of di-hadron FFs}
\label{sec:definition_dihadron}
We now turn our attention to parton fragmentation into two hadrons.
We begin by specifying the kinematics, where we follow closely the notation of Refs.~\cite{Bianconi:1999cd,Radici:2001na,Bacchetta:2002ux}.
The hadron momenta are denoted by $P_1$ and $P_2$, with $P_1^2 = M_1^2$ and $P_2^2 = M_2^2$.
It is convenient to introduce the total 4-momentum and the relative 4-momentum~\cite{Bianconi:1999cd},
\begin{equation} \label{e:dihadron_momenta}
P_h = P_1 + P_2 \,, \quad R = \frac{P_1 - P_2}{2} \,.
\end{equation}
The total invariant mass $M_h$ of the hadron pair is given by $M_h^2 = P_h^2$.
(Notice that this introduces an overload of the symbols $P_h$ and $M_h$ which we also use in the case of single-hadron fragmentation --- see Eq.~(\ref{e:Ph}), 
However, their meaning should always be clear form the context).
We do not consider polarization of the hadrons.
In fact, a classification of DiFFs including hadron polarization does not yet exist.
We consider a frame of reference in which $P_h$ has no transverse component.
Then the 4-momentum of the parton is given by Eq.~(\ref{e:k}), while $P_1$ and $P_2$ may be written as
\begin{eqnarray}
P_1 & = & \bigg( \frac{M_1^2 + \vec{R}_T^2}{(1 + \zeta) P_h^-}, \frac{1 + \zeta}{2} \,P_h^-, \vec{R}_T \bigg) \,,
\\
P_2 & = & \bigg( \frac{M_2^2 + \vec{R}_T^2}{(1 - \zeta) P_h^-}, \frac{1 - \zeta}{2} \,P_h^-, - \vec{R}_T \bigg) \,,
\end{eqnarray}
where the minus-momenta are characterized by the variable $\zeta$~\cite{Bacchetta:2002ux}.
Note that in~\cite{Bianconi:1999cd} instead of $\zeta$ the variable $\xi = (1 + \zeta)/2$ has been used.
With these definitions one readily finds the relation
\begin{equation} \label{e:RT_Mh}
\vec{R}_T^{\,2} = \frac{1 - \zeta^2}{4} M_h^2 - \frac{1 - \zeta}{2} M_1^2 - \frac{1 + \zeta}{2} M_2^2 \,,
\end{equation}
which implies a lower bound on $M_h$,
\begin{equation}
M_h^2 \ge \frac{2}{1 + \zeta} M_1^2 + \frac{2}{1 - \zeta} M_2^2 \,.
\end{equation}
\begin{figure}[t]
\begin{center}
\includegraphics[width=9.0cm]{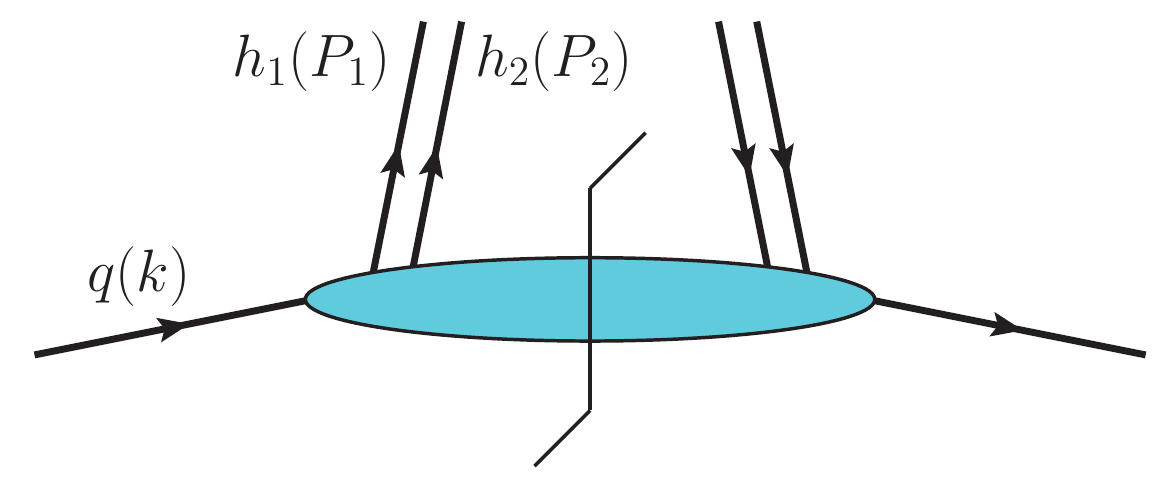} 
\end{center}
\vspace{-0.4cm}
\caption{Representation of the correlator $\Delta^{h_1 h_2/q}$ in~(\ref{e:corr_quark_di-hadron_kT}), which describes the fragmentation of a quark into two hadrons.
The figure contains no reference to the Wilson lines.}
\label{f:quark_frag_correlator_di-hadron}
\end{figure}

The {\it qq} fragmentation correlator for the di-hadron case is illustrated in Fig.~\ref{f:quark_frag_correlator_di-hadron}.
If one keeps the dependence on the transverse quark momentum $k_T$ this correlator, in analogy with Eq.~(\ref{e:corr_quark_kT}), takes the form~\cite{Collins:1993kq,Collins:1994ax,Bianconi:1999cd}
\begin{eqnarray} \label{e:corr_quark_di-hadron_kT}
\Delta^{h_1 h_2/q}(z,\vec{k}_T;P_1,P_2) & = & \sum_X \hspace{-0.5cm} \int \; \int \frac{d \xi^+ \, d^2\vec{\xi}_T}{(2\pi)^3} \, e^{i (k^- \xi^+ - \vec{k}_T \cdot \vec{\xi}_T)} \,
\langle 0 | \, {\cal W}_1(\infty, \xi) \, \psi_q(\xi^+, 0^-, \vec{\xi}_T)  \, | P_1, P_2; X \rangle  
\nonumber \\[0.1cm]
& & \times \; \langle P_1, P_2; X | \, \bar{\psi}_q(0^+, 0^-, \vec{0}_T) {\cal W}_2(0, \infty) \, | 0 \rangle \,,
\end{eqnarray}
where the Wilson lines from~(\ref{e:wilson_1}),~(\ref{e:wilson_2}) are used.
By means of~(\ref{e:corr_quark_trace}) we define traces of the correlator in~(\ref{e:corr_quark_di-hadron_kT}).
The leading-twist traces are parameterized in terms of four DiFFs~\cite{Bianconi:1999cd},
\begin{eqnarray} \label{e:D1_quark_di-hadron_kT}
\Delta^{h_1 h_2/q \, [\gamma^-]}(z, \vec{k}_T; P_1, P_2) & = &  
D_1^{h_1 h_2/q}(z, \zeta, \vec{R}_T^{\, 2}, \vec{k}_T \cdot \vec{R}_T, \vec{k}_T^{\,2}) \,,
\phantom{\frac{1}{1}}
\\[0.1cm]  \label{e:G1_quark_di-hadron_kT}
\Delta^{h_1 h_2/q \, [\gamma^- \gamma_5]}(z, \vec{k}_T; P_1, P_2) & = &  
\frac{\varepsilon_T^{ij} \, R_T^i \, k_T^j}{M_1 \, M_2} \,
G_1^{\perp \, h_1 h_2/q}(z, \zeta, \vec{R}_T^{\, 2}, \vec{k}_T \cdot \vec{R}_T, \vec{k}_T^{\,2}) \,,
\\[0.1cm]  \label{e:H1_quark_di-hadron_kT}
\Delta^{h_1 h_2/q \, [i \sigma^{i-} \gamma_5]}(z,\vec{k}_T; P_1, P_2) & = & 
- \, \frac{\varepsilon_T^{ij} \, R_T^j}{M_1 + M_2} \,
H_1^{\open \, h_1 h_2/q}(z, \zeta, \vec{R}_T^{\, 2}, \vec{k}_T \cdot \vec{R}_T, \vec{k}_T^{\,2})
\nonumber \\
& & - \, \frac{\varepsilon_T^{ij} \, k_T^j}{M_1 + M_2} \,
H_1^{\perp \, h_1 h_2/q}(z, \zeta, \vec{R}_T^{\, 2}, \vec{k}_T \cdot \vec{R}_T, \vec{k}_T^{\,2}) \,.
\end{eqnarray}
While for fragmentation into a single unpolarized hadron only two TMD FFs exist, in the di-hadron case the relative momentum $R_T$ leads to a richer structure of the fragmentation correlator and, hence, to a larger number of FFs.
The DiFFs $G_1^{\perp \, h_1 h_2/q}$ and $H_1^{\open \, h_1 h_2/q}$, like the Collins function $H_1^{\perp \, h_1 h_2/q}$, are na\"ive T-odd objects~\cite{Bianconi:1999cd}.
Here we will not review the definition of subleading twist DiFFs.
For this we refer to the papers~\cite{Bacchetta:2003vn,Gliske:2014wba}.

From Eqs.~(\ref{e:D1_quark_di-hadron_kT})-(\ref{e:H1_quark_di-hadron_kT}) one sees that, when integrating the correlator in~(\ref{e:corr_quark_di-hadron_kT}) upon $k_T$, only two structures survive.
We define the corresponding $k_T$ integrated DiFFs according to
\begin{eqnarray}  \label{e:D1_int_di-hadron}
D_1^{h_1 h_2/q}(z, \zeta, M_h^2) & = & z^2 \int d^2 \vec{k}_T \, \Delta^{h_1 h_2/q \, [\gamma^-]}(z, \vec{k}_T; P_1, P_2) 
\nonumber \\
& = & z^2 \int d^2 \vec{k}_T \, D_1^{h_1 h_2/q}(z, \zeta, \vec{R}_T^{\, 2}, \vec{k}_T \cdot \vec{R}_T, \vec{k}_T^{\,2}) \,,
\\  \label{e:H1_int_di-hadron}
- \, \frac{\varepsilon_T^{ij} \, R_T^j}{M_1 + M_2} \, H_1^{\open \, h_1 h_2/q}(z, \zeta, M_h^2) & = & 
z^2 \int d^2 \vec{k}_T \, \Delta^{h_1 h_2/q \, [i \sigma^{i-} \gamma_5]}(z,\vec{k}_T; P_1, P_2) \,.
\end{eqnarray}
Here we made use of~(\ref{e:RT_Mh}) which allows one to express $|\vec{R}_T|$ in terms of $\zeta$ and $M_h^2$.
The normalization of the $k_T$ integrated DiFFs in~(\ref{e:D1_int_di-hadron}),~(\ref{e:H1_int_di-hadron}) follows the case of single hadron production in Eqs.~(\ref{e:D1_int})-(\ref{e:H1_int}).
Note that both $k_T$ dependent DiFFs in~(\ref{e:H1_quark_di-hadron_kT}) contribute to $H_1^{\open \, h_1 h_2/q}(z, \zeta, M_h^2)$.
If one further integrates the correlator in~(\ref{e:corr_quark_di-hadron_kT}) upon $R_T$ only the DiFF $D_1^{h_1 h_2/q}$ survives.
In analogy with~(\ref{e:D1_quark_def}), the operator definition of this function reads
\begin{eqnarray} \label{e:D1_quark_DiFF_def}
D_1^{h_1 h_2/q}(z_1,z_2) & = & \frac{z}{4} \, \sum_X \hspace{-0.5cm} \int \; \int \frac{d \xi^+}{2\pi} \, e^{i k^- \xi^+} \,
\textrm{Tr} \, \Big[ \langle 0 | \, {\cal W}(\infty^+ , \xi^+)  \, \psi_q(\xi^+, 0^-, \vec{0}_T) \, | P_1, P_2; X \rangle  
\nonumber \\[0.1cm]
& & \times \; \langle P_1, P_2; X | \, \bar{\psi}_q(0^+, 0^-, \vec{0}_T) \, {\cal W}(0^+ , \infty^+) \, | 0 \rangle \, \gamma^- \Big] \,,
\end{eqnarray}
with $z = z_1 + z_2$, and
\begin{equation}
z_1 = \frac{P_1^-}{k^-} = z \, \frac{1 + \zeta}{2} \,, \qquad 
z_2 = \frac{P_2^-}{k^-} = z \, \frac{1 - \zeta}{2} \,.
\end{equation}
Historically, the fully integrated DiFF in~(\ref{e:D1_quark_DiFF_def}) was used first~\cite{Konishi:1978yx,Konishi:1979cb}.

Finally, we briefly discuss $k_T$ dependent DiFFs for gluons.
They are defined through the correlator
\begin{eqnarray} \label{e:corr_gluon_di-hadron_kT}
\Delta^{h_1 h_2/g,ij}(z,\vec{k}_T;P_1,P_2) & = & \sum_X \hspace{-0.5cm} \int \; \int \frac{d \xi^+ \, d^2\vec{k}_T}{(2\pi)^3} \, e^{i (k^- \xi^+ - \vec{k}_T \cdot \vec{\xi}_T)} \,
\langle 0 | \, {\cal W}_{1,ba}(\infty, \xi) \, F_a^{- i} (\xi^+, 0^-, \vec{\xi}_T) \, | P_1, P_2; X \rangle  
\nonumber \\[0.1cm]
& & \times \; \langle P_1, P_2; X | \, F_c^{- j}(0^+, 0^-, \vec{0}_T)  \, {\cal W}_{2,cb}(0, \infty) \, | 0 \rangle \,,
\end{eqnarray}
where we use Wilson lines as specified in~(\ref{e:wilson_1}), (\ref{e:wilson_2}), but in the adjoint representation.
With the decomposition~(\ref{e:corr_gluon_decomp}) of the gluon fragmentation correlator one can identify gluon DiFFs according to
\begin{eqnarray}  \label{e:D1_gluon_di-hadron_kT}
\delta_T^{ij} \, \Delta^{h_1 h_2/g,ij}(z,\vec{k}_T;P_1,P_2) & = & 2 P_h^- \, \bigg[
D_1^{h_1 h_2/g}(z, \zeta, \vec{R}_T^{\, 2}, \vec{k}_T \cdot \vec{R}_T, \vec{k}_T^{\,2}) \bigg] \,,
\\[0.1cm]  \label{e:G1_gluon_di-hadron_kT}
i \varepsilon_T^{ij} \, \Delta^{h_1 h_2/g,ij}(z,\vec{k}_T;P_1,P_2) & = & 2 P_h^- \, \bigg[ \frac{\varepsilon_T^{ij} \, R_T^i \, k_T^j}{M_1 \, M_2} \,
G_1^{\perp \, h_1 h_2/g}(z, \zeta, \vec{R}_T^{\, 2}, \vec{k}_T \cdot \vec{R}_T, \vec{k}_T^{\,2}) \bigg] \,,
\\[0.1cm]  \label{e:H1_gluon_di-hadron_kT}
\hat{S} \, \Delta^{h_1 h_2/g,ij}(z,\vec{k}_T;P_1,P_2) & = & 2 P_h^- \, \hat{S} \, \bigg[
\frac{R_T^i \, R_T^j}{M_1 \, M_2} \, 
H_1^{\open \, h_1 h_2/g}(z, \zeta, \vec{R}_T^{\, 2}, \vec{k}_T \cdot \vec{R}_T, \vec{k}_T^{\,2})
\nonumber \\
&& + \, \frac{k_T^i \, k_T^j}{M_1 \, M_2} \, 
H_1^{\perp \, h_1 h_2/g}(z, \zeta, \vec{R}_T^{\, 2}, \vec{k}_T \cdot \vec{R}_T, \vec{k}_T^{\,2})
\nonumber \\
&& + \, \frac{R_T^i \, k_T^j}{M_1 \, M_2} \, 
H_1^{\open \perp \, h_1 h_2/g}(z, \zeta, \vec{R}_T^{\, 2}, \vec{k}_T \cdot \vec{R}_T, \vec{k}_T^{\,2})
\bigg] \,,
\end{eqnarray}
Like in the quark case we define $k_T$ integrated gluon DiFFs by performing $z^2 \int d^2 \vec{k}_T$ in~(\ref{e:D1_gluon_di-hadron_kT})-(\ref{e:H1_gluon_di-hadron_kT}).
This integration reduces the number of independent structures to two --- one coming from~(\ref{e:D1_gluon_di-hadron_kT}) and another from~(\ref{e:H1_gluon_di-hadron_kT}). 
The latter is proportional to $\hat{S} R_T^i R_T^j$, and the associated $k_T$ integrated DiFF was denoted by $\delta \hat{G}^\open$ in Ref.~\cite{Bacchetta:2004it}, where all three $k_T$ dependent DiFFs in~(\ref{e:H1_gluon_di-hadron_kT}) contribute to $\delta \hat{G}^\open$.
Further integration of the correlator upon $R_T$ leaves one structure only which defines the unpolarized gluon DiFF $D_1^{h_1 h_2 / g}(z_1, z_2)$.

\subsection{Interpretation of FFs}
\label{subs:interpretation}
In the previous section~\ref{sec:definition} we already included some discussion on the interpretation of FFs.
We now give a brief overview of this topic. 
\begin{table}[t]
\parbox{.45\linewidth}{
\centering
\begin{tabular}{|c|c|c|c|}
\hline 
$\textrm{H} \, \big\backslash \, \textrm{q}$ & $\textrm{U}$ & $\textrm{L}$ & $\phantom{aa} \textrm{T}_{\phantom{M_{M}}}^{\phantom{M^{M}}}$ \\
\hline 
$\textrm{U}$ & $D_1^{h/q}$ & & $ H_1^{\perp \, h/q}$ \\
\hline
$\textrm{L}$ &  & $G_1^{h/q}$ & $\phantom{a} H_{1L}^{\perp \, h/q}$ \\
\hline
$\textrm{T}$ & $\phantom{a} D_{1T}^{\perp \, h/q} \phantom{a}$ & $\phantom{a} G_{1T}^{h/q} \phantom{a}$ & $\phantom{a} H_1^{h/q} \quad H_{1T}^{\perp \, h/q} \phantom{a}$ \\
\hline
\end{tabular}
\caption{Interpretation of TMD FFs for quarks. 
The colums indicate the quark polarization --- unpolarized (U), longitudinally polarized (L), transversely polarized (T). 
The rows indicate the hadron prolarization.\label{t:quark_TMDs}}
}
\hfill
\parbox{.45\linewidth}{
\centering
\begin{tabular}{|c|c|c|c|}
\hline 
$\textrm{H} \, \big\backslash \, \textrm{g}$& $\textrm{U}$ & $\textrm{Circ}$ & $\phantom{aa} \textrm{Lin}_{\phantom{M_{M}}}^{\phantom{M^{M}}}$ \\
\hline 
$\textrm{U}$ & $D_1^{h/g}$ & & $ H_1^{\perp \, h/g}$ \\
\hline
$\textrm{L}$ &  & $G_1^{h/g}$ & $H_{1L}^{\perp \, h/g}$ \\
\hline
$\textrm{T}$ & $ \phantom{a} D_{1T}^{\perp \, h/g} \phantom{a}$ & $ \phantom{a} G_{1T}^{h/g} \phantom{a}$ & $\phantom{a} H_1^{h/g} \quad H_{1T}^{\perp \, h/g}$ \\
\hline
\end{tabular}
\caption{Interpretation of TMD FFs for gluons. 
The colums indicate the gluon polarization --- unpolarized (U), circularly polarized (Circ), linearly polarized (Lin). 
The rows indicate the hadron prolarization.\label{t:gluon_TMDs}}
}
\end{table}

It is well known that leading twist FFs (as well as PDFs) have an interprepatation as probability densities, while two-parton higher twist FFs and three-parton FFs have not.
Let us begin with the FF $D_1^{h/q}(z)$.
By using light-front quantization for the quark field operators in~(\ref{e:D1_quark_def}) one can show that this FF is the number density for finding an unpolarized hadron with momentum $P_h^- = z k^-$ inside an unpolarized quark with longitudinal momentum $k^-$~\cite{Collins:1981uw,Collins:2011zzd}.
This interpretation can be generalized to include the transverse momentum $P_{hT}$ of the hadron, that is, to the case of the TMD FF $D_1^{h/q}(z,\vec{P}_{hT}^{\,2})$.
(Discussion on complications of the density interpretation in a gauge theory like QCD, that are related to UV divergences and other issues, can be found in~\cite{Collins:2011zzd} --- see also Refs.~\cite{Brodsky:2002ue,Collins:2003fm}.)
One can further extend the analysis by including the polarization of the parton and of the hadron.
In general, the polarization of the quark is determined by the gamma matrix used in~(\ref{e:corr_quark_trace}).
To be specific, one has the following list:
\begin{itemize}
\item $\Gamma = \gamma^-$: unpolarized quarks
\item $\Gamma = \gamma^- \gamma_5$: longitudinally polarized quarks
\item $\Gamma = i \sigma^{i-} \gamma_5 $: transversely polarized quarks
\end{itemize}
In the case of gluons, the projections in Eqs.~(\ref{e:D1_gluon_z}),~(\ref{e:G1_gluon_z}),~(\ref{e:H1_gluon_z}) correspond to unpolarized gluons, circularly polarized gluons, and linearly polarized gluons, respectively.
This information about parton polarization, combined with the hadron polarization that shows up on the r.h.s.~of Eqs.~(\ref{e:D1_quark_kT})-(\ref{e:H1_quark_kT}) and~(\ref{e:D1_gluon_kT})-(\ref{e:H1_gluon_kT}), one immediately finds the interpretation of the TMD FFs which is summarized in Tab.~\ref{t:quark_TMDs} and Tab.~\ref{t:gluon_TMDs}.
Here one has to keep in mind that if polarization of the parton and/or hadron is involved the respective FF actually corresponds to differences of densities, and therefore it can become negative.
To give just one example, the entire r.h.s.~of Eq.~(\ref{e:D1_quark_kT}) describes the density of transversely polarized spin-1/2 hadrons inside an unpolarized quark, while $D_{1T}^\perp$ describes the difference of two densities with opposite spin orientations of the hadron.
Note that the probability densities also contain the prefactors of the TMD FFs in~(\ref{e:D1_quark_kT})-(\ref{e:H1_quark_kT}) and~(\ref{e:D1_gluon_kT})-(\ref{e:H1_gluon_kT}).

Though three-parton FFs are not probability densities one can still elaborate on their interpretation, at least for some of these functions.
Here we just focus on the function $\hat{H}_{FU}^{h/q, \, \Im}$ discussed in Sec.~\ref{sec:definition_3parton}.
We consider the component $i$ of the average transverse momentum of an unpolarized hadron inside a transversely polarized quark with polarization along the $j$ direction.
In a model-independent way this is given by
\begin{eqnarray}
\langle P_{hT}^i(z)\rangle_{UT}^j & = &
\sum_{S_h} \int d^2\vec{P}_{hT} \, P_{hT}^i \, \Delta^{h/q \, [i \sigma^{j-} \gamma_5]}(z,\vec{k}_T; P_h, S_h)
\nonumber \\
& = & - \, \varepsilon_T^{ij} \, z M_h \, H_1^{\perp (1) \, h/q}(z)
\end{eqnarray}
where we use the correlator from Eq.~(\ref{e:H1_quark_kT}) and $H_1^{\perp (1) \, h/q}(z)$ as defined in~(\ref{e:kT_moment}). 
If one now takes into account the relation in Eq.~(\ref{e:H1perp_twist3}) one obtains a connection between that average transverse momentum and the three-parton FF $\hat{H}_{FU}^{h/q,\, \Im}$.
This connection is similar to relations one has on the side of PDFs.
There, for instance, the average transverse momentum of an unpolarized quark in a transversely polarized spin-$\frac{1}{2}$ target is proportional to the three-parton PDF $T_F$, which is known in the literature as Qiu-Sterman function~\cite{Qiu:1991pp,Qiu:1991wg}.
This relation is also fully compatible with the analysis of the transverse color-force acting on quarks inside a hadron as presented in Ref.~\cite{Burkardt:2008ps}.

Like in the case of fragmentation into a single hadron, the DiFF $D_1^{h_1 h_2 / q}(z_1,z_2)$ is a measure for the probability to find, in an unpolarized quark, the two hadrons $h_1$ and $h_2$ with momentum fractions $z_1$ and $z_2$, respectively.
The corresponding interpretation holds for the gluon DiFF $D_1^{h_1 h_2 / g}(z_1, z_2)$.
Following the discussion for single-hadron FFs one can generalize the interpretation of DiFFs to include parton polarization and a more detailed kinematics of the two hadrons.
If one keeps the transverse momentum $k_T$ the probability density for finding two unpolarized hadrons in a transversely polarized target, according to~(\ref{e:H1_quark_di-hadron_kT}), is not given by just a single DiFF but rather contains the two functions $H_1^{\open \, h_1 h_2/q}$ and $H_1^{\perp \, h_1 h_2/q}$.

It is also worthwhile to mention the relation between DiFFs and so-called jet handedness observables that have been discussed as a tool to measure the polarization of partons~\cite{Nachtmann:1977ek,Efremov:1992pe,Ryskin:1993hu}.
The DiFF $G_1^\perp$, describing the fragmentation of a longitudinally polarized quark into two hadrons, is related to the longitudinal jet handedness, while $H_1^\open$ is related to the transverse jet handedness~\cite{Efremov:1992pe,Boer:2003ya}.
More information on this point can be found in Refs.~\cite{Boer:2003ya,Pisano:2015wnq}.

Some of the DiFFs are nonzero only if there is an interference between at least two different contributions to the amplitude describing the fragmentation process~\cite{Collins:1994ax,Jaffe:1997hf,Jaffe:1997pv,Radici:2001na}.
In the literature they are therefore often denoted as interference fragmentation functions (IFFs)~\cite{Jaffe:1997hf,Jaffe:1997pv}.
Presently, the best studied IFF is $H_1^{\open \, h_1 h_2/q}$.
We point out that the need for an interference between different fragmentation amplitudes is equivalent to what is required for na\"ive T-odd single hadron FFs, as briefly discussed in the paragraph after Eq.~(\ref{e:H3_quark_z}).
As a consequence, for example the Collins function $H_1^{\perp \, h/q}$ might also be called IFF, which however is uncommon.
The nature of $H_1^{\open \, h_1 h_2/q}$ as interference effect is also nicely revealed after a partial wave expansion~\cite{Jaffe:1997hf}.
Such an expansion, which is performed in the {\it cm} frame of the two hadrons~\cite{Bacchetta:2002ux,Bacchetta:2003vn}, is meaningful if $M_h$ is not very large and only few partial waves are needed.
Considering merely $s$-wave and $p$-wave contributions the following expansion has been given~\cite{Bacchetta:2002ux}:
\begin{equation} \label{e:H1open_partial_wave}
H_1^{\open \, h_1 h_2/q}(z,\zeta,M_h^2) \, \sim \, 
H_{1,sp}^{\open \, h_1 h_2/q}(z,M_h) \sin\vartheta + 
H_{1,pp}^{\open \, h_1 h_2/q}(z,M_h) \sin\vartheta \cos\vartheta \,.
\end{equation}
Here the angle $\vartheta$ describes the orientation of $\vec{R}_T$ in the two-hadron {\it cm} frame.
In that frame the variable $\zeta$ is just a linear polynomial in $\cos \vartheta$~\cite{Bacchetta:2002ux}.
After integration upon $\cos \vartheta$ only the $s$-$p$ interference term $H_{1,sp}^{\open \, h_1 h_2/q}$ survives~\cite{Bacchetta:2002ux,Pisano:2015wnq}.
It is precisely this interference term that has been extracted from experimental data~\cite{Courtoy:2012ry,Radici:2015mwa}.
Expansions similar to~(\ref{e:H1open_partial_wave}) exist for the other DiFFs~\cite{Bacchetta:2002ux,Bacchetta:2003vn}.

\subsection{Positivity bounds}
\label{subs:bounds}
Positivity bounds are important model-independent constraints for both PDFs and FFs. 
They can be derived, for instance, by using the analogy between these partonic functions and parton-hadron scattering amplitudes~\cite{Soffer:1994ww,Goldstein:1995ek}.
The simplest bounds for the integrated twist-2 quark FFs are
\begin{equation} \label{e:bound_trivial}
\big| G_1^{h/q}(z) \big| \le D_1^{h/q}(z) \,, \qquad \big| H_1^{h/q}(z) \big| \le D_1^{h/q}(z) \,,
\end{equation}
where the first inequality also holds for the gluon FFs in Eqs.~(\ref{e:D1_gluon_z}),~(\ref{e:G1_gluon_z}).
A stronger bound involving the three leading-twist quark FFs reads
\begin{equation} \label{e:bound_soffer}
\big| H_1^{h/q}(z) \big| \le \frac{1}{2}\big( D_1^{h/q}(z) + G_1^{h/q}(z) \big) \,.
\end{equation}
This constraint was first derived for the corresponding PDFs~\cite{Soffer:1994ww} and is known as Soffer bound.
As pointed out in Ref.~\cite{Goldstein:1995ek} one cannot derive positivity bounds for higher twist FFs, which is related to the lack of a density interpretation of these functions.

It has been argued that the bounds in~(\ref{e:bound_trivial}),~(\ref{e:bound_soffer}) also hold once the transverse momentum of the hadron is taken into account~\cite{Bacchetta:1999kz,Mulders:2000sh}.
For TMD FFs one can derive a number of additional bounds because of the much richer phenomenology.
Here we just list two of them~\cite{Bacchetta:1999kz},
\begin{equation}
\frac{z \, |\vec{P}_{hT}|}{M_h} \, \big| H_{1}^{\perp \, h/q}(z, \vec{P}_{hT}^{\,2}) \big|
\le D_1^{h/q}(z, \vec{P}_{hT}^{\,2})
\,, \qquad
\frac{z \, |\vec{P}_{hT}|}{M_h} \, \big| D_{1T}^{\perp \, h/q}(z, \vec{P}_{hT}^{\,2}) \big|
\le D_1^{h/q}(z, \vec{P}_{hT}^{\,2}) \,,
\end{equation}
which have been used as constraint when fitting the Collins function $H_{1}^{\perp \, h/q}$ and the FF $D_{1T}^{\perp \, h/q}$ to data.
Several bounds for TMD FFs are relatively complicated~\cite{Bacchetta:1999kz,Mulders:2000sh} as they involve a number of different functions which currently are either unknown or just poorly constrained.
Therefore, at present, such bounds are of limited practical use.

It is a nontrivial question whether QCD evolution can spoil positivity bounds for PDFs and FFs.
This point has been studied in some detail for the Soffer bound for PDFs~\cite{Barone:1997fh,Vogelsang:1997ak}, and it was found that the bound is preserved even when considering evolution up to two loops~\cite{Vogelsang:1997ak}.
To the best of our knowledge, especially in the case of bounds for TMD FFs, the potential impact of QCD evolution has not yet been addressed in the literature.

Positivity bounds can also be derived for leading twist DiFFs.
One example is~\cite{Bacchetta:2002ux}
\begin{equation}
\frac{|\vec{R}_T|}{M_h} \, \big|  H_1^{\open \, h_1 h_2/q}(z, \zeta, M_h^2) \big| \le D_1^{h_1 h_2/q}(z, \zeta, M_h^2) \,,
\end{equation}
which is the bound that is most important for the current phenomenology of DiFFs.

\subsection{Constraints from charge conjugation and isospin symmetry}
\label{subs:constraints}
Charge conjugation symmetry and isospin symmetry allow one to drastically reduce the number of independent FFs.
Here we will explicitly discuss just the most important situation, namely the fragmentation of up quarks, down quarks, and gluons into pions.
Using charge conjugation one finds the following exact relations,
\begin{equation} \label{e:charge}
D_1^{\pi^+/u} = D_1^{\pi^-/\bar{u}} \,, \quad
D_1^{\pi^+/\bar{u}} = D_1^{\pi^-/u} \,, \quad
D_1^{\pi^+/d} = D_1^{\pi^-/\bar{d}} \,, \quad
D_1^{\pi^+/\bar{d}} = D_1^{\pi^-/d} \,, \quad
D_1^{\pi^+/g} = D_1^{\pi^-/g} \,. \quad
\end{equation}
where we used that the gluon is its own antiparticle.
In addition, isospin symmetry of the strong interaction provides the two relations
\begin{equation} \label{e:isospin}
D_1^{\pi^+/u} = D_1^{\pi^-/d} \,, \quad
D_1^{\pi^+/d} = D_1^{\pi^-/u} \,,
\end{equation}
which are only broken by (numerically small) electromagnetic effects.
We also mention that all the fits that are discussed below in Sec.~\ref{sec:global_fits} in more detail assume isopsin symmetry, with the exception of the DSS fit~\cite{deFlorian:2007aj,deFlorian:2014xna}.
However, when looking at the change in $\chi^2/{\rm d.o.f.}$ the violation of isospin breaking found in the DSS fit is not significant.
The FFs for the neutral pion are not independent but rather given by~\cite{Collins:2011zzd}
\begin{equation} \label{e:pizero_FF}
D_1^{\pi^0/q} = \frac{1}{2} \big( D_1^{\pi^+/q} + D_1^{\pi^-/q} \big) \,, 
\end{equation}
and likewise for $D_1^{\pi^0/g}$.
The set of relations in~(\ref{e:charge}),~(\ref{e:isospin}), and~(\ref{e:pizero_FF}) reduces the number of FFs from 15 to just 3 independent functions, which are the favored and disfavored FFs,
\begin{eqnarray} \label{e:fav_FF}
D_1^{\rm fav} & \!\! \equiv \!\! & D_1^{\pi^+/u} = D_1^{\pi^+/\bar{d}} = D_1^{\pi^-/d} = D_1^{\pi^-/\bar{u}} \,,
\\  \label{e:dis_FF}
D_1^{\rm dis} & \!\! \equiv \!\! & D_1^{\pi^+/d} = D_1^{\pi^+/\bar{u}} = D_1^{\pi^-/u} = D_1^{\pi^-/\bar{d}} \,,
\phantom{\frac{1}{1}}
\end{eqnarray}
as well as the gluon FF $D_1^{\pi/g}$.
The exact same constraints due to charge conjugation and isospin symmetry hold for all the other single-hadron FFs including higher twist and three-parton FFs.
This discussion may be further extended by considering strange quark fragmentation and different types of hadrons like kaons for example. 
Then one can exploit the ${\rm SU}(3)$ flavor symmetry which is of course less accurate than the isospin symmetry. 

Finally, charge conjugation and isospin symmetry can also be used to establish relations between DiFFs.
In that case one finds for instance~\cite{Jaffe:1997hf,Bacchetta:2006un,Bacchetta:2011ip}
\begin{eqnarray} \label{e:DiFF_D1_relation}
& & D_1^{\pi^+ \pi^- / u} =  D_1^{\pi^+ \pi^- / d}  = D_1^{\pi^+ \pi^- / \bar{u}} = D_1^{\pi^+ \pi^- / \bar{d}} \,,
\\  \label{e:DiFF_H1open_relation}
& & H_1^{\open \, \pi^+ \pi^- / u} = - \, H_1^{\open \, \pi^+ \pi^- / d} = - \, H_1^{\open \, \pi^+ \pi^- / \bar{u}} = H_1^{\open \, \pi^+ \pi^- / \bar{d}} \,,
\phantom{\frac{1}{1}}
\end{eqnarray}
where we just consider fragmentation into a $\pi^+ \pi^-$ pair
The minus sign in~(\ref{e:DiFF_H1open_relation}) appears since the vector $\vec{R}_T$ is reversed when interchanging the $\pi^+$ and the $\pi^-$.

\subsection{Momentum sum rules}
Momentum sum rules represent yet additional important constraints on FFs.
The best known such sum rule exists for the integrated FF $D_1^{h/q}(z)$~\cite{Collins:1981uw},
\begin{equation} \label{e:sum_rule_1}
\sum_h \sum_{S_h} \int_0^1 dz \, z \, D_1^{h/q}(z) = 1 \,,
\end{equation}
which is also valid for the gluon FF $D_1^{h/g}(z)$~\cite{Collins:1981uw,Collins:2011zzd}.
In Eq.~(\ref{e:sum_rule_1}) one sums over all hadrons arising from the fragmentation as well as the (potential) spin orientations of the hadrons.
The sum rule in~(\ref{e:sum_rule_1}) is quite intuitive as it represents conservation of the longitudinal parton momentum.
On the other hand its derivation in field theory requires some work~\cite{Collins:1981uw} (see also Ref.~\cite{Meissner:2010cc}).
Though in this article we have not discussed the difference between bare FFs and ultraviolet renormalized FFs we mention that the sum rule holds in either case.
More details on this point can be found for instance in~\cite{Collins:1981uw,Collins:2011zzd}.

Conservation of the (zero) transverse momentum of a fragmenting quark is expressed through the so-called Sch\"afer-Teryaev sum rule~\cite{Schafer:1999kn,Meissner:2010cc},
\begin{equation} \label{e:sum_rule_2}
\sum_h \sum_{S_h} \int_0^1 dz \, z \, M_h \, H_{1}^{\perp(1) \, h/q}(z) = 0 \,,
\end{equation}
where we use the $k_T$ moment of the Collins function defined in~(\ref{e:kT_moment}).
This constraint has been imposed in several extractions of the Collins function.
Because momentum sum rules for FFs involve a summation over hadron spins no such sum rule exists for any of the FFs that depend on the hadron polarization~\cite{Meissner:2010cc}.
In addition, one can show that for the gluon FF $H_{1}^{\perp \, h/g}$ there is no counterpart of the sum rule in~(\ref{e:sum_rule_2}).

The first measurement of the Collins effect for pion production in SIDIS by the HERMES Collaboration~\cite{Airapetian:2004tw} revealed that the favored and disfavored Collins functions are roughly equal in magnitude but have opposite sign,
\begin{equation} \label{e:Collins_fav_dis}
H_1^{\perp \, {\rm fav}} \approx - \, H_1^{\perp \, {\rm dis}} \,.
\end{equation}
Here we use the definition of favored and disfavored FFs as given in Eqs.~(\ref{e:fav_FF}) and~(\ref{e:dis_FF}).
The relation~(\ref{e:Collins_fav_dis}) is suggested by the Sch\"afer-Teryaev sum rule~(\ref{e:sum_rule_2}) if one assumes that the (light) quarks predominantly fragment into pions~\cite{Vogelsang:2005cs} and that the (magnitude of) favored and disfavored Collins functions have a similar $z$ dependence.
In turn, the relation~(\ref{e:Collins_fav_dis}) can be considered an empirical confirmation of the Sch\"afer-Teryaev sum rule~(\ref{e:sum_rule_2}).

We note in passing that sum rules have also been considered for the unpolarized DiFF $D_1^{h_1 h_2 / q}$.
Here we do not elaborate on this point and just refer to the literature~\cite{Konishi:1979cb,deFlorian:2003cg,Majumder:2004br}.

\subsection{Universality of FFs}
\label{sec:universality}
For leading-twist integrated FFs there is usually no longer any discussion about their universality, i.e., their process independence.
It is rather taken for granted that, for instance, $D_1^{h/q}(z)$ is the same in processes like $e^+ e^-$ annihilation, SIDIS, and hadronic collisions.
However, the question about universality of TMD FFs is nontrivial.
This point becomes immediately obvious if one recalls the universality properties of TMD PDFs, in particular the prediction that the Sivers function~\cite{Sivers:1989cc,Sivers:1990fh} and the Boer-Mulders function~\cite{Boer:1997nt} change sign between SIDIS and the Drell-Yan process~\cite{Collins:2002kn}.
In SIDIS these functions have future-pointing Wilson lines, while for Drell-Yan they have past-pointing Wilson lines.
Though one is dealing with different operator definitions in the two reactions, time-reversal in combination with the parity transformation can be used to relate the definitions, which implies a sign change in the case of na\"ive T-odd TMD PDFs~\cite{Collins:2002kn}.
TMD FFs have future-pointing Wilson lines in $e^+ e^-$ annihilation, while in SIDIS they have, {\it a priori}, past-pointing Wilson lines.
Unlike the case of parton distributions, TMD FFs in the two reactions cannot be related via time-reversal (see also the discussion after Eq.~(\ref{e:H3_quark_z})).
Therefore, initially it seemed that they are not universal and even entirely unrelated when comparing $e^+ e^-$ annihilation and SIDIS. 

The first indication that TMD FFs might nevertheless be universal came from a one-loop spectator model calculation of a certain transverse SSA~\cite{Metz:2002iz}.
That study directly implied universality of the T-odd TMD FF $D_{1T}^{\perp \, h/q}$, and the argument was also extended to the Collins function $H_{1}^{\perp \, h/q}$~\cite{Metz:2002iz}.
The calculation revealed that, due to the specific kinematics in the fragmentation process, one is not sensitive to the direction of the Wilson line.
No statement was made about the universality of T-even FFs.
Also, the model calculation in~\cite{Metz:2002iz} suggests that higher-twist FFs may be non-universal.

A more general analysis was presented in Ref.~\cite{Collins:2004nx} which applies to both T-odd and T-even FFs.
The discussion was made explicit for one-loop graphs in spectator-type models, but the arguments given in that work generalize to higher orders.
The specific aforementioned kinematics allows one to derive factorization such that FFs have future-pointing Wilson lines in both $e^+ e^-$ annihilation and SIDIS and are therefore universal~\cite{Collins:2004nx}.
Alternatively one can in a first step derive factorization with FFs in SIDIS having past-pointing Wilson lines, and then show that the difference between FFs with future-pointing and past-pointing Wilson lines vanishes~\cite{Collins_Metz:unpub}.
The latter step is trivial at one loop, but can systematically be extended to higher orders including more complicated spectator systems.

The arguments leading to universality of TMD FFs also allow one to establish universality of the so-called soft factor which describes leading-twist soft-gluon effects and plays an important role in TMD factorization --- see~\cite{Collins:2011zzd,Echevarria:2011rb} and references therein.
(When deriving TMD factorization one needs to consider leading-twist contributions from (i) hard-gluon emission, (ii) gluon emission that is collinear to the external hadrons, and (iii) soft-gluon emission.
In the case of collinear factorization soft-gluon effects cancel when adding real and virtual diagrams, but there is no such cancellation for TMD factorization.
This is part of the reason why results in collinear factorization are typically simpler than in TMD factorization.)
The soft factor is essentially a vacuum expectation value of four Wilson lines, where in the case of SIDIS two of them, {\it a priori}, are future-pointing and two are past-pointing, while for $e^+ e^-$ annihilation for instance all four Wilson lines are future-pointing. 
Time-reversal does again not allow to relate the two objects, but still one can show that they are identical~\cite{Collins:2004nx}.
A very recent explicit two-loop calculation of the soft factor~\cite{Echevarria:2015usa,Echevarria:2015byo} is fully compatible with the universality of this object.
(See Ref.~\cite{Echevarria:2016scs} for a closely related study.)
It was also argued these fixed-order results might generalize to all orders in perturbation theory~\cite{Echevarria:2012js,Echevarria:2014rua,Echevarria:2015byo}.

Further universality studies considered $k_T$ moments of TMD FFs. 
For instance one can look at the following moment of the correlator in Eq.~(\ref{e:corr_quark_kT})~\cite{Boer:2003cm,Bomhof:2006ra},
\begin{equation} \label{e:kT_moment_univ}
\int d^2\vec{k}_T \, k_T^i \, \Delta^{h/q \,[{\cal U}]}(z,\vec{k}_T;P_h,S_h) = 
{\rm universal} \;\, {\rm term} + C_F^{[{\cal U}]} \, \pi \, \Delta_F^{h/q;\,i}(z,z;P_h,S_h) \,,
\end{equation}
with the superscript $[{\cal U}]$ indicating different paths of the Wilson lines.
According to Eq.~(\ref{e:kT_moment_univ}), this moment is given by a universal term plus a path-dependent term where $C_F^{[{\cal U}]}$ is a calculable factor that gets multiplied by the so-called gluonic pole matrix element, which is the three-parton correlator in~(\ref{e:corr_qgq}) evaluated for the specific case of a vanishing (longitudinal) gluon momentum~\cite{Boer:2003cm,Bomhof:2006ra}.
(If the gluon momentum of the {\it qgq} correlator is zero one hits a pole of a parton propagator from the hard scattering of the process, which is the cause of the name ``gluonic pole matrix element".)
The l.h.s.~of~(\ref{e:kT_moment_univ}) is given by moments of TMD FFs, where for unpolarized hadrons the Collins function moment $H_1^{\perp (1) \, h/q}$ as defined in~(\ref{e:kT_moment}) shows up.
The crucial point of this discussion is that gluonic pole matrix elements for FFs vanish.
This was first shown in a lowest-order spectator model calculation~\cite{Gamberg:2008yt}, and later on in a model-independent way~\cite{Meissner:2008yf,Gamberg:2010uw} (see also the work in~\cite{Belitsky:1997ay}). 
The specific moments of TMD FFs that appear on the l.h.s.~of~(\ref{e:kT_moment_univ}) are therefore universal.
Similar to~(\ref{e:kT_moment_univ}) one may relate higher $k_T$ moments of TMD FFs to certain collinear multi-parton correlators where the number of partons increases with increasing power of $k_T$.
By means of the methods of Ref.~\cite{Meissner:2008yf} or of Ref.~\cite{Gamberg:2010uw} one finds that these multi-parton correlators vanish as well~\cite{Meissner_Metz:unpub,Gamberg:2010uw}.
The benefit of such a (formal) study is, however, not immediately obvious as higher $k_T$ moments of TMD FFs are severely plagued by UV divergences and rapidity divergences.

Several additional works confirmed the universality of TMD FFs.
In Refs.~\cite{Yuan:2007nd,Yuan:2008yv} it was shown, by analyzing Feynman graphs up to two-loop, that the Collins effect in $p^{\uparrow} p \to ({\rm jet} \, h) \, X$ is universal.
Moreover, model-independent calculations of the Collins function $H_1^{\perp \, h/q}$~\cite{Yuan:2009dw} and of the polarizing FF $D_{1T}^{\perp \, h/q}$ at large transverse parton momentum provide universal results.
We finally note that the current phenomenology, in particular for the Collins function, is compatible with universality.

\subsection{Evolution}
\label{subsec:evolution}
Because of QCD dynamics FFs depend on an additional parameter, the renormalization scale $\mu$.
In fact in the case of TMDs FFs, like for TMD PDFs, yet another parameter is needed.
So far we have neglected the dependence on those parameters, which is governed by QCD evolution equations.
Here we give a very brief account of the current status of that field.

\subsubsection{Evolution of integrated leading-twist FFs}
The general structure of the evolution equations for unpolarized twist-2 integrated FFs is given by
\begin{equation} \label{e:FF_evol}
\frac{d}{d \ln \mu^2} D_1^{h/i}(z,\mu^2) 
= \frac{\alpha_s(\mu^2)}{2\pi} \sum_j \int_z^1 \, \frac{du}{u} \, P_{ji}(u, \alpha_s(\mu^2)) \, D_1^{h/j}\Big( \frac{z}{u},\mu^2 \Big) \,,
\end{equation}
which is basically identical with the form of the evolution equations for PDFs.
One just has to keep in mind that the matrix for the time-like splitting functions in~(\ref{e:FF_evol}) is $P_{ji}$, as opposed to $P_{ij}$ in the case of PDFs.
Usually the system of evolution equations in~(\ref{e:FF_evol}) is decomposed into the flavor non-singlet and the flavor singlet sectors.
The splitting functions $P_{ji}$ have a perturbative expansion of the form
\begin{equation} \label{e:splitting}
P_{ji}(u,\alpha_s(\mu^2)) = P_{ji}^{(0)}(u) 
+ \frac{\alpha_s(\mu^2)}{2\pi} \, P_{ji}^{(1)}(u)
+ \Big( \frac{\alpha_s(\mu^2)}{2\pi} \Big)^2 P_{ji}^{(2)}(u) + \ldots \,.
\end{equation}
The LO order time-like splitting functions $P_{ji}^{(0)}$ were computed in~\cite{Owens:1978qz,Uematsu:1978yw,Georgi:1977mg}. 
They agree with the well-known LO space-like DGLAP splitting functions~\cite{Gribov:1972ri,Gribov:1972rt,Altarelli:1977zs,Dokshitzer:1977sg}, which is known in the literature as Gribov-Lipatov relation~\cite{Gribov:1972ri,Gribov:1972rt}. 
This relation can also be traced back to the so-called Drell-Levy-Yan relation between structure functions in DIS and in $e^+ e^- \to h X$~\cite{Drell:1969jm,Drell:1969wd,Drell:1969wb}.
In Ref.~\cite{Blumlein:2000wh} this point has been discussed in some detail. 
The NLO splitting functions $P_{ji}^{(1)}$ were computed in~\cite{Curci:1980uw,Furmanski:1980cm,Floratos:1981hs,Kalinowski:1980ju,Kalinowski:1980wea,Stratmann:1996hn}.
Though they differ from their space-like counterparts one can still relate them by a suitable analytical continuation~\cite{Curci:1980uw,Furmanski:1980cm,Floratos:1981hs,Stratmann:1996hn,Blumlein:2000wh,Dokshitzer:2005bf}.
In the meantime even the NNLO time-like splitting functions have been studied.
Specifically, the function $P_{qq}^{(2)}$ that is needed for the non-singlet evolution was computed in~\cite{Mitov:2006ic}, while $P_{gg}^{(2)}$ was obtained in~\cite{Moch:2007tx}.
The off-diagonal splitting functions $P_{qg}^{(2)}$ and $P_{gq}^{(2)}$ were derived in~\cite{Almasy:2011eq} by making also use of the momentum sum rule in~(\ref{e:sum_rule_1}) and the super-symmetric limit.
An uncertainty still exists for $P_{qg}^{(2)}$ which, however, is not very important numerically~\cite{Almasy:2011eq}.
We also mention that extensive numerical studies of NNLO time-like evolution effetcs were carried out recently in Ref.~\cite{Bertone:2015cwa}.

The structure of the evolution equations for the polarized twist-2 FFs $G_1^{h/i}$ and $H_1^{h/i}$ agrees with equation~(\ref{e:FF_evol}), where for $H_1^{h/i}$ there is no mixing with gluons due to the chiral-odd nature of that function.
The splitting functions for $G_1^{h/i}$ are known up to NLO accuracy~\cite{Stratmann:1996hn}.
In the case of $H_1^{h/i}$ NLO splitting functions do not yet exist, but one should be able to derive them from their space-like counterparts~\cite{Vogelsang:1997ak}.

\subsubsection{Evolution of TMD FFs}
\label{sec:tmd_evolution}
As already mentioned in Sec.~\ref{sec:definition_TMD}, strictly speaking TMD FFs as given by the correlator~(\ref{e:corr_quark_kT}) are undefined as they contain light-cone singularities.
This problem can be solved by including in the definition a soft factor --- basically the same object we talked about above in Sec.~\ref{sec:universality}.
Here we will not present any equations for such new definitions and just refer to the literature --- see~\cite{Collins:2003fm,Ji:2004wu,Cherednikov:2009wk,Collins:2011zzd,Aybat:2011zv,Echevarria:2011rb,Echevarria:2012pw} and references therein.
We merely mention that the soft factor alone contains a light-cone singularity which, in the mentioned modified TMD definition, just cancels the corresponding singularity of the correlator in~(\ref{e:corr_quark_kT}).
It is important that the presence of the soft factor in the definition of TMD FFs (and TMD PDFs) is actually well motivated from a different perspective. 
Cross sections for processes that are sensitive to transverse parton momenta typically also contain uncancelled leading-twist contributions from soft-gluon emission --- see also the related discussion in Sec.~\ref{sec:universality}.
In a proper definition of TMD correlators the soft factor enters in such a way that there is no double-counting of soft-gluon contributions. 
Therefore, this factor cures two problems at the same time.

The inclusion of the soft factor implies that TMD FFs depend on one additional variable, which is often denoted as $\zeta$~\cite{Collins:1981uk, Collins:2011zzd}. 
Let us add a couple of details on this point.
{\it A priori}, all the Wilson lines that appear in the soft factor are light-like, but this leads to an additional divergence, which can be regularized by taking one of them somewhat off the light-cone.
The scale $\zeta$ is directly related to the direction (rapidity) of that Wilson line.
The resulting $\zeta$ dependence is governed by an evolution equation that is typically written for the Fourier transform of the FFs~\cite{Collins:1981uk},
\begin{equation} \label{e:CS_evolution}
\frac{d}{d \ln \sqrt{\zeta}} \, D_1^{h/q}(z,b_T,\zeta,\mu) = K(b_T,\mu) \, D_1^{h/q}(z,b_T,\zeta,\mu) \,,
\end{equation}
where the variable $b_T$ is Fourier-conjugate to $k_T$ and represents the transverse distance between the two quark fields in~(\ref{e:corr_quark_kT}).
(In most of the literature on TMD evolution that distance is denoted by $b_T$ rather than $\xi_T$ used in~(\ref{e:corr_quark_kT}).
Therefore we adapt this convention here.)
The TMD evolution equation does not show a mixing between a certain quark type and other quark flavors or  gluons.
(The l.h.s.~of~(\ref{e:CS_evolution}) is basically just a derivative with respect to the direction of a Wilson line, which does not introduce mixing.)
This is different from the more familiar case of DGLAP evolution in Eq.~(93) for forward FFs.
However, mixing between different partons comes in when deriving final solutions for the TMD evolution equation.
Here one has to keep in mind that for small distances $b_T$ the TMDs can be computed perturbatively, which introduces parton mixing.
In order to maximize the information one can obtain from pQCD one includes such small $b_T$ results in the solution for TMDs.
The large $b_T$ part of TMDs has to be fitted to data.
Let us also add a few words on the scale $\mu$ as it appears in DGLAP evolution and on the scale $\zeta$.
While in a first place $\mu$ is arbitrary, it must be chosen of the same order as $Q$, with $Q$ indicating here the large scale of a physical process.
Otherwise one obtains large logarithms that spoil the convergence of the perturbative expansion.
For instance at one-loop one finds terms of the type $\alpha_{\rm s} \ln \frac{\mu}{Q}$ which would become large once $\mu$ and $Q$ are very different even if $\alpha_{\rm s}$ were small.
The same reasoning applies to $\zeta$, for which one typically picks $\sqrt{\zeta} \sim Q$~\cite{Collins:2011zzd,Aybat:2011zv}
in order to have a well-behaved perturbation series.

The $\mu$ dependence of TMDs follows from a renormalization group equation, 
\begin{equation} \label{e:mu_evolution}
\frac{d}{d \ln \mu} \, D_1^{h/q}(z,b_T,\zeta,\mu) = \bigg( \gamma_D(\alpha_s(\mu)) - \gamma_K(\alpha_s(\mu)) \, \ln \frac{\sqrt{\zeta}}{\mu} \bigg) D_1^{h/q}(z,b_T,\zeta,\mu) \,,
\end{equation}
which differs from DGLAP evolution for integrated FFs.
Expressions for the quantity $K(b_T,\mu)$ as well as the anomalous dimensions $\gamma_D$ and $\gamma_K$ can be found, for instance, in~\cite{Collins:2011zzd,Aybat:2011zv}.
The Fourier transforms of all three correlators in Eqs.~(\ref{e:D1_quark_kT})-(\ref{e:H1_quark_kT}) obey the evolution equations~(\ref{e:CS_evolution}) and~(\ref{e:mu_evolution}).
The equations are solved in $b_T$ space, and then one transforms back to $k_T$ space.
Numerically the evolution of the various TMD FFs differs, not the least because they enter in~(\ref{e:D1_quark_kT})-(\ref{e:H1_quark_kT}) with different prefactors in $k_T$.
For gluon TMD FFs one has the same structure of the evolution equations, but the evolution kernels are different.
\begin{figure}[t]
\begin{center}
\includegraphics[width=0.48\textwidth]{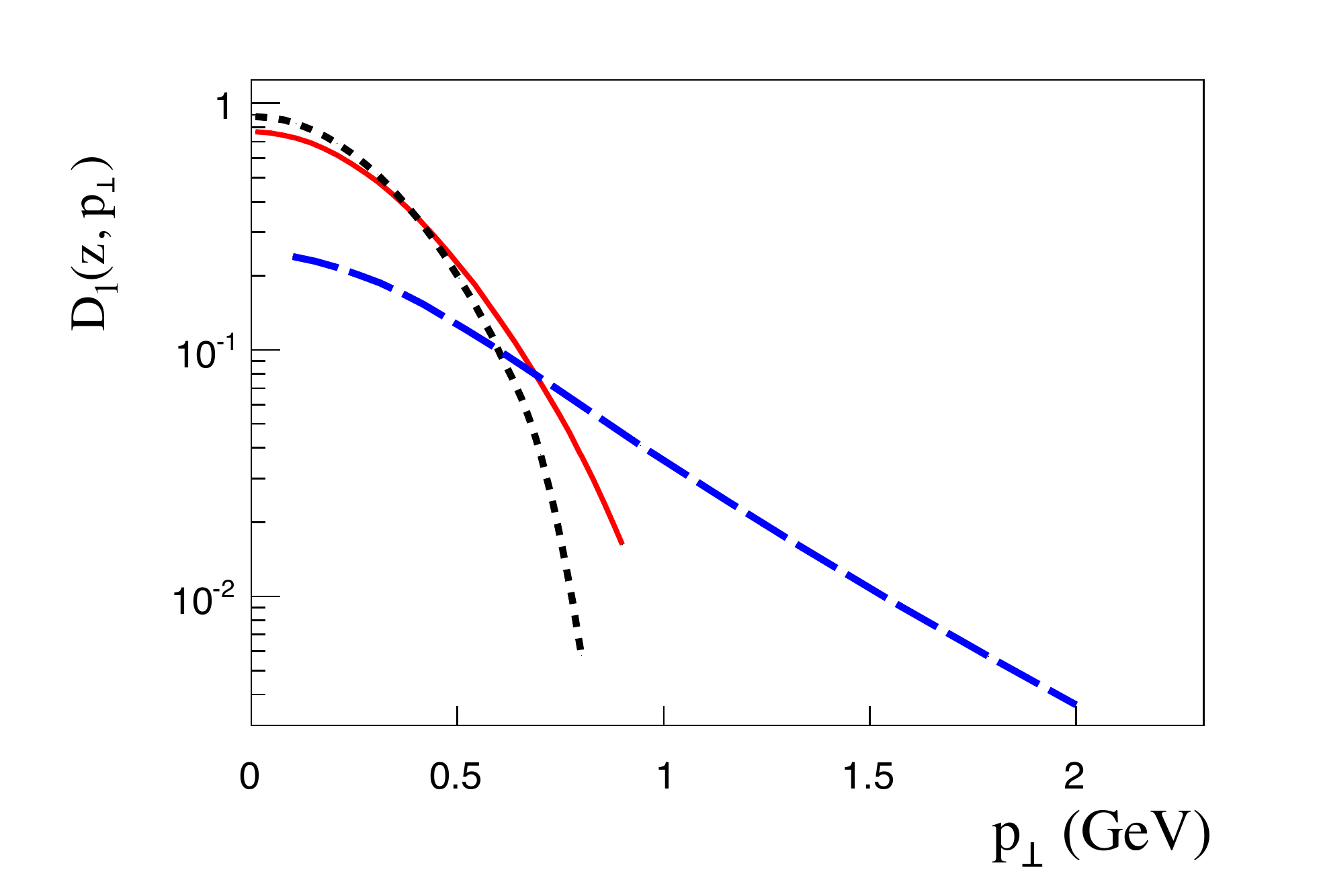} 
\includegraphics[width=0.48\textwidth]{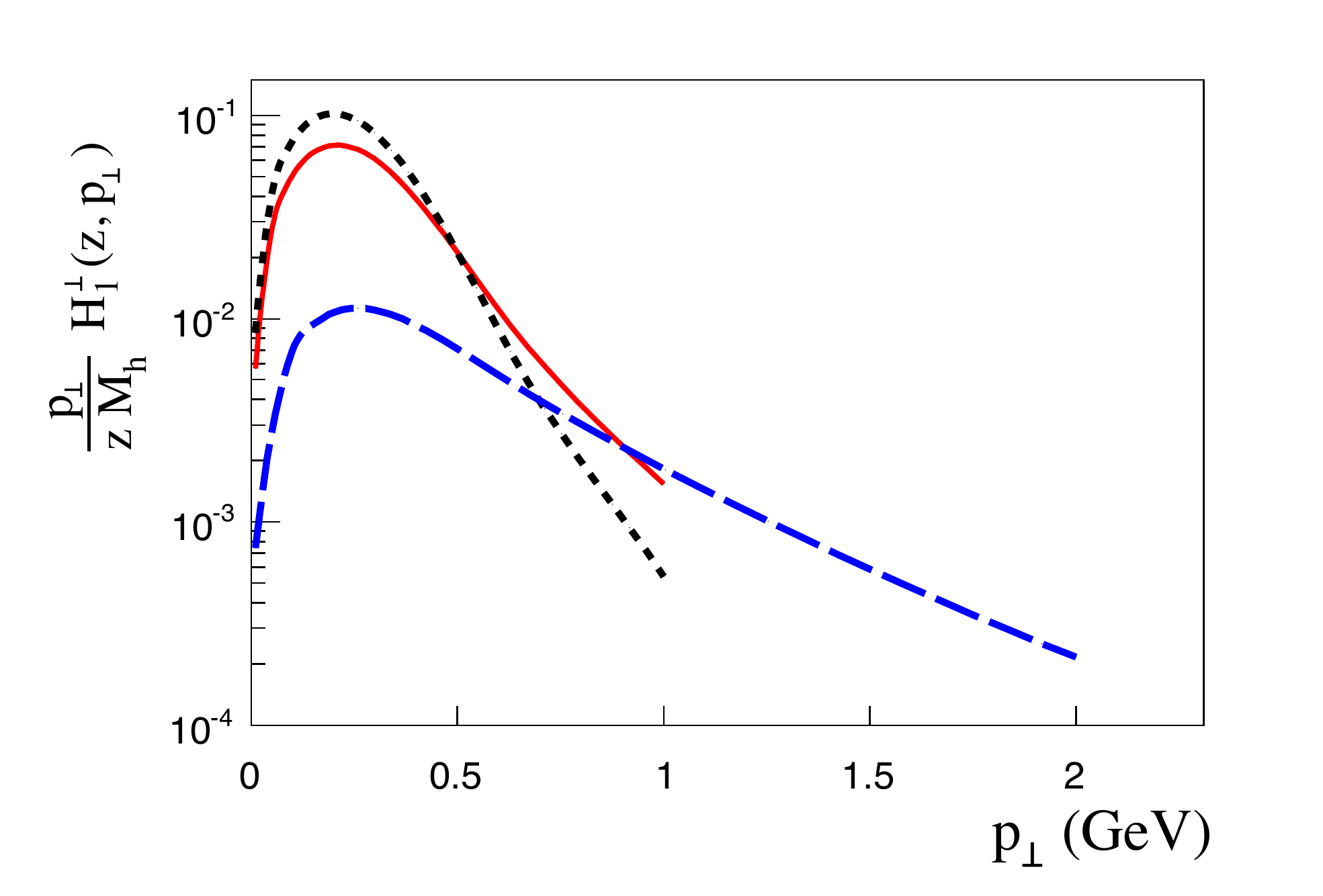}
\end{center}
\vspace{-0.4cm}
\caption{TMD evolution of unpolarized FF and Collins function.
Shown are $D_1^{u/p}(z,z^2 \vec{k}_T^{\, 2})$ (left panel) and $-\frac{1}{z} H_1^{\perp \, u/p}(z,z^2 \vec{k}_T^{\, 2})$ (right panel) in the notation of the present work. Note that $p_\perp |_{\rm plots} = |\vec{k}_T|$. Choosing $Q^2 = \zeta = \mu^2$ the functions are displayed at three different scales: $Q^2 = 2.4 \,\textrm{GeV}^2$ (dotted lines), $Q^2 = 10 \,\textrm{GeV}^2$ (solid lines), $Q^2 = 1000 \,\textrm{GeV}^2$ (dashed lines). 
Figures reprinted with permission from~\cite{Kang:2015msa}~\href{http://dx.doi.org/10.1103/PhysRevD.93.014009}{Z.-B.~Kang, et al., Phys.~Rev.~D93~(2016)~014009}. Copyright (2016) by the American Physical Society.}
\label{f:TMD_evolution}
\end{figure}

Evolution of TMD FFs has been studied in a number of recent works~\cite{Collins:2011zzd,Aybat:2011zv,Anselmino:2012aa,Sun:2013dya,Sun:2013hua,Anselmino:2013lza,Aidala:2014hva,Echevarria:2014xaa,Echevarria:2014rua,Kang:2014zza,Boglione:2014oea,Kang:2015msa,Bacchetta:2015ora,Li:2016ctv,Echevarria:2016scs}.
Since the Fourier transform from $b_T$ space to $k_T$ space also involves the region of large distances, the final result of the evolution depends on non-perturbative physics which needs to be parameterized.
At present the largest numerical uncertainty for TMD evolution is related to the freedom of modelling the non-perturbative part of TMDs in $b_T$ space --- see~\cite{Landry:2002ix,Kang:2011mr,Sun:2013hua,Echevarria:2014xaa,Aidala:2014hva,Sun:2014wpa,Collins:2014jpa} and references therein.
In Fig.~\ref{f:TMD_evolution} TMD evolution is shown for the favored unpolarized TMD FF and the Collins function.
The general feature of TMD evolution is that TMD distributions get broadened as one goes to larger scales.
This holds for all the different studies in the literature, but the magnitude of this broadening can differ considerably.

Some of the recent works related to TMD evolution used the so-called Collins-Soper-Sterman (CSS) formalism~\cite{Collins:1984kg}.
The CSS approach is largely equivalent to the TMD approach, meaning that in either case effects due to soft gluon emission are resummed to all orders.
It is beyond the scope of this review to further elaborate on this point, but more details can be found in Refs.~\cite{Bacchetta:2008xw,Collins:2011zzd} for instance.

\subsubsection{Evolution of higher-twist FFs}
There exists a considerable amount of recent work on the evolution of twist-3 PDFs --- see~\cite{Kang:2008ey,Zhou:2008mz,Vogelsang:2009pj,Braun:2009mi,Ma:2011nd,Schafer:2012ra,Ma:2012xn,Kang:2012em,Schafer:2013wca,Schafer:2013opa,Ratcliffe:2014nla,Dai:2014ala,Zhou:2015lxa} and references therein for earlier papers.
In comparison, the literature on the evolution of twist-3 FFs is sparse.
The scale dependence of the two-parton function $E^{h/q}$ in~(\ref{e:E_quark_z}) was studied in Ref.~\cite{Belitsky:1996hg}, while that of the functions $G_T^{h/q}$ in~(\ref{e:GT_quark_z}) and $H_L^{h/q}$ in~(\ref{e:HL_quark_z}) was given in~\cite{Belitsky:1997by}.
A summary of the main results of these two papers can be found in~\cite{Belitsky:1997ay}.
In a more recent work the evolution of the (twist-3) $k_T$ moments $H_1^{\perp (1) \, h/q}$ and $D_{1T}^{\perp (1) \, h/q}$ (see Eq.~(\ref{e:kT_moment}) for their definition) was discussed~\cite{Kang:2010xv}.
All these functions mix under evolution with three-parton FFs.
We refrain from explicitly listing any of the evolution kernels as they are quite lengthy.
Further work on the evolution of twist-3 FFs is needed in order to improve the phenomenology of certain observables like the SSA $A_N$ in hadronic collisions.

\subsubsection{Evolution of di-hadron FFs}
The first evolution equations for DiFFs were derived for the integrated functions $D_1^{h_1 h_2/q}(z_1,z_2)$~\cite{Konishi:1979cb,Sukhatme:1980vs}.
For a parton $i$ the LO evolution of these objects is given by~\cite{Konishi:1979cb,Sukhatme:1980vs,deFlorian:2003cg}
\begin{eqnarray} \label{e:DiFF_evol}
\frac{d}{d \ln \mu^2} D_1^{h_1 h_2/i}(z_1,z_2,\mu^2) 
& = & \frac{\alpha_s(\mu^2)}{2\pi} \sum_j \int_{z_1 + z_2}^1 \, \frac{du}{u^2} \, P_{ji}(u) \, D_1^{h_1 h_2/j}\Big( \frac{z_1}{u},\frac{z_2}{u},\mu^2 \Big)
\nonumber \\
& & + \frac{\alpha_s(\mu^2)}{2\pi} \sum_{jk} \int_{z_1}^{1 - z_2} \, \frac{du}{u(1-u)} \, \hat{P}_{ji}^k(u) \, D_1^{h_1/j}\Big( \frac{z_1}{u},\mu^2 \Big) \, D_1^{h_2/k}\Big( \frac{z_2}{1-u},\mu^2 \Big) \,. \phantom{aaa}
\end{eqnarray}
It is remarkable that, in contrast to the evolution equation~(\ref{e:FF_evol}) for single-hadron FFs, one finds two terms on the r.h.s.~of Eq.~(\ref{e:DiFF_evol}).
The first term corresponds to the usual homogenous evolution of $D_1^{h/i}$, where the same splitting functions appear.
However, a transition from a parton into two hadrons can also arise when the parton first splits into two partons with each of them afterwards fragmenting into a single hadron.
This mechanism gives rise to the second (inhomogeneous) term in~(\ref{e:DiFF_evol}).
The object $\hat{P}_{ji}^k$ is just the contribution from real emission to the splitting function with $k$ labeling the third parton of the vertex for $i \to j \,k$.
Evolution equations for the DiFFs $D_1^{h_1 h_2/i}(z_1,z_2)$ were also discussed in Refs.~\cite{Majumder:2004wh,Majumder:2004br}, along with numerical solutions of the equations.

In Ref.~\cite{Ceccopieri:2007ip} evolution equations were considered for DiFFs that depend on the relative transverse momentum $|\vec{R}_T|$ between the two hadrons, in addition to their dependence on $z_1$ and $z_2$.
It was argued that for these objects only the homogeneous term contributes.
Specifically, it was proposed that the evolution equation for $D_1^{h_1 h_2/q}(z_1,z_2,\vec{R}_T^{\,2})$ is identical to~(\ref{e:DiFF_evol}) but one just keeps the first term on the r.h.s.~of that equation.
The evolution equation for $H_1^{\open \, h_1 h_2/q}(z_1,z_2,\vec{R}_T^2)$ would then look alike~\cite{Ceccopieri:2007ip}, where one however takes the splitting functions for the evolution of the transversity distribution~\cite{Artru:1990zv}.
This type of evolution equations was used in a number of phenomenological studies~\cite{Bacchetta:2008wb,Bacchetta:2011ip,Courtoy:2012ry,Bacchetta:2012ty,Radici:2015mwa}.

\section{Observables for Light Quark Fragmentation Functions}
\label{sec:observables}
This section gives an overview of the observables that are sensitive to the FFs defined above in Sec.~\ref{sec:definition}.
Unless stated otherwise, it is implied that the produced hadrons are unpolarized.
Relevant processes that either have been measured or for which data can be expected soon are summarized in Tab.~\ref{tbl:observables}, along with the FFs to which they are sensitive.
\LTcapwidth=\textwidth
\begin{longtable}{|c|c|c|}
\hline \multicolumn{1}{|c|}{\textbf{Process}} & \multicolumn{1}{c|}{\textbf{
Quantity}} & \multicolumn{1}{c|}{\textbf{Remarks}} \\ \hline 
\endfirsthead
\multicolumn{3}{c}%
{{\bfseries \tablename\ \thetable{} -- continued from previous page}} \\
\hline \multicolumn{1}{|c|}{\textbf{Process}} &
\multicolumn{1}{c|}{\textbf{Quantity}} &
\multicolumn{1}{c|}{\textbf{Remarks}} \\ \hline 
\endhead

\hline \multicolumn{3}{|r|}{{Continued on next page}} \\ \hline
\endfoot

\hline \hline
\endlastfoot

\hline
\multicolumn{3}{|c|}{Integrated FF $D_1(z)$}\\
\hline
$e^+ e^-\rightarrow h X $ & $\sum_q e^2_q D_{1}^{h/q}(z)$ & \\
$e^+ e^- \to h_a h_b X$ & $\sum_q e^2_q D_{1}^{h_a/q}(z_a) D_1^{h_b/\bar{q}}(z_b) + \{q\leftrightarrow \bar{q}\} $ &\begin{tabular}{@{}c@{}} back-to-back production\\ of hadron pair\end{tabular}\\
$\ell p \rightarrow \ell h X $ & $\sum_q e_q^2 f_1^{q/p}(x) D_1^{h/q}(z)$ & \\
$p p \rightarrow h X $ & $\sum_{i,j,k}  f_1^{i/p_a}(x_a) \otimes f_1^{j/p_b}(x_b) \otimes D_1^{h/k}(z)$ & cannot access $z$ \\
$p p \rightarrow \gamma h  X $ & $\sum_{i,j,k}  f_1^{i/p_a}(x_a) \otimes f_1^{j/p_b}(x_b) \otimes D_1^{h/k}(z)$ & \begin{tabular}{@{}c@{}} back-to-back production \\of hadron with direct $\gamma$ \end{tabular} \\
$p p \rightarrow (h,\textrm{jet}) X $ & $\sum_{i,j,k}  f_1^{i/p_a}(x_a) \, f_1^{j/p_b}(x_b) \, D_1^{h/k}(z)$ &\begin{tabular}{@{}c@{}} hadron in jet:\\
  \begin{tabular}{@{}c@{}} can access $z$  \end{tabular}\end{tabular}\\

\hline
\hline
\multicolumn{3}{|c|}{TMD FF $D_1(z,k_T)$}\\
\hline
$e^+ e^- \to h_a h_b X $ & 
\begin{tabular}{@{}c@{}}
$\sum_q e^2_q D_{1}^{h_a/q}(z_a,k_{aT}) \otimes D_1^{h_b/\bar{q}}(z_b,k_{bT})$ \\ $ + \{q\leftrightarrow \bar{q}\} $ 
\end{tabular}
&\begin{tabular}{@{}c@{}} back-to-back production\\ of hadron pair\end{tabular}\\
$p p \to h_a h_b X $ & \begin{tabular}{@{}c@{}}
$\sum_{i,j,k,l} f_1^{i/p_a}(x_a,p_{aT}) \otimes f_1^{j/p_b}(x_b,p_{bT})$ \\ $\otimes D_{1}^{h_a/k}(z_a,k_{aT}) \otimes D_1^{h_b/l}(z_b,k_{bT})$ \end{tabular} &\begin{tabular}{@{}c@{}} back-to-back production \\of hadron pair\end{tabular}\\
$pp \rightarrow \gamma h X $ & \begin{tabular}{@{}c@{}}
$\sum_{i,j,k} f_1^{i/p_a}(x_a,p_{aT}) \otimes f_1^{j/p_b}(x_b,p_{bT})$ \\ $\otimes D_{1}^{h/k}(z,k_T) $ \end{tabular} 
& \begin{tabular}{@{}c@{}} back-to-back production \\of hadron with direct $\gamma$ \end{tabular} \\
\begin{tabular}{@{}c@{}}$e^+e^-\rightarrow$\\$ (h, \textrm{jet/thrust axis}) X $\end{tabular} & $\sum_q e_q^2 D_{1}^{h/q}(z,k_T)$ & can access $z$, $k_T$ \\
$\ell p \rightarrow \ell h X $ & $\sum_q e_q^2 f_1^{q/p}(x,p_T) \otimes D_1^{h/q}(z,k_T)$ & \\
$pp \rightarrow (h, \textrm{jet}) X $ & $\sum_{i,j,k} f_1^{i/p_a}(x_a) \, f_1^{j/p_b}(x_b) \, D_{1}^{h/k}(z,k_T) $
 & \begin{tabular}{@{}c@{}}hadron in jet:\\ can access $z$, $k_T$\end{tabular}\\
\hline
\hline
\multicolumn{3}{|c|}{TMD FF $H_1^\perp(z,k_T)$}\\
\hline
$e^+ e^- \to h_a h_b X $ & \begin{tabular}{@{}c@{}}
$\sum_q e^2_q H_1^{\perp \, h_a/q}(z_a,k_{aT}) \otimes H_1^{\perp \, h_b/\bar{q}}(z_b,k_{bT})$ \\ $ + \{q\leftrightarrow \bar{q}\} $ 
\end{tabular} &
\begin{tabular}{@{}c@{}} back-to-back production\\ of hadron pair\end{tabular}\\
$\ell p^\uparrow \rightarrow \ell h X $ & $\sum_q e_q^2 h_1^{q/p}(x,p_T) \otimes H_1^{\perp \, h/q}(z,k_T)$ & \\
$p^\uparrow p \rightarrow (h, \textrm{jet}) X $ & $\sum_{i,j,k} h_1^{i/p_a}(x_a) \, f_1^{j/p_b}(x_b) \, H_1^{\perp \, h/k}(z,k_T) $
&\begin{tabular}{@{}c@{}} hadron in jet:\\ can access $z$, $k_T$ \end{tabular}\\
\hline
\hline
\multicolumn{3}{|c|}{Twist-3 FFs}\\
\hline
$\ell p^{\uparrow} \to h X$ & $\sum_q e_q^2 h_1^{q/p}(x) \, \{ H^{h/q}, \; H_1^{\perp (1) \, h/q}, \hat{H}_{FU}^{h/q,\,\Im} \} + \ldots$ & \\
$\vec{\ell} p^{\uparrow} \to h X$ & $\sum_q e_q^2 h_1^{q/p}(x) \, E^{h/q} + \ldots$ & \\
$p^{\uparrow} p \to h X$ & \begin{tabular}{@{}c@{}} $\sum_{i,j,k} h_1^{i/p_a}(x_a) \otimes f_1^{j/p_b}(x_b)$ \\
$\otimes \, \{ H^{h/k}, \; H_1^{\perp (1) \, h/k}, \hat{H}_{FU}^{h/k,\,\Im} \} + \ldots$ \end{tabular} & \\
\hline
\hline
\multicolumn{3}{|c|}{Di-hadron FFs}\\
\hline
$e^+e^-\rightarrow (h_1,h_2) X $ &
$\sum_q e^2_q D_{1}^{ h_1 h_2/q}(z,M_h)$ &
\begin{tabular}{@{}c@{}}
for large $M_h$ \\ also $D_1^{h/i}(z)$ contribute
\end{tabular}\\
\begin{tabular}{@{}c@{}}
$e^+e^-\rightarrow$\\ $ (h_{a1},h_{a2}) \, (h_{b1},h_{b2}) X $ \end{tabular}& 
\begin{tabular}{@{}c@{}}
$\sum_q e^2_q D_1^{h_{a1} h_{a2}/q}(z_a,M_{ha}) \, D_1^{h_{b1} h_{b2}/\bar{q}}(z_b,M_{hb}) $  \\
$+ \{q\leftrightarrow \bar{q}\}$\\
$\sum_q e^2_q H_1^{\open \, h_{a1} h_{a2}/q}(z_a,M_{ha}) \, H_1^{\open \, h_{b1} h_{b2}/\bar{q}}(z_b,M_{hb}) $  \\
$+ \{q\leftrightarrow \bar{q}\}$\\
$\sum_q e^2_q G_1^{\perp \, h_{a1} h_{a2}/q}(z_a,M_{ha}) \, G_1^{\perp \, h_{b1} h_{b2}/\bar{q}}(z_b,M_{hb}) $  \\
$+ \{q\leftrightarrow \bar{q}\}$ \end{tabular} &
\begin{tabular}{@{}c@{}} back-to-back production \\of di-hadron pair\end{tabular}\\
$\ell p \rightarrow \ell (h_1,h_2) X $ & $\sum_q e_q^2 f_1^{q/p}(x) \, D_1^{h_1,h_2/q}(z,M_h)$ & \\
$\ell p^{\uparrow} \rightarrow \ell (h_1,h_2) X $ & $\sum_q e_q^2 h_1^{q/p}(x) \, H_1^{\open \, h_1,h_2/q}(z,M_h)$ & \\
$p p \rightarrow (h_1,h_2) X $ & $\sum_{i,j,k}  f_1^{i/p_a}(x_a) \otimes f_2^{j/p_b}(x_b) \otimes D_1^{h_1 h_2/k}(z,M_h)$ & \\
$p^\uparrow p \rightarrow (h_1,h_2) X $ & $\sum_{i,j,k}  h_1^{i/p_a}(x_a) \otimes f_2^{j/p_b}(x_b) \otimes H_1^{\open \, h_1 h_2/k}(z,M_h)$ & \\
\hline
\caption{Processes and respective observables for which experimental data either exist or will be avialble soon.
We list the LO expressions for the observables in terms of FFs and PDFs, where also the unpolarized PDF $f_1^{i/p}$ and the transversity PDF $h_1^{i/p}$ of a parton $i$ in a proton show up.
For ease of notation we use a simplified notation for TMDs like $D_1^{h/i}(z,k_T)$ instead of $D_1^{h/i}(z,z^2 \vec{k}_T^{\,2})$.
Depending on the observable the symbol $\otimes$ signifies convolution of longitudinal momenta or intrinsic transverse momenta.
Some of the remarks only apply in a LO treatment.
Several of the equations are generic expressions.
In particular for {\it pp} scattering a sum over partonic channels is understood and the hard factors are left out.
For twist-3 observables we just show the twist-3 fragmentation part.} 
\label{tbl:observables}
\end{longtable}

We will first discuss the complementarity of the different experimental configurations: $e^+e^-$ annihilation, SIDIS, and {\it pp} scattering.
(Note that in the case of SIDIS we typically mention lepton-proton scattering only, even though often we also have in mind lepton-scattering off deuteron or $^3$He. 
Likewise, when we talk about {\it pp} scattering often a process like $p\bar{p}$ scattering is implied.)
Then we will define specific processes in which we can access the different FFs. 
We will start with the simplest quantity, the unpolarized integrated leading-twist single-hadron FF $D_1^{h/i}(z)$, and then treat TMDs FFs (including dependence on parton polarization), higher twist FFs, and di-hadrons FFs.

The cleanest access to FFs is through $e^+ e^-$ annihilation, where the final state partons fragment into hadrons.
Compared to SIDIS and $pp$ collision, the advantage is that the FFs are the only non-perturbative objects in the cross-section.
On the other hand, $e^+ e^-$ annihilation also has several general limitations which can be addressed by other processes. 
Here we list the most important such limitations:
\begin{itemize}
\item Extraction of flavor separated FFs, or alternatively favored and disfavored fragmentation as defined in Eq.~(\ref{e:fav_FF}) for instance, is difficult in SIA. 
\item Sensitivity to the gluon FF only comes in through higher order pQCD corrections.
\item The range of scales at which FFs are probed is very narrow and basically fixed by the {\it cm} energy of the measurement. 
In principle this can be addressed by looking at initial state radiation events, but the  limitation in the \textrm{red}{experimental} acceptance for the radiated photon makes this difficult.
\end{itemize}
Below we discuss some experimental techniques which were used to address in particular the flavor and gluon tagging challenges. 
However, none of those provides sufficient information that can be used to obtain a clean separation. 
This is where SIDIS and $pp$ experiments are needed most.
On the other hand SIDIS, and even more so $pp$ measurements, suffer from additional uncertainties due to PDFs and possible nuclear corrections in fixed target experiments. 
Yet data from these experiments are crucial to get sufficiently accurate flavor-separated FFs and gluon FFs.
While SIDIS data provides relatively clean access to FFs, in particular a direct measurement of the energy fraction $z$ of the quark carried by the hadron (in a LO analysis), and allows one to study the flavor structure of FFs with different targets, $pp$ is more challenging but is indespensable in order to measure the gluon FFs. 
Recent theoretical and experimental advances using hadron-in-jet measurements discussed below could help to reconstruct the partonic kinematics in hadronic collisions as well. 
Finally, $pp$ experiments span a range in $\sqrt{s}$ that is currently unmatched by $e^+e^-$ annihilation and SIDIS.

\subsection{Observables for integrated FF $D_1$}
\label{sec:observables_integrated}
The FF $D_1^{h/i}(z)$ enters the cross sections for SIA, SIDIS, and {\it pp} scattering.
For $e^+ e^- \to h X$ and $\ell p \to \ell h X$ the cross section can be expressed through structure functions which contain the FFs.

\subsubsection{Observables for integrated FF $D_1$ in $e^+ e^-$}
For SIA the cross-section can be written as~\cite{Altarelli:1979kv}
\begin{equation} \label{e:xs_SIA}
\frac{1}{\sigma_\textrm{tot}} \, \frac{d\sigma^{e^+e^- \to h X}}{dz} = F^h(z,Q^2) \,,
\end{equation}
where the structure function $F^h(z,Q^2)$ has the meaning of a multiplicity, that is, the number of hadrons of type $h$ per event.
The observable $z=\frac{2 \, E_h}{\sqrt{s}}$ is the hadron energy scaled to half the {\it cm} energy and $Q^2 = s$.
At NLO the total hadronic cross section in~(\ref{e:xs_SIA}) is given by $\sigma_\textrm{tot} = \frac{4\pi\alpha_{em}}{Q^2} \sum_q e_q^2 \, (1 + \frac{\alpha_s}{\pi})$.
The multiplicity $F^h$ is decomposed in terms of two structure functions $F_1$ and $F_L$,
\begin{equation}
\label{eq:SIA_Multiplicities}
F^h=\frac{1}{\sum_q e^2_q} \big( 2F_1^h(z,Q^2) + F_L^h(z,Q^2) \big) \,,
\end{equation}
which, at NLO accuracy, take the form
\begin{eqnarray}
\label{eq:SIA_F1}
2 F_1^h(z,Q^2) & = & \sum_q e^2_q \bigg( D_1^{h/q}(z,Q^2) 
+ \frac{\alpha_s(Q^2)}{2\pi} \Big( C_1^q \otimes D_1^{h/q} + C_1^g \otimes D_1^{h/g} \Big)(z,Q^2) \bigg) \,,
\\
\label{eq:SIA_FL}
F_L^h(z,Q^2) & = & \frac{\alpha_s(Q^2)}{2\pi}\sum_q e^2_q \Big( C_L^q \otimes D_1^{h/q} +C_L^g \otimes D_1^{h/g} \Big)(z,Q^2) \,.
\end{eqnarray}
The coefficient functions $C_1^i$, $C_L^i$ depend on $z$, $\alpha_s$ and the ratio $\frac{Q^2}{\mu^2}$, where $\mu$ here represents the factorization scale. 
The symbol $\otimes$ denotes convolution in longitudinal momentum fractions. 
The NLO coefficient functions can be found for example in~\cite{Kretzer:2000yf}. 
Currently they are known up to NNLO.
As the gluon FF $D_1^{h/g}$ only enters at order $\alpha_s$ its contribution is small, in particular at large $\sqrt{s}$. 
Similar to the access to gluon PDFs from scaling violations, $D_1^{h/g}$ can also be addressed via its contribution to the evolution of the FFs --- see Eq.~(\ref{e:FF_evol}). 
Given the weak (logarthmic) scale dependence one is left with large uncertainties.
Information on gluon FFs can also be extracted by considering three jet events which, however, requires a more complicated theoretical apparatus.
The other issue that one encounters when using Eq.~(\ref{eq:SIA_F1}) is that, at leading order, the object accessed is $\sum_q e_q^2 D_1^{h/q}$, i.e., the charge weighted sum of the FFs. 
In particular, all $q\bar{q}$ pairs with masses below $\sqrt{s}$ can be created. 
This means that the cross section can receive significant contributions from heavy quark production.
In the following we outline some methods that allow one to achieve, to some extent, a separation of FFs for different flavors and for which experimental results are available.
\begin{itemize}
\item The most common way to separate heavy quark fragmentation from light quark fragmentation is to tag heavy quark production by reconstructing mesons containing the respective heavy quark, such as charmed or B-mesons in the event (see. e.g.~\cite{Bevan:2014iga}). 
However, the interpretation of such a non-inclusive observable is non-trivial and care has to be taken not to bias the phase space of the FF measurement.
\item In $e^+e^-$ annihilation at $\sqrt{s} = m_Z$ it is possible to get some separation of quark and antiquark FFs by using polarized beams. 
Since the parity violating weak decay of the $Z^0$ is coupling differently to left- and right-handed quarks, quarks and antiquarks have different preferred directions leading to different angular distributions of the produced hadrons.
The SLD experiment for example claims to have achieved a quark vs antiquark purity of 73\%~\cite{Kalelkar:2000ig}.
\item Some flavor information can be gained by comparing data from $e^+e^-\rightarrow \gamma^*$ with $e^+e^-\rightarrow Z^0$ and taking advantage of the different coupling constants of the quarks to the $\gamma^*$ and the $Z^0$.
\item Another way to access the flavor dependence of FFs in $e^+e^-$ data is to use back-to-back hadron pairs in the process $e^+e^- \to h_1 h_2 X$. 
The cross-section for this process takes the schematic form~\cite{deFlorian:2003cg}
\begin{equation}
\label{eq:diHadB2B}
\sum_{i,j} \hat{\sigma}_{ij}\otimes D_1^{h_1/i} \otimes D_{1}^{h_2/j} \,,
\end{equation}
where $\hat{\sigma}_{ij}$ is the partonic cross section to produce partons $i$ and $j$, which at LO will be a $q\bar{q}$ pair.
In a global fit, using the information of different charge and flavor combinations in the final state, this observable allows one to gain information about the differences of the favored vs disfavored fragmentation process.
Equation~(\ref{eq:diHadB2B}) is only valid if the two hadrons are well separated, so e.g.~are produced in back-to-back jets. 
For a di-hadron system with a small invariant mass $M_h$, the di-hadron production is described by DiFFs~\cite{Boer:2003ya}. 
In the $M_h$ integrated cross-section the single-hadron FFs and DiFFs mix~\cite{deFlorian:2003cg}.
\end{itemize}

\subsubsection{Observables for integrated FF $D_1$ in SIDIS}
The cross section for SIDIS, written in terms of structure functions, takes on a similar form as the one for SIA in $e^+e^-$ annihilation~\cite{deFlorian:1997zj}, 
\begin{equation}
\frac{d^3 \sigma^{\ell p \to \ell h X}}{dx \, dy \, dz} =\frac{2\pi \alpha_{\rm em}^2}{Q^2} 
\bigg( \frac{1+(1-y)^2)}{y} \, 2F_1^h (x,z,Q^2) + \frac{2(1-y)}{y} \, F_L^h(x,z,Q^2) \bigg) \,.
\label{eq:SIDISInclusiveHadronXSect}
\end{equation}
With $P$ and $q$ denoting the 4-momentum of the proton and the exchanged gauge boson, respectively, we use common DIS variables: $Q^2 = - q^2$, the Bjorken scaling variable $x = \frac{Q^2}{2 P \cdot q}$, $y = \frac{P \cdot q}{P \cdot l}$ describing the momentum transfer from the initial lepton to the gauge boson, and $z = \frac{P \cdot P_h}{P \cdot q}$.  
Neglecting target mass corrections one has the well-known relation $Q^2=sxy$.
Note that the cross section in~(\ref{eq:SIDISInclusiveHadronXSect}) is integrated upon the transverse momentum $\vec{P}_{h\perp}$ of the hadron.
Below in Sec.~\ref{sec:observables_TMDFF_SIDIS} we keep the dependence on $\vec{P}_{h\perp}$ which gives sensitivity to TMD FFs.
Also, we consider hadron production in the current fragmentation region.
In an experiment this is usually ensured by a cut on the Feynman variable $x_F = \frac{P_{hL}}{2\sqrt{s}}$, which is the fractional longitudinal {\it cm} momentum of the hadron. 
Otherwise, the cross-section receives contributions from target fragmentation as well. 
Such contributions are described by fracture functions which is a different type of non-perturbative objects~\cite{Trentadue:1993ka,Anselmino:2011ss} (see also the very brief discussion in Sec.~\ref{sec:fracture_functions}).
Like in the $e^+e^-$ case described in Eqs.~(\ref{eq:SIA_F1},\ref{eq:SIA_FL}), the SIDIS structure functions can be expressed in terms of FFs.
At NLO accuracy one has
\begin{eqnarray}
2F_1^h (x,z,Q^2) & = & \sum_{q}e^2_q \bigg( f_1^{q/p} D_1^{h/q} + \frac{\alpha_s(Q^2)}{2\pi} \Big( f_1^{q/p} \otimes C_1^{qq} \otimes D_1^{h/q} 
\nonumber \\
\label{eq:SIDIS_F1}
& &  \hspace{1.0cm} + \, f_1^{q/p} \otimes C_1^{gq} \otimes D_1^{h/g} + f_1^{g/p} \otimes C_1^{qg} \otimes D_1^{h/q} \Big) \bigg] \,,
\\
\label{eq:SIDIS_FL}
F_L^h(x,z,Q^2) & = & \frac{\alpha_s(Q^2)}{2\pi} \sum_q e_q^2 \Big( f_1^{q/p} \otimes C_L^{qq} \otimes D_1^{h/q} + f_1^{q/p} \otimes C_L^{gq} \otimes D_1^{h/g} + f_1^{g/p} \otimes C_L^{qg} \otimes D_1^{h/q} \Big) \,,
\phantom{aaa}
\end{eqnarray}
where the unpolarized integrated PDFs $f_1^{i/p}$ in the proton enter in the convolutions.
The NLO coefficient functions can be found in \cite{deFlorian:1997zj}.
Similar to the SIA cross section, the gluon FF only contributes at order $\alpha_s$. 
For brevity we have omitted the arguments of the PDFs, FFs, and coefficient functions.
Just considering the charge factors, the SIDIS cross section is most sensitive to the u-quark fragmentation. 
Using an effective neutron target (deuterium or $^3$He) will enhance the sensitivity to the d-quark.
Compared to $e^+e^-$ annihilation, heavy quarks basically do not play a role in the SIDIS. 
SIDIS experiments also often report the cross section normalized to the total hadronic cross section $\sigma_\textrm{tot}$, i.e., the multiplicity mentioned above. 
The multiplicity is the more appropriate observable for experiments where the instantaneous luminosity is not precisely known, but that can efficiently trigger on hadronic events. 
This is true for many SIDIS experiments discussed in Sec.~\ref{sec:experiments}.

\subsubsection{Observables for integrated FF $D_1$ in $pp$}
\label{sec:ppScattering}
While in SIDIS one can achieve some flavor-separation by coupling the FFs to different PDFs, $pp$ scattering is needed to directly access the gluon FF.
At sufficiently high transverse momentum of the observed hadron the cross section of the inclusive production of a hadron with energy $E_h$ and momentum $\vec{P}_h$ can be written in the factorized form~\cite{Aversa:1988vb}
\begin{equation}
\label{eq:ppXSection}
\frac{E_h \, d^3\sigma^{pp \rightarrow h X}}{d^3 P_h} =\sum_{i,j,k,l} \int \frac{dx_a}{x_a} \int \frac{dx_b}{x_b} \int \frac{dz}{z^2} f_1^{i/p_a}(x_a) \, f_1^{j/p_b}(x_b) \, D_{1}^{h/k}(z) \, \hat{\sigma}^{ij \to kl} \, \delta(\hat{s} + \hat{t} + \hat{u}) \,,
\end{equation}
where we restricted ourselves to the LO expression.
Through the $\delta$-function containing the partonic Mandelstam variables one has a convolution of longitudinal momentum fractions.
The cross section for $pp \to hX$ is known at NLO accuracy~\cite{Aversa:1988vb,Jager:2002xm}, which is also state of the art in the global fits.

At the high energies reached at RHIC or the LHC for example, quark-gluon or glue-glue scattering dominates, which allows one to access the gluon FFs. 
Also, compared to SIDIS, in {\it pp} scattering there is no u-quark dominance.
On the other hand, the process $pp \to hX$ is not free from challenges.
First, the cross section contains three non-perturbative objects and, depending on the kinematical region, uncertainties from the PDFs can be sizeable. 
Second, many partonic channels contribute to the cross section already at LO.
Third, a number of available data sets are in a kinematical regime in which higher order perturbative corrections which would require resummation, and/or contain higher twist effects are large.
These points complicate a quantitative description and restrict the available datasets for most practical purposes to the ones taken at RHIC, the Tevatron, and the LHC.
Fourth, due to the convolution of longitudinal momenta in~(\ref{eq:ppXSection}), {\it pp} data do not give direct access to the $z$ dependence of FFs even when analyzed in a LO framework.

An alternative approach that provides more direct information on the $z$ dependence of FFs was recently proposed in Refs.~\cite{Procura:2009vm,Jain:2011xz,Procura:2011aq,Arleo:2013tya,Kaufmann:2015hma,Kang:2016mcy,Kang:2016ehg,Dai:2016hzf}.
Instead of the inclusive hadron production cross section, the cross section for hadrons in jets is used.
Using the narrow jet approximation, the energy of the initial parton can be accessed, and the $z$ dependence of the FFs can be measured in processes where hadrons in jets with relatively narrow cone sizes are detected~\cite{Kaufmann:2015hma}.
Following Ref.~\cite{Kaufmann:2015hma} the cross-section can schematically be written as
\begin{equation}
\label{eq:D1FromPPJets}
\frac{d^3\sigma^{pp \to (h,\textrm{jet}) X}}{dP_T^\textrm{jet} \, d\eta^\textrm{jet} \, dz} \propto \sum_{i,j,k} f_1^{i/p_a}(x_a) \otimes f_1^{j/p_b}(x_b) \otimes \hat{\sigma}^{ij \to k \textrm{jet}} \otimes D_{1}^{h/p_k}(\frac{z}{z_k}) 
\end{equation}
In this equation the hard partonic cross section $\hat{\sigma}^{ij \to k \textrm{jet}}$ describes the hard scattering of partons $i$ and $j$ to produce a parton $k$ in a jet with pseudorapidity $\hat{\eta}$. 
The partonic cross section depends on the jet definition, in addition to its dependence on kinematical factors. 
This parton $k$ is allowed at NLO to form another parton, taking $z_k$ of the energy of $p_k$, such that the observed $z = P_{hT}/P_T^\textrm{jet}$ is $z/z_k$ of the energy of the fragmenting parton. 
Independent of the complexity of the cross section at NLO, where one has to deal with parton splitting, the approach is conceptually easy to describe: by reconstructing a jet, an estimate of the momentum of the fragmenting parton is available. 
Measuring then the momentum of a hadron inside this jet, gives access to the $z$ dependence of the FFs. 
Another advantage is that fixing the $P_T^\textrm{jet}$ and its pseudorapidity, one has also some information about the fractional partonic momenta $x_a$, $x_b$.


\subsection{Transverse single-spin asymmetries}
\label{sec:TSSA}
As discussed below in more detail transverse SSAs play an important role for the measurement of TMD FFs, higher-twist FFs, and DiFFs.
Therefore we first give here a few general features of such asymmetries.
Schematically a transverse SSA is defined as
\begin{equation} \label{e:SSA_definition}
A_N = \frac{\sigma^{\uparrow} - \sigma^{\downarrow}}{\sigma^{\uparrow} + \sigma^{\downarrow}} \,.
\end{equation}
Let us first focus on the specific case of processes like $p^{\uparrow} p \to h X$, $\bar{p}^{\uparrow} p \to h X$, and $\ell p^{\uparrow} \to h X$.
These reactions have in common that just one particle is observed in the final state.
It is easy to understand why in such a case $A_N$ is necessarily related to transverse spin.
To see this let us consider for instance {\it pp} collisions, i.e., the process $p(P_a,S_a) + p(P_b) \to h(P_h) + X$, where we have indicated the 4-momenta of the particles and the 4-compenent spin vector $S_a$ of one of the protons.
Neglecting parity-violating interactions the only allowed correlation involving the spin vector $S_a$ is
\begin{equation} \label{e:SSA_correlation}
\varepsilon_{\mu\nu\rho\sigma} P_a^{\mu} P_b^{\nu} P_h^{\rho} S_a^{\sigma} \sim \vec{S}_a \cdot (\vec{P}_a \times \vec{P}_h) \,,
\end{equation}
where we use the 4-dimensional Levi-Civita tensor.
Modulo pre-factors that are irrelevant for the sake of the argument, in~(\ref{e:SSA_correlation}) we also give the result one obtains when boiling down this correlation such that only ordinary 3-vectors show up.
This expression implies that the correlation is nonzero only if the spin vector $\vec{S}_a$ is perpendicular (normal) to the reaction plane which is given by the momenta $\vec{P}_a$ and $\vec{P}_h$.
If one considers in addition parity-violating effects also a longitudinal SSA exists in the aforementioned reactions.
Moreover, parity-conserving longitudinal single-spin effects are allowed in processes with more identified particles such as SIDIS.

A second general feature of the transverse SSA defined in~(\ref{e:SSA_definition}) is the need for an imaginary part in the scattering amplitude.
To discuss this point in a bit more detail let us, for simplicity, look at elastic proton-pion scattering.
In Pauli space, the scattering amplitude of this process takes the familiar form
\begin{equation} \label{e:amplitude_ppi}
{\cal M} = \chi_f^{\dagger} \, (A + i \vec{\sigma} \cdot \vec{B}) \, \chi_i \,,
\end{equation}
with the non-flip amplitude $A$ and the spin-flip amplitude $\vec{B}$.  
In Eq.~(\ref{e:amplitude_ppi}), $\chi_i$ and $\chi_f$ denote the 2-component Pauli spinors of the nucleon in the initial and the final state, respectively.
For polarization of the incoming proton along the $\pm y$-direction, $A_N$ is given by
\begin{equation} \label{e:SSA_ppi}
A_N = \frac{{\rm Tr} \, (A + i \vec{\sigma} \cdot \vec{B}) \, \sigma_y \, (A^{\ast} - i \vec{\sigma} \cdot \vec{B}^{\ast})} {{\rm Tr} \, (A + i \vec{\sigma} \cdot \vec{B}) \, (A^{\ast} - i \vec{\sigma} \cdot \vec{B}^{\ast})}
= \frac{2 \, {\rm Im} \, (A B_y^{\ast} + B_x B_z^{\ast})}{|A|^2 + |B_x|^2 + |B_y|^2 + |B_z|^2} \,.
\end{equation}
This result shows that a nonzero $A_N$ requires the interference between different contributions to the full scattering amplitude and, in particular, an imaginary part in at least one of these contributions.
We emphasize that this result holds for any reaction, irrespective of the number of identified final-state particles.
\vspace{0.5cm}

\subsection{Observables for TMD FFs}
\label{sec:observables_TMD}
There is considerable interest beyond the integrated FFs $D_1^{h/i}(z)$ and towards exploring the dependence of FFs on the intrinsic transverse momentum acquired during the fragmentation process.  
Measuring the transverse momentum dependence of FFs provides deeper insight into the process of hadronization. 
In addition, in a process like SIDIS the transverse momentum dependence of the fragmentation process has to be known in order to extract the transverse momentum dependence of PDFs.

Getting precise information on the transverse momentum dependence of FFs from experimental data is not trivial.
The most detailled recent works in this area exclusively used data on SIDIS --- either cross sections or multiplicities that depend on the transverse momentum of the final state hadron.
However, even in a LO treatment the total hadron's transverse momentum is generated by both the transverse momentum of the active quark relative to the proton and the transverse momentum of the hadron relative to the quark.
The same complication in principle applies to the production of two back-to-back hadrons in $e^+ e^-$ annihilation, where the total transverse momentum of one hadron relative to the other is given by the transverse momentum dependence of the two involved FFs.
However, given that essentially the same TMD FF appears twice in the cross section one may get new valuable information from this observable, especially when looking at different combinations of hadrons.
At present there exists no suitable data that allow one to carry out such an analysis, but work along those lines is in progress.
Another way to address the transverse momentum dependence of FFs is by observing hadrons plus the jet from which they originate, similar to what we discussed above in Sec.~\ref{sec:ppScattering} for the integrated FF $D_1^{h/i}(z)$~\cite{Kaufmann:2015hma} --- see for instance~\cite{Jain:2011xz} for a related discussion.
In a LO calculation such an observable gives direct access to TMD FFs, while the situation is less clean once higher order corrections are included.
Further developments in this area can be expected.
On the experimental side, there has been some effort to correct measured jet quantities on partonic quantities that can be calculated by using simulations from a program like PYTHIA~\cite{Sjostrand:2006za}, but this introduces model dependence and theoretical uncertainties since it is difficult to match results from Monte-Carlo event generators onto a fixed-order calculation.

Keeping the transverse momentum dependence of the fragmentation process also gives rise to new correlations and therefore additional TMD FFs as discussed Sec.~\ref{sec:definition_TMD}.
For unpolarized hadrons there is one such function, namely the Collins function $H_1^{\perp \, h/q}$ for quarks~\cite{Collins:1992kk}. 
It can serve as ``quark polarimeter" as it measures the strength of the triple scalar product $\vec{s}_q \cdot (\vec{k} \times \vec{P}_h)$ containing the spin vector $\vec{s}_q$ of the quark.
This correlation obviously vanishes if the quark and the hadron move collinearly.
Note that only the transverse quark spin matters, that is, the component of $\vec{s}_T$ which is perpendicular to the quark momentum $\vec{k}$.
In this section, we describe observables related to the Collins FF in $e^+ e^-$ annihilation, SIDIS, and {\it pp} scattering. 
Since the Collins function is chiral-odd, in an observable it can only appear in combination with another chiral-odd non-perturbative function.
While the chiral-odd partner can be $H_1^\perp$ itself in back-to-back hadron production in $e^+e^-$ annihilation, in SIDIS and $pp$ the Collins FF appears together with chiral-odd PDFs, which themselves are only poorly known. 
Of these combinations presently the most relevant are observables in which $H_1^{\perp \, h/q}$ couples to the transversity PDF $h_1^{q/p}$.

Finally, the evolution of TMD FFs and integrated FFs are very different --- see Sec.~\ref{sec:tmd_evolution} for a brief introduction.
The quantitative treatment of TMD evolution has a strong dependence on non-perturbative physics and is an active area of research.
Observables involving TMD FFs could provide further insight into this topic.

In the rest of this section we will only present parton model results which is sufficient to show the sensitivity to the FFs.
Proofs of TMD factorization are available for $e^+ e^- \to h_a h_b X$ and SIDIS.
On the other hand, TMD factorization breaks down for back-to-back hadron production in hadronic collisions~\cite{Rogers:2010dm}, even though it is not known how important the factorization breaking is numerically.
For TMD-sensitive hadron production inside jets in {\it pp} scattering the status of TMD factorization is presently unclear.

\subsubsection{Observables for TMD FFs in $e^+e^-$}
\label{sec:ObservablesForTMDFFsInSIA}
\begin{figure}[t]
\includegraphics[width=0.49\textwidth]{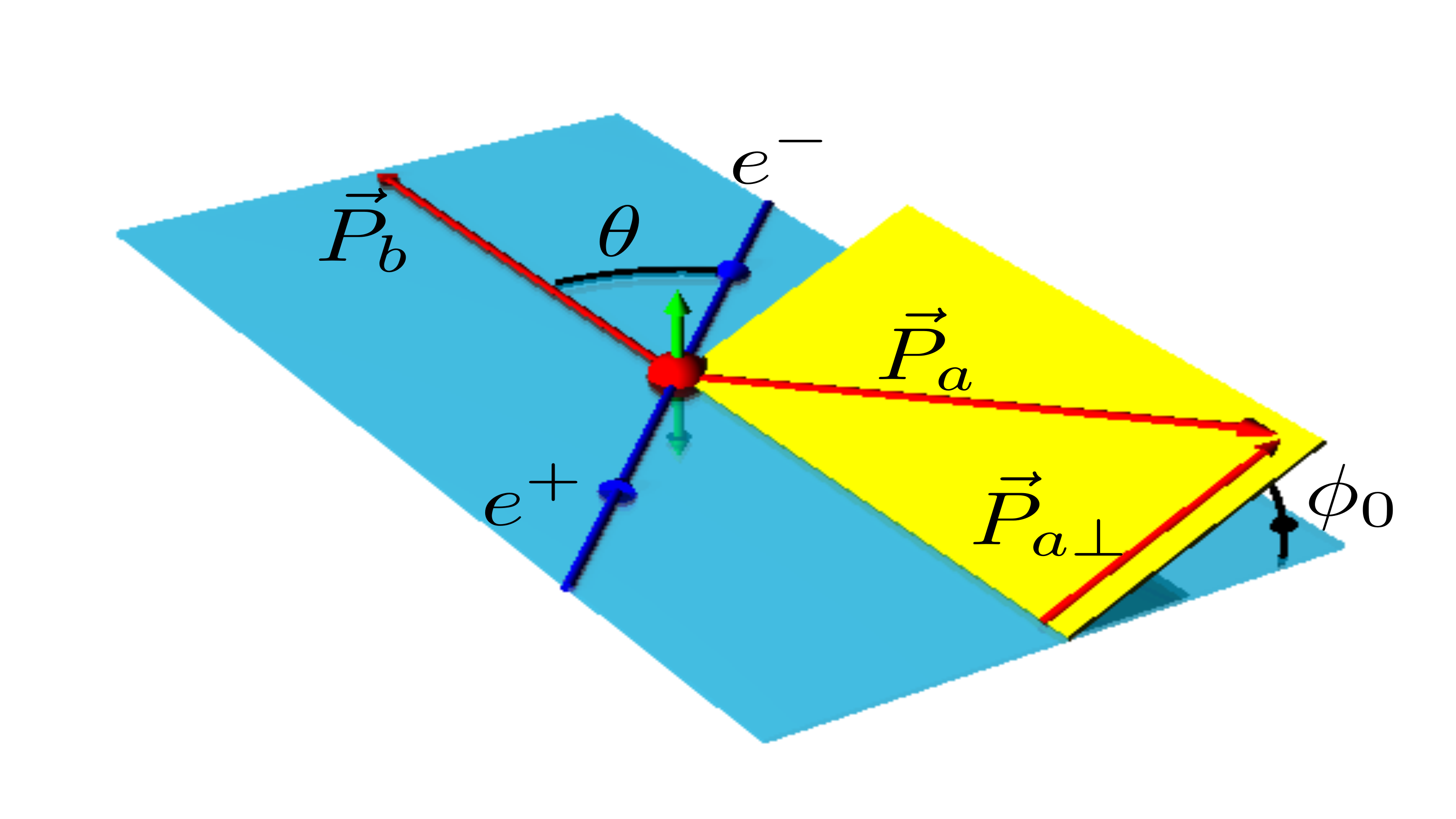}\includegraphics[width=0.49\textwidth]{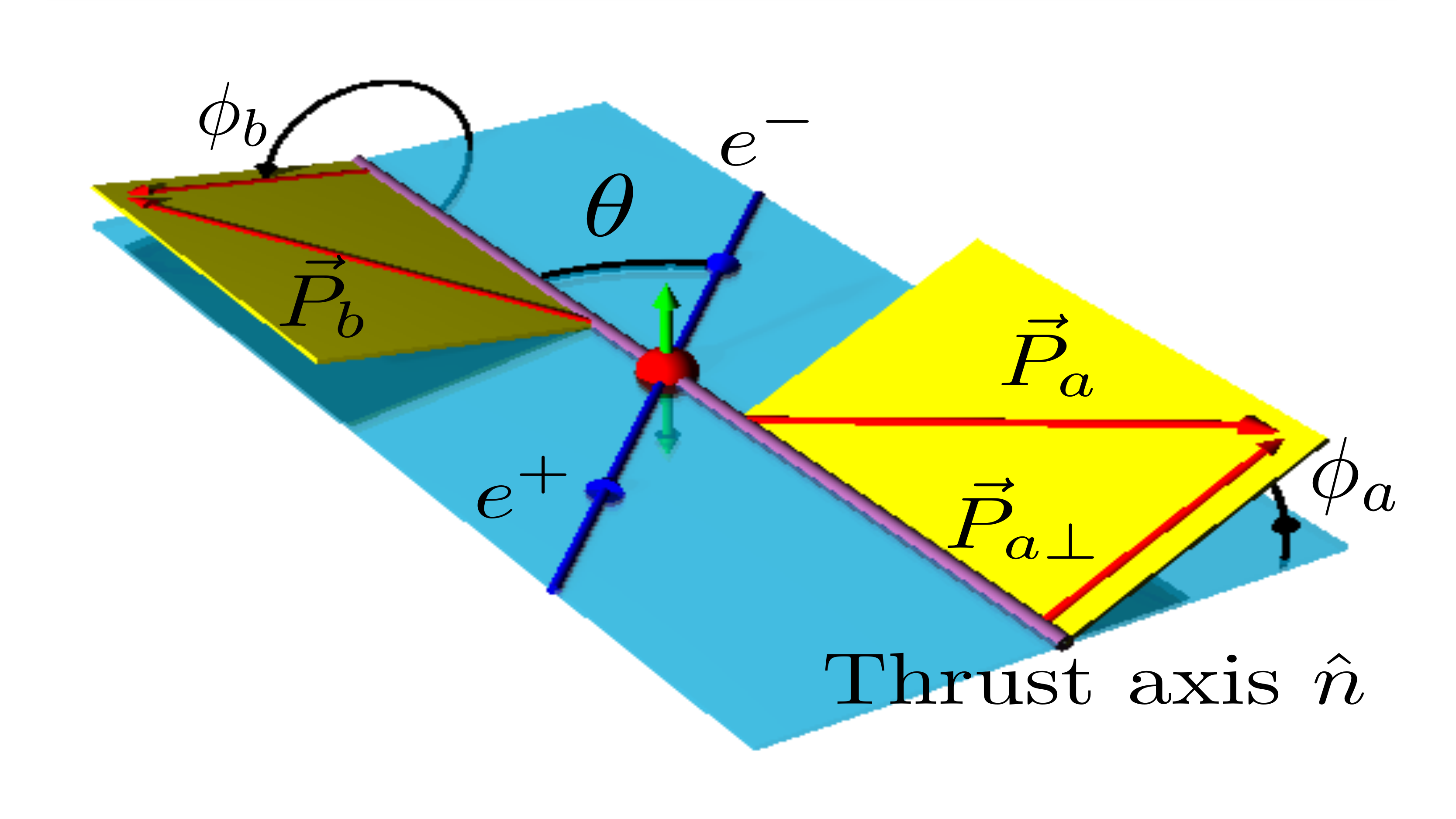}
\caption{Coordinate systems used for the measurement of back-to-back correlations of hadron pairs in $e^+ e^-$ annihilation. On the left an asymmetric coordinate system is shown with the azimuthal angle $\phi_0$. On the right, a symmetric coordinate system is shown, where the azimuthal angles $\phi_1$, $\phi_2$ are measured around the thrust axis. 
Figures adapted with permission from~\cite{Seidl:2008xc}. Copyrighted by the American Physical Society.\label{fig:SIACollinsCoo}}
\end{figure}

For $e^+ e^-$ annihilation we focus on the production of back-to-back hadrons in $e^+ e^- \to h_a h_b X$.
In the case of photon exchange the cross section for this process was worked out in Ref.~\cite{Boer:1997mf}.
For a complete discussion including electroweak and polarization effects we refer to~\cite{Pitonyak:2013dsu} and papers quoted therein.
At leading order in perturbation theory and in $1/Q$ the cross section reads~\cite{Boer:1997mf}
\begin{eqnarray}
\frac{d^6 \sigma^{e^+ e^- \to h_a h_b X}}{d\Omega \, dz_a \, dz_b \, d^2\vec{P}_{a\perp}} & = &
\frac{3 \, \alpha_{\textrm{em}}^2}{Q^2} \, z_a^2 z_b^2 \bigg( A(y) \, \mathcal{C}_{e^+ e^-} \big[ D_1 \bar{D}_1 \Big] 
\nonumber \\
& & + \, B(y) \, \cos(2\phi_0) \,
\mathcal{C}_{e^+ e^-} \bigg[\frac{2\hat{h} \cdot \vec{k}_{aT} \, \hat{h} \cdot \vec{k}_{bT} - \vec{k}_{aT} \cdot \vec{k}_{bT}}{M_a \, M_b} \, H_1^\perp \,\bar{H}_1^\perp \bigg] \bigg) \,,
\label{eq:back2backHadrons}
\end{eqnarray}
where we use the convolution integral
\begin{eqnarray} \label{e:convolution_epem}
{\cal C}_{e^+ e^-}[w D \bar{D}] & = & \sum_q e_q^2 \int d^2\vec{k}_{aT} \, d^2\vec{k}_{bT} \, \delta^{(2)}(\vec{k}_{aT} + \vec{k}_{bT} + \vec{P}_{aT}/z_a) 
\nonumber \\
& & \hspace{1.0cm} \times \, w(\vec{k}_{aT},\vec{k}_{bT}) \, D^{h_a/q}(z_a,z_a^2 \vec{k}_{aT}^{\,2}) \, D^{h_b/\bar{q}}(z_b,z_b^2 \vec{k}_{bT}^{\,2}) 
+ \{ q \leftrightarrow \bar{q} \}\,.
\end{eqnarray}
We are working in the leptonic {\it cm} frame --- see left panel of Fig.~\ref{fig:SIACollinsCoo}.
The momentum $\vec{P}_b$ of hadron $h_b$ is chosen as the axis around which the azimuthal angle $\phi_0$ of the plane spanned by the $e^+e^-$ axis and $\vec{P}_a$ is measured. 
(This reference frame is obviously asymmetric with regard to the two hadrons.
However, a similar symmetric frame can also be used for this process~\cite{Pitonyak:2013dsu}.)
In this frame, the relative transverse momentum of the two hadrons is just $\vec{P}_{a\perp}$ and, according to the convolution integral in~(\ref{e:convolution_epem}), given by the transverse momenta associated with the fragmentation of both the quark and the antiquark.
In Eq.~(\ref{eq:back2backHadrons}) also $\hat{h} = \vec{P}_{a\perp}/|\vec{P}_{a\perp}|$ is used.
The $A(y),B(y)$ are kinematic factors that are defined as 
\begin{equation} \label{e:y_dependence_epem}
A(y) = \frac{1}{2} - y + y^2 \,, \qquad \qquad B(y) = y(1-y) \,,
\end{equation}
where $y=(1 + \cos\theta)/2$, and $\theta$ is the angle between $\vec{P}_b$ and the lepton axis.

Concentrating first on $D_1$, we see that one can extract $\mathcal{C}_{e^+ e^-} [ D_1 \, \bar{D}_1 ]$ by integrating over $\phi_0$.
As already mentioned the downside of the back-to-back measurement is the convolution of the transverse momenta.
However, a fine binning in $z$ should suffice for the extraction of the transverse momentum dependence of the individual FFs.
The advantage of measuring $\sum_q e_q^2 ( D_1^{h_a/q} \otimes D_1^{h_b/\bar{q}} + D_1^{h_a/\bar{q}} \otimes D_1^{h_b/q})$ is that a certain amount of flavor information can be extracted from different combinations of hadrons in the final state.

The second term in~Eq.~(\ref{eq:back2backHadrons}), which can be extracted from the amplitude of the $\cos(\phi_0)$ modulation of the cross-section, is sensitive to the Collins FF $H_1^{\perp \, h/q}$.
Because this function is chiral-odd, it cannot be measured in SIA.
However, the correlation of the spins of the $q\bar{q}$ pair produced in the decay of the virtual photon leads to an azimuthal correlation of two hadrons produced in opposite hemispheres. 
The sign of the Collins function cannot be fixed using this observable. 
On the other hand, like for $D_1^{h/q}$, the freedom in choosing charge and flavor for the hadrons can be used to access different combinations of favored and disfavored FFs. 
This will be described in more detail in Sec.~\ref{sec:CFFExtraction}.
The term in Eq.~(\ref{eq:back2backHadrons}) containing the Collins function can be extracted by dividing the $\phi_0$ dependent cross section by the azimuthally averaged cross section.
Since Eq.~(\ref{eq:back2backHadrons}) defines an unpolarized cross section one cannot cancel systematic effects in the way done for spin asymmetries.
Instead experiments exploit the flavor dependence of $H_1^{\perp \, h/q}$ and use so-called double ratios between the normalized cross section of different charge combinations, which are described in Sec.~\ref{sec:ObservablesSpinDepFFs}.
The $B(y)$ term in~(\ref{eq:back2backHadrons}) implies that the term containing the Collins function is maximal if the $q \bar{q}$ axis, which approximately coincides with the thrust axis of the event, is perpendicular to the lepton axis.
This is advantageous since measurements in the central region of the detector are much easier than in the forward/backward region.

We already mentioned that one may also work in a symmetric frame for the process $e^+ e^- \to h_a h_b X$ by just detecting the two hadrons.
Another way to define a symmetric frame is by adding also the thrust axis.
Such a frame is depicted on the r.h.s.~of Fig.~\ref{fig:SIACollinsCoo}.
With this choice, and by making an appropriate weighting of the cross section it was found in Ref.~\cite{Boer:phd}:
\begin{equation}
\label{eq:SIACollinsTwoScales}
\frac{d^6\sigma^{e^+ e^- \to h_a h_b X}}{d\Omega \, dz_a \, dz_b \, d\phi_a \, d\phi_b} \propto
\sum_q e_q^2 \, B(y) \cos(\phi_a + \phi_b) \, H_1^{\perp (1) \, h_a/q}(z_a) \, H_1^{\perp (1) \, h_b/\bar{q}}(z_b) + \{ q \leftrightarrow \bar{q} \},
\end{equation}
where we only list the term containing the Collins function.
Now transverse momentum moments of the FFs appear as defined in Eq.~(\ref{e:kT_moment}).
The result in~(\ref{eq:SIACollinsTwoScales}) was obtained starting from the parton model expression for the cross section.
Presently it is not fully understood how to compute higher order corrections for such weighted observables.


We also mention that for $e^+ e^- \to h_a h_b X$ other observables involving the Collins function exist if one detects polarization of final state particles as can be done for example in the case of hyperon production~\cite{Pitonyak:2013dsu}.
At present no data exist for such observables.


\subsubsection{Observables for TMD FFs in SIDIS}
\label{sec:observables_TMDFF_SIDIS}
In SIDIS the transverse momenta of the TMD PDFs and TMD FFs contribute to the transverse momentum spectrum of final state hadrons. 
This spectrum depends on the structure function under consideration.
Taking into account lepton and target polarization, the SIDIS cross section can be decomposed in terms of 18 structure functions, which depend on the variables $x$, $y$, $z$, and $\vec{P}_{h\perp}^{\, 2}$~\cite{Diehl:2005pc,Bacchetta:2006tn}. 
We consider a reference frame in which the exchanged virtual photon and the target nucleon are collinear, and the transverse momentum $\vec{P}_{h\perp}$ of the hadron is measured with respect to the direction of the virtual photon. 
Here we only list the five structure functions that are relevant for the subsequent discussion~\cite{Bacchetta:2006tn,Diehl:2005pc},
\begin{eqnarray} \label{e:SIDIS_xs_PT}
\frac{d^6\sigma^{\ell p \to \ell h X}}{dx \, dy \, d\phi_S \, dz \, d\phi_h \, dP_{h\perp}^2} 
& = &\frac{\alpha_{\rm em}^2}{x y Q^2} \bigg[ \bigg( 1 - y + \frac{y^2}{2} \bigg) F_{UU,T}
+ (2 - y) \sqrt{1 - y} \cos \phi_h F_{UU}^{\cos \phi_h}
\nonumber \\
& & + \, (1 - y) \cos(2\phi_h) F_{UU}^{\cos 2\phi_h}
+ |\vec{S}_\perp| \, (1 - y) \sin(\phi_h + \phi_S) F_{UT}^{\sin(\phi_h + \phi_s)}
\phantom{\bigg( }
\nonumber \\
& & + \, |\vec{S}_\perp| \, (1 - y) \sin(3\phi_h - \phi_S) F_{UT}^{\sin(3\phi_h - \phi_s)} + \ldots \bigg] \,,
\end{eqnarray}
where $\phi_h$ and $\phi_S$ represent the azimuthal angle of $\vec{P}_{h\perp}$ and the transverse spin vector of the target $\vec{S}_\perp$ relative to the lepton plane, respectively.
The first subscript for the structure functions indicates an unpolarized ($U$) lepton, while the second subscript indidates either an unpolarized ($U$) or transversely-polarized ($T$) target.
In the case of $F_{UU,T}$ the additional index $T$ characterizes transverse polarization of the virtual photon.
(Note that there exists also the function $F_{UU,L}$ which, however, is a twist-4 quantity~\cite{Bacchetta:2006tn}).
In the region of small $|\vec{P}_{h\perp}|$ four of the structure functions in~(\ref{e:SIDIS_xs_PT}) receive leading twist contributions in TMD factorization.
With the definition 
\begin{equation} \label{e:convolution_SIDIS}
{\cal C}_{\rm SIDIS}[w f D] = x \sum_q e_q^2 \int d^2\vec{p}_T \, d^2\vec{k}_T \, \delta^{(2)}(\vec{p}_T - \vec{k}_T - \vec{P}_{h\perp}/z) \,
w(\vec{p}_T,\vec{k}_T) \, f^{q/p}(x,\vec{p}_T^{\;2}) \, D^{h/q}(z,z^2\vec{k}_T^{\,2})
\end{equation}
for the convolution of TMD PDFs and TMD FFs in SIDIS one finds the parton model results~\cite{Collins:1992kk,Mulders:1995dh,Bacchetta:2006tn} 
\begin{eqnarray} \label{e:F_UU_SIDIS}
F_{UU,T} & = & {\cal C}_{\rm SIDIS} \Big[f_1 D_1 \Big] \,,
\\  \label{e:F_cos2phi_SIDIS}
F_{UU}^{\cos (2\phi_h)} & = & {\cal C}_{\rm SIDIS} \bigg[ - \frac{2 \, \hat{h} \cdot \vec{k}_T \, \hat{h} \cdot \vec{p}_T - \vec{k}_T \cdot \vec{p}_T}{M \, M_h} \, h_1^\perp H_1^\perp \bigg] \,,
\\  \label{e:F_Collins_SIDIS}
F_{UT}^{\sin (\phi_h + \phi_s)} & = & {\cal C}_{\rm SIDIS} \bigg[ - \frac{\hat{h} \cdot \vec{k}_T}{M_h} \, h_1 H_1^\perp \bigg] \,,
\\  \label{e:F_pretelosity_SIDIS}
F_{UT}^{\sin (3\phi_h - \phi_s)} & = & {\cal C}_{\rm SIDIS} \bigg[ \frac{\hat{h} \cdot \vec{k}_T \, \vec{p}_T^{\;2} + 2 \, \hat{h} \cdot \vec{p}_T \, \vec{k}_T \cdot \vec{p}_T - 4 \, \hat{h} \cdot \vec{k}_T \, (\hat{h} \cdot \vec{p}_T)^2} {2 M^2 \, M_h} \, h_{1T}^\perp H_1^\perp \bigg] \,,
\end{eqnarray}
where here $\hat{h} = \vec{P}_{h\perp} / |\vec{P}_{h\perp}|$.
As discussed in Sec.~\ref{sec:data_TMD_D1} and Sec.~\ref{sec:fit_TMDFF_D1} in more detail, it is the structure function $F_{UU,T}$ in~(\ref{e:F_UU_SIDIS}) through which presently we have most of the information about the transverse momentum depedence of $D_1^{h/q}$.
In Sec.~\ref{sec:fit_TMDFF_D1} we will also elaborate more on the structure function $F_{UU}^{\cos \phi_h}$ in~(\ref{e:SIDIS_xs_PT}) and its relation to TMD FFs.
This function describes a twist-3 contribution to the SIDIS cross section.

The structure function $F_{UT}^{\sin (\phi_h + \phi_S)}$ in~(\ref{e:F_Collins_SIDIS}) describes the Collins effect in SIDIS that was proposed to get information on the twist-2 integrated transversity PDF $h_1^{q/p}$~\cite{Collins:1992kk}.
In order to observe this effect one needs transverse target polarization.
Since the spin-dependent part in~(\ref{e:SIDIS_xs_PT}) is only a small correction to the total cross section, experiments combine the information from measuring the cross sections with spin up $\sigma^\uparrow$ and spin down $\sigma^\downarrow$ to cancel the unpolarized part of the cross section as well as systematics that are spin-independent, e.g~the detector acceptance. 
The simplest approach is to use the transverse SSA as defined in Eq.~(\ref{e:SSA_definition}) 

which in SIDIS retains the angular dependence of the structure functions. 
The amplitude of the $\sin(\phi_h + \phi_S)$ modulation is then
\begin{equation} \label{e:Collins_SSA_SIDIS}
A_{UT}^{\rm Collins} 
\propto \frac{\sum_q e_q^2 \, h_1^{q/p}(x,\vec{p}_T^{\;2}) \otimes H_1^{\perp \, h/q}(z,\vec{P}_{hT}^{\,2})}
{\sum_q e_q^2 \, f_1(x,\vec{p}_T^{\;2}) \otimes D_1^{h/q}(z,\vec{P}_{hT}^{\,2})} \,.
\end{equation}
We return to the phenomenology of $A_{UT}^{\rm Collins}$ in Sec.~\ref{sec:ObservablesSpinDepFFs} and Sec.~\ref{sec:CFFExtraction}.
The general principle of extracting the Collins asymmetry can be translated to other SSAs measured in SIDIS and $pp$ scattering. 
One of the earliest works on the extraction of transverse SSAs can be found in~\cite{Ohlsen:1973}. 
Modern experiments usually use an unbinned maximum likelihood fit to the full cross section to minimize systematic effects due to interference with the acceptance, see e.g.~Ref.~\cite{Airapetian:2010ds}.

According to~(\ref{e:F_cos2phi_SIDIS}) the Collins function also matters for the unpolarized structure function $F_{UU}^{\cos 2\phi}$, where it couples to the Boer-Mulders function $h_1^\perp$~\cite{Boer:1997nt}, one of the eight leading twist TMD PDFs.
Currently that structure function cannot be used for getting information on the Collins function, not the least because $h_1^\perp$ is only poorly known.
In fact one can try the reverse, namely learning about the Boer-Mulders function from this azimuthal modulation of the cross section using input for the Collins function~\cite{Barone:2008tn,Barone:2009hw,Barone:2015ksa}.
The same applies to the structure function $F_{UT}^{\sin(3\phi_h - \phi_S)}$ through which the pretzelosity TMD PDF $h_{1T}^{\perp}$ could be constrained~\cite{Avakian:2008dz,Lefky:2014eia}.

\subsubsection{Observables for TMD FFs in $pp$}
\label{sec:observables_TMDFF_pp}
In $pp$ scattering one can also look at hadrons inside jets in order to study TMD FFs~\cite{Yuan:2007nd,D'Alesio:2010am,D'Alesio:2013jka}.
At LO the cross section in that case can be written as~\cite{D'Alesio:2010am,D'Alesio:2013jka}
\begin{eqnarray}
\label{eq:upolPPTMD}
\frac{d^6\sigma^{pp \to (h,\textrm{jet})X}}{d^3\vec{P}_{\textrm{jet}} \, dz \, d^2\vec{P}_{hT}} & \propto & 
\sum_{i,j,k,l} \int \frac{dx_a}{x_a} \, d^2 \vec{p}_{aT} \int \frac{dx_b}{x_b} \, d^2 \vec{p}_{bT} \, \delta(\hat{s} + \hat{t} + \hat{u}) 
\nonumber \\
& & \times \; \Big( \hat{\sigma}_{\textrm{unp}}^{ij \to kl} \, f_1^{i/p}(x_a,\vec{p}_{aT}^{\,2}) \, f_1^{j/p_b}(x_b,\vec{p}_{bT}^{\,2}) \, D_1^{h/k}(z,\vec{P}_{hT}^{\, 2})
\nonumber \\
& & \hspace{0.6cm} + \, \sin(\phi_S - \phi_h) \, \hat{\sigma}_{\textrm{pol}}^{ij \to kl} \, h_1^{i/p}(x_a,\vec{p}_{aT}^{\,2}) \, f_1^{j/p_b}(x_b,\vec{p}_{bT}^{\,2}) \, H_1^{\perp \, h/k}(z,\vec{P}_{hT}^{\, 2}) + \ldots \Big) \,,
\end{eqnarray}
where $\phi_S$ and $\phi_h$ are the azimthal angles of the transverse spin vector of the proton and of the transverse momentum of the hadron $\vec{P}_{hT}$ relative to the jet --- see~\cite{Yuan:2007nd,D'Alesio:2010am} for details on the kinematics.
In a leading order approach one can directly access the transverse momentum dependence of the FFs, but higher order corrections change this situation.
As already stated above, at present there exists no proof of TMD factorization for this process.
It is defintely worthwhile to investigate this point carefully.
Ideally hadron production inside jets in $pp$ scattering can be used for example in global fits of the Collins function and the poorly known transversity PDF.
In addition to the two terms shown in~(\ref{eq:upolPPTMD}), there exist other angular modulations in the cross section through which one could study other TMDs~\cite{D'Alesio:2010am,D'Alesio:2013jka}. 
They also include effects due to linearly polarized gluons which attracted quite some interest recently.

Note that hadrons inside jets could also be explored in $e^+ e^-$ annihilation and SIDIS in order to study TMDs.
It may actually we worthwhile to look into those cases from both an experimental and a theoretical point of view. 
On the theory side for example, one interesting aspect would be trying to compute NLO corrections in such simpler processes first before moving to the $pp$ case.

\subsection{Observables for higher-twist FFs}
\label{sec:observables_twist3}
Twist-3 FFs are not just of academic interest.
They are needed for the QCD description of several key observables, where in some cases data exist as will be discussed in more detail in Sec.~\ref{sec:experiments}.
Our primary focus here is on the transvserse SSA $A_N$, defined in Eq.~(\ref{e:SSA_definition}), that has been observed in processes like $p^{\uparrow} p \to h X$, $\bar{p}^{\uparrow} p \to h X$, and $\ell p^{\uparrow} \to h X$.

In the first attempt to compute $A_N$ for such reactions in pQCD, exclusively twist-2 parton correlation functions were used~\cite{Kane:1978nd}.
It was found that $A_N \sim \alpha_s \, m_q / |\vec{P}_{h\perp}|$, where $\alpha_s$ arises from 1-loop corrections that are needed to generate an imaginary part on the amplitude level~\cite{Kane:1978nd}.
Another suppression comes form the quark mass $m_q$.
Since this result was much too small for explaining data on $A_N$ available at that time, the authors of Ref.~\cite{Kane:1978nd} even speculated that (perturbative) QCD may be incorrect.
However, that conclusion was not justified.
The calculation in~\cite{Kane:1978nd} had just shown that $A_N$, unlike the unpolarized cross section, is a genuine twist-3 observable since there is a contribution proportional to the quark mass.
(One can readily understand why $A_N$ cannot be a twist-2 observable. 
At twist-2, necessarily the transversity PDF $h_1$ of the transversely-polarized proton and the unpolarized PDF $f_1$ of the unpolarized proton enter the factorized cross section.
Since $h_1$ is chiral-odd one needs another non-perturbative chiral-odd parton correlator in order to get a nonzero result.
This could only be provided by the fragmentation process.
However, at twist-2 there is no chiral-odd (collinear) FF for fragmentation into unpolarized hadrons.
Put differently, the helicity structure of the hard scattering cross section for $A_N$ does not allow a twist-2 effect.)
In fact it was pointed out early on that three-parton correlations have to be taken into account for a proper description of $A_N$ in QCD~\cite{Efremov:1981sh,Efremov:1983eb}.
The framework used in~\cite{Efremov:1981sh,Efremov:1983eb} is collinear twist-3 factorization.
This approach was discussed in detail in Ref.~\cite{Ellis:1982wd,Ellis:1982cd}, and later also successfully applied to $A_N$ in hadronic collisions --- see, e.g., Refs.~\cite{Qiu:1991pp,Qiu:1998ia,Eguchi:2006qz,Kouvaris:2006zy,Koike:2007rq,Koike:2009ge,Kanazawa:2010au}.

Here we will just discuss the overall outcome of a calculation of $A_N$ in collinear twist-3 factorization and skip all technical details.
For the process $p(P_a,S_a) + p(P_b) \to h(P_h) + X$ the spin-dependent cross section takes the generic form~\cite{Metz:2012ct}
\begin{eqnarray} \label{e:sigma_generic}
d\sigma(\vec{S}_{\perp}) & = & H \otimes f_{(3)}^{i/p_a} \otimes f_{(2)}^{j/p_b} \otimes D_{(2)}^{h/k} 
\nonumber \\
& + & H' \otimes f_{(2)}^{i/p_a} \otimes f_{(3)}^{j/p_b} \otimes D_{(2)}^{h/k}
\nonumber \\
& + & H'' \otimes f_{(2)}^{i/p_a} \otimes f_{(2)}^{j/p_b} \otimes D_{(3)}^{h/k} \,,
\end{eqnarray} 
where $f_{(t)}^{i/p_a}$ ($f_{(t)}^{j/p_b}$) denotes the parton distribution of parton $i$ ($j$) in proton $a$ ($b$), and $D_{(t)}^{h/k}$ is the fragmentation function of hadron $h$ in parton $k$.  
The twist of the functions is indicated by $t$.  
The hard factors for each term are given by $H$, $H'$, and $H''$, while the symbol $\otimes$ represents convolutions in the appropriate momentum fractions.  
In Eq.~(\ref{e:sigma_generic}) a sum over partonic channels and parton flavors in each channel is implicit.
The expression in~(\ref{e:sigma_generic}) shows that the twist-3 effect can be associated with either the transversely polarized proton, or the unpolarized proton, or the parton fragmentation.
The imaginary part of the first two lines of~(\ref{e:sigma_generic}) comes from a pole in the hard partonic scattering coefficient, while in the 3rd line it arises from the fragmentation correlator.
As we argue below in Sec.~\ref{sec:fit_higher_twist}, the twist-3 fragmentation effect may be the most important piece for the phenomenology of $A_N$ in processes like $p^{\uparrow} p \to h X$.
Its contribution to the spin-dependent cross section schematically reads~\cite{Metz:2012ct}
\begin{equation} \label{e:sigma_generic_frag}
d\sigma(\vec{S}_{\perp}) |_{{\rm frag}} = H'' \otimes h_1^{i/p_a} \otimes f_1^{j/p_b} 
\otimes \{ H^{h/k} , \; H_1^{\perp (1) \, h/k} , \; \hat{H}_{FU}^{h/k, \, \Im} \} \,.
\end{equation}
This expression shows, in particular, that the transversity $h_1^{q/p_a}$ of the polarized proton enters as well as, {\it a priori}, three twist-3 FFs:
the (two-parton) FF $H^{h/q}$ defined in~(\ref{e:H_quark_z}), the moment $H_1^{\perp (1) \, h/q}$ of the Collins function defined in~(\ref{e:kT_moment}), and the (imaginary part of the) three-parton FF $\hat{H}_{FU}^{h/k, \, \Im}$ defined in~(\ref{e:DeltaF_H}).
However, due to the relations~(\ref{e:H_twist3}),~(\ref{e:H1perp_twist3}), the final result can be expressed entirely through the three-parton FF~\cite{Kanazawa:2015ajw}.
The part in~(\ref{e:sigma_generic_frag}) that is related to the Collins function was computed in~\cite{Kang:2010zzb}, while the complete result was obtained in~\cite{Metz:2012ct}. 
The leading order calculation of all the terms in~(\ref{e:sigma_generic}) includes several hundred Feynman graphs.
The corresponding evaluation of $A_N$ in $\ell p^{\uparrow} \to h X$ is much simpler in comparison, and a full analytical result can be found in~\cite{Gamberg:2014eia}.
In that case one of course only has twist-3 effects associated with the transversely polarized nucleon and with the parton fragmentation.
The transverse SSA $A_N$ also exists for polarization in the final state and has been studied in processes like $p p \to \Lambda^{\uparrow} X$.
A brief discussion of the existing twist-3 calculations for this observable will be given in Sec.~\ref{sec:frag_pol_hadron}.

We also mention that alternative theoretical frameworks were used to describe $A_N$ for both hadron-hadron collisions and lepton-nucleon collisions.
In particular, many phenomenological studies exist in the generalized parton model which consistently works with TMD parton correlators --- see~\cite{Anselmino:2009pn,Anselmino:2012rq,Anselmino:2013rya,Anselmino:2014eza,Aschenauer:2015ndk} and references therein. 
Other approaches involve, for instance, more complicated multi-parton dynamics or instanton-induced effects~\cite{Hoyer:2006hu,Hoyer:2008fp,Kang:2011ni,Qian:2011ya,Kovchegov:2012ga,Troshin:2012fr,Kochelev:2013zoa,Qian:2015wyq}. 

The longitudinal-transverse double-spin asymmetry (DSA) $A_{LT}$ for reactions such as $\vec{p} p^{\uparrow} \to h X$ and $\vec{\ell} N^{\uparrow} \to h X$ is also sensitive to twist-3 FFs.
The generic structure of the numerator of the asymmetry coincides with the expression in~(\ref{e:sigma_generic}).
For lepton-nucleon scattering the analytical result for $A_{LT}$ can be found in Ref.~\cite{Kanazawa:2014tda}, where the final expression for the twist-3 fragmentation part only contains the FF $E^{h/q}$ defined in~(\ref{e:E_quark_z}).
A measurement of $A_{LT}$ using a polarized $^3$He-target was performed at Jefferson-Lab~\cite{Zhao:2015wva}.
At present, it is not possible to draw any strong conclusion from comparing the data of~\cite{Zhao:2015wva} with a numerical estimate given in~\cite{Kanazawa:2014tda}.
In the case of {\it pp} scattering a basically complete analytical result for $A_{LT}$ is available as well~\cite{Metz:2012fq,Koike:2015yza,Koike:2016ura}.
Currently, no data exist for this observable, and RHIC would be the only facility in the world that could perform such a measurement.
Given the pioneering character of a potential experimental study, plus the fact that in the past measurements of the ``partner" twist-3 observable $A_N$ gave rise to strong surprises, it would be worthwhile to pursue activities along those lines.

Twist-3 FFs also enter certain structure functions in SIDIS (and in $e^+e^- \to h_1 h_2 X$) for large transverse hadron momenta~\cite{Yuan:2009dw,Kanazawa:2013uia}.
While at present we have no suitable data for such observables, a future electron-ion collider~\cite{Boer:2011fh,Accardi:2012qut} would be ideal to fill this gap.

\subsection{Observables for di-hadron FFs}
\label{sec:observables_dihadron}
Compared to the single hadron case, the phenomenology of DiFFs is richer due to the existence of two hadron 4-momenta $P_1$, $P_2$ (see Sec.~\ref{sec:definition_dihadron}).
Instead of working with $P_1$ and $P_2$ one may also use their sum $P_h$ and (half of) their difference $R$ as defined in Eq.~(\ref{e:dihadron_momenta}).
Similar to the single-hadron TMD FFs one can also consider the transverse momentum dependence of DiFFs.
However, in the following we will restrict ourselves to observables that do not vanish when integrated over $\vec{k}_T$. 
The DiFFs contributing to these observables are functions of the fractional energy $z$ and the invariant mass of the hadron pair $M_h$. 
At large $M_h$, the DiFFs can be computed in pQCD and expressed in terms of single hadron FFs~\cite{Zhou:2011ba} (see also Ref.~\cite{deFlorian:2003cg}).
Here we are mainly concerned with dihadron production for small $M_h$.

In addition to the spin-averaged DiFF $D_1^{h_1 h_2/q}$, the IFF $H_1^{\open \, h_1 h_2/q}$, describing the fragmentation of a transversely polarized quark into two unpolarized hadrons, has attracted considerable interest recently.
As we summarize in this section, it can be accessed in $e^+e^-$ annihilation, SIDIS and $pp$ scattering.
(For an overview of the observables we refer to Tab.~\ref{tbl:observables}.)
Like the Collins function, $H_1^{\open \, h_1 h_2/q}$ is chiral-odd and does couple to the transversity distribution of the nucleon~\cite{Collins:1993kq}.
It can also serve as ``quark polarimeter" as it measures the strength of the triple scalar product $\vec{s}_q \cdot (\vec{R} \times \vec{P}_h)$ where $\vec{s}_q$ is the spin vector of the quark.
Here the (transverse) vector $\vec{R}_T$ plays basically the same role of $\vec{k}_T$ in the Collins effect.
The above correlation survives integration upon the transverse momentum of the quark relative to the two hadrons, and therefore one can use collinear factorization which generally is simpler than TMD factorization.
The price to pay is that one needs to keep the dependence of the DiFFs on $\vec{R}^2_T$, or in other words $M_h$ (see Eq.~(\ref{e:RT_Mh})).

As already mentioned in Sec.~\ref{sec:definition_dihadron} we do not consider twist-3 DiFFs and their observables.
More information on this topic can be found in~\cite{Pisano:2015wnq} and references therein.

\subsubsection{Observables for di-hadron FFs in $e^+e^-$}
\label{sec:DiFF_epem}
In $e^+ e^-$ annihilation one can access $D_1^{h_1h_2/q}(z,M_h)$ through $e^+ e^- \to (h_1,h_2) X$.
The cross section of this reaction is sensitive to $\sum_q e^2_q D_{1}^{h_1 h_2/q}(z,M_h)$~\cite{deFlorian:2003cg}.
However, like in the single hadron case, it is worth considering the back-to-back production of a di-hadron pair, $e^+ e^- \to (h_{a1},h_{a2}) (h_{b1},h_{b2}) X$, to access the product of quark and antiquark DiFFs and gain some sensitivity to the favored vs disfavored FFs.
In order to get any information on the IFF $H_1^{\open \, h_1h_2/q}(z,M_h)$ one actually needs to look at this more complicated final state.

\begin{figure}[t]
\begin{center}
\includegraphics[width=0.55\textwidth]{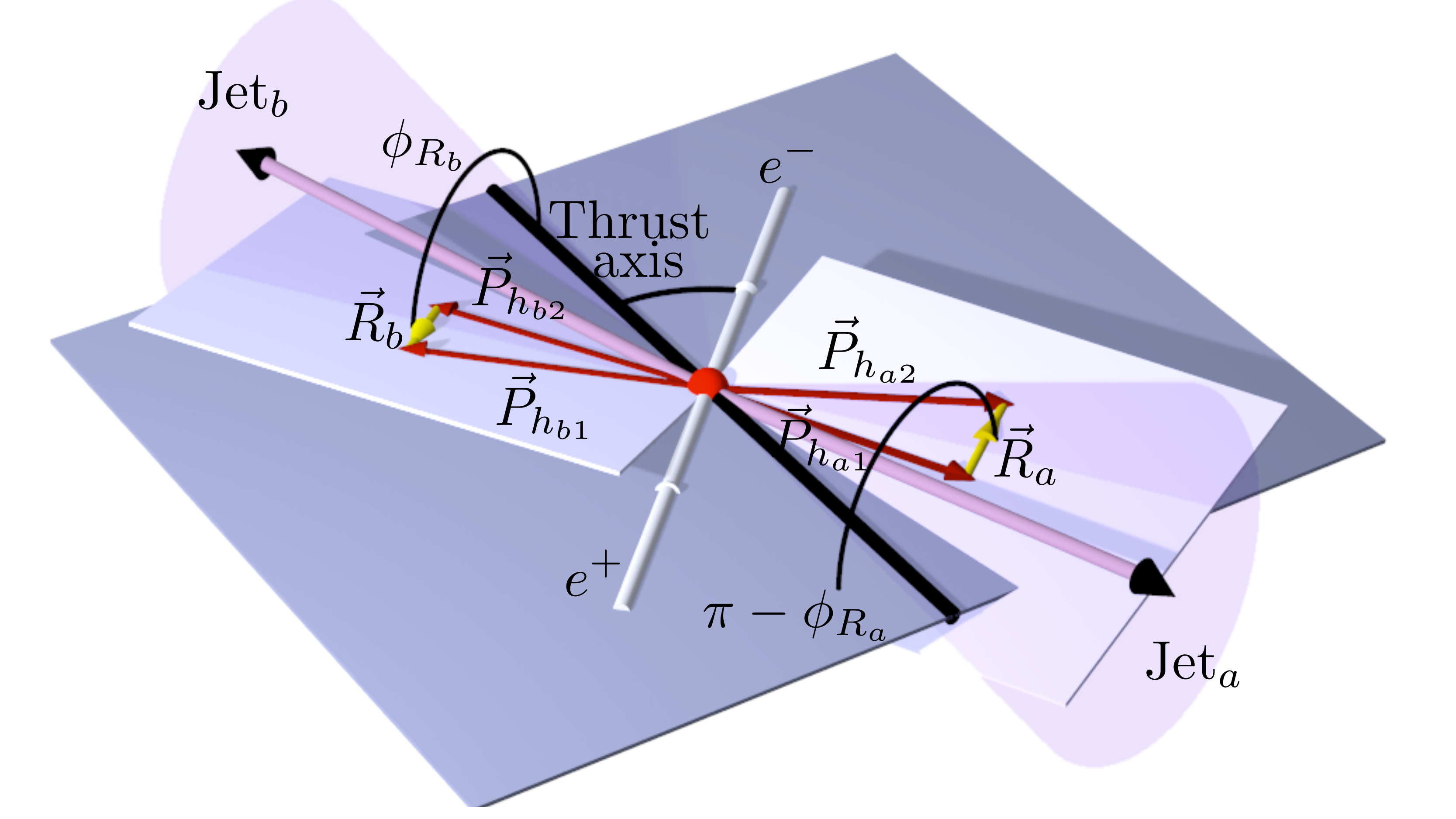}
\end{center}
\caption{Coordinate system used for the Artru-Collins asymmetry for di-hadron pairs. For convenience, the jets are depicted to lie in the event plane which in turn is spanned by the beam axis and the jets.
\label{fig:IFFCoo}}
\end{figure}

Historically, different coordinate systems have been proposed to define the azimuthal angles in the back-to-back correlation of di-hadron pairs.
Originally, the authors of Ref.~\cite{Artru:1995zu} used the coordinate system shown in Fig.~\ref{fig:IFFCoo}, where the azimuthal angles of the vectors $\vec{R}_a$, $\vec{R}_b$ are measured around the axis given by the jet momentum vectors. 
This axis is approximately equal to the thrust axis, so this situation basically corresponds to using the coordinate system for the single hadron asymmetry shown on the right of Fig.~\ref{fig:SIACollinsCoo}. 
Instead in~\cite{Boer:2003ya} a coordinate system is employed in which $\vec{P}_{hb} = \vec{P}_{b1} + \vec{P}_{b2}$ defines the axis around which the azimuthal angles are measured. 
This asymmetric coordinate system thus corresponds to the situation on the left of Fig.~\ref{fig:SIACollinsCoo}.
In both cases, using the parton model approximation, one can write the cross section for the back-to-back production of hadron pairs as
\begin{eqnarray}
\label{eq:diFFInEE}
\frac{d^6 \sigma^{e^+ e^- \to (h_{a1} h_{a2}) (h_{b1} h_{b2}) X}}{d\Omega \, dz_a \, dz_b \, d\phi_{Ra} \, d\phi_{Rb}} & = &
\frac{3\alpha_{\textrm em}^2}{4 \, Q^2} \, z_a^2 z_b^2 \, \sum_q e_q^2 \, \bigg( A(y) D_1^{h_{a1} h_{a2}/q}(z_a,M_{ha}) \, D_1^{h_{b1} h_{b2}/\bar{q}}(z_b,M_{hb})
\nonumber \\
& & + \, B(y) \cos(\phi_{Ra} + \phi_{Rb}) H_1^{\open \, h_{a1} h_{a2}/q}(z_a,M_{ha}) \, H_1^{\open h_{b1} h_{b2}/\bar{q}}(z_b,M_{hb})  \bigg) \,,
\end{eqnarray}
with $A(y)$, $B(y)$ from~(\ref{e:y_dependence_epem}).
The term containing $H_1^\sphericalangle$ can then be extracted from the $\cos(\phi_{Ra} + \phi_{Rb})$ modulation of the crosr section.
To this end the cross section is usually normalized to the azimuthally averaged cross section for di-hadron pair production,
\begin{equation}
\frac{1}{\sigma_\textrm{tot}} \,
\frac{d^6 \sigma^{e^+ e^- \to (h_{a1} h_{a2}) (h_{b1} h_{b2}) X}}{d\Omega \, dz_a \, dz_b \, d\phi_{Ra} \, d\phi_{Rb}} = 
1 + \cos(\phi_{Ra} + \phi_{Rb}) \, A^{\cos(\phi_{Ra} + \phi_{Rb})} \,,
\end{equation}
where $A^{\cos(\phi_{Ra}+\phi_{Rb})}$ is known as the Artru-Collins asymmetry,
\begin{equation}
\label{eq:ArtruCollins_Asymmetry}
A^{\cos(\phi_{Ra} + \phi_{Rb})}= 
\frac{B(y) \, \sum_q e^2_q \, H_1^{\open \, h_{a1} h_{a2}/q}(z_a,M_{ha}) \, H_1^{\open \, h_{b1} h_{b2}/\bar{q}}(z_b,M_{hb}) }
{A(y) \, \sum_q e^2_q \, D_1^{h_{a1} h_{a2}/q}(z_a,M_{ha}) \, D_1^{h_{b1} h_{b2}/\bar{q}}(z_b,M_{hb})} \,.
\end{equation}

\subsubsection{Observables for di-hadron FFs in SIDIS}
As in the case of single hadron production, SIDIS data is important to achieve a better flavor separation of DiFFs.
The unpolarized SIDIS cross section for $\ell p \to \ell (h_1 h_2) X$ is sensitive to $D_1^{h_1 h_2/q}(z,M_h)$ whereas in the cross section off a transversely polarized target the IFF $H_1^{\open \, h_1 h_2/q}(z,M_h)$ enters.
One finds~\cite{Bianconi:1999cd,Pisano:2015wnq,Martin:2014wua} 
\begin{eqnarray}
\label{eq:diffInSIDIS}
\frac{d^5\sigma^{\ell p \to \ell (h_1 h_2) X}}{dx \, dy \, dz \, d\phi_R \, dM_h^2} 
& = & \frac{\alpha_{\textrm em}}{xyQ^2} \sum_q e^2_q \, \bigg( 2 \, A(y) \, f_1^{q/p}(x) D_1^{h_1h_2/q}(z,M_h)  
\nonumber \\
& & - \, |\vec{S}_T| \, B(y) \, \sin(\phi_R + \phi_S) \, \frac{\pi |\vec{R}_T|}{2\, M_h} h_1^{q/p}(x) \, H_1^{\open \, h_1 h_2/q}(z,M_h)\bigg) \,,
\end{eqnarray}
where $\phi_R$ and $\phi_S$ are the azimuthal angles of $\vec{R}_T$ and $\vec{S}_T$, respectively. 
The $\sin(\phi_R+\phi_S)$ dependent term can be extracted by constructing a transverse SSA.
Note the similarity to the Collins asymmetry in $\ell p \to \ell h X$ in~(\ref{e:SIDIS_xs_PT}) which is given by the $\sin(\phi_h + \phi_S)$ modulation of the cross section.

\subsubsection{Observables for di-hadron FFs in $pp$}
Production of di-hadron pairs in hadronic collisions like $p p \to (h_1,h_2)X$ is important to get direct access to gluon DiFFs.
This process can also serve as a tool to address the tranversity distribution of the nucleon in {\it pp} scattering, since in a transverse SSA $h_1^{q/p}$ couples to the IFF $H_1^{\open \, h_1 h_2/q}$.
In comparison to the Collins effect in {\it pp} scattering discussed in Sec.~\ref{sec:observables_TMDFF_pp} one just needs to measure the momenta of the two hadrons and there is no need to reconstruct for example a jet axis.
One can define a reference frame through the momentum vector $\vec{P}_a$ of one of the incoming proton and $\vec{P}_h$~\cite{Bacchetta:2004it}. 
If proton $p_a$ is transversely polarized the spin-dependent part of the cross section takes the schematic form~\cite{Bacchetta:2004it}
\begin{equation}
d\sigma_{UT}^{p_a^{\uparrow} p_b \to (h_1,h_2)X} \propto \sin(\phi_R - \phi_{S_a}) \, \sum_{i,j,k,l} \int dx_a \int dx_b \int dz \, h_1^{i/p_a}(x_a) \, f_1^{j/p_b}(x_b) \, H_1^{\open \, h_1 h_2/k}(z,M_h^2) \, \Delta \hat{\sigma}^{ij \to kl} \,,
\end{equation}
where $\Delta \hat{\sigma}^{ij \to kl}$ are the spin-dependent partonic cross sections. 
The corresponding transverse SSA reads
\begin{equation} \label{eq:diffInPP}
A_{UT} \propto \frac{\sum_{i,j,k} h_1^{i/p_a}(x_a) \, f_1^{j/p_b}(x_b) \, H_1^{\open \, h_1 h_2/k}(z,M_h)}{\sum_{i,j,k} f_1^{i/p_a}(x_a) \, f_1^{j/p_b}(x_b) \, D_1^{h_1 h_2/k}(z,M_h)}
\end{equation}
Very recently this asymmetry was computed using input from $e^+ e^-$ annihilation and SIDIS~\cite{Radici:2016lam}, and good agreement was reported with data from the STAR Collaboration that will be discussed in Sec.~\ref{sec:data_DiFFs}.

\subsubsection{Observables for TMD di-hadron FFs}
\label{sec:diHadTMDs}
If the dependence on the transverse momentum $\vec{k}_T$ of the fragmenting quark relative to $\vec{P}_h$ is kept, the structure of the di-hadron correlator becomes richer than in the collinear case and the single hadron TMD case --- see Eqs.~(\ref{e:D1_quark_di-hadron_kT})--(\ref{e:H1_quark_di-hadron_kT}). 
It has been argued that the $k_T$ dependence of $D_1^{h_1h_2/q}$ allows one to project out an angular dependence that is sensitive to the Sivers effect~\cite{Kotzinian:2014lsa,Kotzinian:2014gza,Kotzinian:2014uya}.
In addition, two spin-dependent TMD DiFFs appear that are of interest: the quark helicity dependent function $G_1^{\perp h_1 h_2 /q}$ and the chiral-odd transverse spin dependent function $H_1^{\perp h_1 h_2/q}$ which could be considered the equivalent of the Collins function.

The DiFF $G_1^{\perp \, h_1 h_2/q}$ describes a correlation between the quark helicity, $\vec{k}_T$, and $\vec{R}_T$.
According to Ref.~\cite{Boer:2003ya} in the process $e^+ e^- \to (h_{a1},h_{a2}) (h_{b1},h_{b2}) X$, the $\cos(2(\phi_{Ra} - \phi_{Rb}))$ modulation of the cross section, where the angles $\phi_{R_i}$ are defined as for $H_1^\sphericalangle$ in $e^+e^-$ in Sec.~\ref{sec:DiFF_epem}, survives even upon integration over $\vec{k}_T$. 
This allows one to define an asymmetry analogous to the Artru-Collins asymmetry in Eq.~(\ref{eq:ArtruCollins_Asymmetry})~\cite{Boer:2003ya},
\begin{equation}
\label{eq:jetHandedness}
A^{\cos(2(\phi_{Ra}-\phi_{Rb}))} = \frac{\sum_q e_q^2 \tilde{G}_1^{\perp \, h_{a1} h_{a2}/q}(z_a,M_{ha}) \, \tilde{G}_1^{\perp \, h_{b1} h_{b2}/\bar{q}}(z_b,M_{hb})} {\sum_q e^2_q D_1^{h_{a1} h_{a2}/q}(z_a,M_{ha}) \, D_1^{h_{b1} h_{b2}/\bar{q}}(z_b,M_{hb})} \,,
\end{equation}
where 
\begin{equation}
\label{eq:g1TInt}
\tilde{G}_1^{\perp \, h_1 h_2/q}(z,M_h)= \int d\zeta \int d\vec{k}_T \, \vec{k}_T \cdot \vec{R}_T \, G_1^\perp(z,\zeta, \vec{R}_T^{\,2},\vec{k}_T \cdot \vec{R}_T, \vec{k}_T^{\,2}) \,.
\end{equation}
Within the partial wave expansion for the DiFFs, which can be performed for $G_1^{\perp \, h_1 h_2/q}$ similar to the $H_1^{\open \, h_1 h_2/q}$ case in Eq.~(\ref{e:H1open_partial_wave}), the integration in Eq.~(\ref{eq:g1TInt}) can be seen as projecting on a specific partial wave term~\cite{Pisano:2015wnq}.
The asymmetry defined in Eq.~(\ref{eq:jetHandedness}) is also of interest, since it is related to the longitudinal jet handedness~\cite{Efremov:1992pe,Boer:2003ya}. 
A non-zero jet-handedness might be related to local strong CP violation~\cite{Boer:2003ya}.

Following Refs.~\cite{Radici:2001na,Pisano:2015wnq} $G_1^{\perp \, h_1 h_2/q}$ can also be measured in SIDIS in longitudinal or transverse SSAs, however in that case the $k_T$ dependence has to be kept for otherwise the asymmetry vanishes.

Finally, the DiFF $H_1^{\perp \, h_1h_2/q}$ can be observed in the same processes like the ordinary Collins function, where in the equations one needs to replace the momentum of the single hadron with the sum of the momenta of $h_1$ and $h_2$. 

\section{Experiments and Datasets}
\label{sec:experiments}
As noted in Sec.~\ref{sec:observables}, to extract precise information on the kinematic and flavor dependence of FFs the data from semi-inclusive $e^+e^-$ annihilation, SIDIS, and {\it pp} collision experiments is needed. 
This section gives an overview of the experiments that are most relevant for the discussion of the datasets used in the fits described below in Sec.~\ref{sec:global_fits}. 
The experiments, their capabilities, and the analyses that have been performed with the respective datasets are summarized in Tab.~\ref{tbl:experiments}.

\begin{table}
\begin{center}
\begin{tabular}{|c|c|c|c|c|c|}
\hline
Experiment & Process & $\mathcal{L}[pb^{-1}]$ &$Q^2 [GeV^2]$ &  Final States \\
\hline
TPC~\cite{Aihara:1984mk,Aihara:1986fz,Aihara:1988fc,Aihara:1988su} & $e^+e^-$ &  140 & 29 & $\pi^\pm, K^\pm,p/\bar{p}$\\
\begin{tabular}{c}
TASSO\\
\cite{Brandelik:1980iy,Althoff:1982dh,Braunschweig:1988hv}\\\cite{Althoff:1984iz,Braunschweig:1989wg,Brandelik:1981ta,Braunschweig:1988wh}\end{tabular} & $e^+e^-$ &  34& 34,44&$\pi^\pm, K^\pm,p/\bar{p},K_S^0,\Lambda/\bar{\Lambda}$  \\
SLD~\cite{Abe:1998zs,Abe:2003iy}    & $e^+e^-$ &20 & $M_Z$&$\pi^\pm, K^\pm, p,K_S^0,\Lambda/\bar{\Lambda}$ \\
ALEPH~\cite{Buskulic:1994ft,Barate:1996fi}  & $e^+e^-$ &800 & $M_Z$&$\pi^\pm, K^\pm,p,K_S^0,\Lambda/\bar{\Lambda}$\\
DELPHI~\cite{Abreu:1998vq,Abreu:2000gw,Abreu:1994rg,Abreu:1993mm}& $e^+e^-$  &800& $M_Z$&$\pi^\pm, K^\pm,p,K_S^0,\Lambda/\bar{\Lambda}$\\
OPAL~\cite{Abbiendi:1999ry,Akers:1994ez,Abbiendi:2000cv,Alexander:1996qj}   &$e^+e^-$   &800& $M_Z$&$\pi^\pm, K^\pm,p,K_S^0,\Lambda/\bar{\Lambda}$\\
H1~\cite{Aaron:2007ds, Aaron:2009ae,Nowak:2011zz} & $e +p$ & 500 & 27.5 (e) + 920 (p) &$h^\pm$, $K_0^S$\\
ZEUS~\cite{Breitweg:1999nt,Chekanov:2009kb,Abramowicz:2010rz} & $e +p$ & 500 & 27.5 (e) + 920 (p) & $h^\pm$\\
BELLE~\cite{Leitgab:2013qh,Seidl:2015lla}  &   $e^+e^-$ & $10^6$& near 10.58 & $\pi^\pm,K^\pm,p/\bar{p}$\\
BaBar~\cite{Anulli:2004nm,Lees:2013rqd}   &$e^+e^-$ & $557\cdot 10^3 $ & near 10.58&$\pi^\pm,K^\pm,\eta, p/\bar{p}$\\
 
HERMES~\cite{Airapetian:2001qk,Airapetian:2012ki}   &$e+p(d)$  & 272 (p) 329(d)& 27.6 fixed target & $\pi^{\pm,0}, K^\pm$\\
COMPASS~\cite{Adolph:2013stb}   &$\mu+p(d)$ & 775 & 160 GeV fixed target & $h^\pm$\\
\begin{tabular}{c}
PHENIX\\
\cite{Adare:2007dg,Adare:2012nq,Adare:2010cy}\\ \cite{Adare:2014wht,Adare:2014qzo,Adare:2015ozj} \end{tabular}  &  $pp$ &\begin{tabular}{@{}c@{}} $16 \times 10^{-3}$\\2.5\\128\end{tabular} & \begin{tabular}{@{}c@{}}62.4\\ 200 \\ 510\end{tabular}&$\pi^{\pm,0},\eta$\\
 \begin{tabular}{@{}c@{}}
STAR\\ \cite{Adams:2006nd,Abelev:2006cs,Adams:2006uz}\\ \cite{Abelev:2009pb,Adamczyk:2012xd,Adamczyk:2013yvv} \end{tabular}    & $pp$ & 8 & 200 &$\pi^{0,\pm},\eta,p/\bar{p},K_S^0,\Lambda/\bar{\Lambda}$\\ 

ALICE~\cite{Abelev:2012cn}   &pp & $6 \times 10^{-3} $ &$7\times 10^3$&$\pi^0, \eta$\\
TOPAZ~\cite{Itoh:1994kb} & $e^+e^-$ & 278 &52-61.4 & $\pi^\pm, K^\pm,K_S^0,$\\
CDF~\cite{Abe:1988yu,Acosta:2005pk} &$ p+\bar{p}$ & n/a & 630 (1800) & $h^\pm,K_S^0\Lambda^0$ \\
\hline

\end{tabular}
\caption{
Selected datasets on hadron production in $e^+e^-$, $pp(\bar{p})$ and SIDIS. For the $e^+e^-$ experiments the integrated luminosity numbers reflect the overall collected statistics. For the SIDIS and $pp(\bar{p})$ the integrated luminosity of the pion cross section measurement.
\label{tbl:experiments}}
\end{center}
\end{table}

\subsection{$e^+e^-$ facilities}
Electron-positron annihilation experiments play a special role in the study of FFs since they provide clean access to light-quark FFs. 
Up until recently, most global extractions of FFs relied mainly on $e^+e^-$ data --- see Sec.~\ref{sec:global_fits} for more details.
Since the first $e^+ e^-$ collider AdA at Frascati in the early 1960s, there has been a long history of $e^+e^-$ experiments, which have been instrumental in fundamental discoveries in particle physics. 
In particular the data taken at facilities at the SLAC National Accelerator Laboratory in the US, at the Japanese High Energy Accelerator Research Organization (KEK), at DESY in Germany, as well as by the LEP experiments at CERN, allowed the precision study of FFs. 
A comprehensive review of the data collected up to 1995 at PETRA, PEP, TRISTAN, SLC, and LEP can be found in~\cite{Lafferty:1995}, while a collection of datapoints from later measurements can be found on the Durham HEP database~\cite{durham}.

The first collider ring at SLAC was SPEAR operating at 7.4~GeV, which provided the first collisions at the MARK-I experiment in 1973~\cite{O'Shaughnessy:1990ek} leading to the discovery of the $J/\Psi$ and the $\tau$ lepton. 
It was followed by the Positron-Electron-Project (PEP) collider ring, which collected data at $\sqrt{s}=29 \; \rm{GeV}$ from 1980 to 1990. 
PEP in turn was succeeded by the Stanford Linear Collider (SLC), which came online in 1987 and which delivered $20 \; \rm{pb}^{-1}$ at the $Z^0$ mass between 1989 and 1998. 
A unique feature of SLC was that the electron beam could be longitudinally polarized, which allowed the separation of the fragmentation products from quarks and antiquarks coming from the decay of the $Z^0$ due to the parity violation of the weak force. 
More recently, PEP-II was in operation at SLAC, which was a B-factory operating at or near the $\Upsilon(4S)$ resonance from 1998 until 2008.

At KEK, the TRISTAN $e^+e^-$ storage ring operated starting in 1986 at {\it cm} energies between $50 \; \rm{GeV}$ and $64 \; \rm{GeV}$. 
Similar to the development at SLAC, it was succeeded by the B-factory KEKB which started data taking in 1999 and achieved world record instantaneous luminosities of more than $2.11 \times 10^{34} \; \rm{cm}^{-2} \; \rm{s}^{-1}$. 
KEKB delivered its last beam in 2010 and is currently being upgraded to SuperKEKB with a target luminosity of 40 times that of KEKB.
The B-factories operated at SLAC and KEK and delivered record amounts of integrated luminosity at an energy relatively close to the relevant scales of the SIDIS and {\it pp} experiments. 
Together with the precision instrumentation --- the Belle and BaBar experiments described below --- operated at these machines, the B-factories are arguably the best places to study FFs in $e^+e^-$ and have provided the most important results over the last decade. 

At DESY in Hamburg, Germany, DORIS ran from 1974 to 2012 collecting data at $\sqrt{s} = 3-5 \; \rm{GeV}$ (DORIS I) and $\sqrt{s} = 7-10 \; \rm{GeV}$ (DORIS II). 
The larger storage ring PETRA at DESY operated at {\it cm} energies between $14 \; \rm{GeV}$ and $46 \; \rm{GeV}$ in the period between 1978 and 1986. 
A limited dataset on charged particle production cross sections that is sensitive to $D_1^{h/q}$ is also available from CLEO at CESR~\cite{Alam:1982ue} which started datataking in 1979.
However, as with the data taken at SPEAR, the limited precision and the analysis techniques and documentation of these older datasets usually means that they are not included in modern fits where they are superseded by newer datasets.

Similar to the later use of CLEO as a charm factory (CLEO-c)~from 2003 onward, the Beijing Electron-Positron Collider (BEPC)-II has been taking data at $\sqrt{s}$ below $4.6 \; \rm{GeV}$ since 2005. 
As the name suggests, BEPC-II is itself an upgrade to BEPC which ran at lower instantaneous luminosities from 1988 to 2005.
For the study of FFs of light quarks, the recent analysis of off-resonance datasets collected by the Beijing Spectrometer (BES)-III Collaboration around $\sqrt{s}=3.65 \; \rm{GeV}$ has been important to extend the energy range of fragmentation measurements at the lower end. 

With the exception of TRISTAN and the SLC, the facilities described previously collected data well below $\sqrt{s}=M_Z$, thus accessing a flavor-mix determined by the strength of the electromagnetic coupling to the light flavors.
To have a handle on flavor separation and increase the lever arm in $\sqrt{s}$ that is needed to access gluon FFs through scale dependence, the large datasets collected by the Large Electron Positron Collider (LEP) experiments at CERN around $\sqrt{s} = M_Z$ are important. 
LEP operated from 1989 to 2000 (LEP2 from 1996 onward) and collected over $800 \; \rm{pb}^{-1}$ at the $Z^0$ mass, an order of magnitude more than is available from TRISTAN and the SLC. 
Therefore LEP provided the dominant dataset in its respective energy regime.

\subsection{$e^+e^-$ experiments}
Now that we introduced the facilities, we want to give a short introduction into the experiments that were taking data at these facilities. 
We start with the TPC/two-gamma experiment~\cite{Aihara:1988fc}, which collected $140 \; \rm{pb}^{-1}$ at PEP. 
It was the first experiment to employ a Time Projection Chamber (TPC) for tracking and particle identification (PID). 
All following general purpose $e^+e^-$ experiments used similar instrumentation with a TPC or drift chamber as the main tracking and PID detector, together with electromagnetic calorimetry, muon identification and, as technology progressed, silicon vertex detectors. 
Other experiments at PEP from which data relevant for the study of FFs are available are HRS and MARK-II, where the later already collected data at SPEAR and would be used at SLC as well.
Operating at the SLC were the MARK-II~\cite{O'Shaughnessy:1990ek} and SLD experiments~\cite{Abe:1995iq} collecting $20 \; \rm{pb}^{-1}$ at the Z mass. 
Part of the data was taken with a 75\% linearly polarized electron beam to separate quark and antiquark fragmentation.
It is worth mentioning that in particular SLD had good PID capabilities with its central drift chamber and its Cherenkov ring imaging detector.
Finally, the BaBar collaboration was operating the experiment~\cite{TheBABAR:2013yha} with the same name at the B-factory PEP-II, accumulating $557 \; \rm{fb}^{-1}$ of data.

At KEK, the TRISTAN $e^+e^-$ storage ring has been hosting the TOPAZ experiment~\cite{Enomoto:1987rd,Tsukamoto:1990pc,Imanishi:1987zc,Enomoto:1987rd,Kamae:1986jd,Fujimoto:1986wc,Kawabata:1987fj,Noguchi:1988qq,Fujii:1987ez}, as well as the AMY experiment~\cite{Ko:1988ud}. 
TOPAZ collected $278 \; \rm{pb}^{-1}$ $e^+e^-$ annihilation events at $\sqrt{s}$ between $52 \; \rm{GeV}$ and $61.4 \; \rm{GeV}$ with an average $\sqrt{s}$ of $58 \; \rm{GeV}$. 
More recently, the Belle experiment~\cite{BelleDetector} at the B-factory KEKB~\cite{KEKB} collected more than $1 \; \rm{ab}^{-1}$ between 1999 and 2010, mainly at the $\Upsilon(4S)$ resonance at $\sqrt{s} = 10.58 \; \rm{GeV}$, but also at the $\Upsilon(1S)$ to $\Upsilon(5S)$ resonances, and at $\sqrt{s} = 10.52 \; \rm{GeV}$.
Currently Belle is upgraded to Belle II~\cite{Abe:2010gxa} which will take data with the full detector at SuperKEKB starting in 2018. 
Together with BaBar, Belle provides arguably the most important datasets to study FFs. 
The record-setting integrated luminosity, together with the precision instrumentation allowed for the first determination of polarization-dependent FFs as well as the most precise measurements of pion and kaon multiplicities. These results will be discussed in Sec.~\ref{sec:results}. 
From DESY we have datasets relevant for the extraction of FFs from the PLUTO~\cite{Hepp:1980ha,Berger:1980zb} and ARGUS experiments~\cite{Albrecht:1992qf} at DORIS-II, and the TASSO experiment~\cite{Brandelik:1979cj} at $28 - 46.8 \; \rm{GeV}$~\cite{Brandelik:1979bv} at PETRA. 
TASSO collected data in the range $\sqrt{s} = 13 - 44 \; \rm{GeV}$, but the experiment was only instrumented with PID for the data subset at $34 \; \rm{GeV}$ and $44 \; \rm{GeV}$, where $34 \; \rm{pb}^{-1}$ were collected~\cite{Braunschweig:1988hv}.


Moving on to LEP, four intersection regions were hosting the ALEPH~\cite{Decamp:1990jra}, DELPHI~\cite{Aarnio:1990vx}, L3~\cite{L3:1989aa} and OPAL~\cite{Ahmet:1990eg} experiments. 
These experiments collected the integrated luminosity delivered by LEP of $160 \; \rm{pb}^{-1}$, and with the upgrade to LEP-II another $700 \; \rm{pb}^{-1}$. 
All experiments, with the exception of L3, had PID capabilities that allowed them to separate pions and kaons. 
OPAL had a jet chamber that could be used for $dE/dx$ measurements, ALEPH had a TPC, and DELPHI had a TPC as well as a ring imaging Cherenkov counter.

A description of the BES III experiment, which used $62 \; \rm{pb}^{-1}$ off-resonance data for FF analyses and was operating at BEPC, can be found in Ref.~\cite{Ablikim:2009aa}.


\subsection{SIDIS facilities and experiments}
Data from SIDIS experiments with polarized targets is in particular important for the extraction of polarization-dependent FFs. 
Additionally, data taken with effective proton or neutron targets provides information on the flavor dependence of the fragmentation process in the polarized as well as the spin-averaged case. 
Heavier target materials can be used to study in-medium fragmentation. 
Because the longitudinal momentum of the fragmenting quark at leading order can be reconstructed in lepton-nucleon scattering, data from SIDIS experiments is also valuable to extract the $z$ dependence of FFs.
Even though the transverse momentum of the initial state cannot be accessed, measuring the $P_{h\perp}$ dependence in SIDIS can provide valuable information to extract the transverse momentum dependence of PDFs and FFs when combined with $e^+e^-$ data.
We will concentrate on the experiments at DESY and CERN, since the data taken up to now at JLab with a $6 \; \rm{GeV}$ electron beam is not used for the extraction of FFs due to the low {\it cm} energy. 
HERA was the only accelerator in the world able to collide protons with either electrons or protons. 
It was operational between 1992 and 2007, providing a proton beam up to $920 \; \rm{GeV}$ and electron or positron beams at $27.5 \; \rm{GeV}$. 
It hosted the H1~\cite{Abt:1996hi} and ZEUS~\cite{ZeusDetector1993} experiments which were located at the interaction points of the lepton and hadron beams, whereas the HERMES~\cite{Ackerstaff:1998av} experiment used internal fixed gaseous polarized and unpolarized targets that could be brought into the lepton beam.
At CERN, the M2 beamline provided a secondary muon beam to a dynasty of experiments since the late 70's starting with the European Muon Collaboration (EMC), the New Muon Collaboration (NMC) the Spin Muon Collaboration (SMC) to the current generation, the COMPASS experiment.
For the study of FFs, in particular precision data from COMPASS and HERMES are of importance, since H1 and ZEUS lacked precise PID capabilities and thus did not make the study of FFs a focus of their research program. 
However, ZEUS and H1 published data on unidentified charged hadron multiplicities, as well as fragmentation into $K_S^0$ and heavy quarks. 
With $\sqrt{s} = 318 \; \rm{GeV}$, ZEUS and H1 represent the energy frontier for SIDIS experiments. 
In contrast to this HERMES operated at $\sqrt{s} = 5.2 \; \rm{GeV}$ only, which implies that the applicability of pQCD calculations to this data has to be carefully evaluated even with the $Q^2 > 1 \; \rm{GeV}$ cut that is usually applied. 
The same argument applies to a lesser degree to the COMPASS and EMC data.
The COMPASS data which are relevant for this review was taken with a $160 \; \rm{GeV}$ muon beam. 
There is some data from EMC on hadron multiplicities available that has been taken at even higher muon-beam energies of up to $280 \; \rm{GeV}$.

HERMES consisted of a forward spectrometer providing good tracking capability with relative momentum resolutions between 1.5\% and 2.5\%. 
A dual radiator ring imaging Cherenkov counter was used to provide good particle separation for momenta between $2 \; \rm{GeV}$ and $15 \; \rm{GeV}$~\cite{Shibata:2014dra}. 
Additionally, a pre-shower detector was used for hadron/lepton separation as well as a transition radiation detector and a lead-glass calorimeter.
For the extraction of light-quark FFs the data collected with polarized and unpolarized proton and deuterium targets are of interest. 
Average polarizations of around 85\% were achieved with both targets.
On the other hand, the COMPASS~\cite{Abbon:2007pq} setup was using the $160 \; \rm{GeV}$ muon beam on solid polarized deuterium ($^6$LiD) and proton ($\rm{NH}_3$) targets. 
The two-stage spectrometer covers a wide momentum range with $Q^2$ values up to about $100 \; \rm{GeV}^2$. 
In addition to a muon wall and electromagnetic and hadronic calorimetery, PID is provided by a RICH detector that enables charged $\pi/K$ separation up to momenta of $40 \; \rm{GeV}$. 
The availability of transversely polarized targets, good PID, and both proton and deuteron targets allowed both collaborations to study the spin and flavor dependence in the fragmentation of pions, kaons, and neutral mesons.
As noted above, there are some results from COMPASS predecessor experiments (SMC and EMC) available. 
However, the newer COMPASS results supersede these, so that they do not play a large role in current analysis of FFs with the exception of a small dataset from EMC that will be discussed in the results section.

\subsection{$pp$ facilities and experiments}
There exists a long history of experiments studying $pp$ collisions, in particular with fixed targets such as the experiments UA2, AFS, WA70, UA1 at CERN's ISR and SPS. 
For the study of FFs, the data with the most impact comes from experiments at the Tevatron, RHIC and, more recently, the LHC. 
As with the $e^+e^-$ and SIDIS configurations, the modern experiments used analysis techniques which are more amenable to modern global fits.
They also collected far more data, and the experimental capabilities are much improved. 
In particular, the experiments that focus on heavy-ion physics, like STAR~\cite{Ackermann:2002ad} at RHIC and ALICE~\cite{Aamodt:2008zz} at the LHC, have very good PID. 
In addition to poorer instrumentation, older (fixed-target) $pp$ data collected at lower energies pose problems for QCD fits, since the large $x_F$ and $x_T=\frac{P_{hT}}{2\sqrt{s}}$ accessed can imply significant higher-order corrections as already mentioned in Sec.~\ref{sec:observables}.
Therefore, below we will only describe the Tevatron, RHIC and LHC experiments.
At {\it cm} energies of about $1 \; \rm{TeV}$ (Tevatron), 62.4, 200 and $500 \; \rm{GeV}$ (RHIC) and $7 \; \rm{TeV}$ (LHC), these $pp$ machines provide the necessary phase space to extract FFs in a factorized QCD framework.
RHIC is the only machine that is able to provide polarized, both longitudinally and transversely, protons. 
Both RHIC and LHC collected data in proton-nucleus collisions allowing to probe nuclear effects in fragmentation, which are not a focus of this review.
The main experiments at the Tevatron were CDF~\cite{Acosta:2004hw,Abulencia:2005ix} and D0~\cite{Abazov:2005pn}. 
Both of these were optimized for jet physics and were lacking good PID capabilities for pions and kaons.
Therefore the data was mostly used for studies of heavy-quark fragmentation and jet fragmentation, both of which are not covered in this review.
At RHIC, the main experiments are PHENIX~\cite{Adcox:2003zm} and STAR~\cite{Ackermann:2002ad}. 
In particular, STAR with a central TPC is well equipped to identify charged hadrons and reconstruct jets in its barrel region, which covers a pseudorapidity region of about $-1 < \eta < 1$. 
PHENIX is well suited to detect electromagnetic energy from neutral mesons ($\pi^0$ and $\eta$) in the central region $|\eta|<0.35$. 
In the forward region, both experiments have electromagnetic calorimeters which allows the reconstruction of neutral mesons and purely electromagnetic jets. 
PHENIX covers roughly the range $3 < \eta < 4$, whereas STAR's coverage is $2 < \eta < 4$.
At the LHC, the ATLAS~\cite{Aad:2008zzm} and CMS~\cite{Chatrchyan:2008aa} experiments are again optimized for physics with jets and leptons at mid-rapidity, whereas the ALICE experiment which focuses on heavy-ion physics, similar to STAR, has a TPC at its center which allows to study the fragmentation of identified hadrons. 
However, so far ALICE recorded relatively little $pp$ data~\cite{LHCPerformance} so that the precision of the data collected at high enough $p_T$ is small compared to RHIC. 
In the future, data relevant for the study of FFs might also come from the LHCb~\cite{Alves:2008zz} experiment, which covers the region $2 < \eta < 5$ with high-precision tracking, PID and calorimetry.
The LHC experiments also provided a wealth of data on heavy-flavor production which, however, is not a focus of this review.

\subsection{Experimental data}
\label{sec:results}
After the introduction of the observables in Sec.~\ref{sec:observables} and the experiments in the first part of this section, we are now ready to give an overview of the available experimental results. 
We follow the same order as we have for the observables, i.~e., for each observable we start with results from $e^+e^-$ annihilation, continued with SIDIS and $pp$ results. 
However, we have split up the spin-dependent and spin-averaged TMD FF results, since the simultaneous experimental study of these two classes of TMDs is not common.

\subsubsection{Experimental data on integrated FF $D_1$}
\label{sec:expDataD1}
\begin{figure}[t]
\begin{center}
\includegraphics[width=0.65\textwidth]{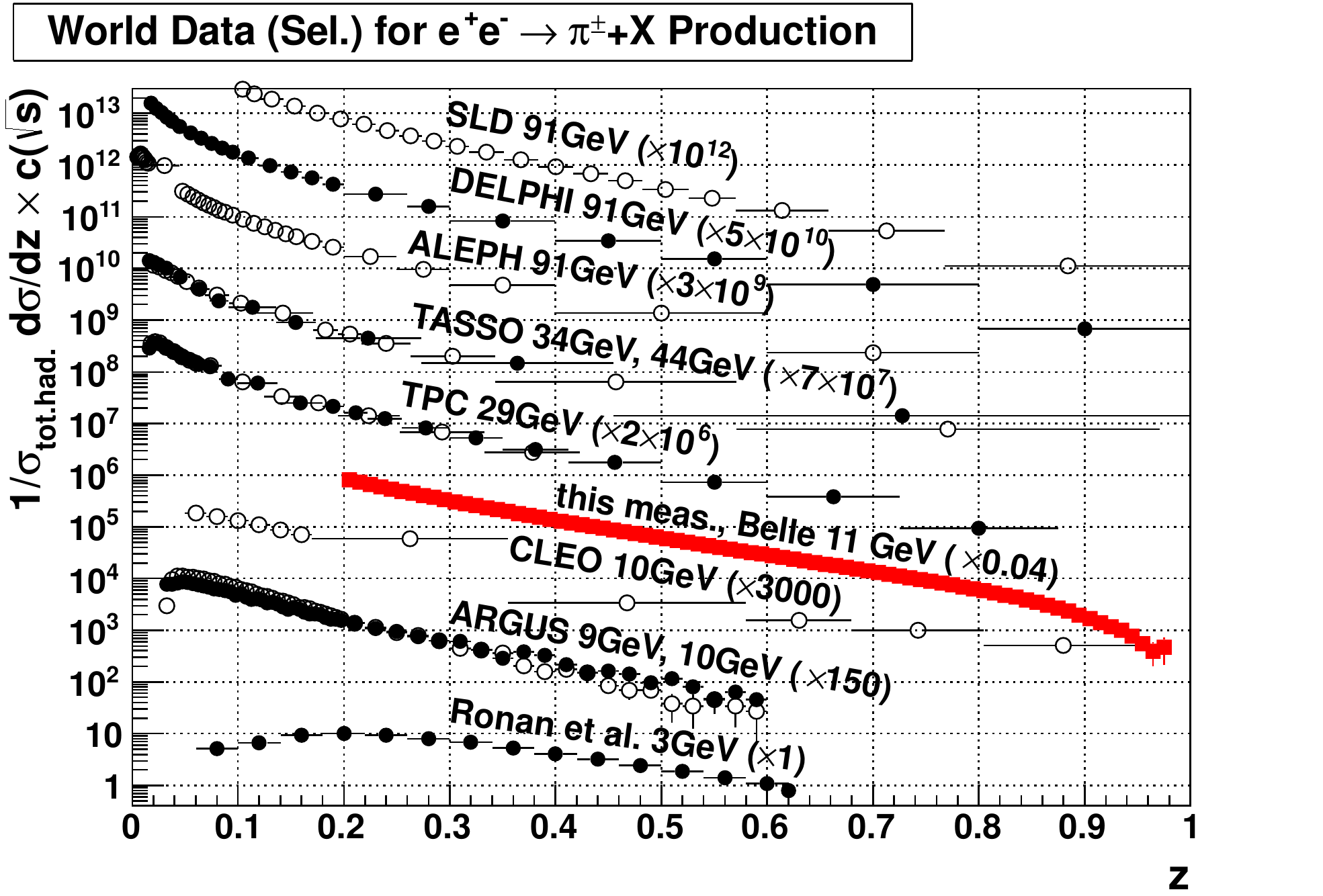}
\end{center}
\caption{World data for identified-pion cross sections in SIA. 
Figure reprinted with permission from~\cite{Leitgab:2013qh}~\href{http://dx.doi.org/10.1103/PhysRevLett.111.062002}{M.~Leitgab, et al., Phys.~Rev.~Lett.~111~(2013)~062002}. Copyright (2013) by the American Physical Society.
\label{fig:pionSIAXsection}}
\end{figure}

\paragraph{$e^+e^-$ experiments:}
The observables for which we have the most data on $D_1^{h/i}$ available are cross sections and multiplicities for unidentified charged hadrons as well as identified pion and kaon cross sections in $e^+e^-$ annihilation. 
In particular, we consider the SIA process described in Eq.~(\ref{e:xs_SIA}). 
Here we have results for unidentified charged hadrons available that were taken over a wide range of energies ($3 \; \rm{GeV} \le \sqrt{s} \le 183 \; \rm{GeV}$) at the facilities described above. 
A compilation  of inclusive particle production data taken till the mid '90s, so excluding the B-factory data, can be found in~\cite{Lafferty:1995}.
Significant data that is used in modern extractions of FFs of non-identified charged hadrons and neutral mesons is available from the TPC, TASSO, SLD, TOPAZ experiments~\cite{Lafferty:1995}.
Looking for identified hadrons, the number of available datasets shrinks somewhat, but since older experiments are anyway not that important for the extraction of FFs for the reasons outlined earlier this is not very concerning.
At the highest energies, we have pion and kaon multiplicities or cross sections from the LEP experiments DELPHI~\cite{Abreu:1998vq}, ALEPH~\cite{Buskulic:1994ft} and OPAL~\cite{Akers:1994ez} which had charged-particle identification.
Similar data exists from SLD~\cite{Abe:1998zs} which, as mentioned above, also took data with a longitudinally polarized beam. 
At intermediate energies, we have datasets on identified pions and kaons from the TPC/two-gamma experiment\cite{Aihara:1988su,Aihara:1986fz,Aihara:1986mv} and from TASSO~\cite{Braunschweig:1988hv}.
The most precise measurements by far on identified pion, kaon and proton cross sections have been done at the B-factories by the Belle~\cite{Leitgab:2013qh,Seidl:2015lla} and BaBar experiments~\cite{Lees:2013rqd}. 
Looking at datasets for neutral light mesons, which means for us $\pi^0$ and $\eta$, we have results from the LEP experiments ALEPH~\cite{Buskulic:1992hn,Barate:1999gb,Heister:2001kp}, L3~\cite{Adriani:1992hd,Acciarri:1994gza} and OPAL~\cite{Ackerstaff:1998ap}. 
From BaBar we also have available preliminary results on $\eta$ meson multiplicities~\cite{Anulli:2004nm}. 
Fig.~\ref{fig:pionSIAXsection} shows a selection of world data on the cross section for identified pions measured at SLD, DELPHI, ALEPH, TASSO, TPC/Two-gamma, Belle, CLEO, ARGUS and MARK-II~\cite{Ronan:1979cz}. 
The data covers a wide range in $\sqrt{s}$, and it illustrates the role the B-factories play in the study of fragmentation measurements with an improvement of the achieved precision by several orders of magnitude. 
This, in particular, allows to map out FFs also at high $z$, a kinematic region previously not accessible.
While generally in SIA it is challenging to address FFs for individual quark flavors, the lever arm in energy  also helps in order to exploit the different coupling strengths of $\gamma^*$ and $Z^0$ and thus gain some flavor sensitivity. 

There have been other approaches to provide some flavor sensitivity from $e^+e^-$ single-particle production.
We already mentioned the SLD analysis~\cite{Kalelkar:2000ig} with a polarized electron beam using the parity-violating coupling of the weak interaction as described in Sec.~\ref{sec:observables_integrated}.
Another approach is to tag the flavor of the jet, either based on the leading hadron in the jet to differentiate between different light quark flavors or tag heavy-quark fragmentation by identifying mesons containing heavy quarks, like $D$ and $B$ mesons, using displaced vertices and/or their reconstruction from decay products.
An example for flavor tagging using the leading hadron is the OPAL analysis in Ref.~\cite{Abbiendi:1999ry}, whereas the SLD analysis in~\cite{Abe:1998zs} is an example for a heavy-flavor-tagged analysis. 
The performed flavor tagging, in particular using the leading hadron, can only be interpreted at the lowest order in pQCD calculations and therefore does not play a role in current fits of FFs~\cite{deFlorian:2014xna}. 

A measurement that gives some information on the flavor of the fragmenting quark and can be interpreted in current pQCD frameworks is the observation of back-to-back hadrons. 
Recently, the Belle collaboration showed results for the observable described in Eq.~(\ref{eq:diHadB2B}) for back-to-back charged pion, kaon and proton pairs~\cite{Seidl:2015lla} which exploits the different flavor contributions to the different charge combinations. 
 
This result also includes the measurement of the cross section  of hadron pairs in the same hemisphere, which is sensitive to the DiFF $D_1^{h_1h_2/q}(z_1,z_2)$. 
As described earlier in Sec.~\ref{sec:observables_dihadron}, the caveat of this observable is that at NLO it also receives contributions from the convolution of single-hadron FFs~\cite{deFlorian:2003cg}. 
The (perturbative) single-hadron FF contribution and DiFF contributions might be separated if the invariant mass of the two-hadron system is measured~\cite{Zhou:2011ba}. 
This second scale then could allow a separation of the perturbative contribution (for large $M_h$) and the non-perturbative contribution.
In addition to tagging specific quark flavors, there was also an effort to understand differences in quark and gluon jets at LEP using three-jet events where two of the jets could be identified as coming from quarks by tagging bottom quarks~\cite{Abreu:2000nw,Abbiendi:2000cv,Buskulic:1995sw,Barate:1998cp}.
We also want to mention that the observation that gluon jets are in general broader than quark jets has implications for the study of gluons in, e.g..  $pp$ collisions.

\paragraph{SIDIS experiments:}
Switching to SIDIS, the relevant observables often are cross sections for hadron production off fixed targets which are sensitive to $D_1^{h/q}$ coupled to the respective PDFs as seen in Eqs.~(\ref{eq:SIDIS_F1}) and (\ref{eq:SIDIS_FL}).
Note that instead of the hadron production cross section, experiments often measure multiplicities, where the SIDIS cross section is normalized to the total DIS cross section. 
In this ratio, quantities that might not be known precisely, like the delivered luminosity, cancel.
\begin{figure}[t]
\begin{center}
\includegraphics[width=0.60\textwidth]{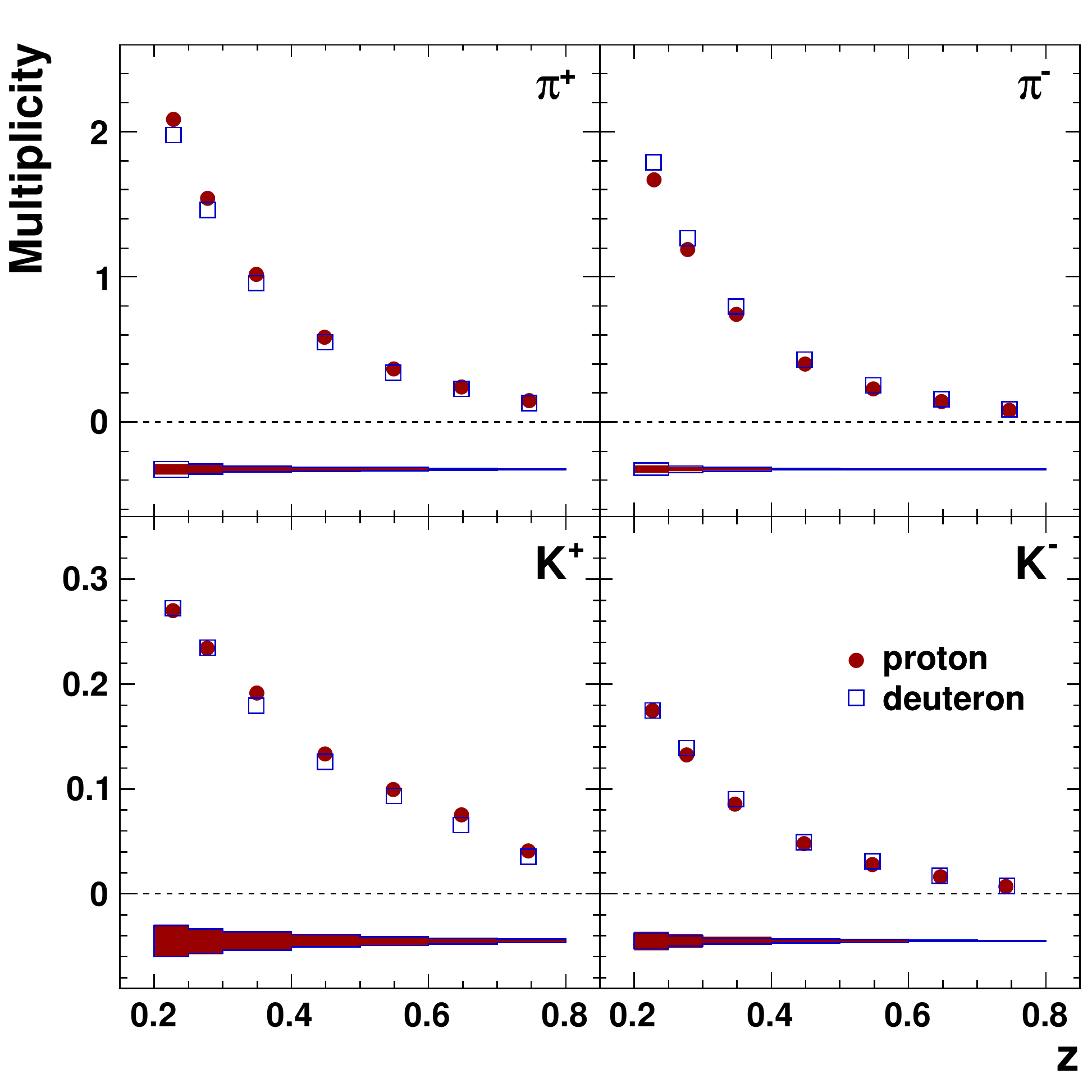}
\end{center}
\caption{Data on identified pion and kaon multiplicities vs.~$z$ from the HERMES experiment.  
Figure reprinted with permission from~\cite{Airapetian:2012ki}~\href{http://dx.doi.org/10.1103/PhysRevD.87.074029}{A.~Airapetian, et al., Phys.~Rev.~D87~(2013)~074029}. Copyright (2013) by the American Physical Society.
\label{fig:HermesMultiplicities}}
\end{figure}

In SIDIS, data for multiplicities of unidentified hadrons, integrated over their transverse momentum, is available at the highest $Q^2$ from the HERA experiments H1~\cite{Aaron:2007ds, Aaron:2009ae} and ZEUS~\cite{Breitweg:1999nt,Chekanov:2009kb,Abramowicz:2010rz}, and from the fixed-target experiments EMC~\cite{Ashman:1991cj} and COMPASS~\cite{Adolph:2012nm} (COMPASS in this case actually reports the cross section). 
The HERA experiments reach low scaled momenta in the current fragmentation region which allows the test of models for the so-called modified leading-log approximation (MLLA) --- see Sec.~\ref{sec:mlla} for more details --- and comparison to $e^+e^-$ experiments.
EMC published results for both $\mu p$ and $\mu d$ scattering, whereas COMPASS only showed the results obtained with their isoscalar target. 
Identified pions and kaons have been measured by EMC~\cite{Aubert:1985zd}, HERMES~\cite{Airapetian:2012ki}, and most recently COMPASS~\cite{Adolph:2016bga}.
Like the earlier EMC results, HERMES used both proton and deuteron targets, whereas COMPASS shows again only results from an isoscalar target. 
COMPASS, like the earlier EMC analysis, also performs a LO extraction of the favored and disfavored pion FFs which are in reasonable agreement with NLO extractions using the DSS, HKNS and LSS frameworks that will be discussed in detail in Sec.~\ref{sec:fits_integrated_D1}.
The HERMES results are shown in Fig.~\ref{fig:HermesMultiplicities}.

\paragraph{$pp$ experiments:}
Data from $pp$ collisions is essential to constrain the gluon FF because constraints derived from scaling violations in SIA datasets at different energies are usually not strong enough to determine the gluon FF with a similar precision as the light-quark FFs. 
This is quite impressively illustrated by comparing the extractions that do not use $pp$ data with those that do. 
Section~\ref{sec:global_fits} will discuss two examples: HKNS~\cite{Hirai:2007cx} which relies solely on SIA data, and DSS~\cite{deFlorian:2014xna} which also includes SIDIS and $pp$ data. 
The extracted gluon FFs are shown in Figs.~\ref{fig:HKNS} and~\ref{fig:dss}. 
Similar to $e^+ e^-$ annihilation, there is a long history of $pp$ scattering experiments, in particular with fixed targets. 
In fact, the first fragmentation models from Feynman and Field~\cite{Feynman:1977yr} were developed using $pp$ data from CERN's ISR. 
However, most older fixed-target $pp$ data are at relatively low transverse momenta and/or high $x_F$.
In this kinematic regime, higher-order corrections are normally substantial, and therefore the data is not used in modern global fits since higher-precision data from RHIC and the LHC became available, which also provide good PID. 
In particular, high-precision data on $\pi^0$ and $\eta$ production is available from PHENIX.
\begin{figure}[t]
\begin{center}
\includegraphics[width=0.60\textwidth]{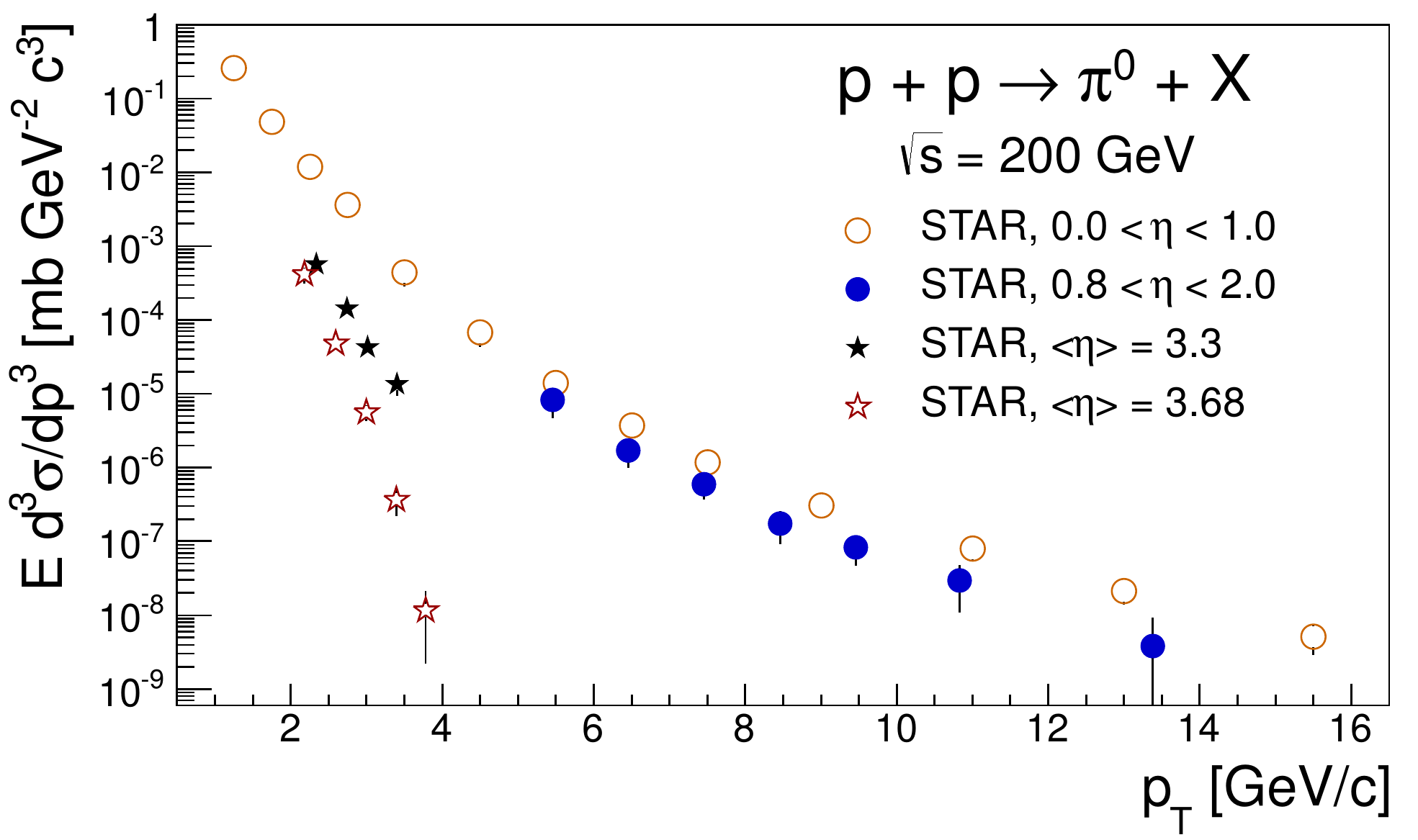}
\end{center}
\caption{Cross section for $\pi^0$ production measured by the STAR experiment at different values of mean pseudorapidity. 
Note that $p_T|_{\rm figure} = |\vec{P}_{h\perp}|$.
Figure reprinted with permission from~\cite{Adamczyk:2013yvv}~\href{http://dx.doi.org/10.1103/PhysRevD.89.012001}{L.~Adamczyk, et al., Phys.~Rev.~D89~(2014)~012001}. Copyright (2014) by the American Physical Society.
\label{fig:STAR_pi0}}
\end{figure}
 
The PHENIX experiment is especially well suited to measure $\pi^0$ and $\eta$ mesons at mid-rapidity ($|\eta| < 0.35$) and forward rapdity ($3 < \eta < 4$). 
Results were published for the cross section of $\pi^0$ at mid-rapidity at {\it cm} energies $62.4 \; \rm{GeV}$~\cite{Adare:2008qb}, $200 \; \rm{GeV}$~\cite{Adare:2007dg}, and $510 \; \rm{GeV}$~\cite{Adare:2015ozj}, as well as for $\eta$ meson cross sections at mid-rapidity at $200 \; \rm{GeV}$~\cite{Adare:2010cy}. 
In the forward region, PHENIX analyzed $\eta$ cross sections at 200~GeV~\cite{Adare:2014qzo}.
In addition to the neutral mesons, PHENIX also measured charged-hadron cross sections at $62.4 \; \rm{GeV}$~\cite{Adare:2012nq} and charged-pion cross sections at $200 \; \rm{GeV}$~\cite{Adare:2014wht}, both at mid-rapidity. 
STAR also published results on the cross section at $200 \; \rm{GeV}$ of $\pi^0$'s at mid-rapidity~\cite{Abelev:2009pb}, intermediate rapidity~\cite{Adamczyk:2013yvv}, as well as forward rapidity~\cite{Adamczyk:2012xd}. 
The later result also contains the cross section for $\eta$.
It should be mentioned that the forward results from both STAR and PHENIX are at relatively large $x_F$ (up to $x_F=0.75$), which again makes it difficult to analyze them using fixed-order pQCD calculations.
Compared to PHENIX, STAR has superior charged particle tracking and identification capabilities at midrapidity   ($|\eta|<1$) provided by its TPC and published results on multiplicities for identified pions, protons and kaons~\cite{Adams:2006nd,Agakishiev:2011dc}.
A compilation of the STAR results of $\pi^0$ cross sections, taken at different mean pseudorapidity $\eta$, are shown in Fig.~\ref{fig:STAR_pi0}.

\begin{figure}[t]
\begin{center}
\includegraphics[width=0.46\textwidth]{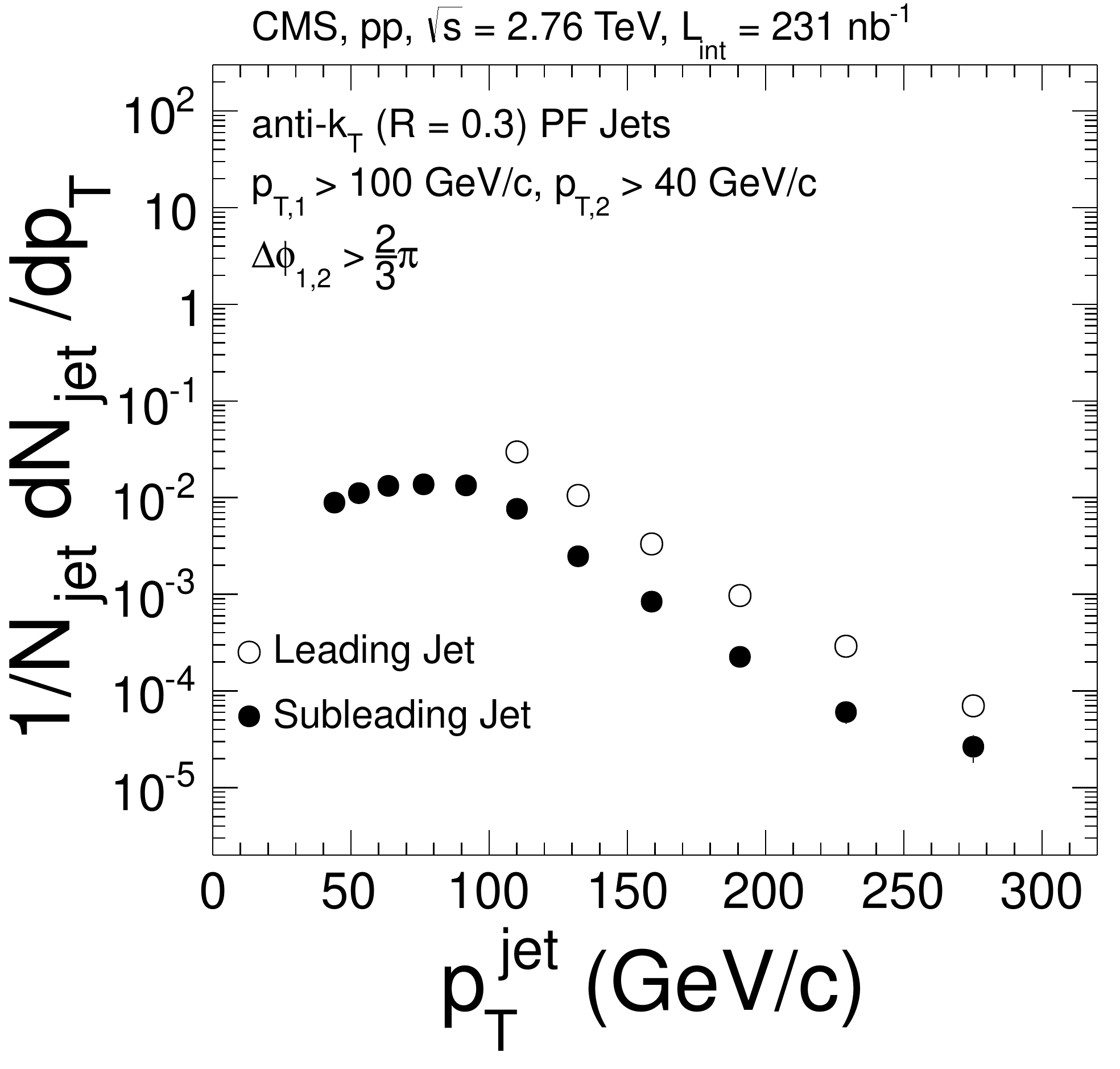}
$\hspace{0.5cm}$
\includegraphics[width=0.46\textwidth]{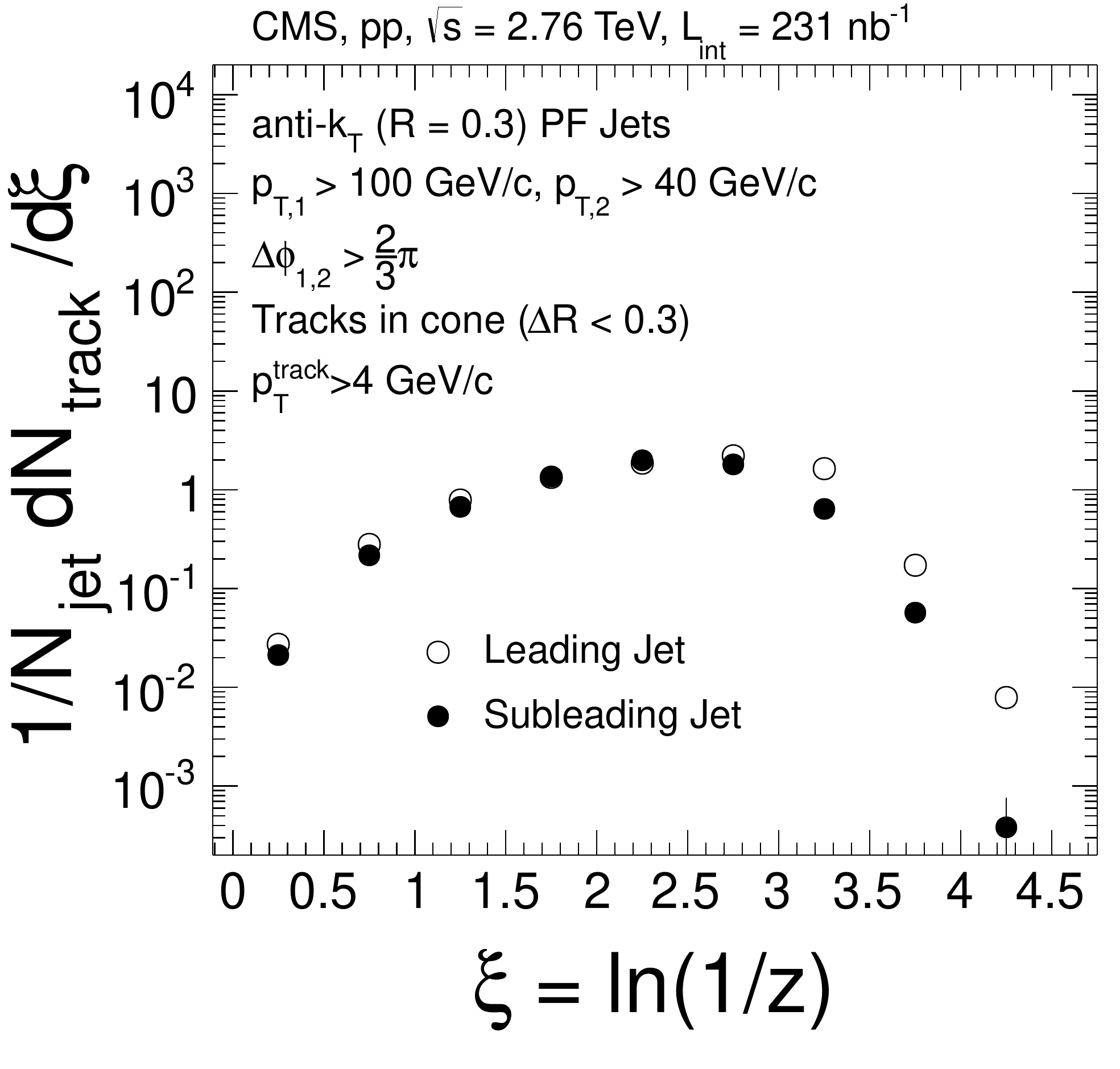}
\end{center}
\caption{Results on the measurement of hadrons in jets by CMS. Left panel: transverse momentum distribution of the leading and subleading jet. Right panel: dependence of the hadrons within the jet on the variable $\zeta$, which is related to $z$ by $\zeta=\ln(\frac{1}{z})$. In this case $z$ designates the fractional momentum of the hadron along the jet axis with respect to the jet momentum. 
Figure reprinted from~\cite{Chatrchyan:2012gw}.
\label{fig:cmsHadronJet}}
\end{figure}

From the LHC we have a number of new results. 
ALICE measured the cross section of $\pi^0$ and $\eta$ production at $\sqrt{s} = 0.9 \; \rm{GeV}$ and $7 \; \rm{TeV}$~\cite{Abelev:2012cn}, as well as unidentified charged hadrons~\cite{Adam:2015pza}.
There are also preliminary results on identified charged pions and kaons~\cite{Volpe:2015tda}.
Given the low integrated luminosity, the precision at high transverse momenta is quite low though.
The ATLAS experiment published multiplicities of unidentified charged hadrons at $900 \; \rm{GeV}$, $2.76 \; \rm{TeV}$, and $7 \; \rm{TeV}$ from significantly larger datasets~\cite{Aad:2010ac}.
A new development is the measurement of hadrons in jets at the LHC which, in a leading-order framework, allows a direct access to the $z$ dependence of FFs in $pp$. 
Fig.~\ref{fig:cmsHadronJet} shows the result on this observable measured at $\sqrt{s} = 2.76 \; \rm{TeV}$ by the CMS experiment~\cite{Chatrchyan:2012gw}. 
ATLAS did a similar analysis at $7 \; \rm{TeV}$~\cite{Aad:2011sc}.
Both RHIC and LHC are $pp$ colliders. 
From $p\bar{p}$ machines we also have some results on charged-particle multiplicities. 
At the Tevatron, CDF has results from $\sqrt{s} = 630 \; \rm{GeV}$ and $1.8 \; \rm{TeV}$~\cite{Abe:1988yu}, and from the UA1 and UA2 experiments, operating at CERN's SPS, results on unidentified hadron multiplicities have been published for {\it cm} energies between $200 \; \rm{GeV}$ and $900 \; \rm{GeV}$~\cite{Bocquet:1995jr,Albajar:1989an,Banner:1984wh}.

\subsubsection{Experimental data on TMD FF $D_1$}
\label{sec:data_TMD_D1}
\begin{figure}[t]
\begin{center}
\includegraphics[width=0.70\textwidth]{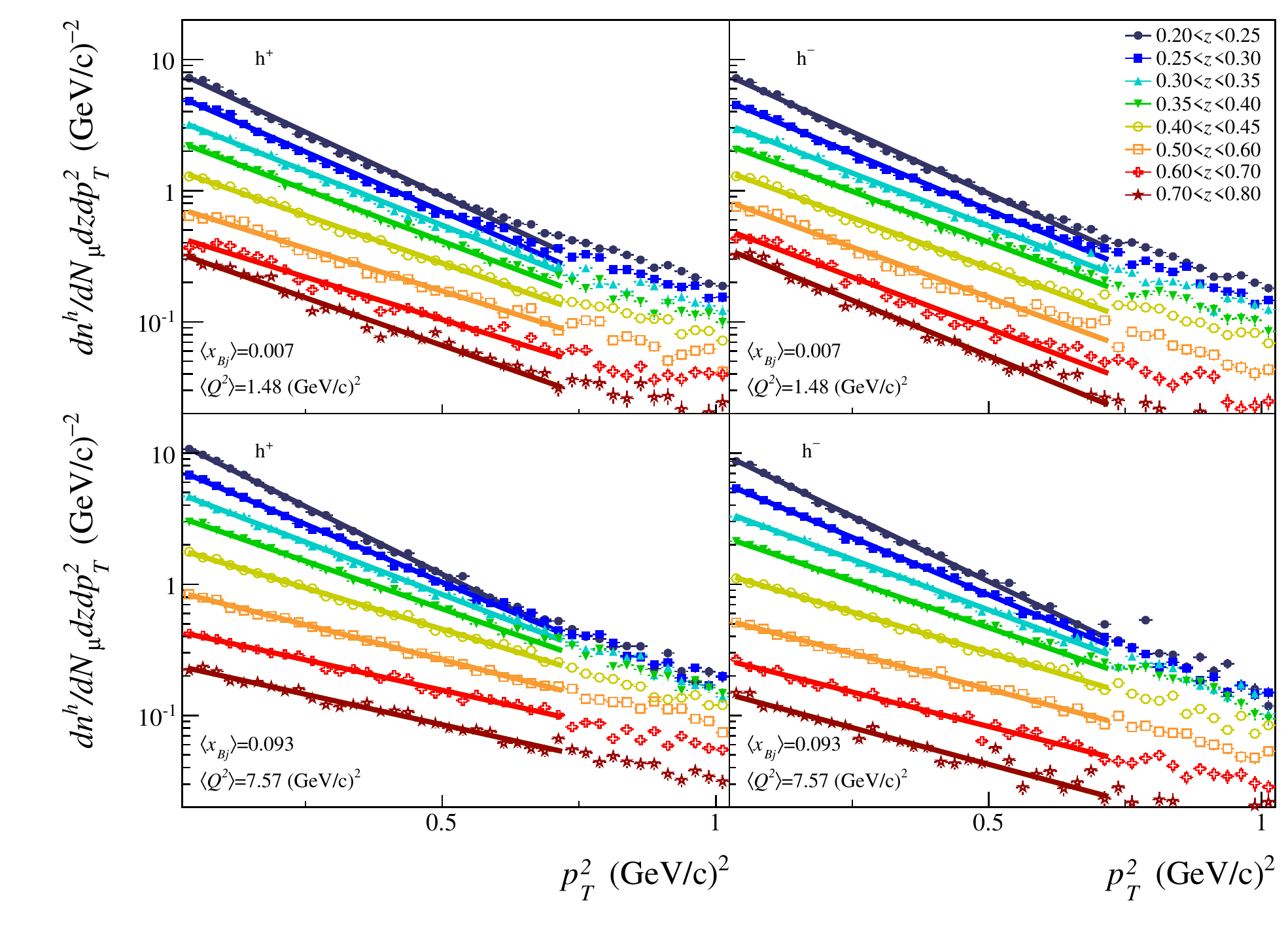}
\end{center}
\caption{Transverse-momentum dependence of unidentified hadrons for different $z$ bins measured by COMPASS~\cite{Adolph:2013stb}. 
Note that $p^2_T|_{\rm figure} = \vec{P}^2_{h\perp}$. 
Figure reprinted from~\cite{Adolph:2013stb} with permission from The European Physical Journal (EPJ). 
\label{fig:Compass_PtMultiplicities}}
\end{figure}

We discussed in Sec.~\ref{sec:observables_TMD} a number of observables that are sensitive to the intrinsic transverse-momentum dependence of FFs. 
As in the transverse-momentum integrated case, the cleanest access can be achieved in $e^+e^-$ annihilation, where there is no contribution from the poorly known $k_T$ dependence of PDFs. 
However, at present no data from $e^+e^-$ annihilation on the transverse-momentum dependence of spin-averaged FFs is available. 
But analyses from Belle data of back-to-back hadrons and hadrons in jets are underway.  
We do have though measurements of the transverse-momentum dependence of ratios of the spin-dependent FF $H_1^\perp$ and of $D_1$ like the one shown in Eq.~(\ref{eq:DR}) observed in so-called double ratios which are discussed in Sec.~\ref{sec:ObservablesSpinDepFFs}.
In SIDIS, the transverse-momentum dependence of FFs can be accessed via the $P_{h\perp}$ dependence of hadron multiplicities $M^h$ which were measured recently by the HERMES Collaboration~\cite{Airapetian:2012ki} and COMPASS Collaboration~\cite{Adolph:2013stb}.
The definitions used in both experiments differ slightly and are given by
\begin{equation}
M^h(x_B,Q^2,z_h,P_{h\perp}^2) \big|_{\rm COMPASS} = 
\frac{1}{2P_{h\perp}} \, M^h \big|_{\rm HERMES} = 
\frac{1}{\frac{d^2\sigma_{\rm DIS}(x_B,Q^2)}{dx_B \, dQ^2}} \, \frac{d^4\sigma_{\rm SIDIS}(x_B,Q^2,z_h,P_{h\perp}^2)}{dx_B \, dQ^2 \, dz_h \, dP_{h\perp}^2} \,.
\end{equation}
Results for this observable from COMPASS~\cite{Adolph:2013stb} are shown in Fig.~\ref{fig:Compass_PtMultiplicities}, again for unidentified hadrons off a deuteron target. 

The data is in good agreement with the Gaussian hypothesis for the transverse-momentum dependence up to intermediate $|\vec{P}_{h\perp}|$ below 1~GeV. 
At higher $|\vec{P}_{h\perp}|$, the transverse-momentum dependence flattens out, which is consistent with the broadening expected from pQCD. 
There is also a $z$ dependence, with higher $z$ having a broader $|\vec{P}_{h\perp}|$ distribution.
The transverse-momentum dependent multiplicities are also available from HERMES for identified pions and kaons~\cite{Airapetian:2012ki}.

In $pp$ collisions, PHENIX measured the $\vec{p}_T$ imbalance of charged hadrons with a trigger $\pi^0$ or direct photon in the opposite hemisphere~\cite{Adare:2010yw}. 
At LO it can be assumed that the direct $\gamma$ balances the fragmenting parton, and thus the $\gamma-h^\pm$ correlations can be used to explore the $k_T$ dependence of $D_1^{h/q}$. 
The PHENIX results are consistent with a Gaussian $k_T$ dependence but also show a tail at large $\vec{k}_T$ which might be consistent with what is expected from TMD evolution. 
A word of caution is in order here.
TMD factorization is expected to be broken for the $\gamma - h$ final state, like is the case for the production of back-to-back hadrons in hadronic collisions~\cite{Rogers:2010dm} --- see also the brief discussion in Sec.~\ref{sec:observables_TMD}. 
On the other hand, at present hardly anything is known about the numerical impact of TMD factorization breaking.
In that regard such correlations in $pp$ can actually be very helpful.

\subsubsection{Experimental data on TMD FF $H_1^\perp$}
\label{sec:ObservablesSpinDepFFs}
Observables for FFs that are sensitive to the polarization of the fragmenting parton as described in Sec.~\ref{sec:observables_TMD} can be roughly divided in two categories. 
One category is comprised of observables where the FF couples to a spin-dependent PDF, which describes the probability that a quark with a given polarization carries a certain momentum.
Examples are observables in SIDIS (see Eq.~(\ref{e:Collins_SSA_SIDIS})) and $pp$ where the transverse-polarization dependent FFs couple to the transversity distribution, or its gluon equivalent. 
However, these (spin-dependent) PDFs itself are poorly known, and their knowledge is in turn dependent on a precise extraction of the FFs.
This makes the second category where the observable is sensitive to the product of two polarization-dependent FFs essential to study the polarization dependence. 
In particular, precision measurements in unpolarized $e^+e^-$ annihilation, where polarization-dependent FFs can be accessed in correlation measurements of back-to-back hadrons or hadron pairs, made possible the first extraction of polarization-dependent FFs. 

\begin{figure}[t]
\begin{center}
\includegraphics[width=0.45\textwidth]{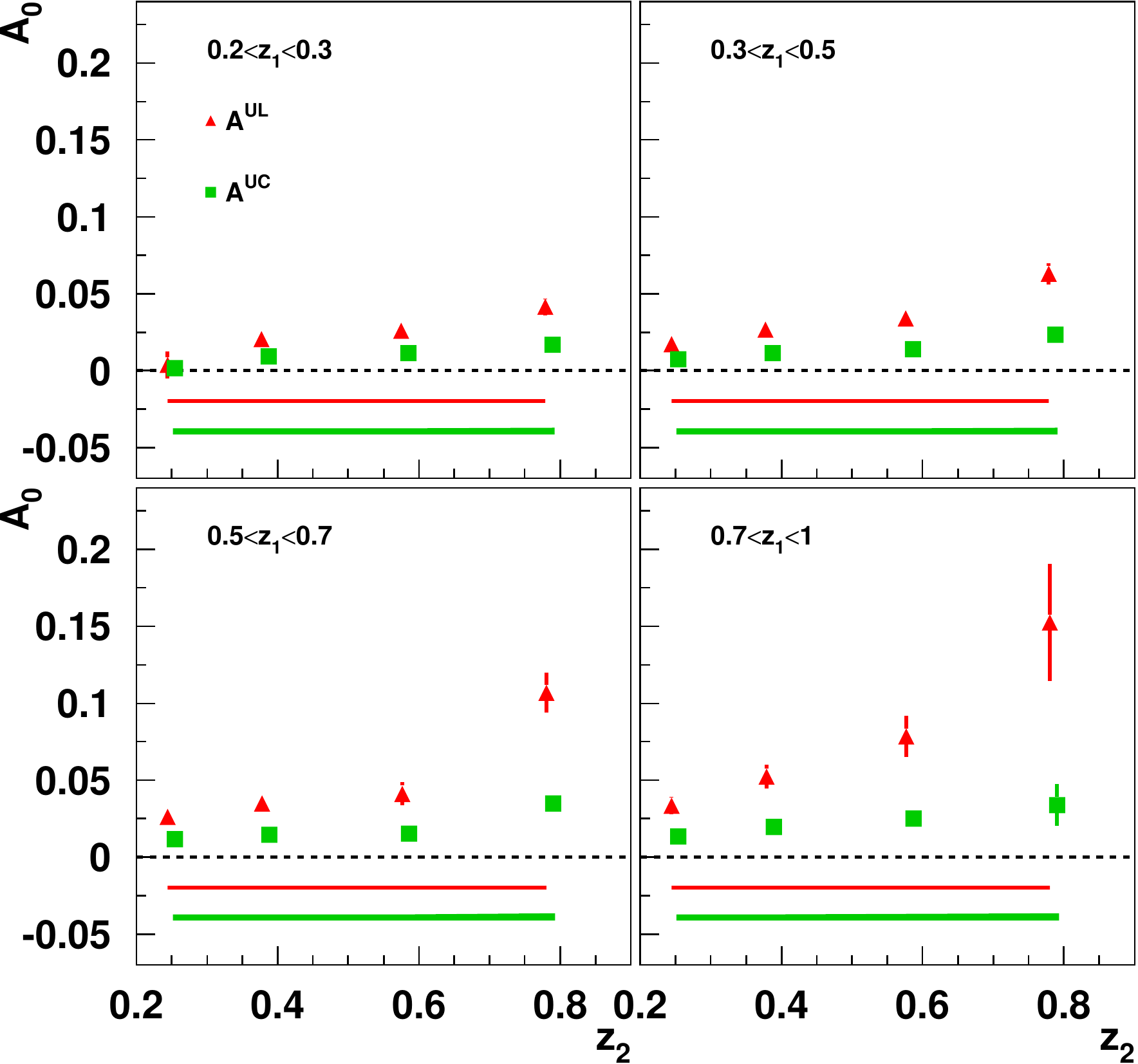}
$\hspace{0.5cm}$
\includegraphics[width=0.45\textwidth]{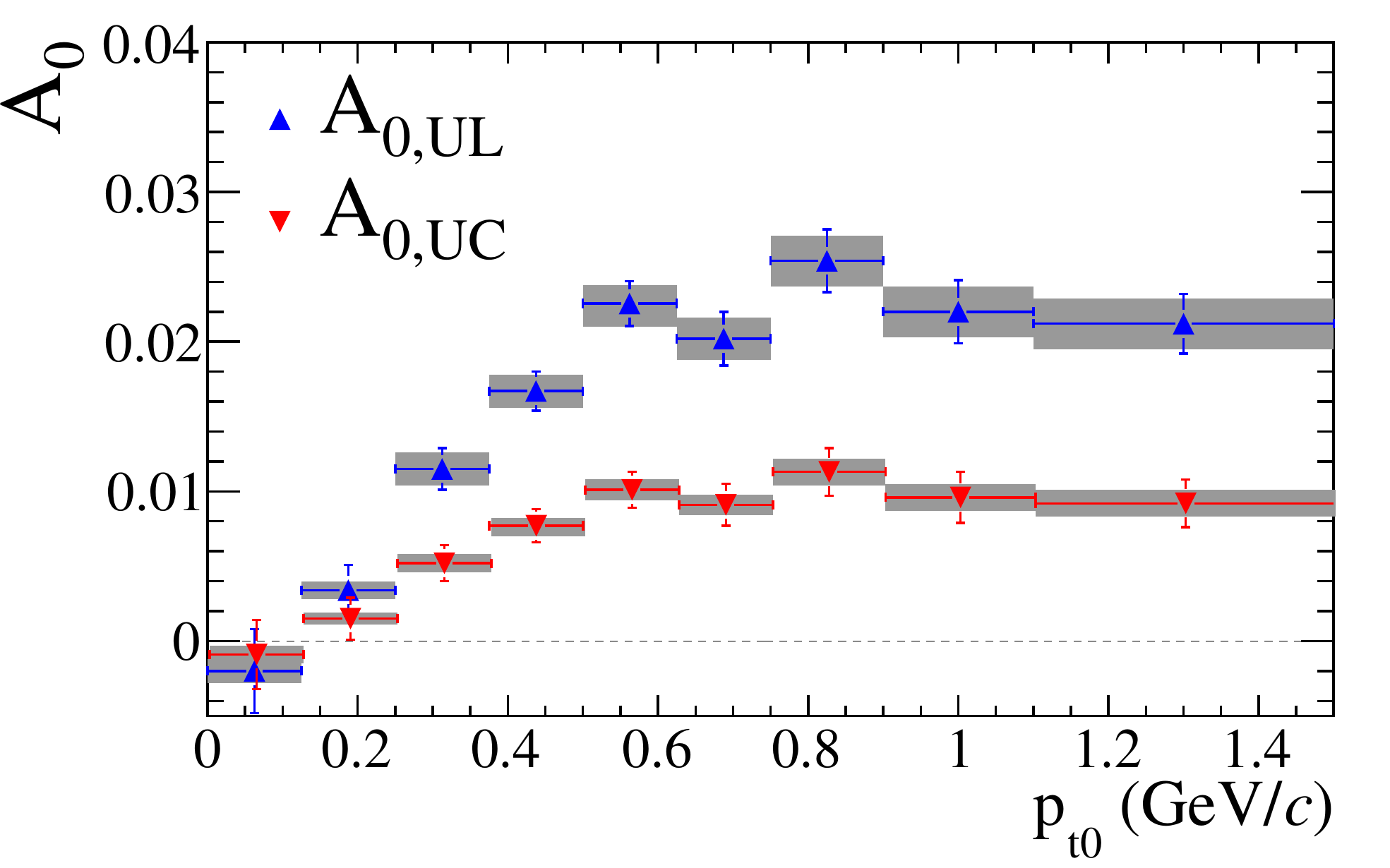}
\end{center}
\caption{Results on Collins asymmetries vs $z$ from Belle (left) and vs $|\vec{P}_{h\perp}|$ (denoted by $p_{t0}$ in the figure) from BaBar (right)\label{fig:SIA_Collins}. The observable $A_0$ shown corresponds to the amplitude of the cosine modulation of the cross section in Eq.~(\ref{eq:back2backHadrons}) in the angle $\phi_0$. Both experiments show double ratios for unlike sign pairs over like sign ($A_{UL}$) and unlike sign over charge-integrated pairs ($A_{UC}$). 
Left figure reprinted with permission from~\cite{Seidl:2008xc}~\href{http://dx.doi.org/10.1103/PhysRevD.78.032011}{R.~Seidl, et al., Phys.~Rev.~D78~(2008)~032011}. Copyright (2008) by the American Physical Society.
Right figure reprinted with permission from~\cite{TheBABAR:2013yha}~\href{http://dx.doi.org/10.1103/PhysRevD.90.052003}{J.~Lees, et al., Phys.~Rev.~D90~(2014)~052003}. Copyright (2014) by the American Physical Society.}
\end{figure}

\paragraph{Results on $H_1^\perp$ in $e^+e^-$:}
The observables in $e^+e-$ annihilation described in Eqs.~(\ref{eq:back2backHadrons}) and~(\ref{eq:SIACollinsTwoScales}) in Sec.~\ref{sec:observables_TMD} are sensitive to the product of the quark and anti-quark FFs of a $q\bar{q}$ pair which is produced in the decay of the virtual photon leading to correlated spins.
Since the product of two potentially small effects is measured, the respective experiments require more statistics than measurements of angular-averaged cross sections. 
In addition, spin-dependent effects are expected to be largest at high $z$ where the detected hadron(s) carries a large fraction of the fragmenting quark's momentum and spin information but the cross section is small.
For these reasons the B-factories Belle and BaBar played a pioneering role in the measurement of the transverse-spin dependent single-hadron FFs and DiFFs $H_1^\perp$ and $H_1^\open$. 
Belle clearly observed for the first time azimuthal correlations of back-to-back pion pairs related to the Collins effect~\cite{Abe:2005zx,Seidl:2008xc}, after DELPHI reported a first study of this effect~\cite{Efremov:1998vd}. 
The Belle result, which is for so-called double ratios of the asymmetries which are defined in the following, is shown in Fig.~\ref{fig:SIA_Collins}. 
The BaBar Collaboration confirmed these results~\cite{TheBABAR:2013yha}, and in addition provided results differential in the transverse momentum $|\vec{P}_{h\perp}|$ and fractional energy $z$ of the hadrons simultaneously. 
The $|\vec{P}_{h\perp}|$ dependent results by the BaBar Collaboration are also shown in Fig.~\ref{fig:SIA_Collins}. Asymmetries for pairs including charged kaons have also been published by the BaBar Collaboration~\cite{Aubert:2015hha}. 

A key experimental technique to measure spin-dependent quantities with high precision is to form asymmetries between different polarization states in which the spin-independent part of the cross-section as well as detector acceptance effects cancels. 
In unpolarized $e^+e^-$ though, this is not directly possible. 
Instead, the observables that are used are ratios of the normalized azimuthal-dependent cross sections (\ref{eq:back2backHadrons}),~(\ref{eq:SIACollinsTwoScales}) for different charge combinations, so-called double ratios. 
This way, non-charge dependent effects from acceptance and gluon radiation cancel out, but, since the physics effect is flavor-dependent, it persists in the azimuthal dependence of the double ratio. 
Both, Belle and BaBar show amplitudes of the fits to the modulations of the ratios of unlike sign pion pairs over like sign pairs ($A_{UL}$) and unlike sign over charge-integrated pairs ($A_{UC}$) that correspond to the polarization-dependent FF to be extracted. 
For example, using double ratios of the cross section from Eq.~(\ref{eq:back2backHadrons}), the $\cos(2\phi_0)$ modulation would be fitted to extract the values for $A_{0,UL}$ and $A_{0,UC}$ shown in Fig.~\ref{fig:SIA_Collins}.
Since both, denominator and numerator of the double ratio already correspond to charge-summed products of FFs for quark and antiquark, the double ratio contains an even more involved combination of FFs that can be used as an input for the globals fits discussed in Sec.~\ref{sec:global_fits}.
For example, 
\begin{equation}
A_{0,UL} \propto \frac{\sum_q e^2_q \, \Big( H_1^{\perp \, \textrm{fav}} \, \bar{H}_1^{\perp \, \textrm{fav}} 
+ H_1^{\perp \, \textrm{dis}} \, \bar{H}_1^{\perp \, \textrm{dis}} \Big)}
{\sum_q e^2_q \, \Big( H_1^{\perp \, \textrm{fav}} \, \bar{H}_1^{\perp \, \textrm{dis}}
+ H_1^{\perp \, \textrm{dis}} \, \bar{H}_1^{\perp \, \textrm{fav}} \Big)} \,,
\label{eq:DR}
\end{equation}
where we used the notation of favored and disfavored FFs in order to simplify the expression.
The convolution over the transverse momentum dependence of $H_1^\perp$ is implied.

The asymmetries rise with $z$, confirming the naive expectation that high $z$ hadrons carry more of the original polarization information. 
The asymmetries also rise with $|\vec{P}_{h\perp}|$, fulfilling the boundary condition of a vanishing Collins effect at $|\vec{P}_{h\perp}|=0$, up to intermediate $|\vec{P}_{h\perp}|$ where they flatten out. 
For large $|\vec{P}_{h\perp}|$ one would expect the asymmetries to fall again, because at fixed $\sqrt{s}$ large $|\vec{P}_{h\perp}|$ correspond to low $z$, and at $|\vec{P}_{h\perp}| > 1 \; \rm{GeV}$ one would expect the asymmetries to be diluted by the gluon radiation generating the large  transverse momenta. 
However, the observed asymmetries flatten out and the acceptance of the experiment is not enough to observe the full $|\vec{P}_{h\perp}|$ shape.
The asymmetries of charged $\pi K$ pairs measured by BaBar are consistent with those for charged pion pairs. 
Interestingly, the like sign over unlike sign asymmetries for charged kaons are larger than the ones for $\pi K$ or $\pi\pi$ pairs. 
This is not true for the like sign over charge-integrated asymmetries which are consistent again with the combinations involving pions. 
This might point to different amplitudes for the favored and disfavored kaon Collins FF. 
However, the double ratio receives contributions from different combinations of favored and disfavored Collins FFs of up, down and strange quarks which makes the interpretation of the result difficult. 
More constraints can come from SIDIS and $pp$ measurements described below, where kaons also exhibit a different behavior than pions. 

The $|\vec{P}_{h\perp}|$ dependence of the asymmetry is obviously sensitive to the intrinsic transverse-momentum dependence of the TMD Collins FF. 
The scale dependence (TMD evolution) of the intrinsic transverse momentum of TMD quantities has recently been the focus of theoretical interest --- see also Sec.~\ref{sec:tmd_evolution}. 
In this respect the recent measurement by the BES-III Collaboration of the $z$ and $|\vec{P}_{h\perp}|$ dependence of the Collins effect at the lower {\it cm} energy of $\sqrt{s} = 3.65 \; \rm{GeV}$~\cite{Ablikim:2015pta} is important to explore the effects of TMD evolution when compared to the Belle and BaBar results.
Other than at high $z$, where the acceptance of the experiment might be biased towards a small volume in phase space, the BES-III results are consistent with calculations including TMD evolution~\cite{Ablikim:2015pta}.

\paragraph{Results on $H_1^\perp$ in SIDIS:}
Even though they reach very high precision, the measurements in $e^+e^-$ annihilation are not well suited to access the flavor dependence of the Collins function.
Other information, such as the absolute sign, is also not accessed in the $e^+ e^-$ correlation measurements. 
Instead, one can use information from SIDIS experiments with a transversely polarized target and $p^\uparrow p$ scattering, where $H_1^{\perp \, h/q}$ couples to the transversity distribution $h_1^{q/p}$ and can be accessed in transverse SSAs, as discussed in Sec.~\ref{sec:observables_TMD}. 
For SIDIS the relevant transverse SSA (Collins asymmetry) is schematically defined in~(\ref{e:Collins_SSA_SIDIS}).
Both, HERMES and COMPASS performed measurements of Collins transverse SSA on proton targets~\cite{Airapetian:2004tw,Airapetian:2010ds,Alekseev:2008aa,Adolph:2012sn,Adolph:2014zba} for charged pions and kaons. 
\begin{figure}[t]
\begin{center}
\includegraphics[width=0.50\textwidth]{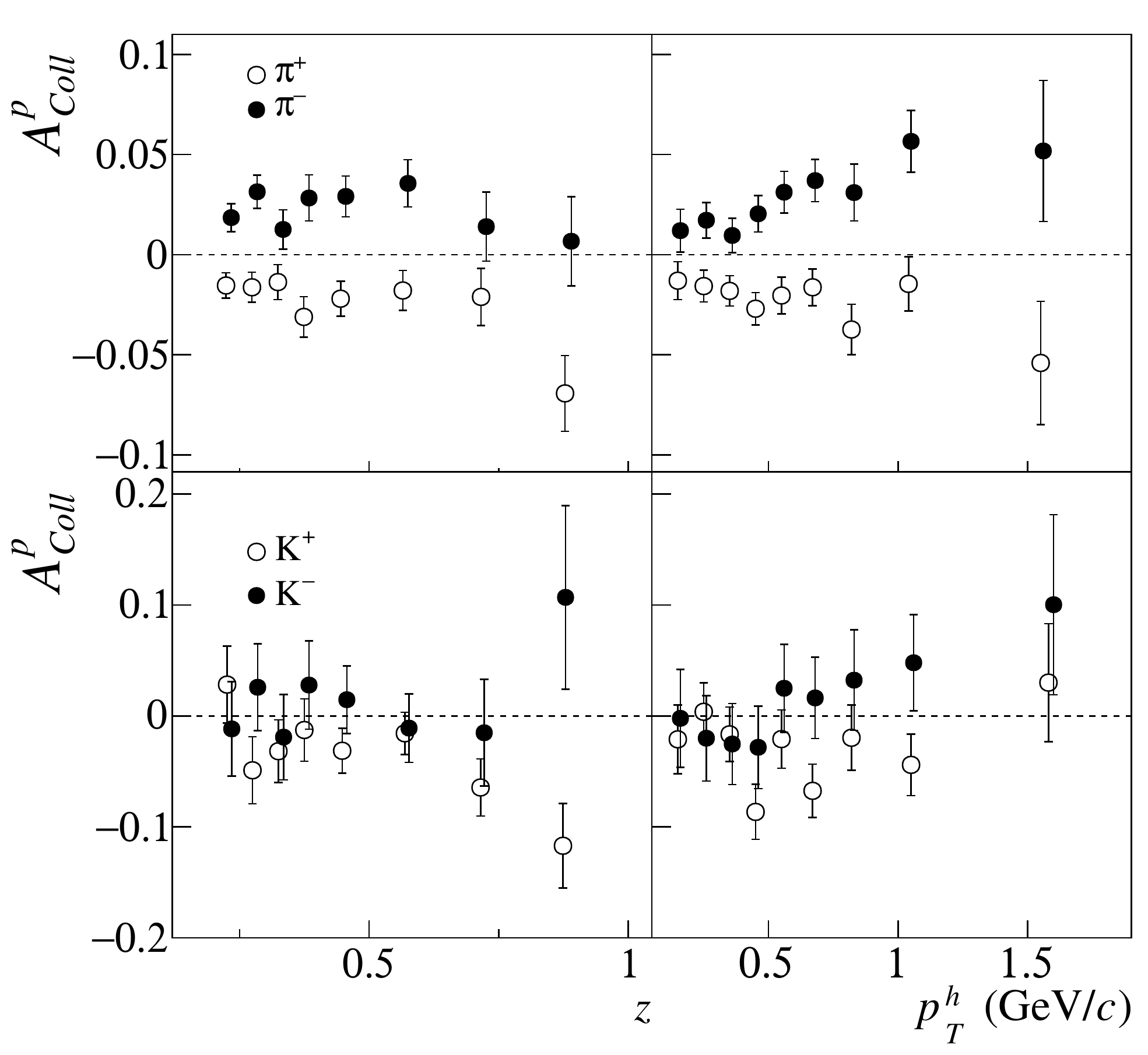}
\end{center}
\caption{Collins asymmetries measured in SIDIS off a proton target at COMPASS\label{fig:CompassCollins}. The results are shown versus $z$ and the transverse momentum $|\vec{P}_{h\perp}|$, denoted by $p_T^h$ in the figure, of the produced hadron. The asymmetry shown is sensitive to the Collins FF coupled to transversity. 
Figure reprinted from~\cite{Adolph:2014zba}.}
\end{figure}

Fig.~\ref{fig:CompassCollins} shows the results from COMPASS for the pion and kaon asymmetries off a proton target. 
The results for charged pions are consistent with the HERMES proton result, which showed the Collins effect for the first time albeit with significantly lower precision. 
Significant asymmetries can be observed for both charges with similar magnitudes that rise with the fractional energy $z$ the hadron is carrying as well as its transverse momentum $|\vec{P}_{h\perp}|$. 
This behavior is consistent with what we saw in the $e^+e^-$ results from Belle and BaBar above. 
The sign of the $\pi^+$ and $\pi^-$ asymmetries are reversed and there is some hint that the $\pi^-$ asymmetries are larger than the signal for $\pi^+$. 
This is consistent with the favored and disfavored Collins FFs being of similar magnitude and opposite sign.
For the charged kaons the precision of the measurements is lower than for the pions. 
For both HERMES and COMPASS, the positive kaons have the same sign as the asymmetries for the $\pi^+$, but the magnitude of the kaon asymmetries seems to be larger. 
COMPASS sees $K^-$ asymmetries that are consistent with the pion asymmetries of the same sign, however at HERMES there are hints that $K^-$ asymmetries have the opposite sign of the $\pi^-$ asymmetries. 
It is not quite clear what causes the behavior, but given the fact that HERMES and COMPASS are measuring at different $\sqrt{s}$, this might be a higher-twist effect.
HERMES also measured asymmetries for neutral pions, whereas COMPASS measured asymmetries for neutral kaons. 
Both results are consistent with zero. 
In particular, the small $\pi^0$  asymmetry is again consistent with favored and disfavored Collins FFs of similar magnitude but opposite sign.
COMPASS additionally reports results off the deuteron target where no significant asymmetries are observed, as well as asymmetries of charged hadrons off proton and deuteron targets~\cite{Alexakhin:2005iw,Ageev:2006da,Alekseev:2010rw}.
This can be interpreted as being caused by the isoscalar nature of the target and the opposite sign for the u-quark and d-quark transversity. 

\paragraph{Results on $H_1^\perp$ in $pp$:}
In $p^\uparrow p$, the measurements of TMD FFs into unpolarized hadrons, like the Collins effect are more challenging since the spin dependence is generally exhibited as an azimuthal correlation of the hadron momentum around the fragmenting quark axis, as described in Eq.~(\ref{eq:upolPPTMD}). 
This means that the outgoing quark axis has to be reconstructed. 
In $pp$, an outgoing quark initiates a jet that can be observed and the jet axis can be taken as  an approximation for the quark axis, even though this simple picture only holds at LO. 
In order to use these observables for an extraction of the FFs, a careful study of the connection of jet observables to quantities in QCD calculations has to be done. 
If one tries to do the LO identification of the jet axis with the quark axis, like is done for the observables in Sec.~\ref{sec:observables_TMDFF_pp}, then among the relevant systematics are the smearing of the jet axis with respect to the outgoing quark axis, the probability that a reconstructed jet matches to an outgoing quark and the change in the underlying partonic process that is caused by triggering on high-energy jets. 
This disadvantage of the study of FFs in $pp$ is balanced by the access to a complementary kinematic regime, in particular at high momentum transfers compared to SIDIS experiments. 
\begin{figure}[t]
\begin{center}
\includegraphics[width=0.55\textwidth]{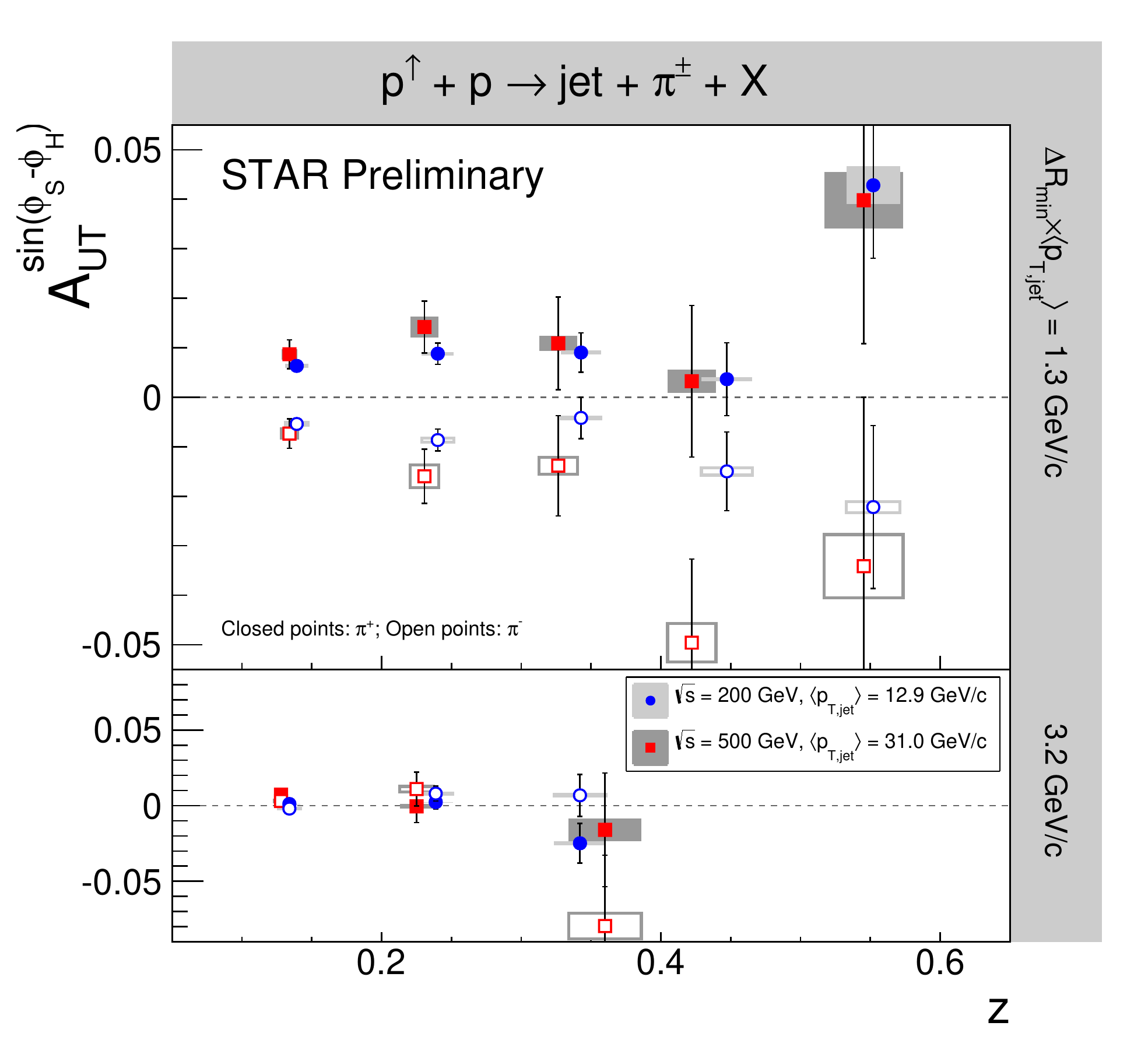}
\end{center}
\caption{Collins asymmetries from the STAR experiment at $\sqrt{s}=200$~GeV and $\sqrt{s}=500$~GeV. Results at forward rapidities (pseudorapidity region $0<\eta<1$ relative to the polarized beam) are shown. The transverse momentum ranges of the jet have been chosen to sample similar parton $x$ values and the angular cuts have been chosen to sample similar $|\vec{P}_{hT}|$ values.
Significant asymmetries are observed that do not exhibit a dependence on $\sqrt{s}$.
\label{fig:StarCollins}}
\end{figure}

It is also worth mentioning that in the LO approximation, compared to the Collins effect in SIDIS, there is no convolution of the transverse-momentum dependence of PDF and FF for the hadron-in-jet measurement since the momentum of the outgoing parton is reconstructed.
Up to now, only the STAR experiment has observed the Collins effect for jets in $pp$ collisions at $\sqrt{s} = 200 \; \rm{GeV}$ and $\sqrt{s} = 500 \; \rm{GeV}$~\cite{Aschenauer:2015eha,Drachenberg:2014txa}. 
The results are shown in Fig.~\ref{fig:StarCollins}.

They show a significant effect which is consistent between the two energies. 
So far, the STAR results are not used in global fits to the Collins FF. 
However, one interesting observation is that there seems to be no evidence for strong evolution effects due to TMD evolution if data points with corresponding $x_T$ and $|\vec{P}_{h\perp}|$ (called $j_T$ by STAR) are compared. 
Even though a cancellation between the TMD evolution effects in $H_1^\perp$ and the spin-independent FF $D_1$ might play a role, this behavior is different from the situation at lower energies, where asymmetries measured at Belle/BaBar are significantly smaller than those measured at BES-III. 
However it is not quite clear if this is mainly an effect of the covered $z$ and $|\vec{P}_{h\perp}|$ ranges. 
In addition to the Collins measurement, STAR has results for the "Collins-like" FF $H_1^{\perp \, h/g}$, the analogue of $H_1^{\perp \, h/q}$ for the fragmentation of linearly polarized gluons. 
No significant signal is observed~\cite{Drachenberg:2014jla}, which most likely is caused by the small magnitude of this particular gluon FF~\cite{D'Alesio:2010am}.

\subsubsection{Experimental data on transverse single-spin asymmetry $A_N$}
\label{sec:twist3FF_data}
At present, the most important observable for twist-3 FFs is the single-scale transverse SSA $A_N$ discussed in Sec.~\ref{sec:observables_twist3}. 
It is the simplest transverse SSA that can be defined and has been observed in $l p^{\uparrow} \rightarrow h X$ and $p^\uparrow p\rightarrow h X$ over a wide range of $\sqrt{s}$.
Here we will concentrate on the results in $p^\uparrow p$ where $A_N$ has been measured at high enough value of the relevant scale, that is the transverse momentum $|\vec{P}_{h\perp}|$ of the produced hadron with respect to the beam axis, such that the cross-section can be described in a factorized QCD approach. 
It should be mentioned though, that there are some hints that $A_N$ receives contributions from diffractive processes at RHIC energies~\cite{Aschenauer:2016our}. 
We also mention that results from SIDIS are available as well, e.g. from HERMES~\cite{Airapetian:2013bim}. 

\begin{figure}[t]
\begin{center}
\includegraphics[width=0.98\textwidth]{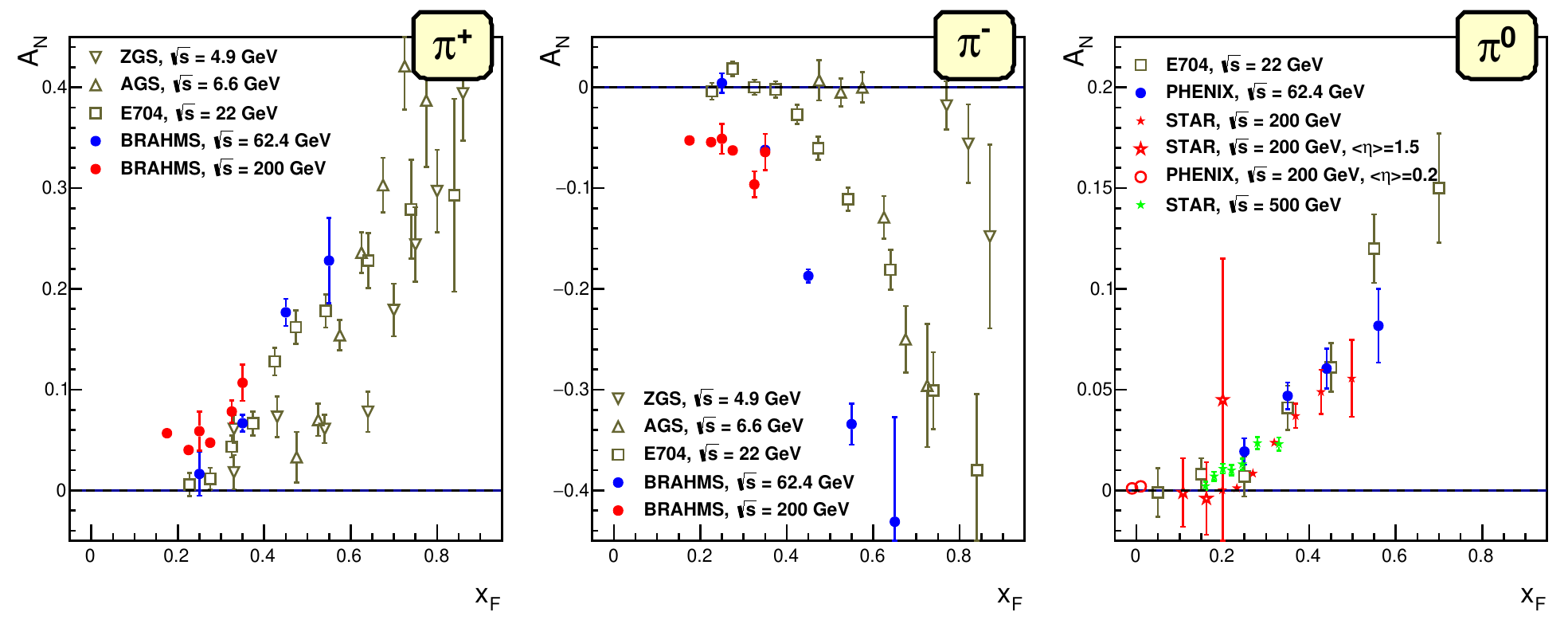}
\end{center}
\caption{Selection of world data on $A_N$ in $pp$ for neutral and charged pions. In particular in the $\pi^0$ case, the so-called $x_F$ scaling is evident, which means that the asymmetry is almost independent of $\sqrt{s}$. In general, the dependence of $A_N$ on $x_F$ is almost linear.
Data compiled by Oleg Eyser~\cite{Aschenauer:2016our}.
\label{fig:A_N}}
\end{figure}

Large asymmetries for $\pi^0$ but also for charged pions and kaons have been observed over a wide range of energies in the forward region where the valence quarks of the polarized protons are probed~\cite{Klem:1976ui,Adams:1991cs, Adams:1991ru,Allgower:2002qi,Adams:2003fx,Lee:2007zzh,Arsene:2008aa,Abelev:2008af,Adamczyk:2012xd,Adare:2013ekj}.
Fig.~\ref{fig:A_N} shows a selection of world data for $A_N$ in $p^\uparrow p$ for charged and neutral pions. 
There are several key features of the data.
Especially for the charged pions at large $x_F$ one has very significant effects, with $A_N$ becoming as large as $\pm$40\%.
Also, when comparing $A_N$ for $\pi^+$ and $\pi^-$ one finds, for the same kinematics, (almost) equal magnitudes but reversed signs.
Another remarkable feature of the data is the virtual independence of the neutral-pion asymmetry of $\sqrt{s}$ of the experiment at a fixed value of $x_F$, even though one would assume that the underlying processes might be quite different. 
or instance for high $x_F$, at low $\sqrt{s}$ the mean $|\vec{P}_{h\perp}|$ are too low for reliable pQCD calculations.
Instead the kinematic dependence of the asymmetries can be expressed solely by the Feynman variable $x_F$, a feature that is known as $x_F$ scaling. 
A trend for the same scaling holds for the charged pions, however some $\sqrt{s}$ dependence can be observed, which seems to average out for the $\pi^0$. 
Most recently, results from STAR show a persistence of the asymmetries at $\sqrt{s} = 500 \; \rm{GeV}$~\cite{Heppelmann:2013ewa}, even at transverse momenta up to about $7 \; \rm{GeV}$.
(Note that in the plots in Fig.~\ref{fig:A_N} the transverse hadron momentum is integrated over.)
In fact, no experiment observed a significant fall-off of the asymmetries with $|\vec{P}_{h\perp}|$, which na\"ively would be expected for a higher-twist observable. 
However, the reach in $|\vec{P}_{h\perp}|$ of existing experiments might not be high enough to observe this behavior. 
Asymmetries for kaons have been measured by the BRAHMS experiment at $62.4 \; \rm{GeV}$ and $200 \; \rm{GeV}$ at forward pseudorapidities~\cite{Videbaek:2008zz}. 
The magnitude of the kaon asymmetries are similar to the pion asymmetries, however both kaon charges exhibit the same sign. 
This behavior poses interesting questions about the production and fragmentation process, since the $K^+$ contains a valence quark of the parent proton, whereas the $K^-$ does not.
Another asymmetry sensitive to favored strange-quark fragmentation is the $A_N$ for $\eta$ mesons. 
Precise measurements of this quantity at $200 \; \rm{GeV}$ have been done by PHENIX~\cite{Adare:2014qzo} and STAR~\cite{Adamczyk:2012xd}. 
PHENIX sees asymmetries consistent with the $\pi^0$, whereas STAR sees a slight enhancement of the asymmetry at large $x_F$, however the precision is not sufficient to make a definite statement.

Let us finally come back to the large magnitude of the (twist-3) $A_N$, especially in the case of charged pions.
These effects are clearly larger than, for instance, the observed twist-2 Collins asymmetry in SIDIS.
They are also larger than the twist-2 longitudinal double-spin asymmetry $A_{LL}$, which was measured in the same process, i.e., $\vec{p} \vec{p} \to h X$, but for different proton polarizations.
Based on these results one may question the convergence of the twist expansion.
However, from a phenomenological point of view the twist expansion can only be studied in a meaningful way when considering leading-twist and subleading-twist effects for the same process and the same observable.

\subsubsection{Experimental data on di-dadron FFs $D_1$, $H_1^\open$, $G_1^\perp$}
\label{sec:data_DiFFs}

\paragraph{Spin-averaged di-hadron FF $D_1^{h_1h_2/q}$:} 
There is no data available on $D_1^{h_1h_2/q}(z,M_h)$, i.e. sensitive to the $z$ and $M_h$ dependence of $D_1^{h_1h_2/q}$ simultaneously. 
However, Belle recently published results on the cross-section of di-hadron pair production (see Sec.~\ref{sec:DiFF_epem}) for pions, kaons and protons in $e^+e^-$~\cite{Seidl:2015lla}. 
However, the dependence on $M_h$ has not been considered yet. 
Instead the measurement is differential in $z_1,z_2$, where the $z_i$ are the fractional energies of the two hadrons making up the hadron pair. 
As mentioned several times before, in this observable the DiFF and the single-hadron FFs mix at NLO, and much of the information about $D_1^{h_1h_2/q}$ is contained in the $M_h$ dependence which is not measured. 
For the $z$ integrated $M_h$ spectra of charged pion pairs, we have some data available from $e^+e^-$ annihilation from LEP~\cite{Acton:1992sa,Abreu:1992xx,Buskulic:1995gm} and Cornell~\cite{Cohen:1980zg}, from SIDIS (EMC)~\cite{Aubert:1983un,Arneodo:1986tc}, and from $pp$ scattering~\cite{Blobel:1973wr,AguilarBenitez:1991yy,Adams:2003cc}.
However, these results focus on $\rho^0$ production and the shape of the di-pion spectrum. 
They lack a measurement of the $z$ dependence and are therefore not sufficient to extract the DiFFs with sufficiently high precision.

\begin{figure}[t]
\begin{center}
\includegraphics[width=0.98\textwidth]{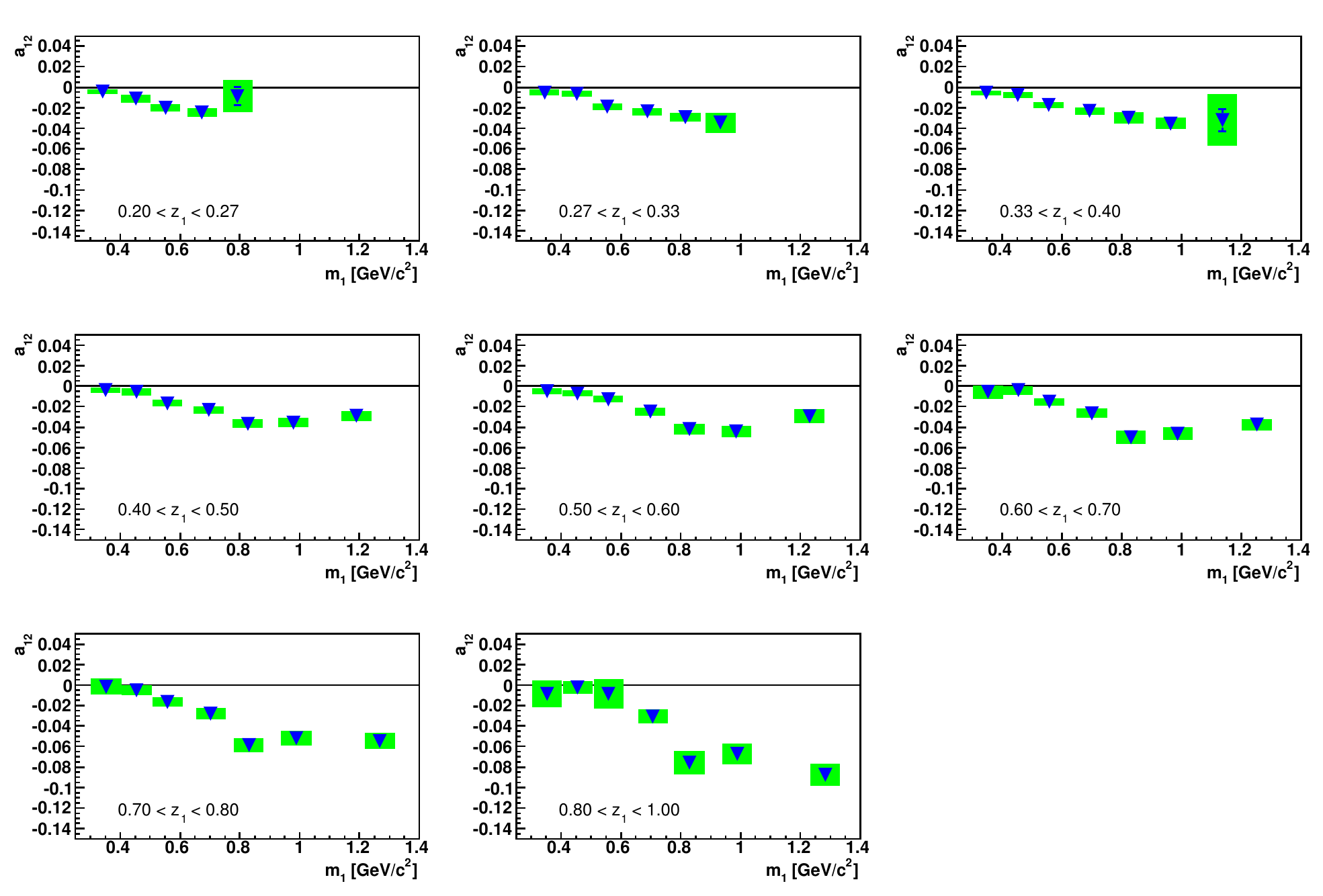}
\end{center}
\caption{Azimuthal correlations of di-pion pairs measured by Belle. The observable denoted by $a_{12}$ in the plots corresponds to the amplitude of the $\cos(\phi_{Ra}+\phi_{Rb})$ modulation of the cross section in Eq.~(\ref{eq:diFFInEE}). The results are shown as a function of the invariant mass $M_{ha}$, denoted by $m_1$ in the figure, of the hadron pair in one hemisphere in each panel, and each panel shows a different bin $z_a$, which is denoted by $z_1$ in the figure. As explained in the text, the observed asymmetry is sensitive to the product of $H_1^\open$ for the quark and antiquark. 
Figure reprinted with permission from~\cite{Vossen:2011fk}~\href{http://dx.doi.org/10.1103/PhysRevLett.107.072004}{A.~Vossen, et al., Phys.~Rev.~Lett.~107~(2011)~072004}. Copyright (2011) by the American Physical Society.
\label{fig:SIA_IFF}}
\end{figure}

\paragraph{Spin-dependent di-hadron FFs $H_1^\open$, $G_1^\perp$ in $e^+e^-$:}
Similar to the Collins effect in $e^+e^-$, the spin-dependent di-hadron IFF $H_1^\open$ can also be measured in the correlation of back-to-back di-hadron pairs. 
Belle measured for the first time the corresponding observable described in Eq.~(\ref{eq:diFFInEE}). 
The results are shown in Fig.~\ref{fig:SIA_IFF}.

Different from the results for the Collins FF, the acceptance effects are sufficiently averaged out by looking at two hadrons in the final state on each side, such that no double ratios are needed to extract the asymmetry. 
This means that the results directly correspond to the amplitude of the modulation of the normalized cross section in Eq.~(\ref{eq:diFFInEE}), i.e., the product $H_1^\open \bar{H}_1^\open$. 
Since $H_1^\open$ is not a TMD, there is also no convolution of intrinsic transverse momenta.
Therefore, the results are not shown vs $|\vec{P}_{hT}|$, but instead the second degree of freedom introduced by the second hadron is expressed in the invariant mass of the hadron pair $M_h$.
As is the case for the Collins effect, the asymmetries rise generally with $z$ and exhibit almost a linear dependence on $z$. 
As function of $M_h$ the asymmetries exhibit a bit more structure. 
As described in Sec.~\ref{subs:interpretation} and Eq.~(\ref{e:H1open_partial_wave}), it is expected that the signal for $H_1^\open$ is enhanced where amplitudes for pion pairs produced in different partial waves interfere. 
Some models suggest that this is the case for $M_h\approx M_\rho$ --- see also Sec.~\ref{sec:spectator_models}.
This is what is observed at Belle. 
However, the asymmetries do not significantly fall for $M_h>M_\rho$, but rather flatten out. 
This is different from the results in SIDIS and $pp$ which are shown below. 
One potential explanation is that at Belle energies there is a significant fraction of charm-quark production, and the resonance structure is richer with larger contributions from partial waves with non-zero orbital angular momentum at large $M_h$.

The Belle and BaBar programs on spin-dependent FFs continue. 
Most recently Belle showed the first measurement of azimuthal correlations in jets that are thought to be sensitive to the helicity-dependent TMD DiFF $G_1^\perp$~\cite{Abdesselam:2015nxn}.
The relevant observable is discussed in Sec.~\ref{sec:diHadTMDs}.

\begin{figure}
\begin{center}
\includegraphics[width=0.98\textwidth]{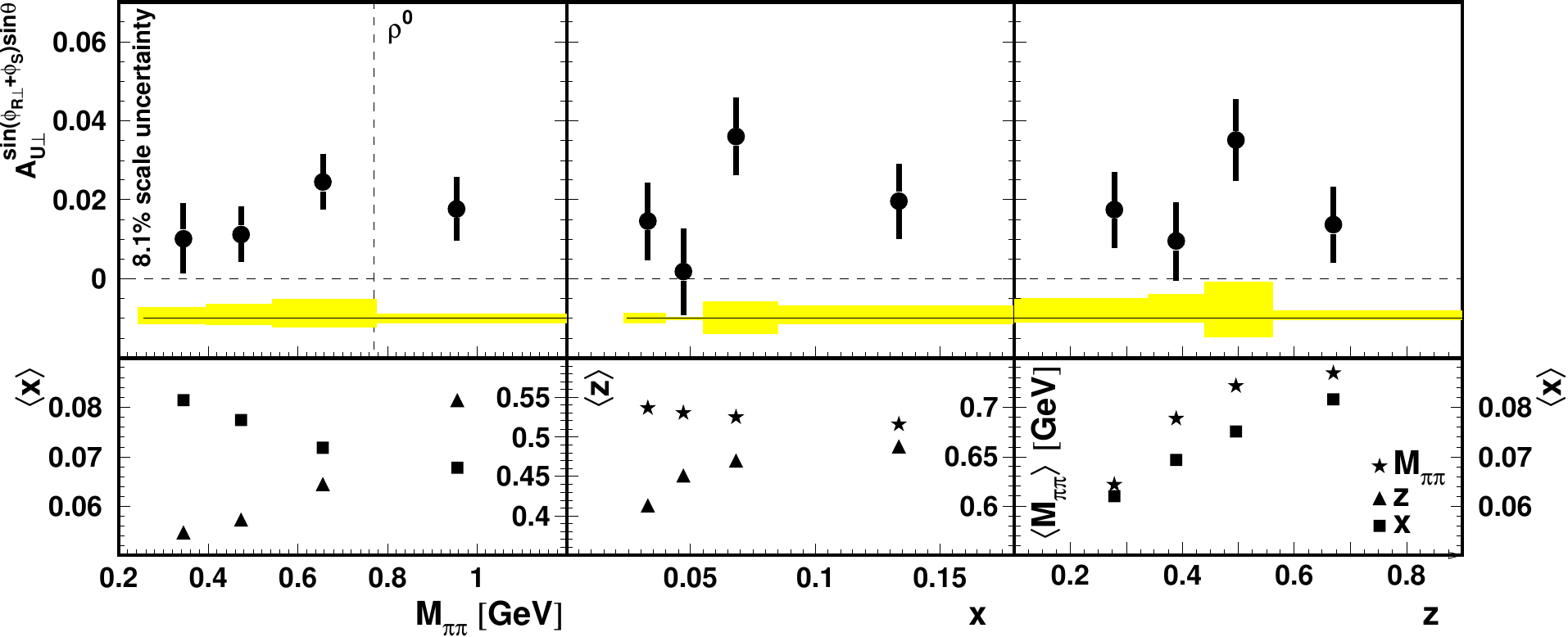}
\end{center}
\caption{Di-hadron asymmetries sensitive to $H_1^\sphericalangle$ times the transversity distribution measured by the HERMES collaboration\label{fig:HermesIFF}. The data is shown vs. the mass of the hadron pair, $x$ and $z$. The asymmetry corresponds to the size of the $\sin(\phi_R+\phi_S)$ modulation of the cross-section~(\ref{eq:diffInSIDIS}), however, the $\sin(\theta)$ dependence of the asymmetry on the decay angle $\theta$ of the hadron pair has already been factored out according to the partial wave decomposition in Eq.~(\ref{e:H1open_partial_wave}). 
Figure reprinted from~\cite{Airapetian:2008sk}.}
\end{figure}

\paragraph{Spin-dependent di-hadron FF $H_1^\open$ in SIDIS:}
The observable for the transverse-polarization dependent DiFF $H_1^\open$ is analogous to the single-hadron case in SIDIS. 
The relevant spin-dependent cross section is shown in Eq~(\ref{eq:diffInSIDIS}), and the experimental observable is again a transverse SSA.
HERMES was the first experiment to observe a significant nonzero asymmetry~\cite{Airapetian:2008sk} off a proton target. 
This result is shown in Fig.~\ref{fig:HermesIFF}. 

The measurement was repeated by COMPASS using both a deuteron and a proton target. 
The result off the proton target is consistent with the HERMES result, but much more precise~\cite{Adolph:2014fjw}. 
At the higher $\sqrt{s}$ of the COMPASS experiment the hadron multiplicities are also higher, which contributes quadratically to the precision of the measurement of hadron pairs. 
Similar to the Collins effect, asymmetries grow with the fractional energy $z$ the pair is carrying. 
The asymmetries are enhanced around $M_h=M_\rho$ and fall again for $M_h > M_\rho$. 
This is at variance with the results seen in $e^+e^-$, which might be due to the different spectrum of partial waves contributing to the pion pair production.
COMPASS also studied the interplay between the di-hadron asymmetries and single hadron asymmetries~\cite{Adolph:2015zwe}.
 
\begin{figure}[t]
\begin{center}
\includegraphics[width=0.55\textwidth]{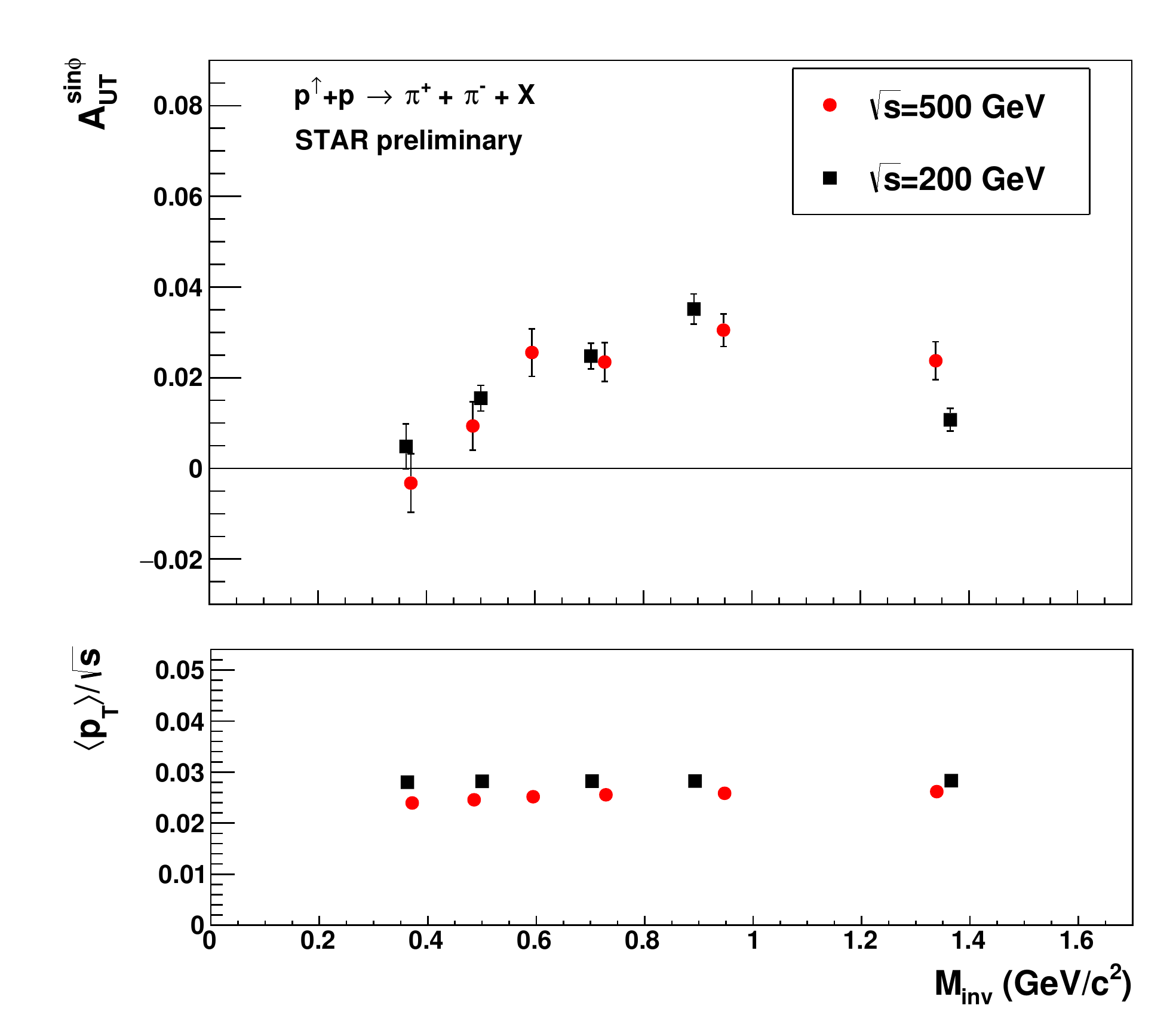}
\end{center}
\caption{Di-hadron asymmetries  from the STAR experiment. Shown are results measured in the pseudorapidity region $0<\eta<1$ relative to the polarized beam for both $\sqrt{s} = 200 \; \rm{GeV}$ and $500 \; \rm{GeV}$. The lower left panel shows the mean transverse momentum of the hadron pair with respect to the beam axis normalized to $\sqrt{s}$. For a fixed pseudorapidity of the hadron pair, this value is proportional to the partonic $x$ at LO. The STAR results therefore show significant asymmetries for both energies that do not show a dependence on $\sqrt{s}$ when probed at similar partonic kinematics. Figure reprinted from~\cite{Aschenauer:2016our}.
\label{fig:StarIFF}}
\end{figure}

\paragraph{Spin-dependent di-hadron FF $H_1^\open$ in $pp$:}
The polarization-dependent DiFF $H_1^\open$ survives an integration over the intrinsic transverse momentum of the hadron pair with respect to the fragmenting quark axis. 
Since the relative transverse momentum to the quark axis therefore does not have to be reconstructed, different compared to measurements sensitive to $H_1^\perp$ discussed earlier, the measurement of this effect does not require the jet reconstruction in $p^\uparrow p$ scattering.
Instead, the second hadron provides another degree of freedom that allows to form two independent vectors, the difference $\vec{R}$ and the sum $\vec{P}_h$ of the two hadron momenta. 
This actually significantly simplifies the di-hadron measurement compared to the single-hadron measurement of the Collins TMD. 
Because the jet axis does not have to be reconstructed, the systematic effects are significantly reduced. 
The observable in Eq.~(\ref{eq:diffInPP}) is the spin-dependent part of the di-hadron cross section that also depends on the azimuthal angle of $\vec{R}$ around $\vec{P}_h$.
The corresponding transverse SSA observable has been measured for $\pi^+/\pi^-$ pairs at the STAR experiment, where significant asymmetries have been observed both at $\sqrt{s} = 200 \; \rm{GeV}$ and $\sqrt{s} = 500 \; \rm{GeV}$~\cite{Adamczyk:2015hri,Aschenauer:2015eha,Skoby:2016jkp}. 
These results are also shown in Fig.~\ref{fig:StarIFF}.
The asymmetries generally rise with $M_h$ and $p_T$. 
Here the $p_T$ is the transverse momentum of the hadron pair with respect to the beam axis, which should not be confused with the intrinsic $|\vec{P}_{hT}|$ relative to the fragmenting quark acquired during the fragmentation process. 
Therefore higher $p_T$ are correlated with higher $z$ and higher values of $x$, the momentum fraction of the active parton in the parent nucleon.  
The $M_h$ dependence is very similar to the dependence observed in SIDIS which is expected since the underlying partonic processes are also similar.
Earlier, PHENIX showed an analysis of the correlations of charged unidentified hadrons, and the correlation between $\pi^0$'s and charged hadrons~\cite{Yang:2009zzr}.
However, no signal was observed in the covered kinematic region.
As discussed in Sec.~\ref{sec:data_DiFFs}, there is only very little data available for the spin-averaged FF $D_1^{h_1 h_2/q}$ which is needed to extract $H_1^\open$ from the observables discussed in this section. 
This data is also needed to extract $h_1$ from $pp$ and SIDIS using the di-hadron channel
where for the $pp$ case knowledge of $D_1^{h_1 h_2/g}$ is necessary.

\section{Global Fits}
\label{sec:global_fits}
A multitude of processes that have been measured over a wide range of scales give information on FFs. 
In order to make use of the constraints posed by each measurement, global fits are essential to extract FFs. 
In addition to the extraction of FFs and their uncertainties, global fits are valuable tools to make predictions at different energies and for different processes.
In this section we discuss global fits for the unpolarized FF $D_1^{h/i}$, the polarization-dependent FFs $H_1^{\perp \, h/q}$ and $H_1^{\open \, h_1 h_2/q}$, as well as twist-3 FFs.
The fits are based on the datasets described in Sec.~\ref{sec:results}.

\subsection{Fits of integrated FF $D_1$}
\label{sec:fits_integrated_D1}
\begin{table}[t]
\begin{center}
\begin{tabular}{|c|c|c|c|}
\hline
Fit & Datasets & Uncertainties & Final States \\
\hline 
EMC~\cite{Aubert:1985zd,Arneodo:1989ic} & SIDIS & no & $\pi^\pm$ \\
CGMG \cite{Chiappetta:1992uh}  &$e^+e^-$ , $pp(\bar{p})$ &no & $\pi^0$ \\
BKK~\cite{Binnewies:1994ju,Binnewies:1995pt,Binnewies:1995kg} & $e^+e^-$& no &$h^\pm,\pi^\pm,K^\pm,K_S^0$\\
DSV~\cite{deFlorian:1997zj} & $e^+e^-
$& no &$\Lambda$ \\
Kretzer~\cite{Kretzer:2000yf} & $e^+e^-$, SIDIS~\cite{Kretzer:2001pz} & no& $h^\pm, \pi^\pm, K^\pm$  \\
BFGW~\cite{Bourhis:2000gs} & $e^+e^-$ & yes & $h^\pm$ \\

KKP~\cite{Kniehl:2000fe} & $e^+e^-$& no & $\pi^\pm, K^\pm, p(\bar{p})$\\
Bourrely, Soffer~\cite{Bourrely:2003wi} &$e^+e^-$ , $pp(\bar{p})$ & no & $\pi^0$ \\


AKK08~\cite{Albino:2008fy} & $e^+e^-$, $pp(\bar{p})$ & no &$\pi^\pm,K^\pm,p/\bar{p},K_S^0,\Lambda/\bar{\Lambda}$\\
HKNS~\cite{Hirai:2007cx} &$e^+e^-$& yes & $\pi^\pm, K^\pm,p/\bar{p}$\\
AESSS~\cite{Aidala:2010bn}  &$e^+e^-$ , $pp(\bar{p})$ & yes & $\eta $\\
DSS~\cite{deFlorian:2007aj,deFlorian:2007ekg,deFlorian:2014xna}  &$e^+e^-$ , $pp(\bar{p})$, SIDIS & yes & $h^\pm,\pi^\pm,K^\pm ,p/\bar{p},\eta $\\

LSS~\cite{Leader:2015hna} & SIDIS & no & $\pi^\pm$ \\
ASR-NNLO~\cite{Anderle:2015lqa} &$e^+e^-$ & no &  $\pi^\pm$ \\
COMPASS-LO~\cite{Adolph:2016bga} & SIDIS & no & $\pi^\pm$ \\

\hline
\end{tabular}
\caption{Selection of global fits to $D_1$. The order of the fits from top to bottom is approximately chronologically.\label{tbl:fits}}
\end{center}
\end{table}
\begin{figure}[t]
\begin{center}
\includegraphics[width=0.50\textwidth]{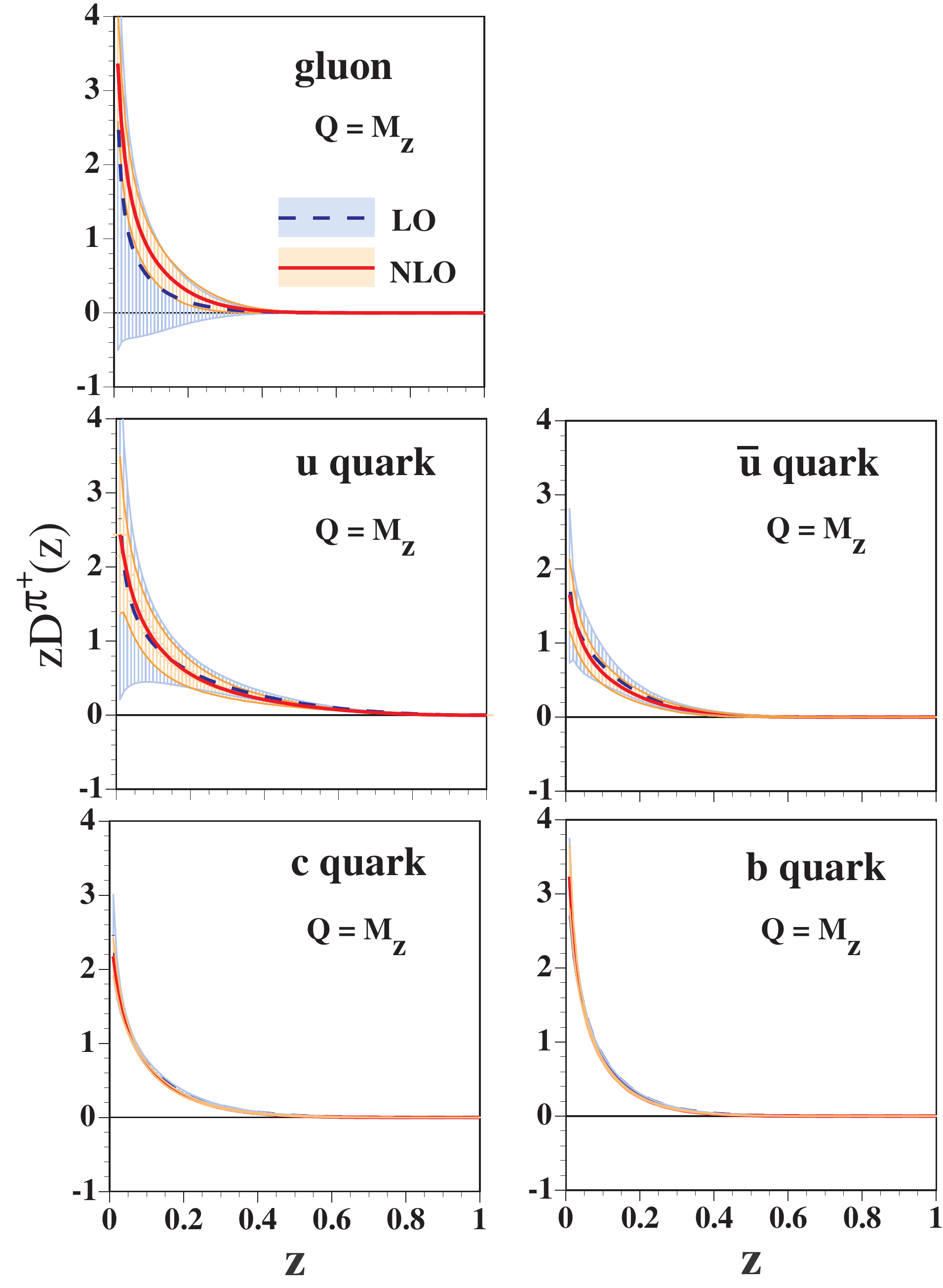}
\end{center}
\caption{HKNS Extraction of $z D_1^{\pi^+/i}$ from $e^+e^-$ data. The light quark FFs can be determined quite well from SIA. Same for the heavy quark FFs where flavor tagged data from LEP can be used. 
Figure reprinted with permission from~\cite{Hirai:2007cx}~\href{http://dx.doi.org/10.1103/PhysRevD.75.094009}{M.~Hirai, et al., Phys.~Rev.~D75~(2007)~094009}. Copyright (2007) by the American Physical Society.
\label{fig:HKNS}}
\end{figure}
There is a rich history of fits that use $e^+e^-$ data and subsequently SIDIS and $pp$ data to extract $D_1$. See Tab.~\ref{tbl:fits} for a summary.
Most fits use a power law parameterization of the form $z^\alpha(1-z)^\beta$, which can be seen as inspired by the Feynman-Field model~\cite{Field:1977fa} or motivated by approaches to fit PDFs. The Drell-Levy Yan-relation~\cite{Drell:1969jm,Drell:1969wd,Drell:1969wb} as well as  modified Gribov-Lipatov~\cite{Gribov:1972rt, Gribov:1971zn,Brodsky:1996cc,Barone:2000tx}  are used to justify using similar approaches for PDFs and FFs. A notable exception from the power law description is the statistical approach by Bourrely and Soffer~\cite{Bourrely:2003wi} which, considering the Gribov-Lipatov relation between PDFs and FF, applies a similar statistical approach that was already applied by the same authors to PDFs. There is not yet an approach based on neural networks similar to the NNPDF extraction~\cite{Ball:2008by} for PDFs. Therefore the uncertainty of all modern fits discussed in this section exhibit some bias due to the functional form used which provides constraints in unmeasured regions.
Depending on the discriminating power of the experimental data used in the fit more constraints are imposed on $SU(3)$ flavor symmetry and/or relations between FFs of different flavors are fixed.
Observables in $e^+e^-$ annihilation provide the cleanest access to FFs and have therefore been used for the first extractions of $D_1$. Among the pioneering first NLO fits to FFs of identified light mesons were BKK~\cite{Binnewies:1994ju}, KKP~\cite{Kniehl:2000fe}, AKK~\cite{Albino:2005me} and Kretzer~\cite{Kretzer:2000yf}. Each of those used $e^+e^-$ data only and do not have a rigorous determination of the uncertainties of the FFs beyond the statistical uncertainties of parameters as they come out of the fit or differences between different extraction methods. BFGW~\cite{Bourhis:2000gs} use some $pp$ data in addition to $e^+e^-$ data. However, they only extract FFs for unidentified hadrons and no updated results with later data is available.
In order to have access to flavor information and the gluon FF, some of the fits (KKP, AKK) use the tagged LEP data described in the previous Sec.~\ref{sec:results}.  
Similarly, the later HKNS and DSS fits discussed below take advantage of the flavor separated LEP and SLD data. As discussed earlier, it is not clear how to treat this tagged data at NLO in a consistent way.
Later in Ref.~\cite{Kretzer:2001pz} SIDIS data were included which allowed to extract flavor separated FFs. 
The first extraction of $D_1$ for light hadrons that provides a rigorous evaluation of the uncertainties of the extracted functions was performed by Hirai, Kumano, Nagai, and Sudoh (HKNS)~\cite{Hirai:2007cx}. 
The results for $D_1^{\pi/q}$ at the Z-mass scale are shown in Fig.~\ref{fig:HKNS}. This extraction only uses $e^+e^-$ data available at the time. 
In particular, the high precision data on pion and kaon multiplicities and cross-sections from the Belle and BaBar experiments are not used. 
Similar to the error estimate for PDFs, two methods are in use to determine the uncertainties for FFs: the Hessian method~\cite{Pumplin:2001ct} and the Lagrange multiplier method~\cite{Stump:2001gu}. 

The HKNS group uses the Hessian method for their extraction, where the $\chi^2$ of the fits to the data is expanded around the minimum. Using only the leading quadratic term, the confidence interval is chosen using a $\Delta \chi^2$ corresponding to the 68\% confidence interval in the given N-dimensional space under the assumption that the $\chi^2$ method behaves analogously to a normal distribution.
The advantage of this method is, that it is computationally less intensive compared to the Lagrange multiplier method and eigenvectors corresponding to the shape of the $\chi^2$ function can be used to easily estimate uncertainties of arbitrary observables. On the other hand, the assumptions regarding the shape of the $\chi^2$ function have to be validated. 
The Lagrange multiplier method does not use those assumptions but is computationally harder and it is more difficult to estimate the uncertainties on arbitrary observables. A detailed comparison between the two methods and its applicability for the extraction of $D_1$ can be found in Ref.~\cite{Epele:2012vg}.
In addition to the light quark FFs for pions, kaons and protons, HKNS also extract charm and bottom FFs.
The tagged data from SLD and DELPHI allow a quite precise determination of these FFs.
In the light quark sector, the restriction to $e^+e^-$ data means that HKNS effectively extracts favored and unfavored FFs for $u$ and $d$, with a slightly different parametrization of the strange quark due to its different mass. A flavor symmetric light sea is assumed as well.
As Fig.~\ref{fig:HKNS} shows, the extracted FFs are quite precise with the exception of the gluon FF. This FF can be determined better by including $pp$ data.
The first simultaneous fit of $pp$ data together with $e^+e^-$ and SIDIS data was performed by de Florian, Sassot, and Stratmann (DSS) in Refs.~\cite{deFlorian:2007aj,deFlorian:2007hc}.
\begin{figure}[t]
\includegraphics[width=0.49\textwidth]{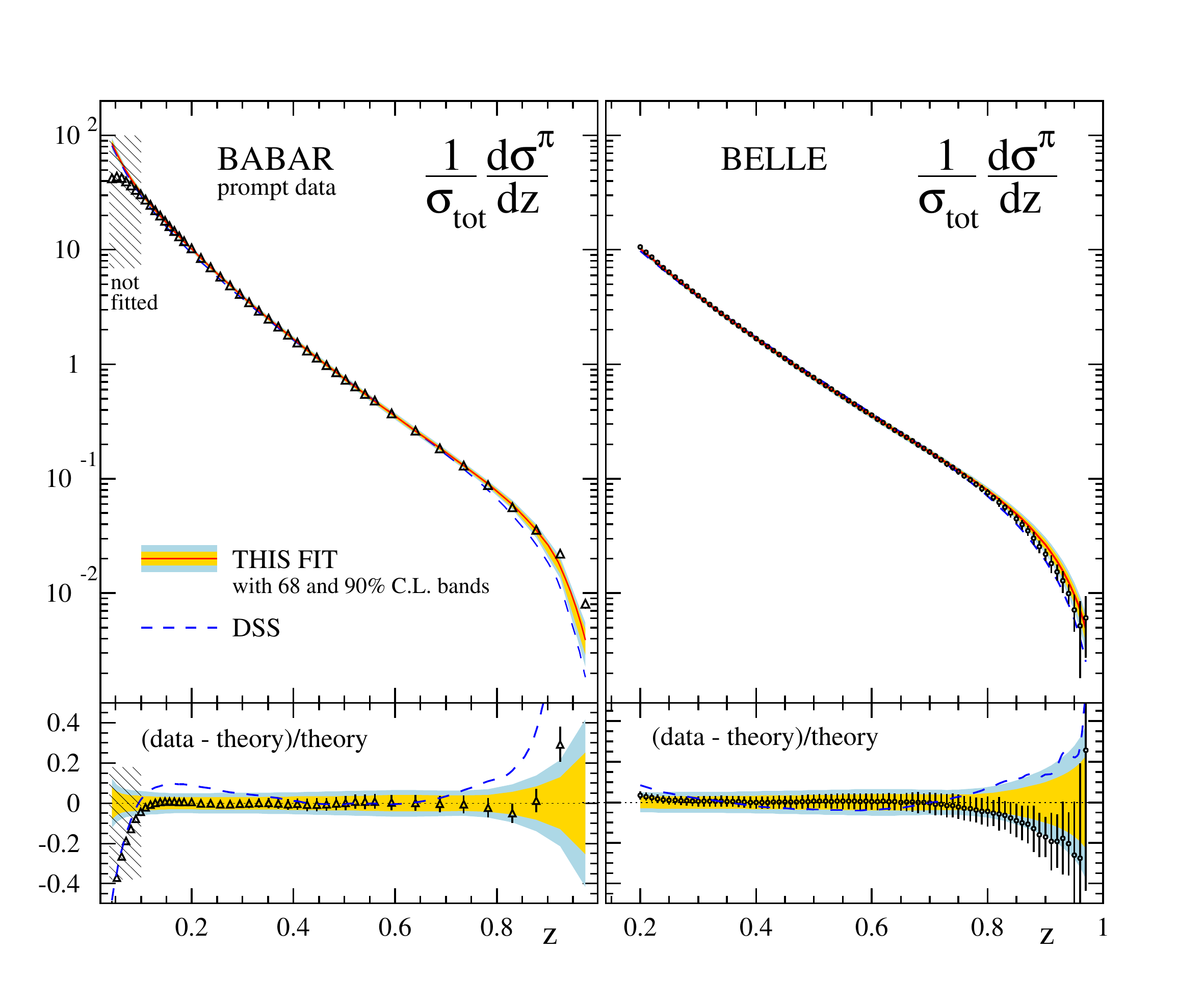}
\includegraphics[width=0.49\textwidth]{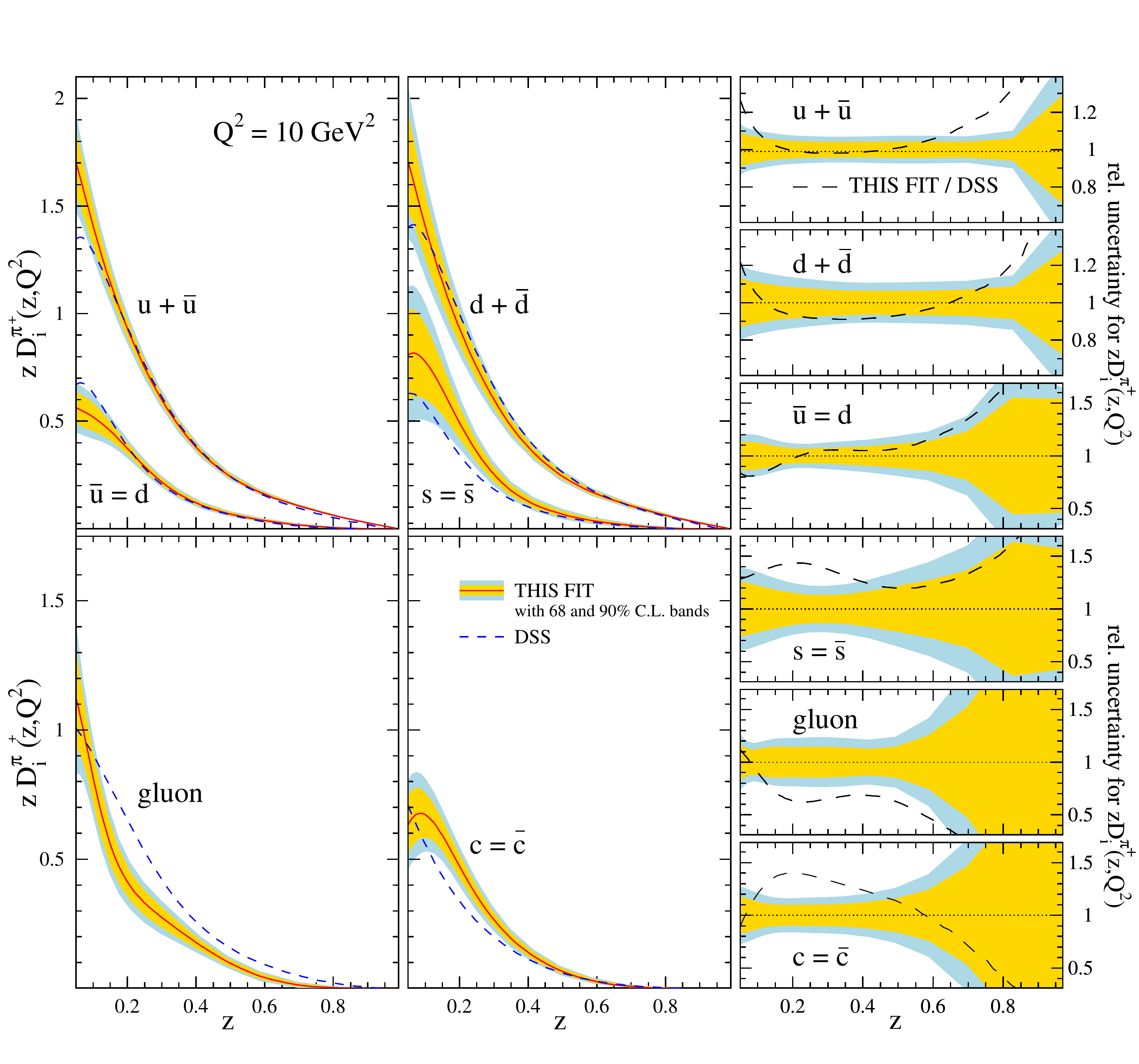}
\caption{DSS Extraction of $z D_1^{\pi^+/i}$ from $e^+e^-$ and $pp$ data. Left panel: fit to Belle and BaBar SIA data. The agreement with the precision data is very good over most of the kinematic regime. At low and high $z$ higher order corrections play a larger role as discussed in the text. In particular at high $z$ the theoretical uncertainties, dominated by the scale dependence, are larger than the experimental uncertainties. Right panel: extracted FFs. The inclusion of $pp$ data leads to uncertainties on the gluon FF which are comparable to the uncertainties of the quark FFs. 
Figure reprinted with permission from~\cite{deFlorian:2014xna}~\href{http://dx.doi.org/10.1103/PhysRevD.91.014035}{D.~de Florian, et al., Phys.~Rev.~D91~(2015)~014035}. Copyright (2015) by the American Physical Society.
\label{fig:dss}}
\end{figure}

Their fit included precise RHIC data to extract the FFs of pions and kaons as well as CDF and UA1 and UA2 data for the extraction of FFs for unidentified charged hadrons and protons.
The latest fit of the pion FFs~\cite{deFlorian:2014xna} also includes LHC data from the ALICE experiment, taken at {\it cm} energies up to $7 \, \textrm{TeV}$, as well as the new precision SIA data from Belle and BaBar.
The extraction of the FFs as well as the fit to the $e^+e^-$ data is shown in Fig.~\ref{fig:dss}. Compared to HKNS, the precision of all FFs is improved. In particular the gluon FF, which was only very poorly constrained in the HKNS fit, is now determined to a similar precision as the light quark FFs. This is due to the $pp$ data which is available from RHIC for $200 \, \textrm{GeV}$ and from the LHC up to $7 \, \textrm{TeV}$, as well as to the precision SIA data which is now also available over a wide energy range and therefore constrains the gluon FF contributing to the evolution of the quark FFs. It should also be noted that the functional form of the DSS approach is more flexible than the previous fits, since the data provides more constraints. In particular, fewer constraints based on flavor symmetry are imposed. Another difference to HKNS is the use of Lagrange multipliers. However, as mentioned above, the consistency with and applicability of the Hessian method was shown which was used for the latest DSS extractions.

Even though there is no strong tensions between the datasets used for the DSS fit, which indicates that the theory is quite well under control, there are still surprises, when new data becomes available. This is in particular true for the gluon FF, for which the NLO analysis of the $pp$ has to be under control. Before LHC, the most stringent contraints came from RHIC data. However, the resulting FFs did a poor job describing LHC data~\cite{d'Enterria:2013vba}. The new DSS fit solves this problem with a refit using the LHC data and removing the PHENIX low $p_T$ points, which cause the most tension, via a minimum $p_T$ cut. 
This illustrates that the FFs extracted from a global fit are only reliable in the kinematic region and for the processes that were used to fit the FFs. Of course, observables where FFs couple to PDFs are particularly challenging. 
One example for the interdependence of PDF and FF extraction is the so-called  
strangeness 'puzzle', the fact that the nucleon's strangeness extracted from SIDIS data can be different from the one extracted from inclusive measurements, a discrepancy that might be traced back to the FFs used~\cite{Leader:2014uua}.

Currently NLO calculations are state-of-the-art and most analysis described in this section are done at this order. Recently~\cite{Anderle:2015lqa} a first NNLO calculation was performed to extract charged pion FFs using the DSS framework. 
Only data from $e^+e^-$ annihilation was used since there are no calculations for the underlying hard process in SIDIS and $pp$ at NNLO available yet.
As expected, the K-factors, which give the ratio between the present calculation and the one performed at NLO, go down. Away from the thresholds at low and high $z$, they are below 10\% down from about 40\% for NLO/LO, giving an estimate of the contribution beyond NNLO. 
The fit is also able to better describe BaBar data on pion production at low $z$. (Belle has a $z$ cut of 0.2, whereas BaBar goes below $z$ of 0.1.) This is expected because at low $z$ higher order terms dominate, making resummation techniques necessary. We come back to this point below.
Most importantly though, the dominating theoretical error from the scale dependence of the extraction goes down from about 10\% to about 5\% at Belle and BaBar energies. This is significant, since this is now close to the experimental uncertainties. At higher {\it cm} energies the theoretical uncertainty is even smaller, reaching 1-2\% at LEP energies.

Data covering kinematic regions where resummation effects or target mass corrections are substantial are often excluded. 
In particular, a minimum on the hard scale of the interaction ($Q^2$ in SIDIS, $P_{h\perp}^2$ in $pp$ and $\sqrt{s}$ in $e^+e^-$) is required. A usual cut of $z>0.1$ is applied to minimize the effect of mass corrections as well as an $x_T=\frac{p_T}{2\sqrt{s}}$ cut. 
At phase space boundaries, e.g. high $z$ and, in $pp$ collisions, at high $x_T$, the importance of higher order terms leads to significant uncertainties in fixed order calculations.
Resummation techniques can be used to address some of these problems. 
Of the fits discussed earlier, the AKK update, AKK08~\cite{Albino:2008fy} already contains resummation at high $z$ and also includes target mass effects for the FFs of heavier particles (See also the work in~\cite{Albino:2008gy}). More recently~\cite{Accardi:2014qda} showed the impact of large $z$ resummation and target mass corrections at low $z$ in the description of the precision data from Belle and BaBar. In comparison, fixed order fits, e.g. DSS, also achieve good fits at high $z$, where resummation would be needed, which is probably because those higher order corrections are fitted as well. 
A special role is played by the low $z$ region, roughly for $z<0.1$. In this region the time-like splitting functions have singularities~\cite{Agashe:2014kda} that dominate at $z$ values that are orders of magnitude higher than the $x$ at which these singularities become important for the corresponding space-like splitting functions of PDFs. The modified leading-log technique described in Sec.~\ref{sec:mlla} can be used to fit this region and conversely the data at low $z$ can be used to test predictions obtained from resummation.



\subsection{Fits of TMD FF $D_1$}
\label{sec:fit_TMDFF_D1}
Here we discuss progress that has been made in understanding the transverse momentum dependence of the unpolarized FF $D_1^{h/i}$.
As already mentioned in Sec.~\ref{sec:observables_TMD} detailled studies in that area only used data from SIDIS, and they are limited to FFs for quarks.
Most of the works are based on the parton model expression for the unpolarized SIDIS cross section --- see Eqs.~(\ref{e:SIDIS_xs_PT}), ~(\ref{e:F_UU_SIDIS}).
The delta-function in~(\ref{e:convolution_SIDIS}) implies
\begin{equation} \label{e:SIDIS_entangle}
\vec{P}_{h\perp} = z \, \vec{p}_T + \vec{P}_{hT} \,,
\end{equation}
and therefore measuring the $P_{h\perp}$ dependence of observables does not provide exclusively information on the transverse momentum dependence of the fragmentation process. 
(For convenience we repeat the meaning of the momenta in Eq.~(\ref{e:SIDIS_entangle}): $\vec{P}_{h\perp}$ is the transverse momentum of the hadron relative to the virtual photon, $\vec{p}_T$ is the transverse momentum of the parton inside the target, while $\vec{P}_{hT}$ indicates the transverse momentum of the produced hadron relative to the fragmenting parton.)
On the other hand, Eq.~(\ref{e:SIDIS_entangle}) also indicates that transverse momentum effects associated with the distribution side and the fragmentation side to some extent may be disentangled through variation of $z$.

Let us now briefly describe the framework that was used in a number of papers to extract information on TMD FFs (and PDFs)~\cite{Anselmino:2005nn,Collins:2005ie,Schweitzer:2010tt,Signori:2013mda,Anselmino:2013lza,Barone:2015ksa}.
Like in the case of integrated FFs one works with a parameterization.
However, for TMD FFs the task in principle is much more complicated since one is dealing with the two variables $z$ and $P_{hT}$.
In order to make some progress it is reasonable to keep in a first step the parameterization as simple as possible.
Arguably the simplest ansatz one can think of for both PDFs and FFS reads
\begin{equation} \label{e:Gauss_ansatz}
f_1^{q/p}(x,\vec{p}_T^{\;2};\mu^2) = f_1^{q/p}(x,\mu^2) \, \frac{e^{-\vec{p}_T^{\;2} / \langle p_T^2 \rangle}}{\pi \, \langle p_T^2 \rangle} \,, \qquad
D_1^{h/q}(z,\vec{P}_{hT}^{\,2};\mu^2) = D_1^{h/q}(z,\mu^2) \, \frac{e^{-\vec{P}_{hT}^{\,2} / \langle P_{hT}^2 \rangle}}{\pi \, \langle P_{hT}^2 \rangle} \,,
\end{equation}
i.e., the dependence on longitudinal and on transverse momenta is separated, where the former is given by the respective integrated partonic functions while the latter is assumed to be Gaussian.
As discussed in Sec.~\ref{sec:spectator_models}, model calculations do not support such a separation, yet the ansatz in~(\ref{e:Gauss_ansatz}) may be a good approximation to model results.
Moreover, one can describe reasonably well many existing data using the simple Gaussian ansatz in~(\ref{e:Gauss_ansatz}) --- see for instance Ref.~\cite{Schweitzer:2010tt}.
The Gaussian widths $\langle p_T^2 \rangle$, $\langle P_{hT}^2 \rangle$ are then fitted to data.
In Eq.~(\ref{e:Gauss_ansatz}) we also show the dependence of the TMDs on the renormalization scale $\mu$ which, in this simplified framework, is given by the (DGLAP) evolution of the integrated partonic functions. 
More sophisticated frameworks include Gaussian widths that depend on the flavor and/or on $z$ $(x)$ for TMD FFs (TMD PDFs). 
Ideally the full machinery of pQCD should be used including higher order corrections in the hard scattering and, in particular, proper TMD evolution.
Below we will summarize what is known about such approaches in the context of $D_1^{h/q}(z,\vec{P}_{hT}^{\,2})$.
\begin{figure}[t]
\begin{center}
\includegraphics[width=0.35\textwidth,angle=-90]{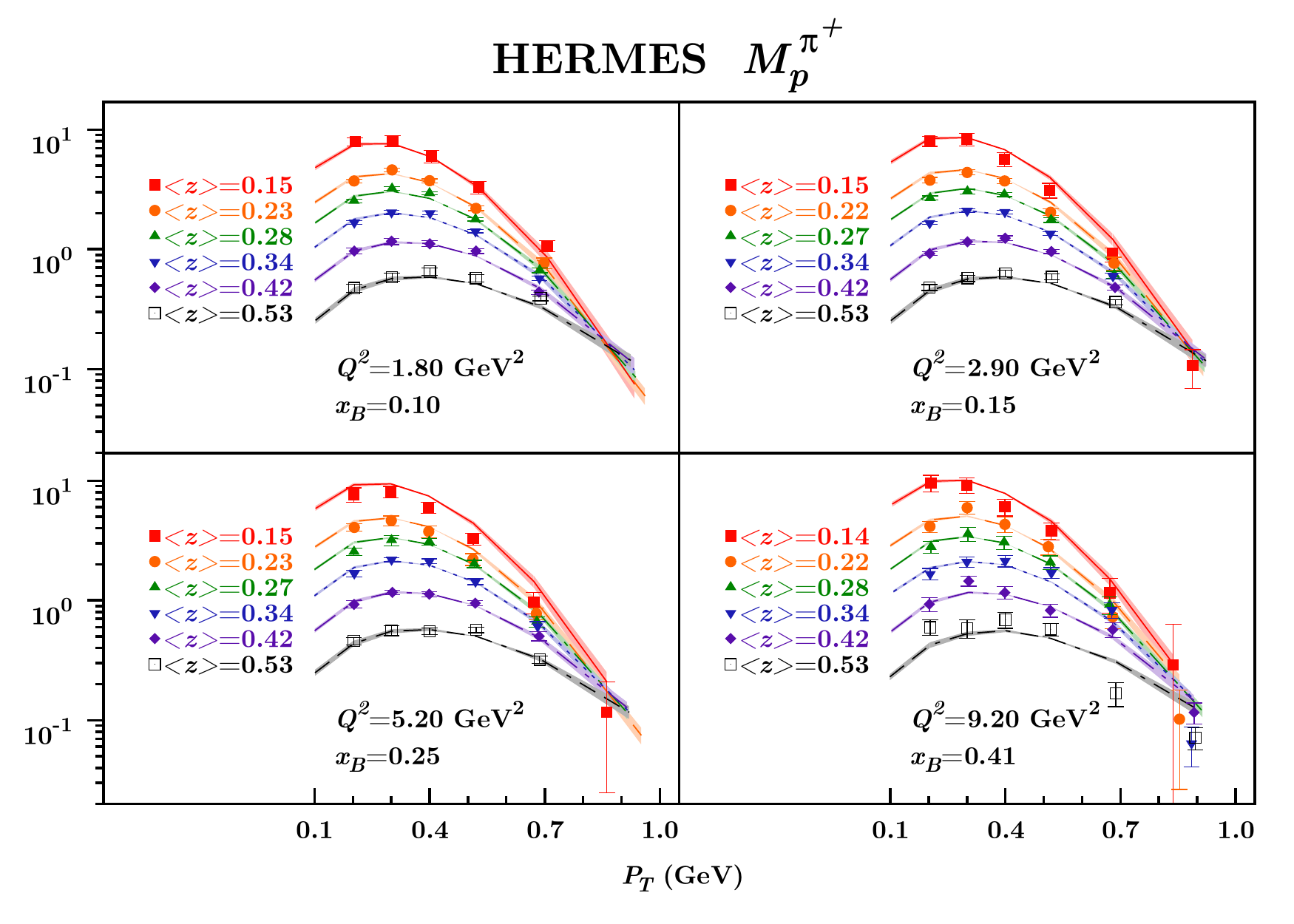} 
\includegraphics[width=0.35\textwidth,angle=-90]{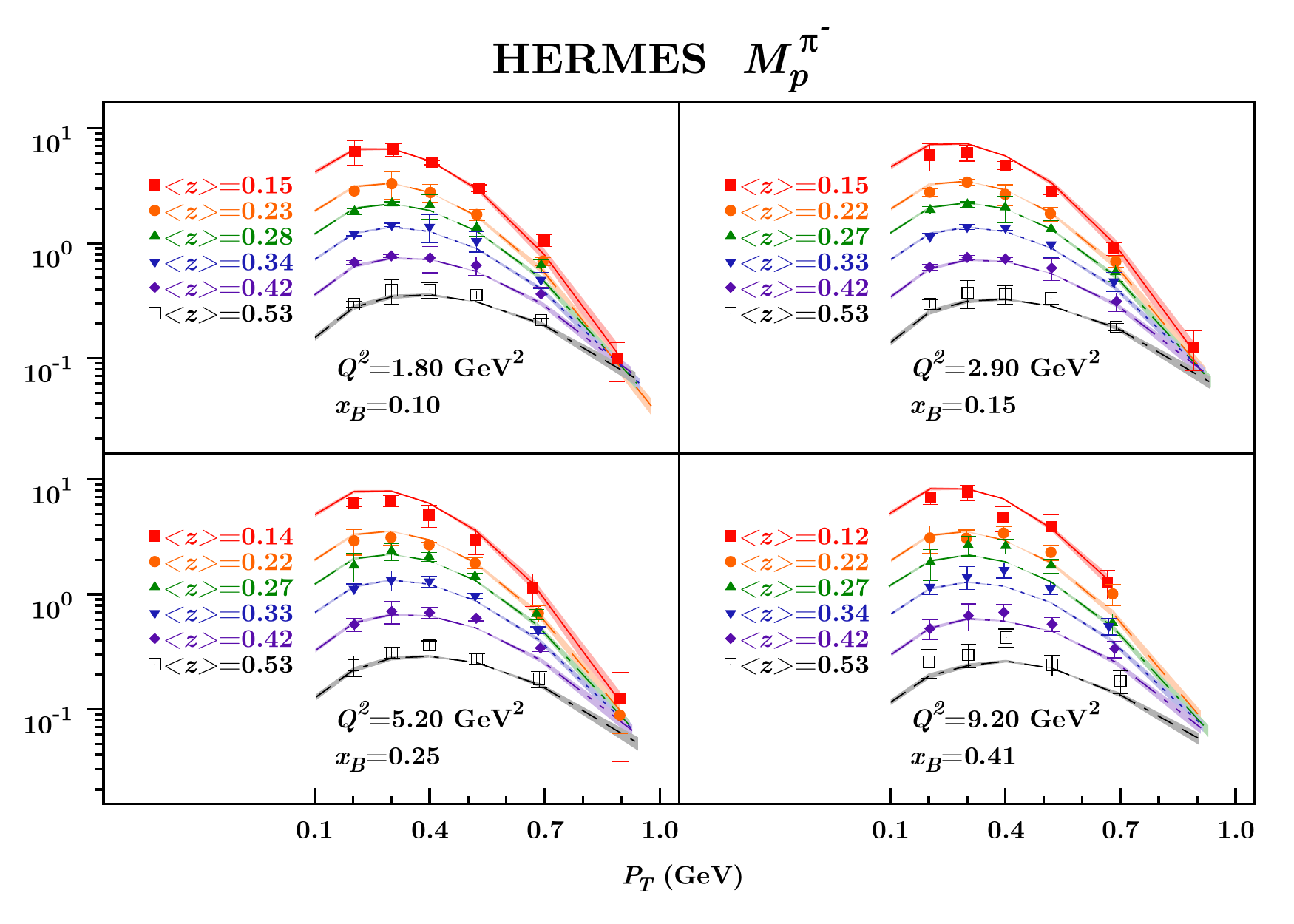}
\end{center}
\vspace{-0.4cm}
\caption{The multiplicities $M_p^{\pi^+}$ obtained compared with HERMES measurements for $\pi^+$ (left panel) and $\pi^-$ (right panel) SIDIS production off a proton target~\cite{Airapetian:2012ki}. 
Note that $P_T|_{\rm figures} = |\vec{P}_{h\perp}|$.
Figures reprinted from Ref.~\cite{Anselmino:2013lza}.}
\label{f:multiplicities_Anselmino}
\end{figure}

Early work on the transverse momentum dependence of $D_1^{h/q}$ can be found in Ref.~\cite{Anselmino:2005nn}.
The authors of that paper considered the azimuthal $(\phi_h)$ dependence of the SIDIS cross section, more precisely the Cahn effect~\cite{Cahn:1978se,Cahn:1989yf} described by the structure function $F_{UU}^{\cos \phi_h}$ in Eq.~(\ref{e:SIDIS_xs_PT}) which is a twist-3 observable.
Using Eq.~(\ref{e:Gauss_ansatz}) one obtains~\cite{Anselmino:2005nn}
\begin{equation} \label{e:Cahn_Gauss}
F_{UU}^{\cos \phi_h} \propto \sum_q e_q^2 \, f_1^{q/p}(x) \, D_1^{h/q}(z) \, \frac{\langle p_T^2 \rangle \, z \, |\vec{P}_{h\perp}|}{\langle P_{h\perp}^2 \rangle \, Q} \, \frac{e^{- \vec{P}_{h\perp}^{\,2} / \langle P_{h\perp}^2 \rangle}}{\pi \langle P_{h\perp}^2 \rangle} \,, \quad \textrm{where} \quad 
 \langle P_{h\perp}^2 \rangle = z^2 \langle p_T^2 \rangle + \langle P_{hT}^2 \rangle \,.
\end{equation}
The expression in~(\ref{e:Cahn_Gauss}) was fitted to data from the EMC Collaboration~\cite{Arneodo:1986cf,Ashman:1991cj} and the E665 Collaboration~\cite{Adams:1993hs} leading to $\langle p_T^2 \rangle = 0.25 \, \textrm{GeV}^2$, $\langle P_{hT}^2 \rangle = 0.20 \, \textrm{GeV}^2$~\cite{Anselmino:2005nn}, i.e., very similar widths were obtained for the PDFs and FFs.
However, for a couple of reasons the analysis in~\cite{Anselmino:2005nn} could be doubted.
First, it did not take into account all the contributions to $F_{UU}^{\cos \phi_h}$ that one finds in TMD factorization~\cite{Mulders:1995dh,Bacchetta:2006tn}.
Second, the status of TMD factorization at twist-3 is presently unclear~\cite{Gamberg:2006ru,Bacchetta:2008xw}.
In fact, in Ref.~\cite{Bacchetta:2008xw} it was precisely the Cahn effect for which a problem was made explicit.
On the other hand, contributions that are neglected in~(\ref{e:Cahn_Gauss}) might be small.
An indication that this could indeed be the case came from Ref.~\cite{Collins:2005ie}, where HERMES data on average transverse momenta of produced hadrons~\cite{Airapetian:2002mf} were fitted.
That work reported the values $\langle p_T^2 \rangle = 0.33 \, \textrm{GeV}^2$, $\langle P_{hT}^2 \rangle = 0.16 \, \textrm{GeV}^2$~\cite{Collins:2005ie}, which are relatively close to the results from~\cite{Anselmino:2005nn}.
Based on later results from the HERMES Collaboration on the transverse momentum dependence of the SIDIS cross section~\cite{Airapetian:2009jy} the following widths were found in Ref.~\cite{Schweitzer:2010tt},
\begin{equation} \label{e:widths_STM}
\langle p_T^2 \rangle \big|_{\textrm{Ref.}\;\scriptsize{\cite{Schweitzer:2010tt}}} = (0.38 \pm 0.06) \, \textrm{GeV}^2 \,, \qquad
\langle P_{hT}^2 \rangle \big|_{\textrm{Ref.}\;\scriptsize{\cite{Schweitzer:2010tt}}} = (0.16 \pm 0.01) \, \textrm{GeV}^2 \,.
\end{equation}
Data from Jefferson Lab~\cite{Mkrtchyan:2007sr,Osipenko:2008aa} are compatible with the numbers in~(\ref{e:widths_STM})~\cite{Schweitzer:2010tt}.
Comparing the results of Refs.~\cite{Anselmino:2005nn,Collins:2005ie,Schweitzer:2010tt} a quite consistent phenomenology seemed to emerge including the twist-3 Cahn effect.
However, a recent analysis of new data from HERMES~\cite{Airapetian:2012yg} and COMPASS~\cite{Adolph:2014pwc} on the Cahn effect gave rise to quite different values for the Gaussian widths~\cite{Barone:2015ksa}.
In particular, very small values ($\langle p_T^2 \rangle  = (0.03 \; \textrm{---} \; 0.04) \, \textrm{GeV}^2$) were obtained for the width of the TMD PDF $f_1^{q/p}$~\cite{Barone:2015ksa}.
This result indicates that the phenomenology of the Cahn effect is actually not under control and suggests that other non-perturbative twist-3 effects may matter for the numerics.
Additionally, higher order pQCD corrections might play a role as they also can generate an azimuthal dependence of the SIDIS cross section~\cite{Georgi:1977tv,Chay:1991nh}.

An important next step in this field were the measurements of multiplicities in SIDIS by HERMES~\cite{Airapetian:2012ki} and COMPASS~\cite{Adolph:2013stb}, which are described in more detail in Sec.~\ref{sec:data_TMD_D1}.
These data were analyzed in Refs.~\cite{Signori:2013mda,Anselmino:2013lza} using essentially the framework described above.
A new element in Ref.~\cite{Signori:2013mda} though were flavor-dependent widhts.
It was argued that the fit suggests a somewhat larger width for disfavored FFs of pions than for favored FFs.
This outcome could be compared to studies of the flavor-dependence of the widths for TMD PDFs.
In that regard a recent work, based on the chiral quark solition model, found the interesting result that TMD PDFs for sea quarks are considerably broader than distributions of valence quarks~\cite{Schweitzer:2012hh}.
Also the authors of~\cite{Anselmino:2013lza} saw an indication of flavor-dependent widths for TMD FFs, but with less significance. 
Additionally, widths that depend on the longitudinal momentum fractions $x$, $z$ were explored in~\cite{Signori:2013mda}, and an improvement of the fits was reported.
(Related older work on TMD FFs with $z$-dependent widths can be found in Refs.~\cite{Boglione:1999pz,D'Alesio:2004up}.
These studies mostly focus on single-inclusive hadron production in hadronic collisions.)
In contrast, according to~\cite{Anselmino:2013lza} it is difficult to constrain the additional parameters of such a scenario.
In Fig.~\ref{f:multiplicities_Anselmino} we show some fit results of Ref.~\cite{Anselmino:2013lza} for multiplicities from HERMES~\cite{Airapetian:2012ki}. 
The following ranges for the Gaussian widths are given in~\cite{Anselmino:2013lza},
\begin{equation} \label{e:widths_Anselmino}
\langle p_T^2 \rangle \big|_{\textrm{Ref.}\;\scriptsize{\cite{Anselmino:2013lza}}} = (0.4 \; \textrm{---} \; 0.8) \, \textrm{GeV}^2 \,, \qquad
\langle P_{hT}^2 \rangle \big|_{\textrm{Ref.}\;\scriptsize{\cite{Anselmino:2013lza}}} = (0.1 \; \textrm{---} \; 0.2) \, \textrm{GeV}^2 \,,
\end{equation}
which suggest that the uncertainties for both TMD PDFs and TMD FFs are still quite large.
In Ref.~\cite{Anselmino:2013lza} also the potential influence of TMD evolution was considered in a simplified manner by allowing for $Q^2$-dependent
widths.
It was found that the COMPASS data (at different $Q^2$) prefer such a scenario over constant widths, while the HERMES data do not. 
This outcome is in qualitative agreement with the expectation based on TMD evolution that transverse momentum distributions broaden when going to higher energies and scales --- see also Sec.~\ref{sec:tmd_evolution}.
For a discussion of this point related to older data from Jefferson Lab, HERMES and COMPASS we refer to~\cite{Schweitzer:2010tt}.

Let us finally address approaches that include higher order QCD corrections.
In that case the parton model result~(\ref{e:F_UU_SIDIS}) for the structure function $F_{UU,T}$ gets replaced by~\cite{Collins:1981uk,Collins:1984kg,Collins:2011zzd}
\begin{eqnarray} \label{e:F_UU_SIDIS_evol}
F_{UU,T} & \propto & \sum_q {\cal H}_q(Q;\mu) \int d^2\vec{p}_T \, d^2\vec{k}_T \, \delta^{(2)}(\vec{p}_T - \vec{k}_T - \vec{P}_{h\perp}/z) 
\nonumber \\
& & \times \; f_1^{q/p}(x,\vec{p}_T^{\;2};\zeta_F,\mu) \, D_1^{h/q}(z,z^2\vec{k}_T^{\,2};\zeta_D,\mu) + Y(Q,\vec{P}_{hT}) \,,
\end{eqnarray}
where ${\cal H}_q$ represents higher order terms in the hard part of the structure functions.
For the PDFs and FFs proper TMD evolution is used, where $\zeta_F$ and $\zeta_D$ represent the relevant scale for the evolution of the PDF and the FF, respectively --- see also Sec.~\ref{sec:tmd_evolution}. 
The so-called $Y$-term on the r.h.s.~of Eq.~(\ref{e:F_UU_SIDIS_evol}) describes the region for large values of $|\vec{P}_{h\perp}|$ where a fixed higher-order calculation is supposed to work.
A numerical result for the TMD evolution of $D_1^{u/p}$, based on Ref.~\cite{Kang:2015msa}, is shown in Fig.~\ref{f:TMD_evolution}.  
As already mentioned in Sec.~\ref{sec:tmd_evolution}, currently there is a lot of discussion about the phenomenology of TMD evolution, which is largely related to the non-perturbative input to the evolution.
One challenge in that regard is the simultaneous description of SIDIS data taken at relatively low scales by HERMES and COMPASS, and data for gauge boson production in hadronic collisions --- see~\cite{Rogers:2015sqa} and references therein.
In Ref.~\cite{Bacchetta:2015ora} a detailled discussion is given on how new data for $e^+ e^- \to h_a h_b X$ might help to put further constraints on TMD evolution.
Though there is still a lot of uncertainty concerning the numbers coming out of such full QCD analyses, we quote here Gaussian widths obtained in one such study.
For example in Ref.~\cite{Echevarria:2014xaa} it was found 
\begin{equation} \label{e:widths_EIKV}
\langle p_T^2 \rangle \big|_{\textrm{Ref.}\;\scriptsize{\cite{Echevarria:2014xaa}}} = (0.25 \; \textrm{---} \; 0.44) \, \textrm{GeV}^2 \,, \qquad
\langle P_{hT}^2 \rangle \big|_{\textrm{Ref.}\;\scriptsize{\cite{Echevarria:2014xaa}}} = (0.16 \; \textrm{---} \; 0.20) \, \textrm{GeV}^2 \,,
\end{equation}
where the numbers refer to the scale $\mu^2 = 2.4 \, \textrm{GeV}^2$.
Obviously these numbers basically agree with the ones in~(\ref{e:widths_STM}) and~(\ref{e:widths_Anselmino}) based on the parton model approximation.
Considerable differences between the two approaches typically appear when the TMDs are evolved to higher scales.
The full QCD machinery for TMDs can be expected to become the standard approach for analyzing TMD-related observables once its phenomenological issues, in particular in the case of SIDIS, have been sorted out.

\subsection{Fits of TMD FF $H_1^\perp$}
\label{sec:CFFExtraction}
\begin{figure}[t]
\begin{center}
\includegraphics[width=0.48\textwidth]{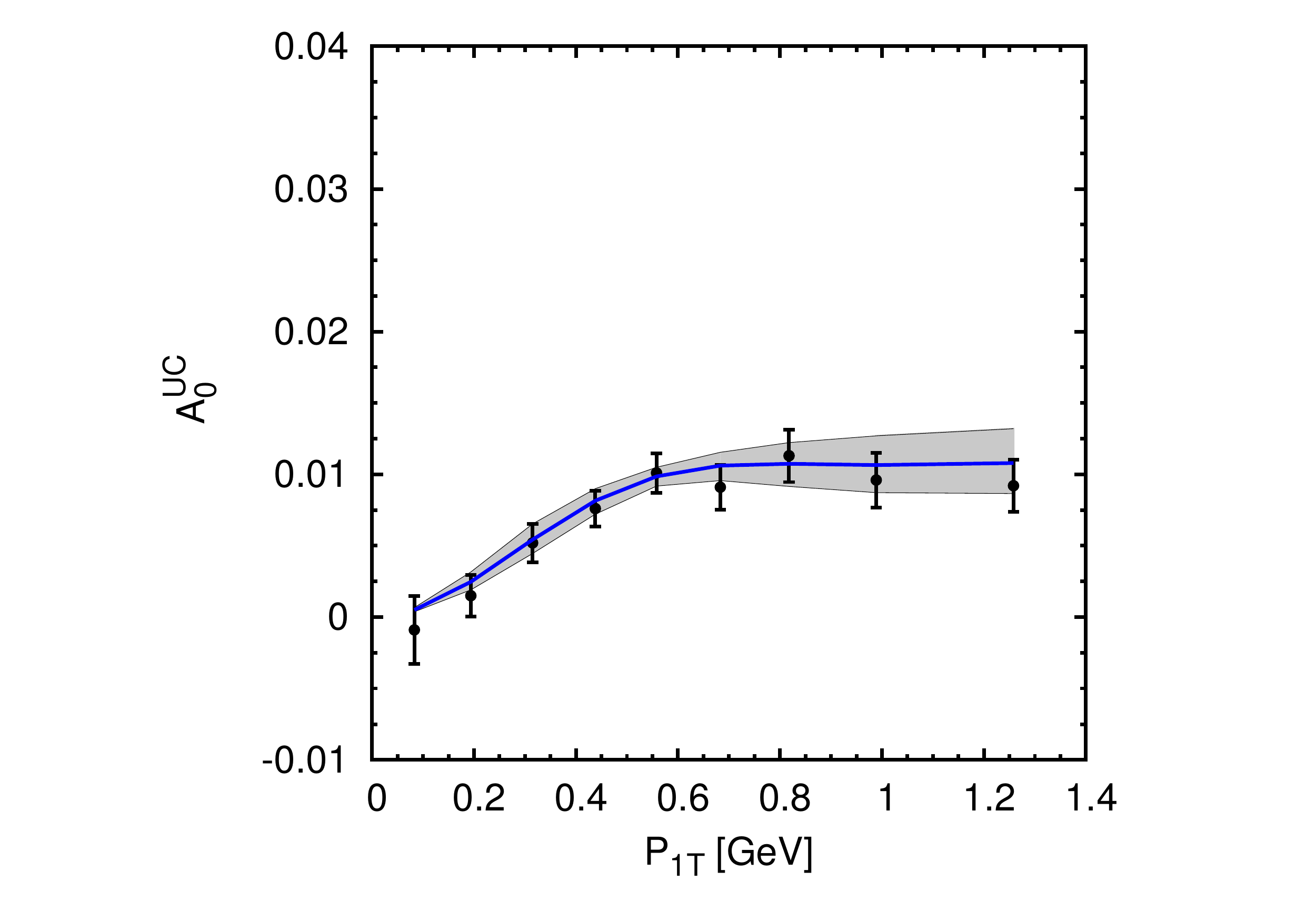} 
\includegraphics[width=0.48\textwidth]{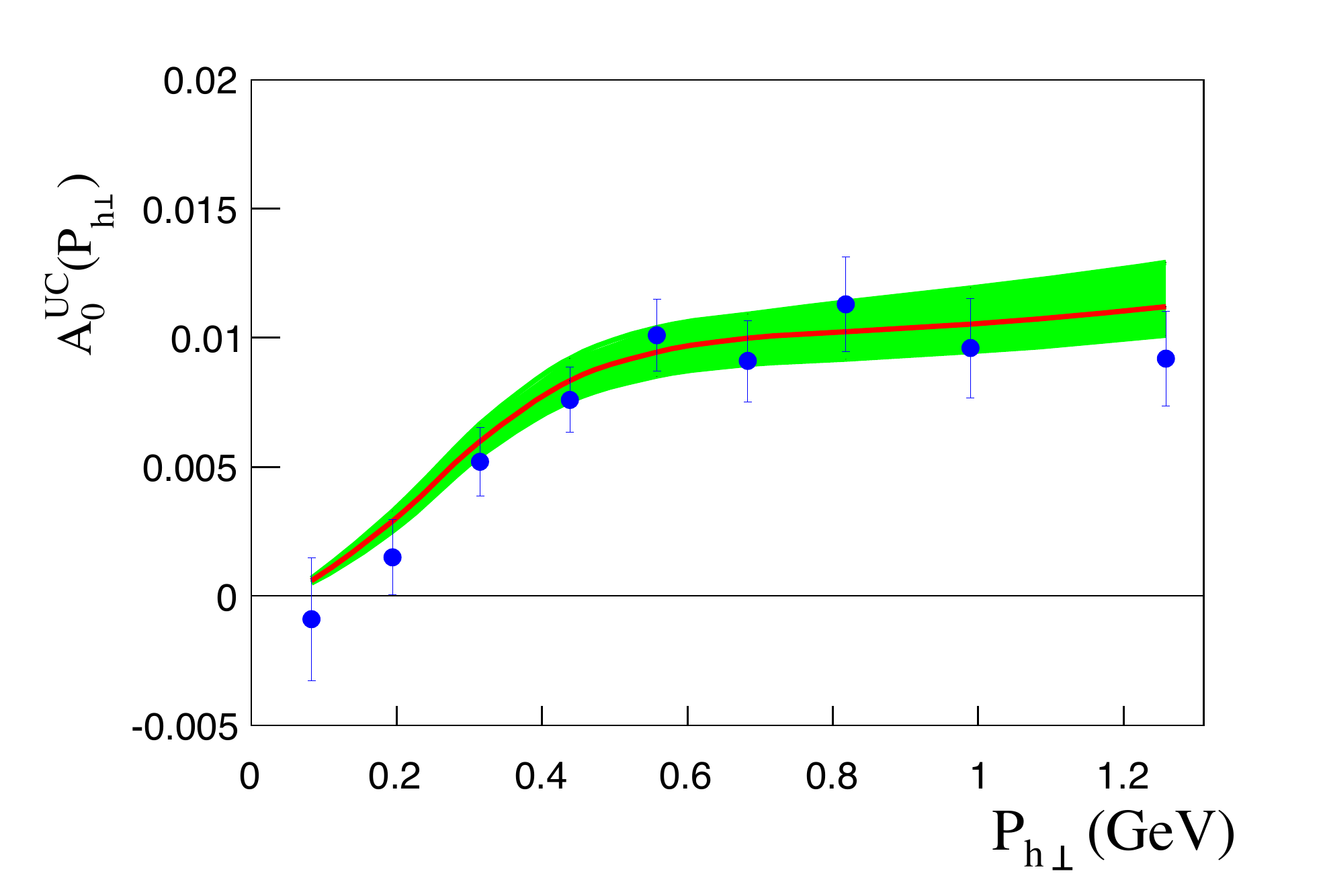}
\end{center}
\vspace{-0.4cm}
\caption{Data on the azimuthal correlation $A_0^{UC}$, as measured by the BaBar Collaboration~\cite{TheBABAR:2013yha}.
They are compared to a fit without TMD evolution (left panel), and with TMD evolution (right panel).
Note that $P_{1T}|_{\rm left \; figure} = |\vec{P}_{h\perp}|$.
Left figure reprinted with permission from~\cite{Anselmino:2015sxa}~\href{http://dx.doi.org/10.1103/PhysRevD.92.114023}{M.~Anselmino, et al., Phys.~Rev.~D92~(2015)~114023}. Copyright (2015) by the American Physical Society.
Right figure reprinted with permission from~\cite{Kang:2015msa}~\href{http://dx.doi.org/10.1103/PhysRevD.93.014009}{Z.-B.~Kang, et al., Phys.~Rev.~D93~(2016)~014009}. Copyright (2016) by the American Physical Society.
\label{fig:extractedCollinsFF_pT}}
\end{figure}
\begin{figure}[t]
\begin{center}
\includegraphics[width=0.48\textwidth]{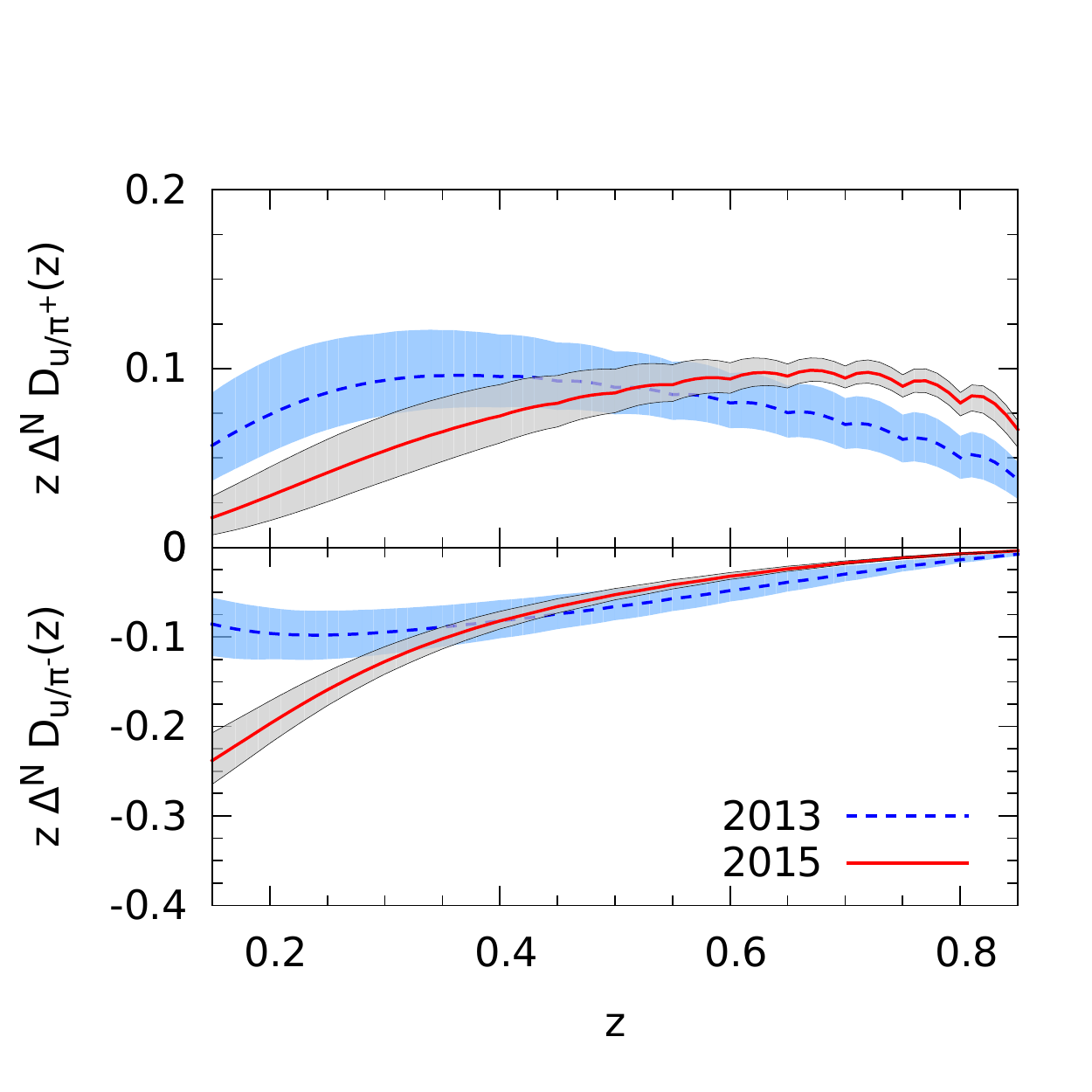} 
\includegraphics[width=0.45\textwidth]{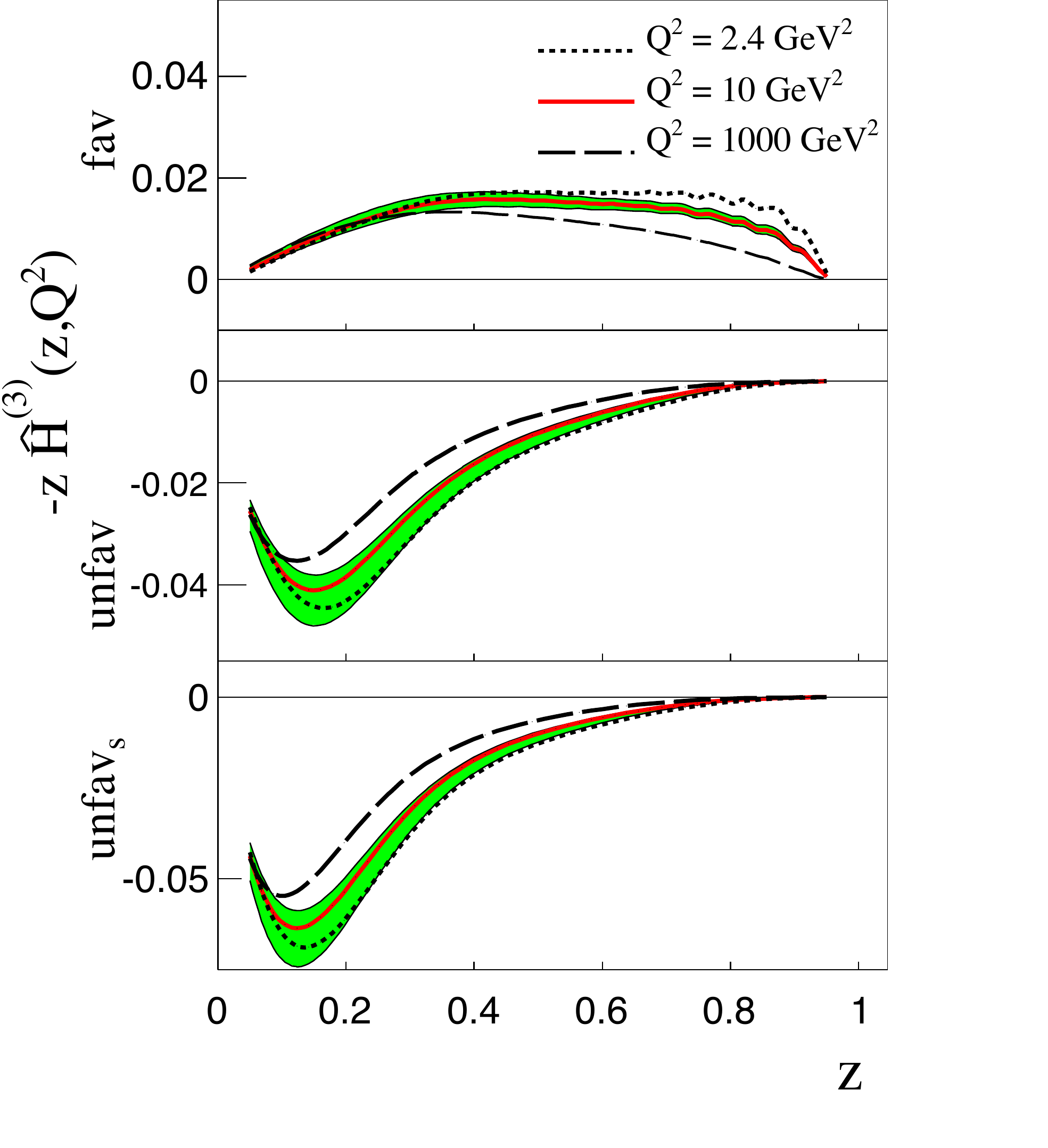}
\end{center}
\vspace{-0.4cm}
\caption{Left panel: Extraction of the Collins function (with error bands) at $Q^2 = 2.4 \; \rm{GeV}^2$ without TMD evolution. 
A comparison to a previous extraction from the same group~\cite{Anselmino:2013vqa} is also given.
Note that $z \, \Delta^N D(z)|_{\rm{plots}} = \frac{2}{M_h} \int d^2\vec{P}_{hT} \, |\vec{P}_{hT}| \, H_1^{\perp}(z,\vec{P}_{hT}^2)$.
Right panel: Extraction of the Collins function. 
The error band is for the curves at $Q^2 = 10 \; \rm{GeV}^2$.
Note that $z \, \hat{H}^{(3)}(z)|_{\rm plots} = - 2 \, z^2 M_h H_1^{\perp \, (1)}(z)$ with the moment definition from~(\ref{e:kT_moment}).
Left figure reprinted with permission from~\cite{Anselmino:2015sxa}~\href{http://dx.doi.org/10.1103/PhysRevD.92.114023}{M.~Anselmino, et al., Phys.~Rev.~D92~(2015)~114023}. Copyright (2015) by the American Physical Society.
Right figure reprinted with permission from~\cite{Kang:2015msa}~\href{http://dx.doi.org/10.1103/PhysRevD.93.014009}{Z.-B.~Kang, et al., Phys.~Rev.~D93~(2016)~014009}. Copyright (2016) by the American Physical Society.
\label{fig:extractedCollinsFF_z}}
\end{figure}

The Collins function $H_1^\perp$ is a spin dependent, chiral odd TMD. In SIDIS (and $pp$) it couples to the similarly poorly known transversity distribution $h_1$. 
Fits that do not include $e^+e^-$ data therefore have to simultaneously extract $h_1$ and $H_1^\perp$. 
Since $H_1^\perp$ describes the spin dependence of the intrinsic transverse momentum in the fragmentation process, the transverse momentum dependence of the FF and the PDF has to be modeled as well. 
Furthermore, the experimental input consists of SSAs measured at SIDIS experiments and ratios measured in $e^+e^-$ annihilation, as described in Sec.~\ref{sec:results}, which contain a nontrivial mix of spin-dependent and spin-averaged PDFs and FFs that both carry transverse-momentum dependence. 
First extractions of $H_1^\perp$ have been performed using SIDIS data~\cite{Vogelsang:2005cs} and, when the Belle data became available, using information from SIDIS and $e^+e^-$ at the same time~\cite{Efremov:2006qm}. 
The first simultaneous extraction of $H_1^\perp$ and $h_1$ from a global fit to SIDIS data from HERMES and $e^+e^-$ data from Belle has been performed by~\cite{Anselmino:2007fs}. This fit has then been subsequently improved as new data became available from COMPASS and BaBar~\cite{Anselmino:2008jk,Anselmino:2013vqa,Anselmino:2015sxa,Anselmino:2015fty}.
The most recent iterations include the $|\vec{P}_{h\perp}|$ dependence and the kaon asymmetries measured by BaBar.
The fit factorizes the collinear part of FF and PDF and the transverse momentum dependent part. Due to the lack of data, only favored and disfavored FFs for (u,d) and s quarks are considered. There is also too little data to extract the transverse momentum dependence, instead a gaussian shape is assumed with a width that is fixed for all $Q^2$. The width is taken to be equal to the widths of unpolarized PDFs and FFs  which are determined as described in Sec.~\ref{sec:fits_integrated_D1} above.
The collinear part of the FF is parametrized in terms of the unpolarized FF with two free parameters. The kaon FF has only one free parameter and is chosen to be proportional to $D_1^{\pi/q}$ and only the favored part of the Collins FF for kaons is constrained in a meaningful way. Only the collinear part is evolved using regular DGLAP.
Even though this approach is rather simple, it achieves a very good description of the available data within their uncertainties. Given that the multiplicity data in a similar kinematic regime can be fitted well with the same approach, this might be expected. 
Fig.~\ref{fig:extractedCollinsFF_pT} shows the fit to the BaBar pion data on the left side compared to the extraction of~\cite{Kang:2015msa} on the right side. 
The later uses TMD evolution in the CSS framework --- see also Sec.~\ref{sec:tmd_evolution}.
Fig.~\ref{fig:extractedCollinsFF_z} shows the resulting extractions of $H_1^\perp$ vs $z$. Even though the Gaussian approach in~\cite{Anselmino:2015fty} is much simpler, the description of the $z$ as well as the $|\vec{P}_{h\perp}|$ dependence of the data is equally good.
Even the prediction for the asymmetries measured at BESIII, which are not part of the fit are in reasonable agreement with the data.
Since $H_1^\perp$ is a TMD, it is expected that the simple model for the $|\vec{P}_{hT}|$ dependence as well as the restriction of the evolution to the collinear part fails for sufficiently precise data or predictions at sufficiently different scales. 
To address this issue, Ref.~\cite{Kang:2015msa} uses TMD evolution.
Here, in addition to the collinear part, which is parametrized in a similar way as in the Gaussian approach of ~\cite{Anselmino:2015fty}, another non-perturbative function is needed, the so-called soft part of the evolution kernel. However, this soft part is universal and a parametrization from literature is choosen, i.e., it does not have to fitted to the data. 
(We also refer to Sec.~\ref{sec:fit_TMDFF_D1} for a brief discussion of present phenomenological problems related to TMD evolution.)
Instead of $|\vec{P}_{hT}|$, the relevant parametrizations are done in $b_T$ space and are not trivial. 
However, the resulting $|\vec{P}_{hT}|$ shape, at least at low to moderate scales, is similar to the Gaussian shape chosen by~\cite{Anselmino:2015fty}.
This is shown in Fig.~\ref{fig:extractedCollinsFF_z} for the $z$ dependence and in Fig.~\ref{fig:extractedCollinsFF_pT} for the $|\vec{P}_{h\perp}|$ dependence of the $e^+e^-$ data which we referred to earlier. 
As already remarked above, the fit to the data and the size of the error bars are comparable to the non-TMD evolved fit shown in the left panels.


\subsection{Fits of higher-twist FFs}
\label{sec:fit_higher_twist}
At present, experimental input on higher twist FFs primarily comes from data on the transverse SSA $A_N$ in $p^{\uparrow} p \to \pi X$~\cite{Adams:2003fx,Lee:2007zzh,Arsene:2008aa,Abelev:2008af,Adamczyk:2012xd} (see also Sec.~\ref{sec:twist3FF_data}).
The general structure of this observable in collinear twist-3 factorization is discussed above in Sec.~\ref{sec:observables_twist3}.
According to Eq.~(\ref{e:sigma_generic}) it has three main contributions.
The first term in~(\ref{e:sigma_generic}) was studied in a number of papers~\cite{Qiu:1998ia, Kouvaris:2006zy, Koike:2007rq, Koike:2009ge,Kanazawa:2010au,Kang:2011hk,Beppu:2013uda}.  
It contains quark-gluon-quark correlations and tri-gluon correlations in the polarized proton, where for the former one distinguishes between so-called soft-gluon pole and soft-fermion pole contributions.
The second term in~(\ref{e:sigma_generic}), which arises from twist-3 effects in the unpolarized proton, was shown to be small~\cite{Kanazawa:2000hz,Kanazawa:2000kp}.
A complete analytical result for the third term in~(\ref{e:sigma_generic}), that is, the twist-3 fragmentation contribution, was obtained in~\cite{Metz:2012ct}.
For some time it was generally believed that the soft-gluon pole contribution, which is associated with the transversely polarized proton and given by the twist-3 Qiu-Sterman function $T_F$~\cite{Qiu:1991pp,Qiu:1991wg}, dominates $A_N$ in hadronic collisions.
The situation changed when people made use of a relation between $T_F$ and the Sivers TMD PDF~\cite{Boer:2003cm}.
The Sivers function enters the SIDIS cross section and had been fitted to available data.
But an attempt failed to simultaneously explain $A_N$ and the Sivers asymmetry in SIDIS~\cite{Kang:2011hk}. 
The $T_F$ extracted from the two reactions had opposite signs~\cite{Kang:2011hk}, which is now known as ``sign-mismatch" problem.
Other scenarios discussed in the literature either could not solve the problem~\cite{Kang:2012xf} or are unlikely to do so~\cite{Koike:2009ge,Beppu:2013uda}.
\begin{figure}[t]
\begin{center}
\includegraphics[width=0.48\textwidth]{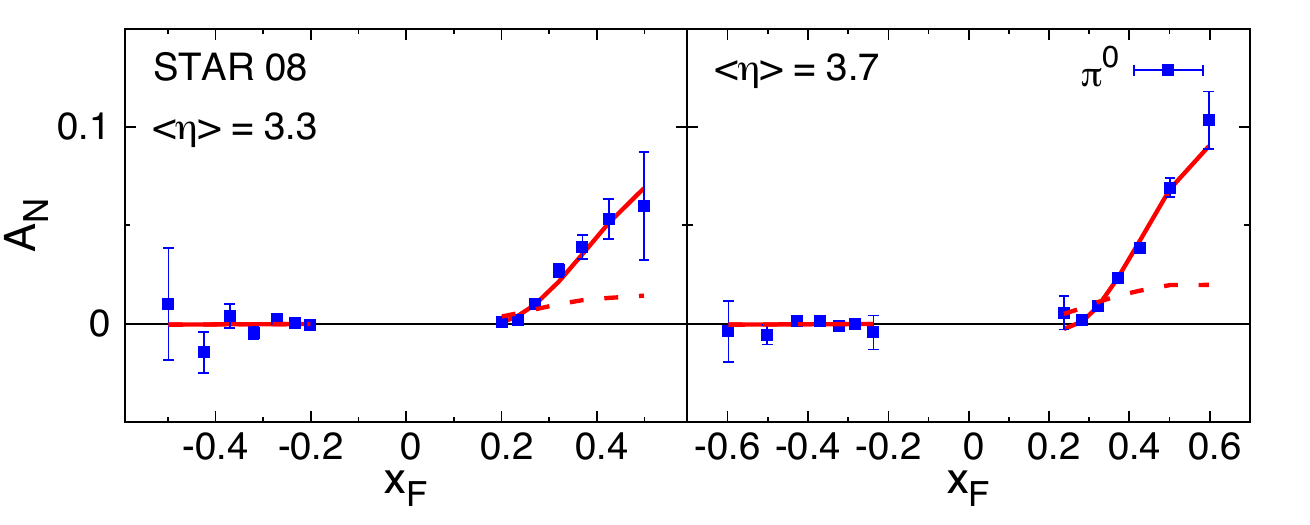} 
\includegraphics[width=0.48\textwidth]{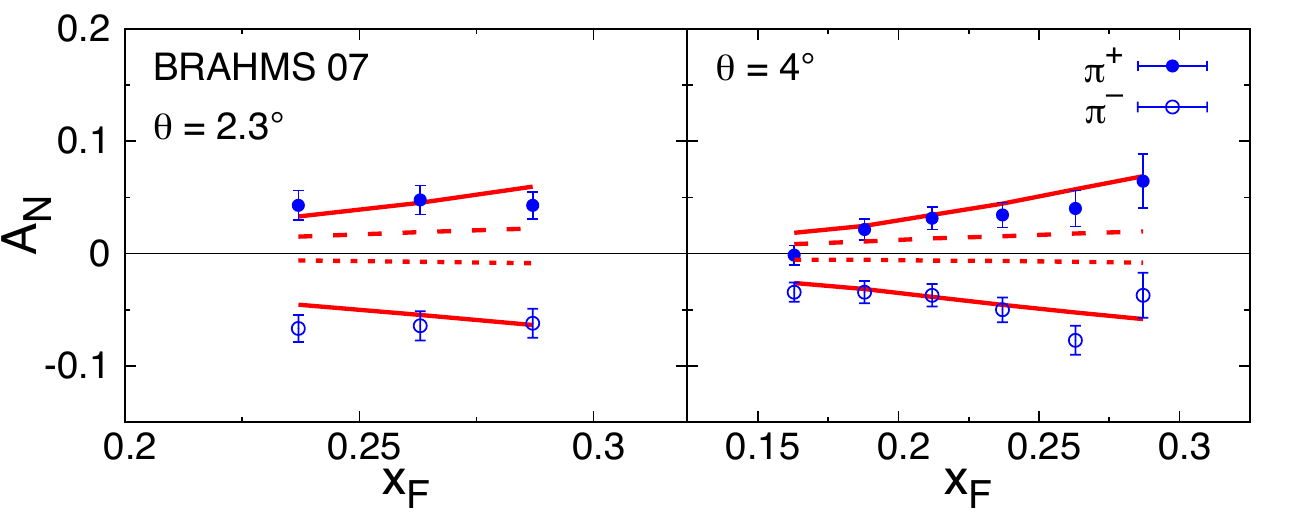}
\end{center}
\vspace{-0.4cm}
\caption{Fit to data on $A_N$ from RHIC for neutral pion production (left panel) from STAR~\cite{Adams:2003fx,Abelev:2008af,Adamczyk:2012xd} and charged pion production (right panel) from BRAHMS~\cite{Lee:2007zzh}.
The fit corresponds to $\chi^2/{\rm d.o.f.} = 1.03$~\cite{Kanazawa:2014dca}.
The dashed line (dotted line in the case of $\pi^-$) shows the result if the three-parton FF $\hat{H}_{FU}^{\Im}$ is switched off.
Figures reprinted with permission from~\cite{Kanazawa:2014dca}~\href{http://dx.doi.org/10.1103/PhysRevD.89.111501}{K.~Kanazawa, et al., Phys.~Rev.~D89~(2014)~111501}. Copyright (2014) by the American Physical Society.}
\label{f:twist3_fit}
\end{figure}

These findings definitely put into question the dominance of the first term in~(\ref{e:sigma_generic}).
Also data on the transverse SSA in fully inclusive DIS~\cite{Airapetian:2009ab,Katich:2013atq} seem to support this point of view~\cite{Metz:2012ui}.
Therefore, in Ref.~\cite{Kanazawa:2014dca} the potential role of the twist-3 fragmentation part of~(\ref{e:sigma_generic}) was studied.
According to Eq.~(\ref{e:sigma_generic_frag}), three twist-3 FFs enter that contribution to the asymmetry, where the analysis in~\cite{Kanazawa:2014dca} took into account the relation in~(\ref{e:relation_twist3}) between these functions.
The moment of the Collins function $H_1^{\perp (1) \, h/q}$ and the transversity distribution $h_1^{q/p}$ were taken from a simultaneous fit of the Collins asymmetry in SIDIS and the azimuthal $\cos(2\phi)$ asymmetry in $e^+ e^- \to h_a h_b X$~\cite{Anselmino:2013vqa}.
The three-parton FF $\hat{H}_{FU}^{h/q \, \Im}$ was fitted to RHIC data for $A_N$ in $p^{\uparrow} p \to \pi X$, taken at $\sqrt{s} = 200 \, \textrm{GeV}$ and at transverse hadron momenta $|\vec{P}_{h\perp}| \ge 1 \, \textrm{GeV}$~\cite{Kanazawa:2014dca}.
The third twist-3 FF $H^{h/q}$ is then fixed through the relation~(\ref{e:relation_twist3}).
The aforementioned soft-gluon pole contribution to $A_N$ was also included in the numerical calculation and fixed through the Sivers function given in Refs.~\cite{Anselmino:2008sga,Anselmino:2013rya} which was extracted from data on the Sivers asymmetry in SIDIS. 
In this framework a good fit was obtained (see Fig.$\,$\ref{f:twist3_fit}), where the analysis also showed that the data cannot be described at all if the three-parton FF $\hat{H}_{FU}^{h/q \, \Im}$ is switched off~\cite{Kanazawa:2014dca}.
As pointed out in~\cite{Kanazawa:2014dca}, the twist-3 approach can also explain the rather flat behaviour of $A_N$ as function of $|\vec{P}_{h \perp}|$ observed at RHIC~\cite{Heppelmann:2013ewa}, which was generally considered a big challenge.
The twist-3 fragmentation contribution was found to be the dominant contribution to $A_N$.
In this context one has to keep in mind that the soft-gluon pole term comes with an opposite sign compared to the data due to the sign-mismatch problem.
The fitted function $\hat{H}_{FU}^{h/q \, \Im}$ of Ref.~\cite{Kanazawa:2014dca} shares similarities with a spectator model calculation~\cite{Lu:2015wja}.
In general, the work~\cite{Kanazawa:2014dca} suggested that one can simultaneously describe $A_N$ in {\it pp} scattering using collinear twist-3 factorization, the Sivers asymmetry in SIDIS, the Collins asymmetry in SIDIS, and the azimuthal $\cos(2\phi)$ asymmetry in $e^+ e^- \to h_a h_b X$.
In particular, twist-3 fragmentation effects may play an important role when trying to answer the longstanding question about the origin of the observed large transverse SSAs in hadronic collisions.

Based on the fit of~\cite{Kanazawa:2014dca}, in Ref.~\cite{Gamberg:2014eia} $A_N$ for $\ell N^{\uparrow} \to h X$ was computed and compared to data from HERMES~\cite{Airapetian:2013bim} and Jefferson-Lab~\cite{Allada:2013nsw}.
Agreement in sign was found, but the calculation typically overshoots the data. 
A number of potential reasons for the discrepancy were discussed in~\cite{Kanazawa:2014dca}.
Most importantly NLO corrections may be very large for $A_N$ in lepton-nucleon scattering for the kinematics of the HERMES and Jefferson-Lab experiments.
For the twist-2 unpolarized cross section of $\ell N \to h X$ the NLO contributions are indeed  very significant~\cite{Hinderer:2015hra}.
It would be important to also push the calculation of $A_N$ for this process to the NLO level. 
Another key point for the phenomenology may be that Ref.~\cite{Kanazawa:2014dca} did not use the constraint among the three tiwst-3 FFs which follows from the LIR relations~\cite{Kanazawa:2015ajw} --- see Eqs.~(\ref{e:H_twist3}),~(\ref{e:H1perp_twist3}) and the discussion in Sec.~\ref{sec:observables_twist3}.
This may reduce the freedom one has when fitting $\hat{H}_{FU}^{h/q \, \Im}$.
Therefore an update of the numerics of Ref.~\cite{Kanazawa:2014dca} is needed.

\subsection{Fits of di-hadron FF $H_1^\open$}
\label{sec:fit_dihadron}

\begin{figure}[!]
\begin{center}
\includegraphics[width=0.48\textwidth]{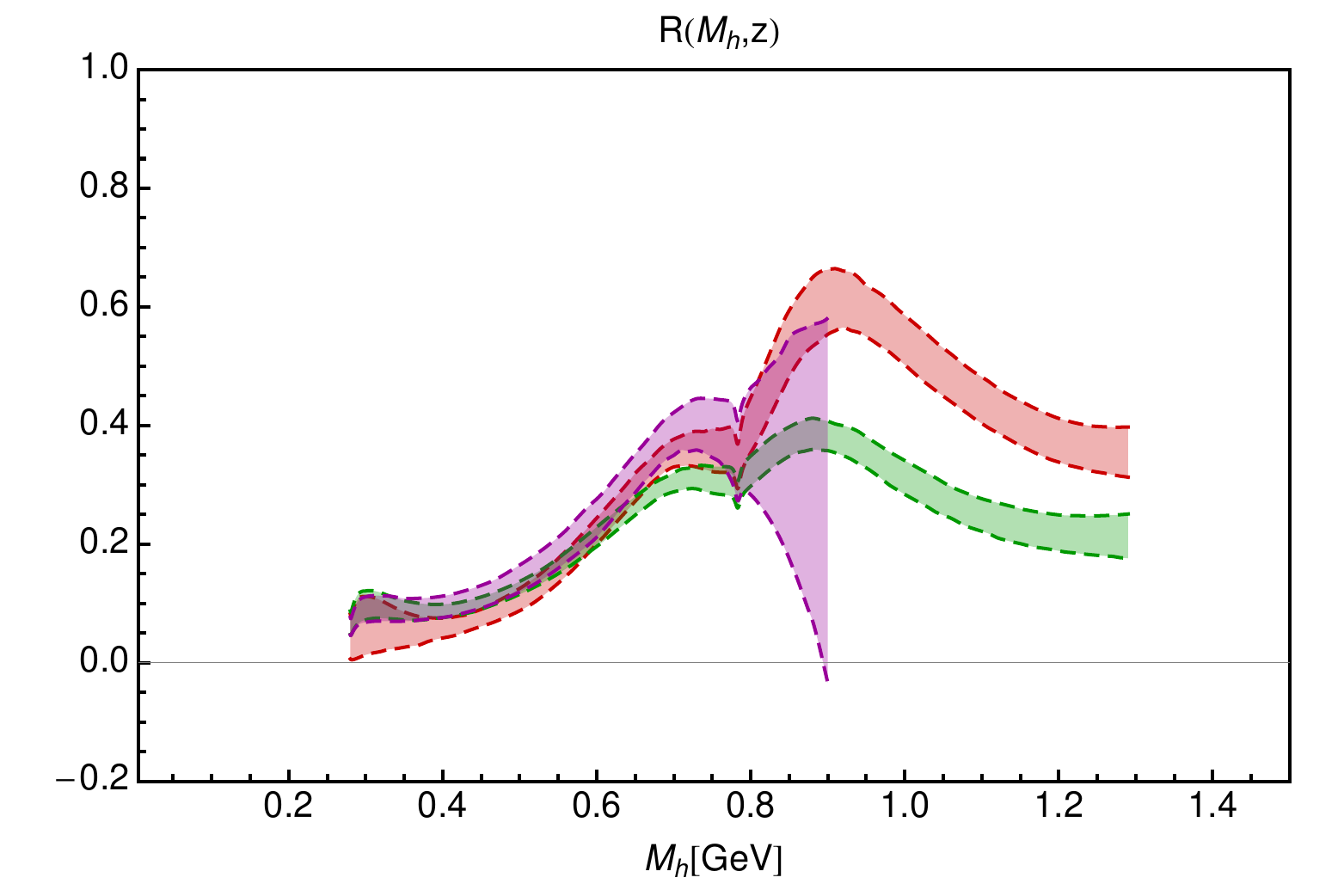} 
\includegraphics[width=0.48\textwidth]{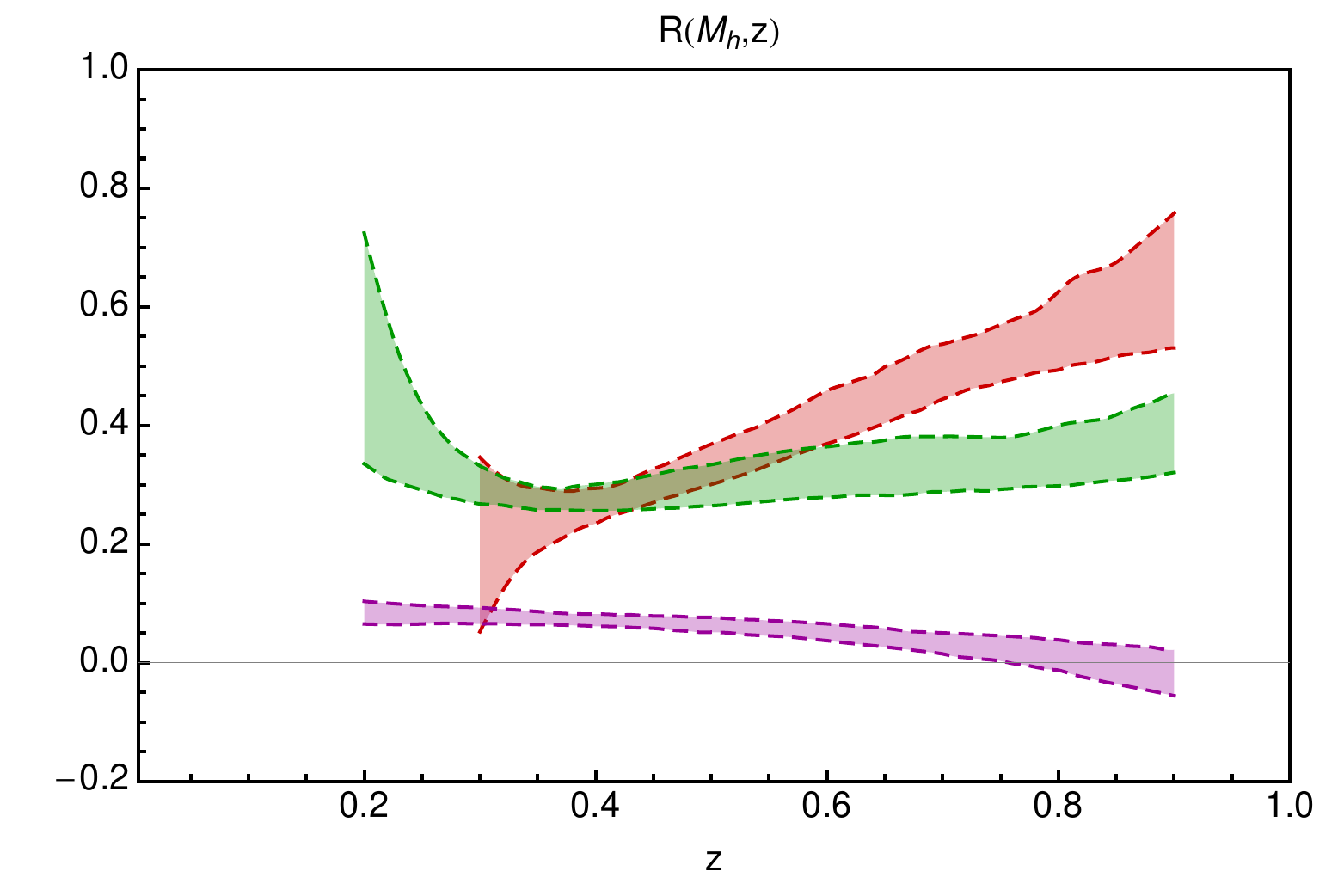}
\end{center}
\vspace{-0.4cm}
\caption{Ratio $R$ between $H_1^{\open h_1 h_2 / q}$ and $D_1^{h_1 h_2 /q}$ as a function of $M_h$ (left panel) and as a function of $z$ (right panel). 
The results are for the scale $Q^2 = 1 \; \rm{GeV}^2$.
Left panel: results for three different values of $z$: $z = 0.25$ (shortest band), $z = 0.45$ (lower band at $M_h \sim 1.2 \; \rm{GeV}$), and $z = 0.65$ (upper band at $M_h \sim 1.2\, \textrm{GeV}$). 
Right panel: results for three different values of $M_h$: $M_h = 0.4 \,\textrm{GeV}$ (lower band at $z \sim 0.8$), $M_h = 0.8\, \textrm{GeV}$ (mid band at $z \sim 0.8$), and $M_h = 1 \, \textrm{GeV}$ (upper band at $z \sim 0.8$). 
All results for $\alpha_s(M_Z^2) = 0.125$.
Figures reprinted from~\cite{Radici:2015mwa}.
\label{fig:IffFit}}
\end{figure}

The integrated di-hadron fragmentation function $H_1^\sphericalangle$ has been extracted from the Belle measurements of di-pion pairs in~\cite{Courtoy:2012ry,Radici:2015mwa}.
Both, unpolarized and  polarized part of $H_1^\sphericalangle$ have been modeled by the product of a $z$ dependent part, a part depending on the invariant mass of the pion pair, and a part with mixed dependence. 
The functional form of the $z$ dependent part has been chosen similar to the functional form used for the Collins single hadron fragmentation function and for the $M_{\pi\pi}$ dependent part a Breit-Wigner function has been chosen. A purely $M_{\pi\pi}-z$ factorized form does not lead to a good fit, so a polynomial mixing $M_{\pi\pi}$ and $z$ has been added. 
In particular the dependence on the mass of the hadron pair is inspired by model calculations where $H_1^\sphericalangle$ receives contributions from the $\pi\pi$ continuum as well as resonant channels, where pion pairs in relative partial waves with different quantum numbers than the continuum are produced. In the kinematic region $M_{\pi\pi}< 1 GeV$, these are mainly pions coming from $\rho$, $\omega$ and $K_S^0$ decays and each of these channels are added to the functional form.
Unlike for the single hadron TMD $H_1^\perp$, there does not exist an extraction of the unpolarized part $D_1^{h_1 h_2/q}$  yet. Therefore the unpolarized part has been taken from fits to Pythia~\cite{Sjostrand:2006za} simulations.
The resulting extraction for the normalized ratio of  $H_1^\sphericalangle/D_1^{h_1 h_2/q}$ is shown in Fig.~\ref{fig:IffFit} vs $M_{\pi\pi}$ and $z$.
It is interesting that the slope of the $z$ dependence changes for different values of the invariant mass. The $M_{\pi\pi}$ dependence shows some $z$ dependence of the contributing di-pion channels.
The fragmentation function $H_1^\sphericalangle$ can also be extracted from SIDIS and $pp$ data in a similar way as the global fits to $H_1^\perp$ described above, with the added advantage that a collinear framework can be used for which factorization and evolution is better understood. However, at this point in time no such truly global extraction is available.


\section{Models}
\label{sec:models}
It is challenging to describe FFs by using non-pertubative approaches to the strong interaction. 
So far FFs have not been computed in lattice QCD.
This can be traced back to the sum over the spectators $X$ in the definition of FFs which can not simply be eliminated using closure because of the (complicated) state $|P_h, S_h; X \rangle$.  
Therefore one has to resort to models.
Roughly speaking two general classes of models have been explored.
One of them is comprised of spectator models in which the parton, in a single step, fragments into a hadron plus a spectator.
The second class comprises models that describe multiple hadron emission, where a number of different versions exist in the literature.
Some recent works aim at combining the two frameworks.
It is impossible to give here a full account of all the existing approaches, and we will mostly focus on the spectator models which recently have been used frequently to calculate (spin-dependent) TMD FFs as well as DiFFs.
In general, at present it is still difficult to make robust predictions for FFs.
Typically one has to fit a number of model parameters to existing data.

\subsection{Spectator models}
\label{sec:spectator_models}
\begin{figure}[t]
\begin{center}
\includegraphics[width=7.0cm]{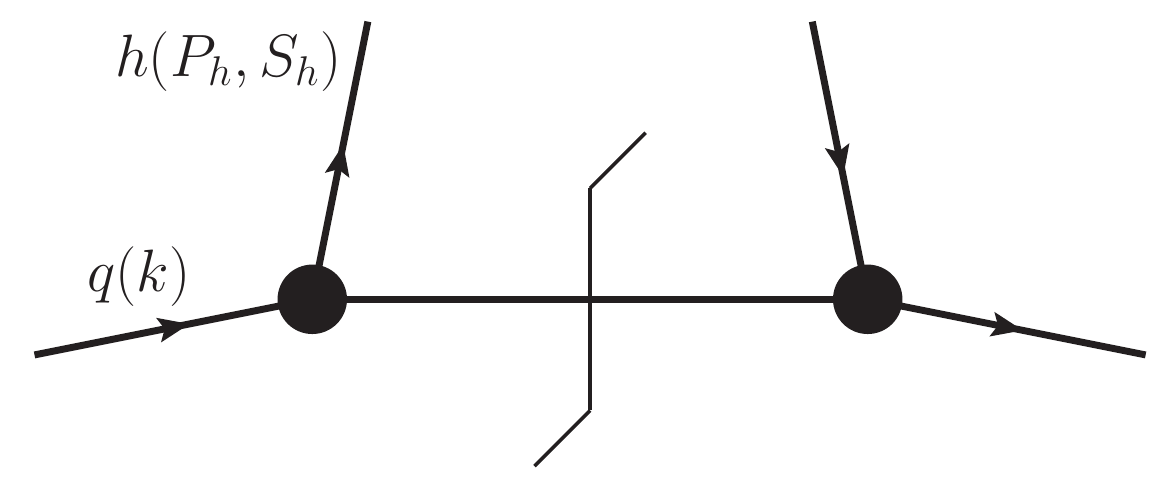} 
\end{center}
\vspace{-0.4cm}
\caption{Representation of the fragmentation correlator $\Delta^{h/q}$ in Eq.~(\ref{e:corr_quark_int}) in a spectator model to leading order in the quark-hadron-spectator coupling.
The inclusive system $X$ is represented by a single spectator particle, which carries 4-momentum $k - P_h$ and is on-shell.}
\label{f:spectator}
\end{figure}

The underlying idea of spectator models for FFs can be seen from the diagram in Fig.$\,$\ref{f:spectator}: a (time-like off-shell) parton fragments into a hadron and a single (real) spectator particle. 
So far such spectator models have been used to describe fragmentation of quarks.
Since the spectator is on-shell one readily finds that the virtuality $k^2$ of the fragmenting quark is fixed once $\vec{k}_T$ (or equivalently $\vec{P}_{h\perp}$) and $z$ are known,
\begin{equation} \label{e:parton_virtuality}
k^2 = \vec{k}_T^{\,2}\frac{z}{1-z} + \frac{M_s^2}{1-z} + \frac{M_h^2}{z} \,,
\end{equation}
where $M_s$ is the spectator mass.
The type of the spectator depends on the fragmentation process.
For instance, if an up quark fragments into a $\pi^+$ the spectator is a down quark.
Or, if an up quark fragments into a proton the spectator is a $\bar{u}\bar{d}$ anti-diquark.
Diquark spectators can either be in a spin-0 state (scalar diquark) or in a spin-1 state (axial diquark)~\cite{Melnitchouk:1993nk,Jakob:1997wg}.

In order to compute the fragmentation of, e.g., light quarks into pions one could use a self-consistent field-theoretic model as defined by the Lagrange density
\begin{equation} \label{e:ps_coupling}
{\cal L}_{\pi qq} = i \, g_{\pi q} \, \bar{q}(x) \, \gamma_5 \, q(x) \, \pi(x) \,,
\end{equation}
which describes a pseudo-scalar quark-pion coupling.
Note that in~(\ref{e:ps_coupling}) we neglect any reference to isospin.
Evaluating the LO diagram in Fig.$\,$\ref{f:spectator} by using~(\ref{e:ps_coupling}) one finds the following favored fragmentation into pions~\cite{Collins:1992kk,Bacchetta:2001di},
\begin{equation} \label{e:ps_D1_kT}
D_1^{\rm fav}(z,z^2\vec{k}_T^{\,2}) = D_1^{\pi^+/u}(z,z^2\vec{k}_T^{\,2}) = \frac{1}{z} \, \frac{g_{\pi q}^2}{8 \pi^3} \, \frac{\vec{k}_T^{\,2} + m^2}{\Big(\vec{k}_T^{\,2} + m^2 + \frac{1-z}{z}m_{\pi} \Big)^2} \,,
\end{equation}
with $m$ representing the quark mass.
Obviously, in this model both the fragmenting parton and the spectator have the same mass.
Nonzero disfavored FFs can be obtained only through higher order graphs.
For instance, in the case of $D_1^{\pi^-/u}$ the up quark first has to convert into a down quark by emitting a $\pi^+$ which becomes part of the (two-particle) spectator system. 

Integrating the expression in~(\ref{e:ps_D1_kT}) upon $k_T$ provides $D_1^{\rm fav}(z)$.
Since for large transverse momenta $D_1^{\rm fav}(z,z^2\vec{k}_T^{\,2})$ in~(\ref{e:ps_D1_kT}) behaves like $1/\vec{k}_T^{\,2}$ --- the same large $k_T$ behaviour that one finds in perturbative QCD --- this integral is UV-divergent.
One way of getting a finite result is through introducing a cutoff on the virtuality of the fragmenting quark~\cite{Bacchetta:2002tk} which, due to Eq.~(\ref{e:parton_virtuality}), also implies a cutoff for $k_T$.
The numerical result shares qualitative features with the phenomenology of $D_1^{\rm fav}(z)$ in that it decreases with increasing $z$, but one cannot expect such a simple model to give quantitative agreement.
In order to improve the phenomenology also the pseudo-vector quark-pion couping, as utilized in the chiral-invariant Georgi-Manohar model~\cite{Manohar:1983md}, was explored~\cite{Bacchetta:2002tk}.
This approach should be more realistic because the pion decouples from the quark in the limit of a vanishing momentum as expected for a Goldstone boson.
In fact one gets a quite good description of the $z$ behavior of $D_1^{\rm fav}(z)$, but the magnitude is just about half of typical fit results~\cite{Bacchetta:2002tk}, unless one uses a value for the axial quark-pion coupling that differs from the original Georgi-Manohar work~\cite{Manohar:1983md}.
Other similar models were used as well.
In the Nambu-Jona-Lasinio model~\cite{Nambu:1961tp,Nambu:1961fr} integrated FFs for pions were obtained in Ref.~\cite{Ito:2009zc} and for kaons in~\cite{Matevosyan:2010hh}.
In the same approach, the effect of vector mesons on pion and kaon FFs, and fragmentation into baryons were studied in Ref~\cite{Matevosyan:2011ey}.
Moreover, in Refs.~\cite{Nam:2011hg,Nam:2012af,Yang:2013cza} pion and kaon FFs were derived in a lonlocal chiral quark model~\cite{Diakonov:2002fq}.
We refer to the original papers for a detailled discussion of the numerical results.
Overall one again just finds qualitative agreement with existing fits.
Such models typically provide (much) better results for PDFs than for FFs which supports the aforementiond statement that modelling parton FFs is a difficult task.
We also point out that in all these model results for FFs the $z$-dependence and the $k_T$-dependence do not decouple, and the $k_T$-dependence is not Gaussian. 
Both assumptions were frequently used in existing extractions of TMD FFs.
On the other hand, for small $k_T$ the transverse momentum dependence obtained in such spectator models can be approximated by Gausssian forms.
(In this context we also refer to~\cite{Boffi:2009sh}.)
Also, existing data do not permit to unambiguously rule out these assumptions.
We note in passing that Refs.~\cite{Ito:2009zc,Meissner:2010cc,Collins:2011zzd} contain discussion on checks of sum rules for FFs in spectator models.
\begin{figure}[t]
\begin{center}
\includegraphics[width=17.0cm]{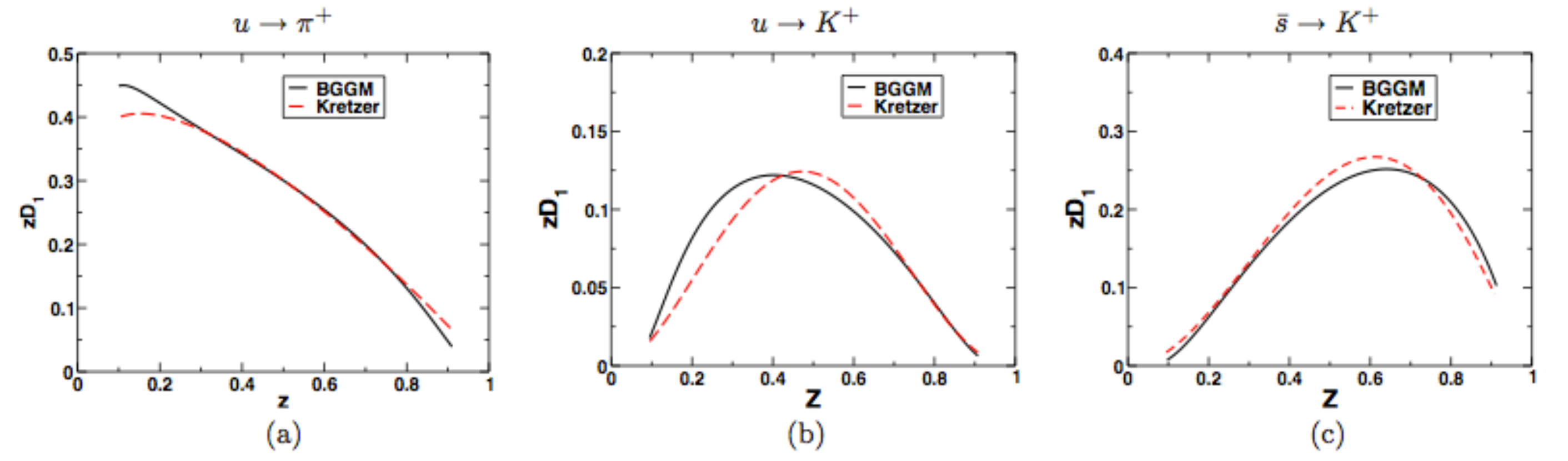} 
\end{center}
\vspace{-0.4cm}
\caption{Spectator model results for $zD_1^{h/q}$ for different transitions: (a) $u \to \pi^+$, (b) $u \to K^+$, (c) $\bar{s} \to K^+$.
The parameters of the model were fixed from a fit to the parameterization of Ref.~\cite{Kretzer:2000yf} (dashed line).
Figures reprinted from ~\cite{Bacchetta:2007wc}.
\label{f:spectator_BGGM}}
\end{figure}

One way to improve the phenomenology of spectator models for FFs is by introducing more free parameters.
Normally two types of modifications are investigated.
First, the mass of the spectator particle is allowed to vary.
Second, the quark-hadron-spectator vertex is decorated by a form factor which depends, in different ways, on the kinematics of the fragmentation process.
Needless to say that such approaches are not related to an underlying Lagrange density.
Studies along these lines can be found, for instance, in Refs.~\cite{Londergan:1996vf,Jakob:1997wg,Kitagawa:2000ji,Kitagawa:2001ig,Yang:2002gh,Bacchetta:2007wc}.
Here we only briefly discuss results for $D_1^{h/q}$ from the model of Ref.~\cite{Bacchetta:2007wc}, where pion and kaon FFs were considered.
The model has a total of five free parameters that were fitted to the FFs from~\cite{Kretzer:2000yf}.
The curves in Fig.$\,$\ref{f:spectator_BGGM} give an indication of the quality of the fit and of the flexibility of the model.
Once the parameters of such a model have been fixed one can predict other FFs.

\begin{figure}[t]
\begin{center}
\includegraphics[width=12.0cm]{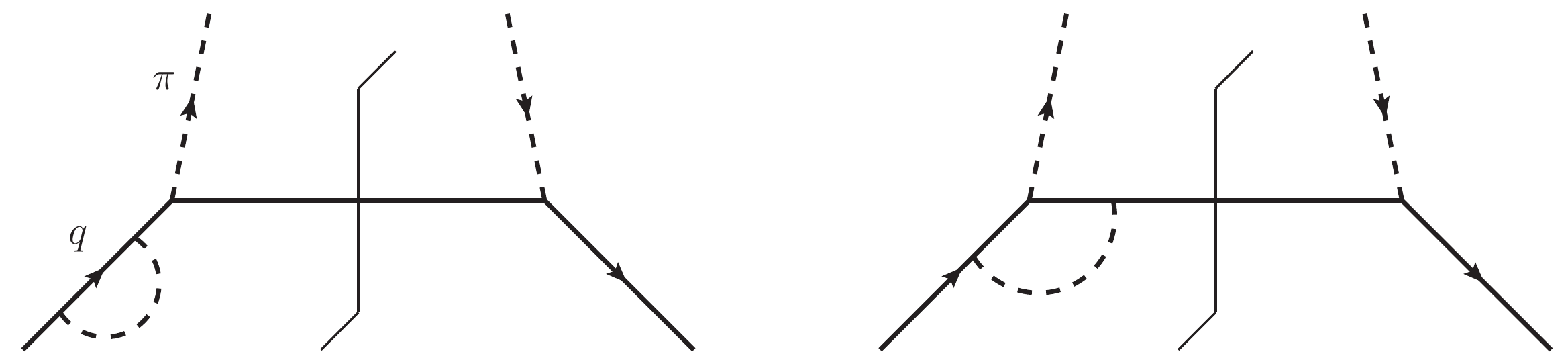} 
\end{center}
\vspace{-0.4cm}
\caption{One-loop corrections to the fragmentation of a quark into a pion in the model defined in Eq.~(\ref{e:ps_coupling}).
Displayed are only the graphs that contribute to the Collins function: quark self-energy (left panel) and vertex correction (right panel).
Other one-loop graphs are irrelevant as they do not provide an imaginary part.
Hermitean conjugate diagrams are not shown.}
\label{f:ps_Collins}
\end{figure}

Spectator models were also used to estimate the Collins function.
A first such calculation was already presented in the original paper by Collins~\cite{Collins:1992kk}.
In that work a nonzero $H_1^\perp$ was obtained by taking into account the imaginary part of the propagator of the time-like fragmenting quark.
Despite this result doubts remained in the community as to whether the Collins function would actually be nonzero or if a cancellation between different imaginary parts may occur in nature --- see for instance Ref.~\cite{Jaffe:1997hf}.
Given that situation, in Ref.~\cite{Bacchetta:2001di} the Lagrange density in~(\ref{e:ps_coupling}) was used for a full field-theoretic model calculation to lowest non-trivial order in the pseudo-scalar coupling $g_{\pi q}$.  
Based on the diagrams in Fig.$\,$\ref{f:ps_Collins} one finds the result~\cite{Bacchetta:2001di}
\begin{equation} \label{e:ps_Collins}
H_1^{\perp \, \pi^+/q}(z,z^2 \vec{k}_T^{\,2}) = - \, \frac{g_{\pi q}^2}{4\pi^3} \frac{m_{\pi}}{1-z} 
\bigg( \frac{m \, {\rm Im} \, \tilde{\Sigma}(k^2)}{(k^2 - m^2)^2} 
+ \frac{{\rm Im} \, \tilde{\Gamma}(k^2)}{k^2 - m^2}\bigg) \bigg|_{k^2 = \vec{k}_T^{\,2}\frac{z}{1-z} + \frac{m^2}{1-z} + \frac{m_{\pi}^2}{z}} \,,
\end{equation}
where ${\rm Im} \, \tilde{\Sigma}$ and ${\rm Im} \, \tilde{\Gamma}$ represent contributions from the imaginary part of the quark self-energy graph and the vertex correction, respectively.
The final result for the Collins function in~(\ref{e:ps_Collins}) is nonzero, which gave support to its existence from the theoretical point of view~\cite{Bacchetta:2001di}.
In the meantime there is of course multiple compelling experimental evidence for a non-vanishing Collins function as discussed in more detail in Sec.$\,$\ref{sec:experiments} and Sec.$\,$\ref{sec:global_fits}.
In Ref.~\cite{Bacchetta:2002tk} the Georgi-Manohar model~\cite{Manohar:1983md} with pion loops was used to calculate $H_1^{h/q \, \perp}$.
Later on also gluon loops were taken into account in different spectator models~\cite{Bacchetta:2003xn,Amrath:2005gv,Bacchetta:2007wc,Matevosyan:2012ga}.
In such approaches even a robust prediction of the sign of the Collins function is nontrivial since individual diagrams can contribute with opposite signs~\cite{Amrath:2005gv}.
The latest phenomenological papers~\cite{Bacchetta:2007wc,Matevosyan:2012ga} used gluon loops only.
The Collins function for pions and kaons was computed in Ref.~\cite{Bacchetta:2007wc} after the parameters of the model were fitted to $D_1^{h/q}$ as discussed in the previous paragraph. 
Disfavored Collins functions were not computed in the spectator model but rather fixed through the Sch\"afer-Teryaev sum rule in~(\ref{e:sum_rule_2})~\cite{Schafer:1999kn,Meissner:2010cc}, combined with a couple of simplifying assumptions~\cite{Bacchetta:2007wc}.
In the case of pions reasonable agreement with parameterizations was obtained, while at that time the paper was written no experimental information was available on the Collins function for kaons~\cite{Bacchetta:2007wc}.

In some papers also higher twist single-hadron FFs were studied in spectator models.
Specifically, calculations of the FF $E^{h/q}$ defined in~(\ref{e:E_quark_z}) can be found in Refs.~\cite{Ji:1993qx,Gamberg:2003pz}.
In a recent work the T-odd functions $H^{h/q}$ in~(\ref{e:H_quark_z}) and the (three-parton) FF $\hat{H}_{FU}^{h/q,\, \Im}$ in~(\ref{e:DeltaF_H}) were computed~\cite{Lu:2015wja}.

Several spectator model calculations are available for DiFFs.
In Ref.~\cite{Collins:1994ax} the unpolarized DiFF $D_1^{\pi\pi/q}$ as well as $H_1^{\open \, \pi\pi/q}$ were computed in the linear sigma model, which was the first estimate ever for the IFF.
In that model the necessary imaginary part comes from the decay width of the sigma meson.
Another approach was used in~\cite{Jaffe:1997hf} with the aim of getting a more realistic phenomenology of $H_1^\open$.
In particular, it was argued that quark fragmentation into two pions could be described as 2-step process:
the quark first fragments into a rho meson, which then decays into two pions.
Based on this picture and on data for $\pi\pi$ phase shifts the DiFFs were estimated.
A remarkable prediction of that study was a sign change of $H_1^{\open \, \pi^+ \pi^-/q}$ around the mass of the rho, which however has not been confirmed by the measurements.
Using a spectator model with several free parameters, in Ref.~\cite{Bianconi:1999uc} the T-odd DiFFs $G_1^\perp$, $H_1^\open$, and $H_1^\perp$ defined in Eqs.~(\ref{e:G1_quark_di-hadron_kT}),~(\ref{e:H1_quark_di-hadron_kT}) were studied for the proton-pion final state.
The calculation includes the production of the Roper resonance, where the decay width of this resonance provides the imaginary part needed for a nonzero T-odd FF.
The spectator model calculation of DiFFs for pions in~\cite{Radici:2001na} is very similar in its spirit to Ref.~\cite{Bianconi:1999uc}, with pion production arising through a (non-resonant) emission from the quark and through the decay of the rho resonance.
The focus was again on the three DiFFs $G_1^\perp$, $H_1^\open$, and $H_1^\perp$.
It is interesting that, in contrast to Ref.~\cite{Jaffe:1997hf}, no sign change was found for the IFF $H_1^{\open \, \pi^+ \pi^-/q}$~\cite{Radici:2001na}. 
In Ref.~\cite{Bacchetta:2006un} another spectator model calculation of DiFFs for pion production was presented.
The parameters of the model were fixed by fitting $D_1^{\pi\pi/q}$ to the output of the PYTHIA event generator for two-pion production in deep inelastic scattering for HERMES kinematics.
Like PYTHIA, the model contains non-resonant pion production, and fragmentation into several heavier mesons plus their decays into pions. 
After having fixed the parameters, the IFF $H_1^{\open \, \pi^+ \pi^-/q}$ was computed, where again no sign changed was found~\cite{Bacchetta:2006un}.
The unpolarized DiFF $D_1^{h_1 h_2/q}$ was also calculated in the Nambu--Jona-Lasinio model by considering the three lightest quarks, and the $\pi\pi$, $\pi K$ and $KK$ final states~\cite{Casey:2012ux}.
Based on the results of Ref.~\cite{Casey:2012ux} the evolution of DiFFs was studied in detail in~\cite{Casey:2012hg}.
Later on the model calculations of~\cite{Casey:2012ux} were extended by including the production of vector mesons~\cite{Matevosyan:2013aka}. 
Through their decay, vector mesons influence to DiFFs of pseudo-scalar mesons, where numerically significant contributions were obtained.
This observation is in line with the output from the PYTHIA event generator and the general philosophy of~\cite{Bacchetta:2006un}.
An independent study of the unpolarized DiFF $D_1^{h_1 h_2/q}$ in the Nambu--Jona-Lasinio model was reported in~\cite{Yang:2014eca}.
The results do not agree with the earlier calculation in Ref.~\cite{Casey:2012ux}.
In the same work also a nonlocal chiral quark model was used.
The numerical results from these two models were found to differ significantly~\cite{Yang:2014eca}. 
We also mention that very recently the unpolarized gluon FF $D_1^{h/g}$ was studied in the Nambu-Jona--Lasinio model~\cite{Yang:2016gnd}.

\subsection{Models for multiple hadron emission}
\label{sec:modelsMultipleHadEmission}
\begin{figure}[t]
\begin{center}
\includegraphics[width=9.0cm]{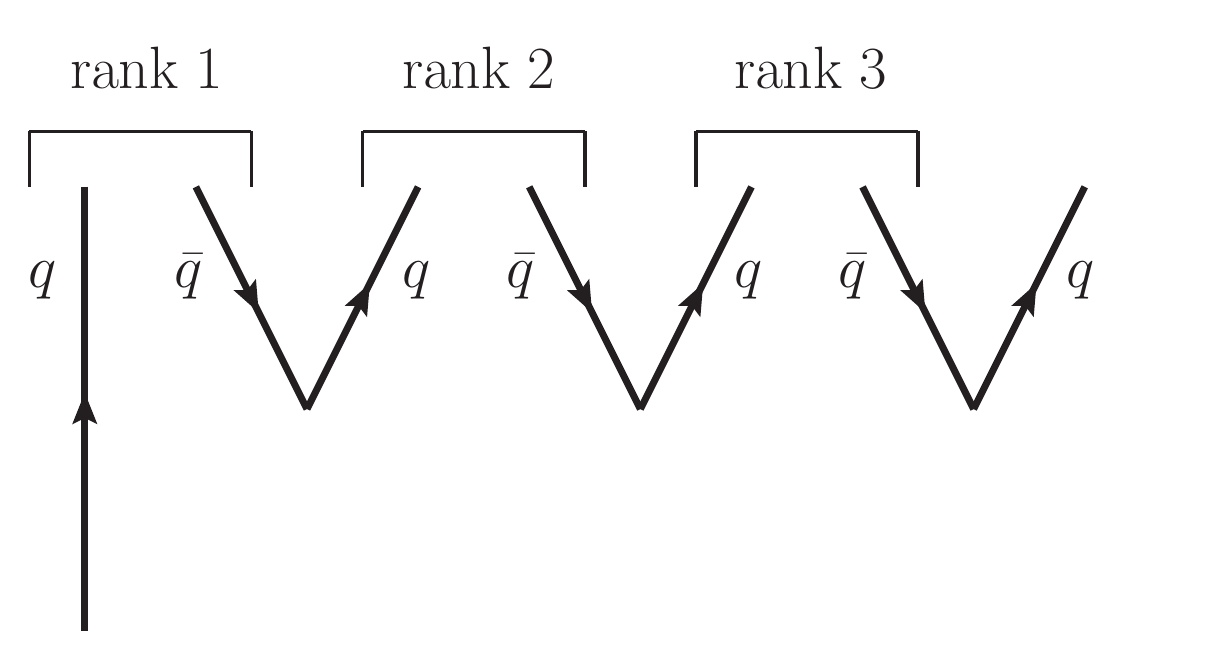} 
\end{center}
\vspace{-0.4cm}
\caption{Hierarchy of mesons formed when an initial quark combines with an antiquark from a produced $q\bar{q}$ pair forming the meson of rank 1.
The resulting quark then combines with an antiquark from another produced $q\bar{q}$ pair forming the meson of rank 2 and so on.}
\label{f:Feynman_Field}
\end{figure}

The second general class of fragmentation models are based on hadron production in a multi-step process.
One representative of that type is the Feynman-Field model for FFs~\cite{Field:1977fa}.
The general principle of that approach is sketched in Fig.$\,$\ref{f:Feynman_Field}: a high-energetic quark is combined with an antiquark from a $q\bar{q}$ pair created out of the vacuum.
This combination process repeats until the leftover energy falls below some cut-off.
A key underlying idea of the Feynman-Field model is that one can describe multiple hadron emission originating from a single parton by just one function $f(\eta)$ characterizing a single emission.
Here $f(\eta)$ describes the probability that the first hierarchy (rank 1) meson leaves fractional momentum $\eta$ to the remaining cascade.
The function is normalized so that~\cite{Field:1977fa}
\begin{equation}
\int_0^1 d\eta \, f(\eta) = 1 \,.
\end{equation}
One can then find the following integral equation for the FF $F(z)$ (generic notation)
\begin{equation} \label{e:FF_integral_rel}
F(z) = f(1-z) + \int_z^1 \frac{d\eta}{\eta} \, f(\eta) \, F \Big( \frac{z}{\eta} \Big) \,.
\end{equation}
The meson might be first in rank, as given by the first term on the r.h.s.~of~(\ref{e:FF_integral_rel}).
The second term calculates recursively the probability for the production of higher rank mesons.
Choosing the simple form
\begin{equation}
f(\eta) = (d + 1) \, \eta^d \,,
\end{equation}
with some parameter $d$, one finds for the fragmentation function~\cite{Field:1977fa}
\begin{equation}
z F(z) = f(1-z) = (d+1)(1-z)^d \,.
\end{equation}
The parameter $d$ then has to be fitted to data.
For simplicity we have not shown any flavor-dependence which was of course incorporated in such a model~\cite{Field:1977fa,Hua:2003ie}.
With a very limited set of parameters people were quite successful in describing data from early $e^+ e^-$ annihilation experiments~\cite{Hoyer:1979ta,Ali:1979em}.
Even though nowadays when fitting FFs to data one uses full QCD dynamics entering through higher order Wilson coefficients and evolution equations, the Feynman-Field model still gives some guidance for the parameterizations of FFs at a given initial scale.

An approach similar to the Feynman-Field model is the string fragmentation model~\cite{Artru:1974hr,Bowler:1981sb,Andersson:1983jt,Andersson:1983ia}, where hadrons are also formed in a hierarchy as indicated in Fig.$\,$\ref{f:Feynman_Field}.
However, the kinematics used in the two models is different. 
Also, in the string model, applied to $e^+ e^-$ annihilation for instance, one does not consider independent fragmentation of the outgoing quark and antiquark as done in the Feynman-Field model, but rather hadronization of the $q\bar{q}$ pair.
The quark and antiquark loose energy to the color field between them, which is supposed to collapse into a stringlike configuration.
The string has a uniform energy per unit length.
If that energy gets too high the string breaks up into hadrons.
Related to the string model are so-called cluster models~\cite{Field:1982dg,Webber:1983if}.
The underlying idea of such models is that in a full event in a first step color-neutral clusters of partoncs form which then hadronize.
A longer discussion of the key features and differences of these three types of models can be found, e.g., in Ref.~\cite{Ellis:1991qj}, along with more references to the original literature.
Such approaches for multiple hadron emission are a key ingredient in Monte-Carlo event generators that describe the complete final state of an event, where the more mordern generators incorporate mostly a string model or a cluster model.

The string fragmentation approach was used in Ref.~\cite{Artru:1995bh} in order to model the Collins function.
To explain the general idea we consider fragmentation of a transversely polarized quark into a spin-0 particle like a pion.
In the string model one assumes that a $q\bar{q}$ pair generated through string breaking carries the quantum numbers of the vacuum, that is, $J^{P} = 0^+$~\cite{Artru:1995bh}. 
This can be achieved if the pair has spin $S = 1$ and orbital angular momentum $L = 1$, with spin and orbital angular momentum pointing in opposite directions. 
In other words, there is a correlation between the spin of the antiquark from the $q \bar{q}$ pair and its orbital angular momentum.
Since the antiquark and the original quark form a spin-0 meson there is also a correlation between the orbital angular momentum of the antiquark, which is transferred to the meson, and the transverse polarization of the fragmenting quark.
This implies a nonzero Collins function, namely that the meson is produced in a preferred direction relative to the plane given by the momentum of the initial quark and its transverse spin.
The model was orginally applied to the transverse SSA in processes like $p^{\uparrow} p \to h X$ and provided the correct sign compared to the data~\cite{Artru:1995bh}.
The sign also agrees with the Collins function extracted from SIDIS data.
A disfavored FF is generated from rank 2 (and higher rank) mesons, and one readily finds that the model predicts opposite signs for the favored and disfavored Collins functions~\cite{Artru:1995bh}, in agreement with phenomenology.
Moreover, the model leads to the following interesting relation between the Collins function for a pion and for a rho meson~\cite{Czyzewski:1996ih},
\begin{equation} \label{e:Artru_rho_pi}
H_1^{\perp \, \rho/q} = - \, \frac{1}{3} \, H_1^{\perp \, \pi/q} \,,
\end{equation}
which has not yet been tested experimentally.
Despite the apparent successes of the model there may be a couple of caveats.
The original model as formulated in Ref.~\cite{Artru:1995bh} is a semi-classical description, which can hardly be compared with (field-theoretic) spectator model calculations based on interference between different amplitudes and the need for an imaginary part in the production amplitude --- see the discussion above in Sec.~\ref{sec:spectator_models}.
Also, since individual contributions in field-theoretic models can give rise to different signs for the Collins function, it is not clear how robust a semi-classical argument about the sign of the Collins functions can be.
In this context we also refer to~\cite{Artru:2010st} which apparently contains interesting further developments of the string fragmentation model including (transverse) spin effects.

Another recent line of research combines spectator model calculations as discussed in Sec.~\ref{sec:spectator_models} with the ideas of the Feynman-Field model.
More precisely, a single emission of a hadron is computed in a field-theoretic approach, and the result is then iterated in the spirit of the Feynman-Field approach.
Calculations along these lines were performed for $D_1^{h/q}(z)$ in the Nambu--Jona-Lasinio model~\cite{Ito:2009zc,Matevosyan:2010hh,Matevosyan:2011ey} and in a non-local chiral-quark model~\cite{Nam:2011hg,Nam:2012af,Yang:2013cza}.
Dedicated studies of TMD FFs can be found in~\cite{Matevosyan:2011vj,Matevosyan:2012ga,Bentz:2016rav}, including a discussion on how model-independent constraints such as the Sch\"afer-Teryaev sum rule~\cite{Schafer:1999kn,Meissner:2010cc} can be satisfied in such a picture.
Related calculations on DiFFs can be found in~\cite{Casey:2012hg,Matevosyan:2013aka,Matevosyan:2013eia,Yang:2014eca}.
The general aim of such studies is to get a quantitative description of FFs which contains as few parameters as possible by computing single-hadron emission based on a Lagrange density.
At present it is not clear if this aim can be reached or if ultimately one needs to work with more flexible parameterizations describing single-hadron production, which of course would increase the number of free parameters.

\section{Other Topics}
\label{sec:otherTopics}

\subsection{FFs for polarized hadrons}
\label{sec:frag_pol_hadron}
Many of the FFs defined in Sec.~\ref{sec:definition} are for polarized hadrons.
Here we just give a very brief discussion of some important points in this field.
At present, for most of these FFs we do not have any information from experiment.
In the majority of cases where experimental data exist, it is for fragmentation into $\Lambda$ and/or $\bar{\Lambda}$ hyperons which are self-analyzing, that is, they reveal their polarization through the angular depedence of the decay products.
The analysis of such experiments is not easy though, since a large fraction of the produced $\Lambda$ particles comes from the decay of heavier hyperons rather than direct production. 

We begin by addressing the integrated twist-2 FF $G_1^{\Lambda/q}(z)$ defined in Eq.~(\ref{e:G1_quark_z}).
This function can be measured in $e^+ e^- \to \vec{\Lambda} X$ on the $Z^0$ resonance~\cite{Burkardt:1993zh}.
The relevant observable is a longitudinal SSA which is allowed for this process due to the parity-violating weak interaction (see also the related discussion in Sec.~\ref{sec:observables_twist3}).
Using a NLO framework and making some simplifying assumptions about the flavor structure of $G_1^{\Lambda/q}(z)$, in Ref.~\cite{deFlorian:1997zj} first information on $G_1^{\Lambda/q}$ for favored fragmentation was obtained on the basis of data from LEP.
This FF is essentially unknown in the case of disfavored and gluon fragmentation. 
In the 1990s people started hoping that the production of longitudinally polarized $\Lambda$'s could also shed light on the spin crisis of the nucleon.
In this context, one question was about the contribution of strange quarks to the nucleon spin.
Since several models predicted that the polarization of the $\Lambda$ is largely associated with a polarized strange quark, one expected in the SIDIS process $\ell \vec{p} \to \ell \vec{\Lambda} X$, in which $G_1^{\Lambda/q}(z)$ enters~\cite{Ellis:1995fc,Jaffe:1996wp}, a strong sensitivity to the strange quark helicity distribution in the proton.  
The FF $G_1^{\Lambda/q}(z)$ also shows up in $\vec{\ell} p \to \ell \vec{\Lambda} X$, a process which was measured by the E665 Collaboration~\cite{Adams:1999px}, the HERMES Collaboration~\cite{Airapetian:1999sh,Airapetian:2006ee}, and the COMPASS Collaboration~\cite{Alekseev:2009ab}.
However, so far there is no data available for $\ell \vec{p} \to \ell \vec{\Lambda} X$.
A future electron-ion collider~\cite{Boer:2011fh,Accardi:2012qut} would be ideal for such a measurement and allow one to get back to the strange quark polarization in the nucleon and the production of polarized hyperons.
The functions $G_1^{\Lambda/i}(z)$ for both quarks and gluons also appear in the QCD description of $\vec{p} p \to \vec{\Lambda} X$~\cite{deFlorian:1998ba}.
A measurement of this double-spin observable was performed by the STAR Collaboration~\cite{Abelev:2009xg} where the result for $\Lambda$'s is compatible with zero, while a small nonzero effect was observed for $\bar{\Lambda}$'s.

The second spin-dependent integrated twist-2 FF is $H_1^{\Lambda/q}(z)$ defined in Eq.~(\ref{e:H1_quark_z}).
In principle, this function could be measured in the reaction $e^+ e^- \to \Lambda^{\uparrow} \bar{\Lambda}^{\uparrow} X$~\cite{Contogouris:1995xc}.
It is of particular interest since, due to its chiral-odd nature, it couples to the transversity distribution of the nucleon in processes like $\ell p^{\uparrow} \to \ell \Lambda^{\uparrow} X$ and $p^{\uparrow} p \to \Lambda^{\uparrow} X$~\cite{Jaffe:1996wp,deFlorian:1998am}. 
From a theoretical point of view the relevant transverse DSAs are very appealing as they are described in standard collinear factorization.
Compared to, for instance, transversity studies through di-hadron production, which are also based on collinear factorization, one deals with a standard integrated FF that only depends on one variable.
Nevertheless, at present no final data of these DSAs are available.

Among the TMD FFs defined in Eqs.~(\ref{e:D1_quark_kT})--(\ref{e:H1_quark_kT}) that depend on the polarization of the hadron so far only the function $D_{1T}^{\perp \, \Lambda/q}$ has been studied in some detail.
This FF describes the fragmentation of an unpolarized quark into a transversely polarized hadron or, in other words, a transverse single-spin effect. 
Using the generalized parton model, in Ref.~\cite{Anselmino:2000vs} a fit of this function was made based on data for the transverse SSA in processes like $p p \to \Lambda^{\uparrow} X$.
In the next paragraph we return to such SSAs.
The result of the fit was then used to estimate transverse SSAs in SIDIS~\cite{Anselmino:2001js}, including neutrino-nucleon scattering $\nu N \to \ell^{\pm} \Lambda^{\uparrow} X$ for which some data are available from the NOMAD Collaboration~\cite{Astier:2000ax}.
In Ref.~\cite{Boer:2010ya} it was discussed how through observables involving $D_{1T}^{\perp \, \Lambda/q}$ one can experimentally check the universality of TMD FFs~\cite{Metz:2002iz,Collins:2004nx}.
A number of SIDIS observables related to (moments of) TMD FFs can be found in~\cite{Boer:1997nt}.
In the case of $e^+ e^- \to h_a h_b X$ a complete list of tree-level results for polarized hadrons is available for both photon and $Z^0$ boson exchange~\cite{Boer:1997mf,Boer:1997qn,Boer:1999mm,Boer:2008fr,Pitonyak:2013dsu}.
Systematic studies of such observables could bring our understanding of hyperon polarization and the fragmentation process in general to the next level.

We now proceed to observables which are sensitive to (integrated) twist-3 FFs for polarized hadrons as defined in Eqs.~(\ref{e:E_quark_z})--(\ref{e:HL_quark_z}) and~(\ref{e:DeltaF_F})--(\ref{e:DeltaF_H}).
The one measured first was the transverse SSA in processes like $pp \to \Lambda^{\uparrow} X$, which in collinear twist-3 factorization involve twist-3 FFs.
For some time it has been observed that hyperons produced in unpolarized hadronic collisions exhibit sizable asymmetries towards large $x_F$ values~\cite{Bunce:1976yb,Heller:1978ty,Erhan:1979xm,Heller:1983ia,Lundberg:1989hw,Yuldashev:1990az,Ramberg:1994tk,Fanti:1998px,Abt:2006da}, similar to the transverse SSAs discussed in Sec.~\ref{sec:twist3FF_data}. 
Some of the earlier data are reviewed in~\cite{Pondrom:1985aw,Panagiotou:1989sv}.
At the LHC, the most recent measurement was performed by the ATLAS experiment~\cite{ATLAS:2014ona}, which did not observe an asymmetry consistent with the small $x_F$ of the measurement. 
Note that at $x_F = 0$ this effect should vanish.
If the transverse momentum of the hyperon is large enough the most rigorous framework to calculate this asymmetry is collinear twist-3 factorization (see also the discussion in Sec.~\ref{sec:observables_twist3}).
The general structure of the spin-dependent cross section is identical with Eq.~(\ref{e:sigma_generic}). 
The available calculations address the twist-3 effect associated with the incoming (unpolarized) hadrons~\cite{Kanazawa:2000cx,Zhou:2008fb,Koike:2015zya}.
It is important to also calculate the twist-3 fragmentation contribution in order to have a complete result in collinear twist-3 factorization.
The same final state asymmetry was also measured in photo-production $\gamma N \to \Lambda^{\uparrow} X$~\cite{Aston:1981em,Abe:1983jy} and quasi-real photo-production in $\ell N \to \Lambda^{\uparrow} X$~\cite{Airapetian:2007mx,Airapetian:2014tyc}.
The complete analytical result for the case of lepto-production was provided in Ref.~\cite{Kanazawa:2015jxa}, while a numerical estimate is still missing.
The final expression contains contributions from the twist-3 FFs $D_{1T}^{\perp (1) \, \Lambda/q}$ (where we use the moment definition in~(\ref{e:kT_moment})), $D_T^{\Lambda/q}$ in~(\ref{e:DT_quark_z}), as well as the three-parton FFs $\hat{D}_{FT}^{\Lambda/q, \, \Im}$ in~(\ref{e:DeltaF_F}) and $\hat{G}_{FT}^{\Lambda/q, \, \Im}$ in~(\ref{e:DeltaF_G}). 
Using the relations presented in~\cite{Kanazawa:2015ajw} the entire final result can be expressed through the three-parton FFs only.
In principle the simplest (twist-3) transverse SSA for polarized hyperons can be measured through $e^+ e^- \to \Lambda^{\uparrow} X$~\cite{Lu:1995rp}.
It is worth noting that according to Ref.~\cite{Boer:1997mf} its analytical result at LO is very compact as it just contains the twist-3 FF $D_T^{\Lambda/q}$.
A measurement by OPAL at LEP energies~\cite{Ackerstaff:1997nh} did not show a significant signal~\cite{Ackerstaff:1997nh}, which could well be due to the high $\sqrt{s}$. It would be very interesting to search for this fundamental asymmetry in lower-energy data from the B-factories.

\subsection{FFs for photons}
\label{sec:FF_photons}
Higher order calculations of cross sections for inclusive photon production necessarily involve the functions $D_1^{\gamma/i}$, that is, FFs of quarks and gluons into photons.
Naively one might think that photon production can be computed in quantum electrodynamics using fixed order perturbation theory.
However, this does not work as soon as photon emission is (quasi-) collinear with respect to a parton.
Three parameterizations of photon FFs are available in the literature that are based on a NLO framework for photon production in $e^+ e^-$ annihilation~\cite{Gluck:1992zx,Bourhis:1997yu}.
No error estimates were given in these papers.
While data from lepton-nucleon scattering and from {\it pp} collisions could further constrain photon FFs~\cite{GehrmannDeRidder:2006vn,Klasen:2014xfa,Kaufmann:2016nux}, at present no global analysis is available.
A perturbative calculation of the Collins function $H_1^{\perp \, \gamma/q}$ for photons was carried out in Ref.~\cite{Gamberg:2012iq}.

\subsection{Modified leading-log approximation}
\label{sec:mlla}
A special role is played by the low-$z$ region, roughly for $z < 0.1$. 
In this region, the time-like splitting functions in Eq.~(\ref{e:splitting}) have singularities that dominate at $z$ values that are orders of magnitude higher than the $x$ values at which these singularities become important for the corresponding space-like splitting functions of PDFs~\cite{Agashe:2014kda}.
Therefore, instead of a fixed-order treatment, for low $z$ often the so-called modified leading-log approximation (MLLA) is used~\cite{Azimov:1984np} --- for reviews of this method see for instance Ref.~\cite{Dokshitzer:1991wu,Khoze:1996dn,Albino:2008gy}.
The MLLA framework allows for important tests of our understanding of pQCD.
However, it is not the focus of the extractions of FFs discussed in this review.
Therefore we here only give a very brief account of this approach and refer to the literature for more details --- see for instance Ref.~\cite{Dokshitzer:1991wu,Khoze:1996dn,Albino:2008gy}.

The MLLA is an extension of the double-log approximation (DLA). 
In the latter one not only resums the large $\ln Q^2$ terms of DGLAP evolution but also leading logarithms in $\xi = \ln \frac{1}{z}$~\cite{Dokshitzer:1991wu}.
In particular, the DLA predicts that the low-$z$ behavior of the FFs can be described by a Gaussian in the variable $log(\frac{1}{z})$ $\xi$, where the width and peak position varies with  the hard scale $Q$. 
Specifically, one finds
\begin{equation}
z \, D_1^{h/q}(z,Q^2) \sim \rm{exp} \, \bigg[ - \frac{1}{2\sigma^2}(\xi - \xi_p)^2 \bigg] \,,
\end{equation}
with
\begin{equation}
\xi_p = \frac{1}{2} \ln \frac{Q}{\Lambda_{\rm QCD}} \,, \qquad
\sigma^2 = \frac{\sqrt{6}}{12} \ln^{\frac{3}{2}} \frac{Q}{\Lambda_{\rm QCD}} \,.
\end{equation}
In the MLLA one takes into account certain subleading terms that go beyond the DLA~\cite{Dokshitzer:1991wu}.
This mainly leads to a shift of the peak position $\xi_p$ towards higher values of $\xi_p$ which typically impoves the phenomenology~\cite{Khoze:1996dn}.
Detailed discussions on the phenomenology of the MLLA can, e.g., be found in the reviews in Ref.~\cite{Khoze:1996dn,Albino:2008gy,Agashe:2014kda}, as well as a recent paper from the Belle Collaboration~\cite{Lees:2013rqd}.

The MLLA approximation is primarily concerned with the low-$z$ behavior of the time-like splitting functions, but resummation of higher-order terms in the coefficient functions that are relevant at small $z$ has also been accomplished~\cite{Albino:2011si,Vogt:2011jv,Albino:2011cm,Kom:2012hd}. 

We also mention that extensions of the MLLA have been proposed in the literature~\cite{Albino:2005gg,Albino:2005gd,Arleo:2007wn,PerezRamos:2007cr}.
For instance, while the MLLA typically is a good approximation to the data in the peak region around $\xi_p$, other approaches provide a better description over a larger kinematical range including, in particular, large-$z$ region~\cite{Albino:2005gg,Albino:2005gd}.


\subsection{Fracture functions}
\label{sec:fracture_functions}
In SIDIS one can use FFs if hadrons are produced in the current fragmentation region.
However, depending on the kinematics, they may also emerge from the target fragmentation region, in particular if the {\it cm} energy of the collision and the fragmentation variable $z$ are not very high.
For such a situation another class of non-perturbative objects was introduced and called fracture functions~\cite{Trentadue:1993ka}. 
They can be considered as a hybrid object sharing similarities with both PDFs and FFs. 
The dependence of fracture functions on spin degrees of freedom and on transverse momenta was discussed in Ref.~\cite{Anselmino:2011ss}.
Afterwards the study in~\cite{Anselmino:2011ss} was extended to the production of two hadrons, one in the current fragmentation region and one in the target fragmentation region which, in principle, would allow one to measure also chiral-odd fracture functions~\cite{Anselmino:2011bb}.
Looking at two hadrons in these regions can also give rise to new observables~\cite{Anselmino:2011aa}.
Related recent work on fracture functions and the target fragmentation region can be found for instance in~\cite{Sivers:2008my,Sivers:2009qj,Ceccopieri:2015kya}.

\section{Conclusions and Outlook}
\label{sec:conclusions}
We have presented an overview of parton FFs in theory and experiment. 
We have concentrated on fragmentation of light quarks in the vacuum. 
We gave emphasis on the emerging fields of TMD, higher twist, di-hadron as well as spin-dependent fragmentation, where 
significant progress has been made over the last decade or so.
In the following we list points which will be important to study or where we expect new developments in the near future.
\paragraph{For the unpolarized FF $D_1$} NNLO fits that include SIDIS data, in addition to the $e^+e^-$ data, should be within reach. 
We recognize that a calculation of $pp \to hX$ at two loop accuracy lies further in the future. 
Looking at hadrons inside jets in hadronic collisions can provide interesting new insights.
Target mass corrections and resummation effects are also under active development. 
This line of research is quite important, in particular for making full use of future data from JLab-12.  
\paragraph{For the TMD FFs} a robust framework QCD framework including TMD evolution is needed, in particular with regard to the non-perturbative part entering the evolution. New data on the transverse momentum dependence from $e^+e^-$ annihilation will be very important to disentangle the contributions of PDFs and FFs from the measured transverse momenta and allow access for example to the flavor dependence of the $|\vec{P}_{hT}|$ width. 
Like for integrated FFs, sudying hadrons inside jets may be a very promising route for TMD FFs as well.
It would be important to have a definite statement about the status of TMD factorization for that case.
\paragraph{For the twist-3 FFs} a new fit to data on the transverse SSA in $p^{\uparrow} p \to h X$ is needed.
Moreover, testing the predictions of the twist-3 FF fits with new data in different kinematic regimes will be important to verify the central finding that they can explain the large $A_N$ observed in $pp$. 
Another important step forward in this area could be a NLO calculation of the transverse SSA in $\ell p^{\uparrow} \to h X$.
\paragraph{For the di-hadron FFs} experimental results that give access to the unpolarized DiFFs are urgently needed, in particular for the gluon FF $D_1^{h_1 h_2/g}$. 
To extract these functions the di-hadron cross-section in $pp$ has to be measured. 
A model independent extraction of transversity from $pp$ data relies on the knowledge of these FFs. 
On the theory side, the calculations for the observables should be pushed to the next order in perturbation theory.
This would allow one to extract the transversity distribution at NLO accuracy in a collinear framework.
\paragraph{For the physics program at the future facilities} JLab-12 and EIC, the precise knowledge of FFs will be crucial. 
This is especially true for the TMD program.
Input from Belle II, which will start taking data with the full detector in 2018, at the next generation B-factory SuperKEKB will be important to provide multi-differential observables for identified light quark FFs. 
\\[1.0cm]
{\bf Acknowledgements:} This work has been supported by the National Science Foundation under Contract No.~PHY-1516088 (A.M.) and PHY-1306942 (A.V.). 



\bibliography{main}

\end{document}